\documentclass{aa}
\usepackage{graphicx}
\usepackage{txfonts}
\usepackage{epstopdf}
\usepackage{color}
\usepackage{multirow}
\usepackage{lscape}
\usepackage{amsmath}
\usepackage{upgreek}
\begin{document}
    \title{NELIOTA: Methods, statistics and results for meteoroids impacting the Moon}
     \author{A. Liakos\inst{1}
             \and
              A. Z. Bonanos\inst{1}
             \and
              E. M. Xilouris\inst{1}
             \and
              D. Koschny\inst{2,3}
             \and
             I. Bellas-Velidis\inst{1}
             \and
             P. Boumis\inst{1}
            \and
             V. Charmandaris\inst{4,5}
            \and
             A. Dapergolas\inst{1}
            \and
             A. Fytsilis\inst{1}
            \and
             A. Maroussis\inst{1}
            \and
            R. Moissl\inst{6}
            }
   \institute{Institute for Astronomy, Astrophysics, Space Applications and Remote Sensing, National Observatory of Athens,\\
              Metaxa \& Vas. Pavlou St., GR-15236, Penteli, Athens, Greece \\
              \email{alliakos@noa.gr}
             \and
              Scientific Support Office, Directorate of Science, European Space Research and Technology Centre (ESA/ESTEC), \\
              2201 AZ Noordwijk, The Netherlands
             \and
              Chair of Astronautics, Technical University of Munich, 85748 Garching, Germany
             \and
              Department of Physics, University of Crete, GR-71003, Heraklion, Greece
              \and
              Institute of Astrophysics, FORTH, GR-71110, Heraklion, Greece
              \and
              European Space Astronomy Centre (ESA/ESAC), Camino bajo del Castillo, Villanueva de la Ca\~{n}ada, 28692 Madrid, Spain
             }
   \date{Received September xx, 2019; accepted March xx, 2019}
  \abstract
   {This paper contains the results from the first 30 months of the NELIOTA project for impacts of Near-Earth Objects/meteoroids on the lunar surface.  Our analysis on the statistics concerning the efficiency of the campaign and the parameters of the projectiles and the impacts is presented.}
   {The parameters of the lunar impact flashes based on simultaneous observations in two wavelength bands are used to estimate the distributions of the masses, sizes and frequency of the impactors. These statistics can be used both in space engineering and science. 
   }
   {
   The photometric fluxes of the flashes are measured using aperture photometry and their apparent magnitudes are calculated using standard stars. Assuming that the flashes follow a black body law of irradiation, the temperatures can be derived analytically, while the parameters of the projectiles are estimated using fair assumptions on their velocity and luminous efficiency of the impacts.}
   {79 lunar impact flashes have been observed with the 1.2~m Kryoneri telescope in Greece. The masses of the meteoroids range between 0.7~g and 8~kg and their respective sizes between 1-20~cm depending on their assumed density, impact velocity, and luminous efficiency. 
   We find a strong correlation between the observed magnitudes of the flashes and the masses of the meteoroids. Moreover, an empirical relation between the emitted energies of each band has been derived allowing the estimation of the physical parameters of the meteoroids that produce low energy impact flashes. 
   }
   {The NELIOTA project has so far the highest detection rate and the faintest limiting magnitude for lunar impacts compared to other ongoing programs. Based on the impact frequency distribution on Moon, we estimate that sporadic meteoroids with typical masses less than 100~g and sizes less than 5~cm enter the mesosphere of the Earth with a rate $\sim108$~meteoroids~hr$^{-1}$ and also impact Moon with a rate of $\sim8$~meteoroids~hr$^{-1}$.}


   \maketitle

%

\section{Introduction}
\label{sec:intro}

The `NEO Lunar Impacts and Optical TrAnsients' (NELIOTA) project has begun in early 2015\footnote{The official observational campaign began in March 2017} at the National Observatory of Athens (NOA) and is funded by the European Space Agency (ESA). Its short-term goal is the detection of lunar impact flashes and the estimation of the physical parameters of the meteoroids (e.g. mass, size) as well as those of the impacts (e.g. temperature, craters on the surface). The mid-term goal concerns the statistics of the frequency and the sizes of the meteoroids and small NEOs to be used by the space industry as essential information for the shielding of space vehicles. For the purposes of the project, a dedicated instrumentation set-up has been installed at the 1.2~m Kryoneri telescope\footnote{\url{http://kryoneri.astro.noa.gr/}} in Greece allowing high resolution observations at a high recording frame rate (30~frames-per-second) simultaneously in two different wavelength bands. This provides the opportunity: a) to validate events using a single telescope, and b) to estimate directly the temperature of the flashes as well as the thermal evolution in time for those that are recorded in consecutive frames. The method used for the determination of lunar impact temperatures as well as the results for the first ten observed flashes have been published in \citet{BON18} (hereafter Paper I). Another significant contribution of this project to the study of lunar impact flashes is the size of the telescope, that permits flash detections up to $\sim$12th magnitude in $R$ filter, i.e. about 2~mag fainter than the previous campaigns. Details about the instrumentation setup and its efficiency/performance on lunar impacts can be found in \citet{XIL18} (hereafter Paper II). The observing campaign started on March 2017 and is scheduled to continue until January 2021. Brief presentations of the NELIOTA project and the methods followed for the derivation of the meteoroid and flash parameters can be found also in \citet{BON15, BON16b, BON16a}, and in \citet{LIA19}.

Although the NELIOTA project was designed mainly to provide information about the meteoroids reaching the atmosphere and the close vicinity of the Earth, it can also contribute to the current and the future space missions to the Moon. During the last decade the interest of many space agencies (CNSA, ESA, ISRO, JAXA, NASA, Roscosmos, SpaceIL) for the Moon has been rapidly increasing with many robotic and crewed missions to be either in progress or scheduled for the near future. It appears that currently there is strong interest of the major funding agencies to establish a lunar base for further exploration and exploitation of the Moon. The recent research works of \citet{HUR17} and \citet{TUC19} showed that meteoroid impacts produce chemical sputtering (i.e. remove $OH$ from the lunar regolith) and along with the solar wind are the most likely source mechanisms supplying $H_2$ to the lunar exosphere. Therefore, continuous and/or systematic monitoring of the lunar surface is considered extremely important. The results from the NELIOTA observations can be also used to calculate the meteoroid frequency distribution on the lunar surface which will provide the means to the space agencies to select an appropriate area (e.g. less likely to be hit by a meteoroid) for establishing the first lunar base. Moreover, estimating the temperatures of the flashes and the kinetic energies of the projectiles will be very important to the structural engineers regarding the armor that should be used for any permanent infrastructure on or beneath the lunar surface.

Near-Earth Objects (NEOs) are defined as asteroids or comets whose orbits cross that of the Earth and potentially can cause damage either on space vehicles (e.g. satellites, space stations, space telescopes) or even on the surface of the Earth (e.g. destroy infrastructure). Meteoroids are tiny objects up to one meter that are mostly asteroidal or cometary debris. The majority of meteoroids are composite of stone (chondrites and achondrites) but there are also such objects of stone-iron and only of iron. They are formed mostly from asteroid collisions on the main asteroids belt (asteroidal debris) and from the outgassing of the comets when they pass close to the Sun (cometary debris). However, some of them can also be formed from asteroid impacts on other planets (e.g. Mars). Asteroidal and cometary debris, which is still close to the orbit of its parent body, impacts the Moon at defined times with defined velocities and directions. The latter are called `meteoroid streams' and give rise to `meteor showers' when entering the atmosphere of the Earth. Objects which cannot be associated with their parent any more are called `sporadics' \citep{KOS19}.

Observing small NEOs and meteoroids entering the Earth's atmosphere has certain difficulties. The observations from ground-based equipment can cover only a very limited surface i.e. $\sim35\times10^3$~km$^{2}$ for an atmospheric height of 75~km (mesosphere area). This results in a very small number of objects of this size or larger to be observed per hour. The idea of using the Moon as a laboratory for impacts is based on the need of systematic observations for detecting indirectly small size NEOs and meteoroids by their impact flashes. Moreover, the surface area of the Moon facing the Earth is $\sim$19$\times10^6$~km$^{2}$, which is $\sim1000$ times greater than the respective available area on Earth's atmosphere for a given site. 

So far, there have been several regular campaigns on lunar impact flashes \citep{ORT06, BOU12, SUG14, MAD14, MAD15a, ORT15, REM15, AIT15a,AIT15b}. Moreover, after 2014, many lunar flashes, produced mostly during meteoroid streams, have been also reported \citep{MAD17, MAD19a}. The similarities of these campaigns are: a) the small-size telescopes (diameter of 30-50~cm) used and b) the unfiltered or single band observations \citep[e.g.][$I$ filter]{MAD19b}. The only multi-filter observations were made for one specific flash ($V$ and $I$ filters) in 2015 by \citet{MAD18}. All these campaigns managed to observe both sporadic and meteoroid stream flashes providing useful constrains on the physical parameters of the impactors. However, due to the small diameter of the telescopes the majority of the flashes are brighter than 10.5~mag \citep[e.g.][]{SUG14}. In addition to the times close to new Moon, impact flashes were reported also during a total lunar eclipse \citep{MAD19b}.

The first peer-reviewed published results for the temperature determination of lunar impact flashes were presented in Paper I based on the NELIOTA observations. Three months later, \citet{MAD18} published a similar peer-reviewed work based on their own multi-filter observations occurred for this purpose in 2015. Recently, \citet{AVD19}, using the online database of NELIOTA, calculated the temperatures of the first 55 validated flashes (until October 2018) and the corresponding masses of the meteoroids. It should be noted that the information given in the NELIOTA online database\footnote{\url{https://neliota.astro.noa.gr/}} is limited (i.e. rounding of values, the frames of the standard stars are not given) and the results based strictly on these data should be considered as fairly approximative.

This paper aims to present in detail all the methods applied in the project and the full statistical analysis of lunar impact flashes from the first 30 months of NELIOTA operations. The method of the flash temperature calculation has been revised in comparison of that of Paper I. The errors for all parameters take into account the scintillation effect, which has been proven as a significant photometric error contributor. Moreover, the association of the projectiles with active meteoroid streams is examined. In Section~\ref{sec:Obs}, the instrumentation and the observational strategy followed are briefly presented. In Sections~\ref{sec:VAL}-\ref{sec:PHOT}, we present in detail our methodology on the validation and the photometry of the flashes. The results for the all the detected flashes and the statistics of the NELIOTA campaign are given in Section~\ref{sec:RES}. In Section~\ref{sec:Calc}, all the methods for the calculation of the parameters of the impacts and the meteoroids are described in detail. In Section~\ref{sec:CorDib}, the distributions and the correlations for the parameters of impact flashes and meteoroids are presented. In Section~\ref{sec:Rates}, we calculate the meteoroid flux and its extrapolation to Earth, while the current results of the campaign are discussed in Section~\ref{sec:Disc}. 


\section{Observations}
\label{sec:Obs}

The NELIOTA observations are carried out at the Kryoneri Observatory, which is located at Mt. Kyllini, Corinthia, Greece at an altitude of $\sim930$~m. The primary mirror of the telescope has a diameter of 1.2~m and its focal ratio is 2.8. Two twin front illuminated sCMOS cameras (Andor Zyla 5.5) with a resolution of $2560\times2160$~pixels and a pixel size of $\sim6.5~\mu$m are separated by a dichroic beam-splitter (cut-off at 730~nm) and they are set at the prime focus of the telescope. Each camera is equipped with one filter of Johnson-Cousins specifications. In particular, the first camera records in the red ($R_{\rm c}$) and the other in the near-infrared ($I_{\rm c}$) passbands, with the transmittance peaks to be $\lambda_{\rm R}=641$~nm and $\lambda_{\rm I}=798$~nm, respectively. The Field-of-View (FoV) of this setup is $\sim 17' \times14.4'$. The cameras record simultaneously at a rate of 30 frames-per-second (fps) in $2\times2$ binning mode. A software pipeline has been developed for the purposes of the project, which splits into four parts: a) Observations (NELIOTA-OBS), b) data reduction and detection of events (NELIOTA-DET), c) archiving (NELIOTA-ARC), and d) information (NELIOTA-WEB).

\begin{figure}
\centering
\begin{tabular}{rl}
\includegraphics[width=4.2cm]{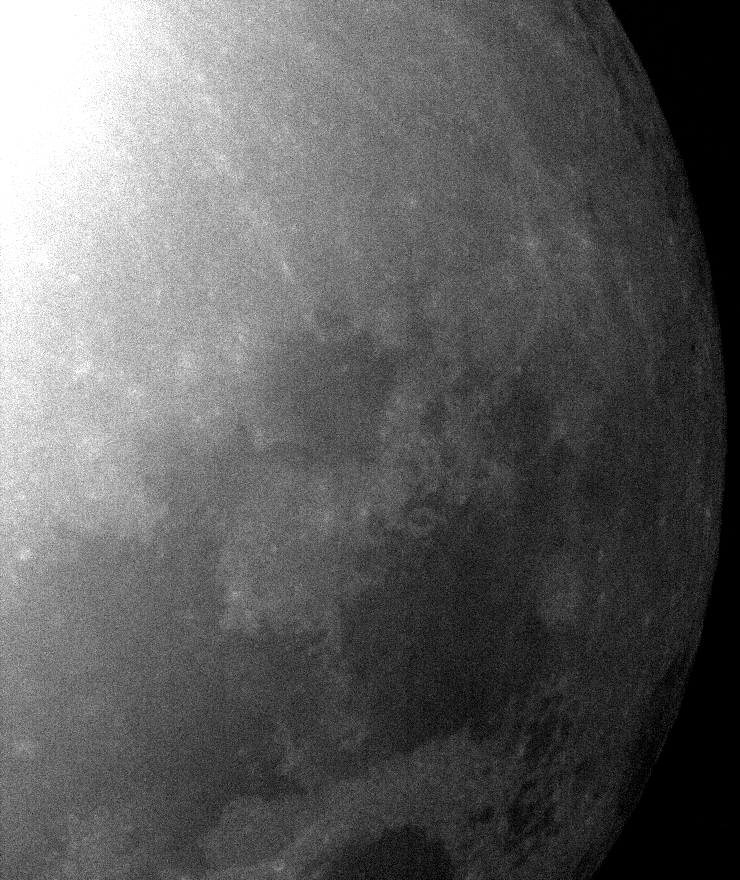}&\includegraphics[width=4.2cm]{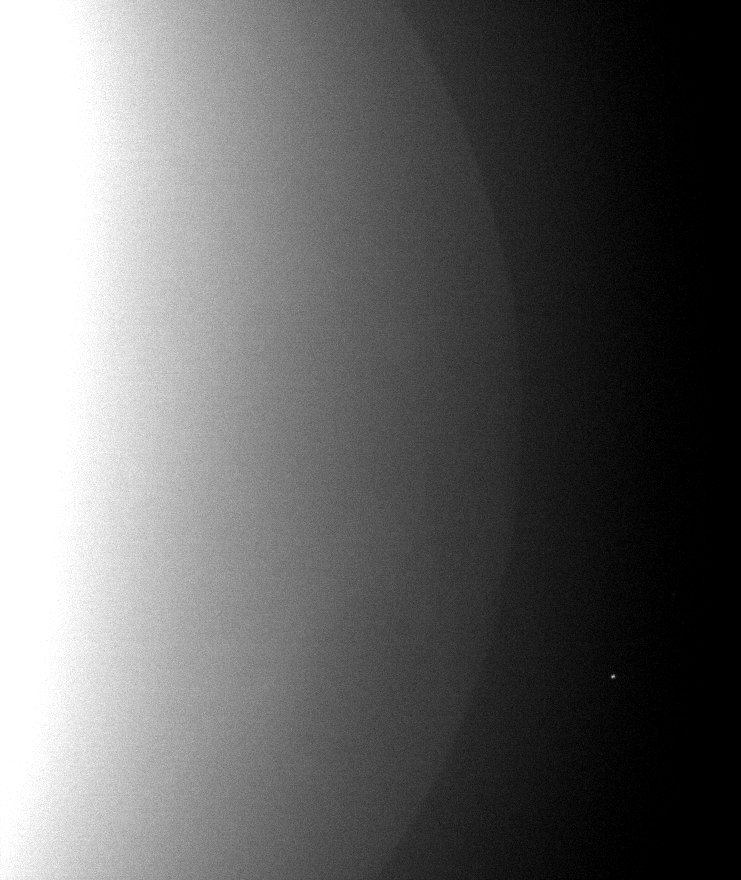}\\
\end{tabular}
\caption{Effect of glare (straylight) from the sunlit part of the Moon on the observations. The left image was obtained during lunar phase 0.14, while the right at 0.44. The contrast of the images is set to a value that allows the lunar surface to be clearly visible. The illuminated part on the left edge of both images is not saturated, but has the highest background values. It should be noted that the terminator is far away from the edges. In the left image, many lunar features can be easily seen. In the right image, the features are very blurred and the total covered area is decreased about $\sim30\%$ with respect to the left one.}
\label{fig:glare}
\end{figure}

Systematic observations are made between lunar phases of $\sim$0.10 and $\sim$0.45 (i.e. before and after new Moon; 5-8 nights per month) at the non-sunlit (nightside) part of the Moon. The upper limit of the lunar phase during which observations can be obtained depends strongly on the intensity of the glare coming from the sunlit part of the Moon. In particular, when the Moon is close to the apogee, i.e. the total observed area increases, the glare is stronger. Therefore, in order to avoid the very high lunar-background noise, the telescope is repositioned towards the lunar limb. With this method, the observed lunar surface at a phase greater than 0.4 is up to $\sim40\%$ less than that observed in less bright phases. So, the upper limit of the lunar phase has been set at $\sim0.44$ during the apogee and $\sim0.46$ during perigee. The effect of glare on lunar images is shown in Fig.~\ref{fig:glare}. It should be noted that the observations near the upper limit are very important because their duration is the longest of all observing nights.
Sky flat-field frames are taken before or after the lunar observations, while the dark frames are obtained directly after the end of them. Standard stars are observed for magnitude calibration reasons every $\sim$15~min. The minimum duration of the observations is $\sim20$~min (at low brightness lunar phases), while the maximum is $\sim$4.5~hr (at lunar phases near $\sim0.45$). It should be noted that the $\sim40\%$ of the total available observing time is lost due to a) the read-out time of the cameras ($\sim30\%$) and b) the repositioning of the telescope for the standard stars observations ($\sim10\%$).

\begin{figure}
\centering
\includegraphics[width=8.5cm]{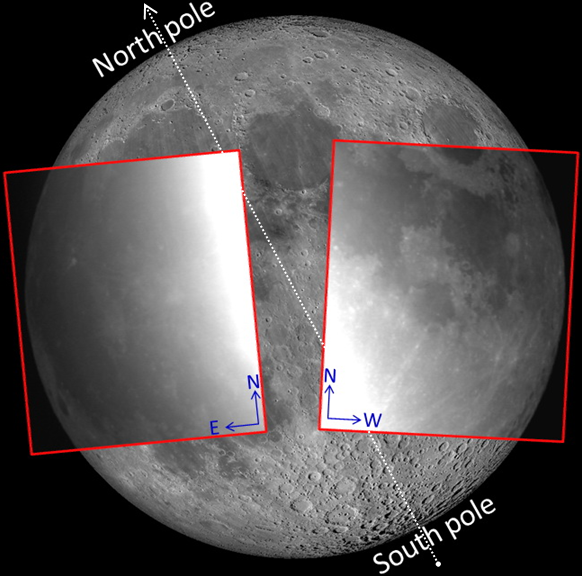}
\caption{Examples of the Field-of-View of the NELIOTA setup (images in rectangles) superimposed on a high detail lunar map. The images were obtained before the first quarter (left) and after the last quarter (right) lunar phases. The celestial orientation is denoted by blue arrows, while the north and south lunar poles are also indicated.}
\label{fig:FoV}
\end{figure}

The orientation of the cameras has been set in such a way that the longer axis is almost parallel to the lunar terminator during the whole year, i.e. it corresponds to the declination equatorial axis (celestial North-South axis). Therefore, only the eastern-western hemispheres of the Moon are observed and not its poles in order to avoid the straylight from the lunar terminator. The libration, the inclination with respect to the orbital plane, and the varying distance of the Moon allow to observe areas with latitudes up to $\pm50\degr$ from the equator. Examples of the covered lunar area during typical NELIOTA observations are illustrated in Fig.~\ref{fig:FoV}.

More details about the individual subsystems, the performance of the NELIOTA setup, the observations strategy, and the software pipeline (e.g. data acquisition, data chuncks) can be found in Paper II.

\begin{figure*}[h!]
\centering
\includegraphics[width=18.5cm]{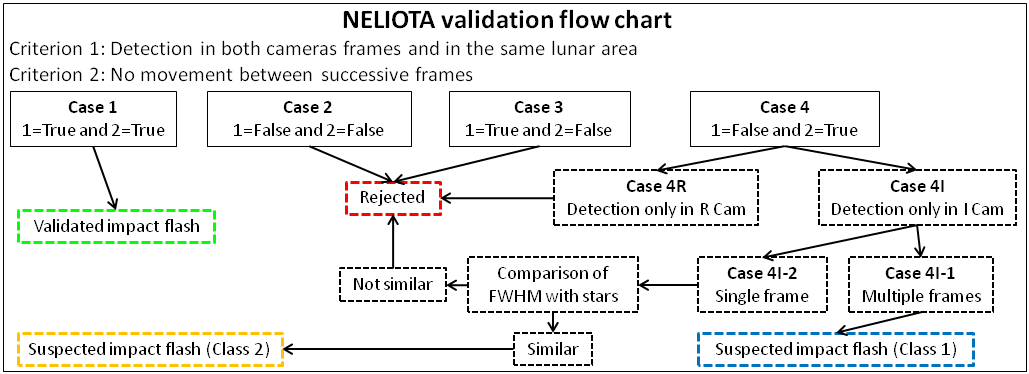}
\caption{Validation flow chart for the events detected by NELIOTA-DET as implemented by the expert user. The criteria are noted on the upper part and all the possible characterizations of the events are given in colored-edge text boxes.}
\label{fig:Valflow}
\end{figure*}

\section{Validation of the events}
\label{sec:VAL}

Before proceeding to the validation procedure, the definition of the frequently used term `event' has to be stated. An `event' is defined as whatever the NELIOTA-DET managed to detect. An event could be a cosmic ray hit, a satellite, an airplane, a distant bird, field stars very close to the lunar limb, and, obviously, a real impact flash. The validation of the events has two steps. The first one concerns the examination of the images, which NELIOTA-DET flagged as including an event, and it is is described in detail in Section~\ref{sec:VAL1}. The second one (Section~\ref{sec:VAL2}) concerns only the events that were either identified as validated or as suspected flashes. Their location on the Moon is compared with orbital elements of satellites in order to exclude the possibility of misidentification.

\subsection{Validation based on data inspection}
\label{sec:VAL1}

\begin{figure}
\centering
\begin{tabular}{cc}
\includegraphics[width=\columnwidth]{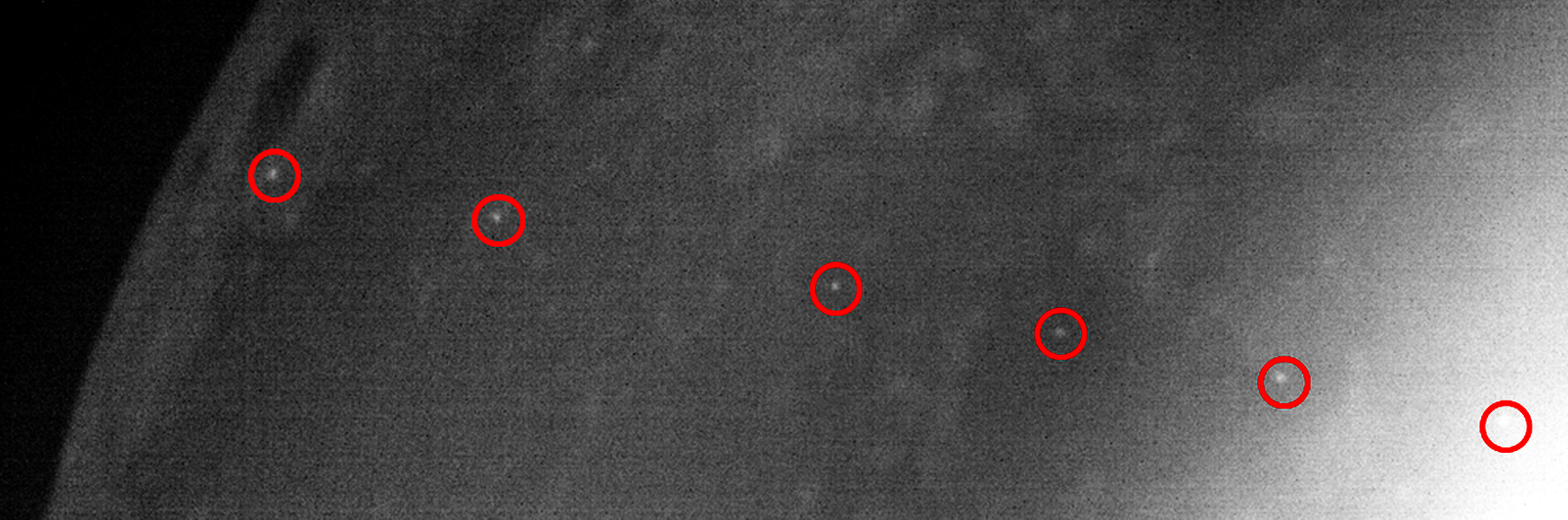}\\
\includegraphics[width=\columnwidth]{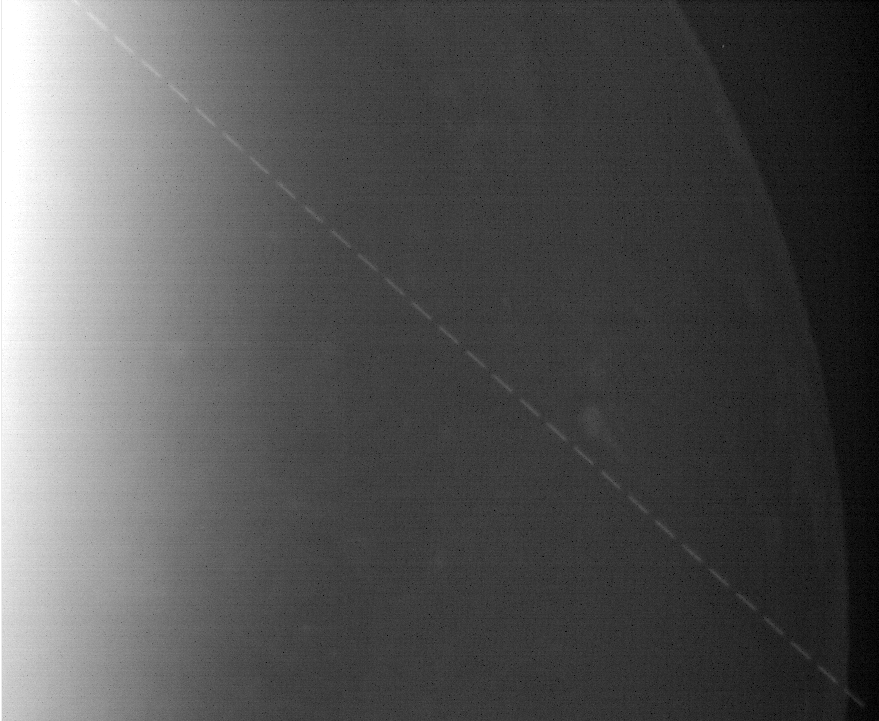}\\
\end{tabular}
\caption{Examples of moving artifacts detected by the NELIOTA-DET. The upper figure is an overlay of seven images with a time difference of 2~sec and shows a slow fly by of a satellite in front of the lunar disk. The rotation period of the satellite was such that could be seen only every 2~sec. The missing point is probably due to temporary bad fluctuation of seeing that did not permit the detection. The lower figure consists of 29 successive images and shows a fast passage of a satellite in front of the lunar disk. The gaps between the lines correspond to the read out time of the camera.}
\label{fig:sat}
\end{figure}

\begin{figure}
\begin{tabular}{rll}
\includegraphics[height=3.1cm]{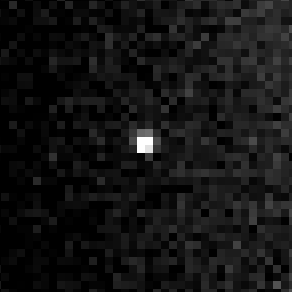}&\includegraphics[height=3.1cm]{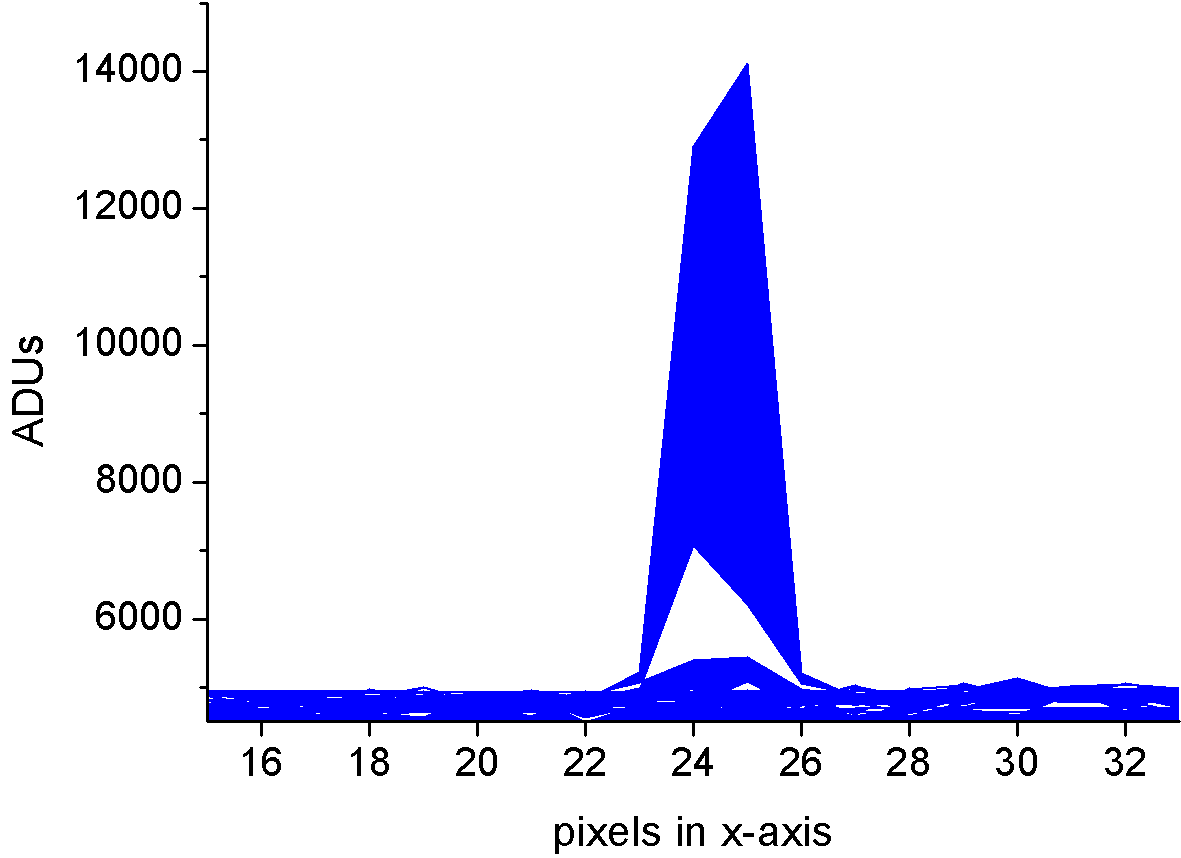}&(a)\\
\includegraphics[height=3.1cm]{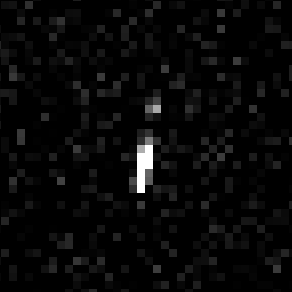}&\includegraphics[height=3.1cm]{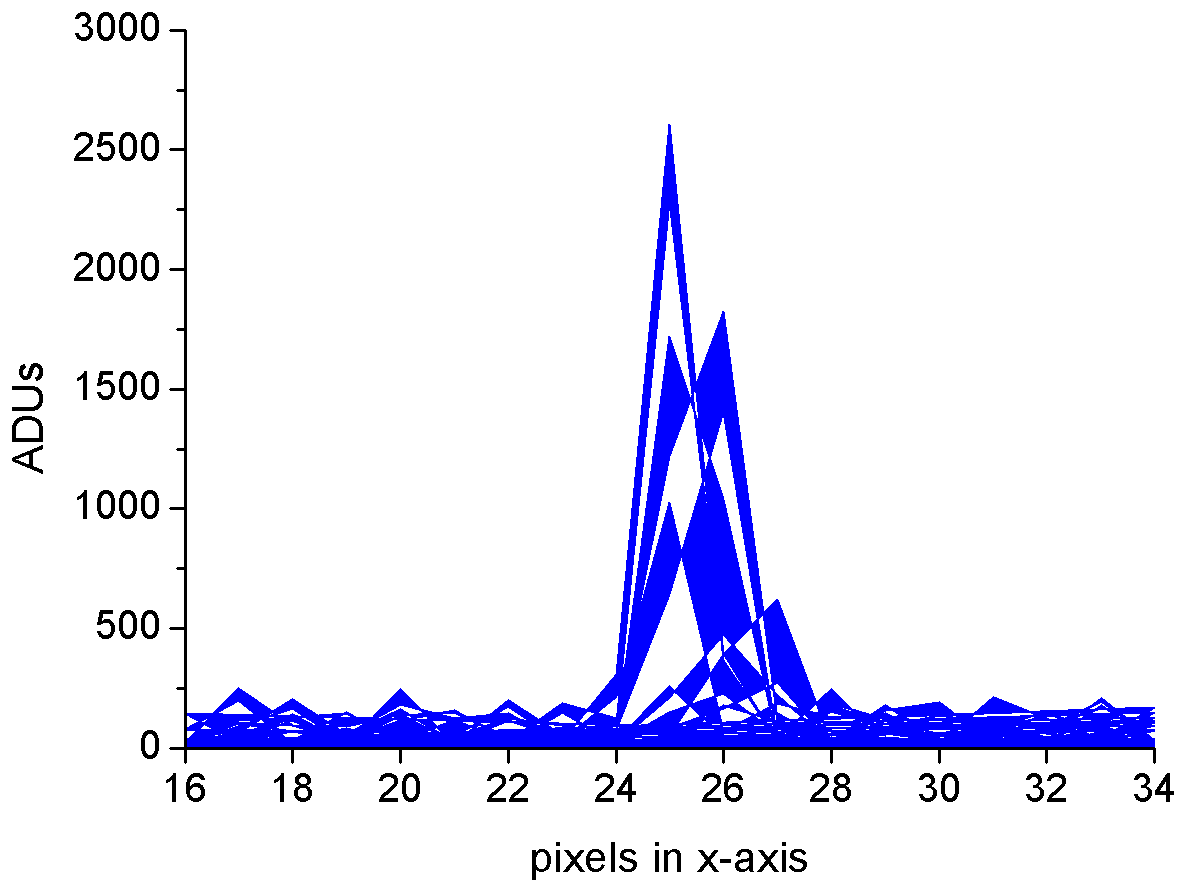}&(b)\\
\includegraphics[height=3.1cm]{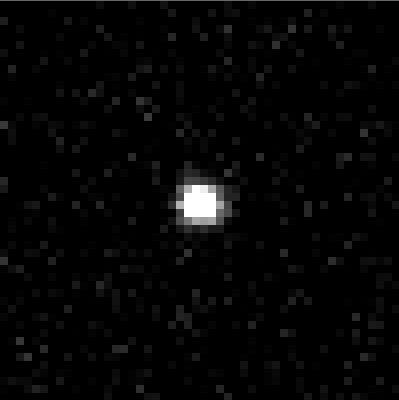}&\includegraphics[height=3.1cm]{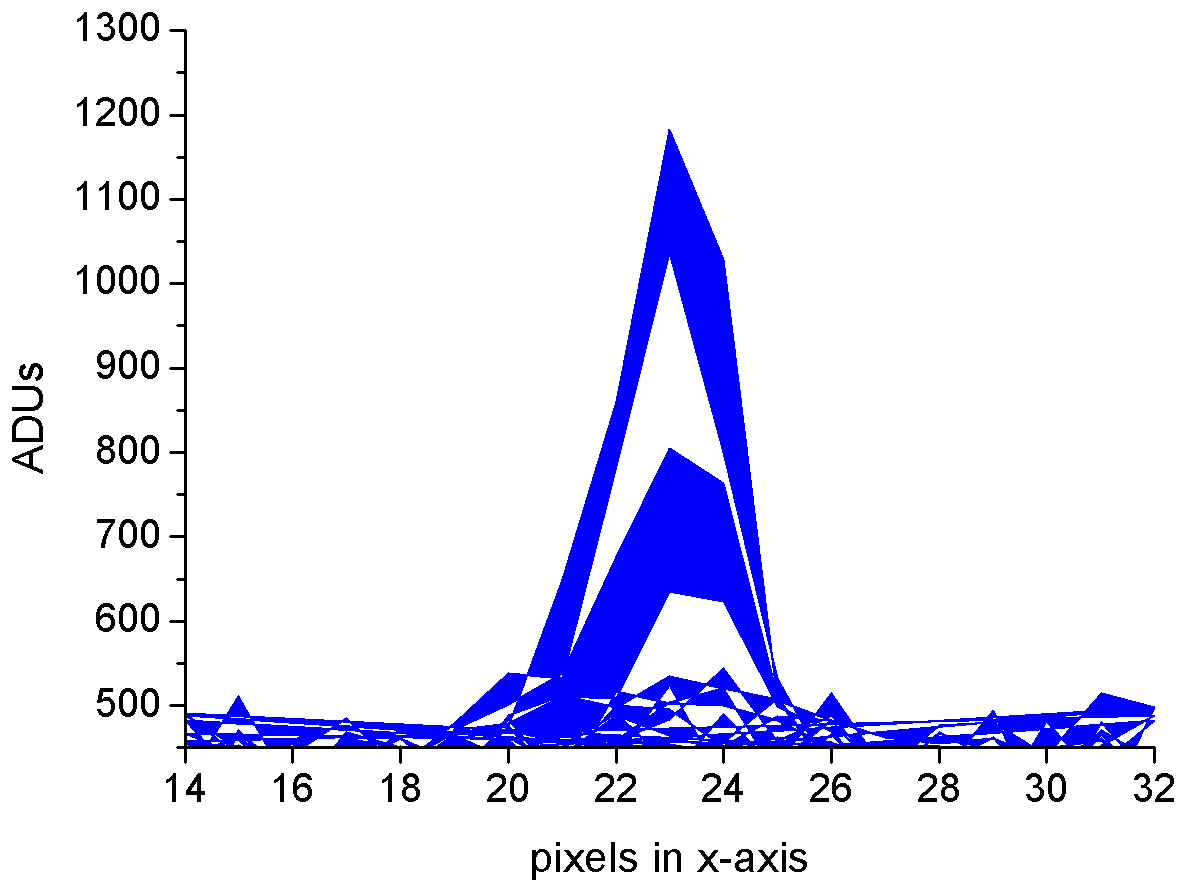}&(c)\\
\includegraphics[height=3.1cm]{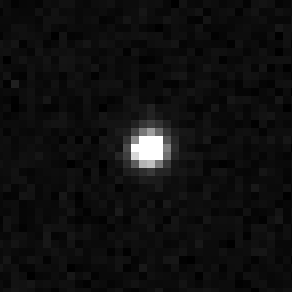}&\includegraphics[height=3.1cm]{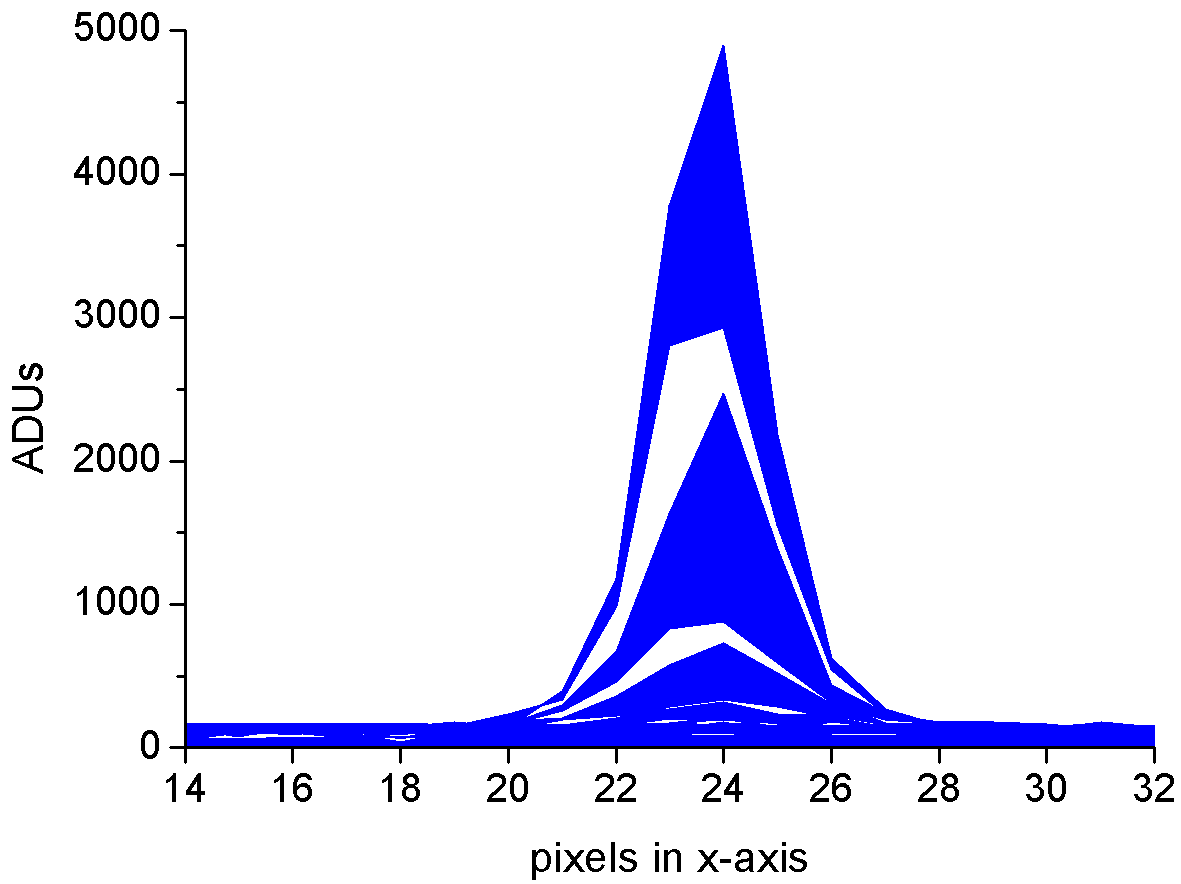}&(d)\\
\end{tabular}
\caption{Photometric profile examples of events. Left: Images of the events (50$\times$50~pixels area). Right: corresponding profiles (18$\times$18~pixels area) of the events. (a): Cosmic ray that hit the sensor almost perpendicularly producing a round PSF. (b): Cosmic ray hit, but this time with an impact angle that produced an elongated profile. (c): Validated impact flash. (d): Standard star used for the photometric calibration. The similarity in terms of FWHM is obvious for (c) and (d) (FWHM difference $\sim0.2$~pixels), while a large difference is seen also between (a) and (c) (FWHM difference) $\sim1.5$~pixels.}
\label{fig:Prof}
\end{figure}

\begin{figure}[h!]
\centering
\includegraphics[width=8cm]{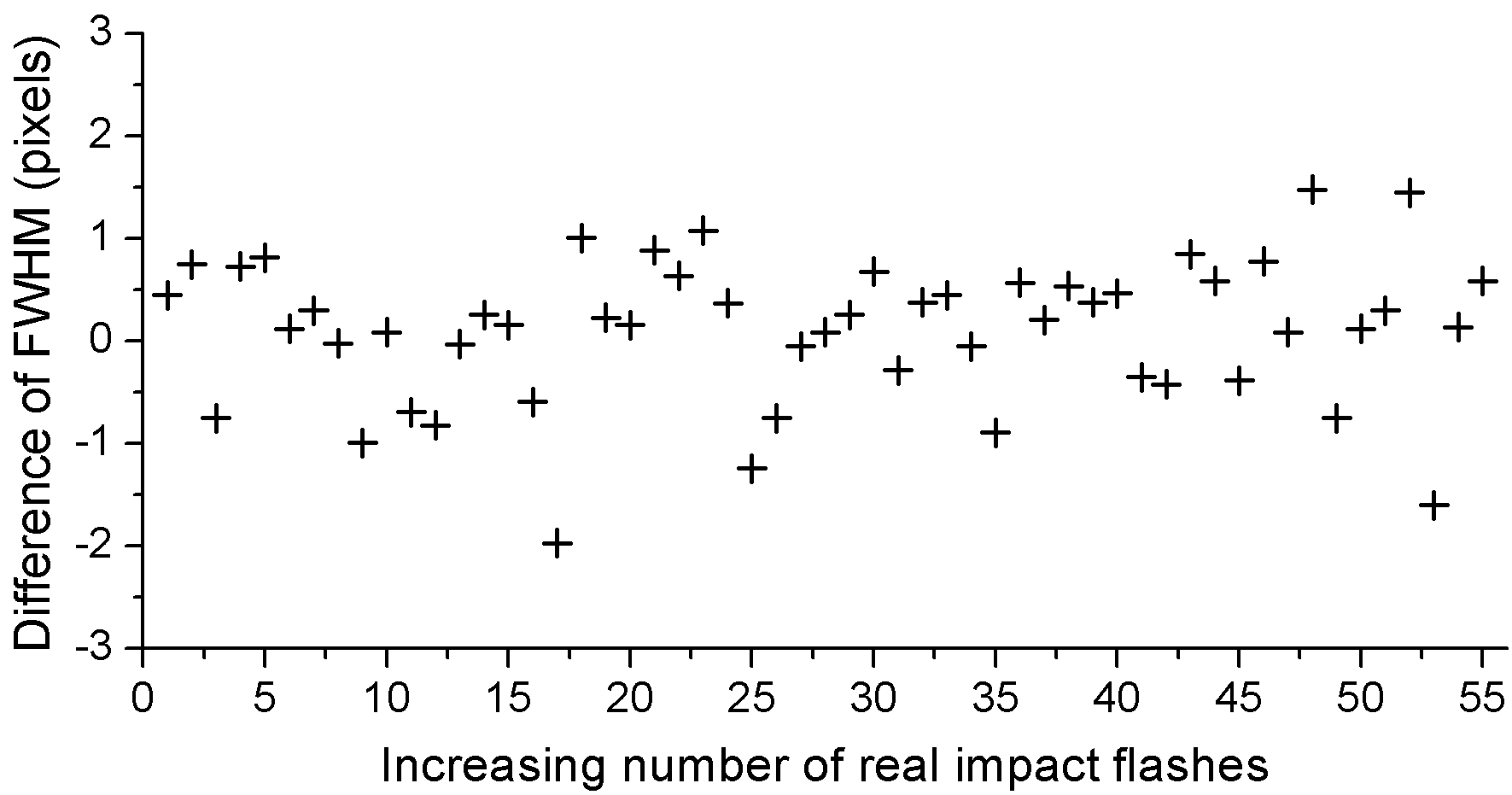}
\caption{Differences between the FWHM of standard stars and validated impact flashes in $I$-band filter for the first 55 cases.}
\label{fig:FWHM}
\end{figure}

Firstly, we set the following two validation criteria for characterizing the events. The first criterion concerns the detection of an event in the frames of both cameras and on exactly the same lunar area (pixels). At this point, it should be noted that the cameras have a small offset between them (plane axes and rotational axis). For this, a pixel transformation matrix for the frames of the cameras has been derived in order to be feasible to search for an event at a specific pixel area in the frames of one camera if it has been detected in the frames of the other. The second criterion is the lack of motion of the event between successive frames. Based on the aforementioned criteria, four possible cases for their (non) satisfaction are produced, which are addressed below, while a schematic flowchart is given in Fig.~\ref{fig:Valflow}.
\\
\\
\textit{Case 1}: Both criteria are fulfilled. The event is characterized as a `validated lunar impact flash'.
\\
\\
\textit{Case 2}: Neither criterion is fulfilled. The event is detected in multiple frames of only one camera. This case happens mostly when satellites (Fig.~\ref{fig:sat}) are detected in the frames of only one of the cameras (depends on the inclination of their solar panels) or when stars are very close to the lunar limb and are faint in one of the two passbands (depends on the temperature of the star). The event is characterized as false.
\\
\\
\textit{Case 3}: The first criterion is fulfilled but the second is not. This case is frequently met when moving objects (satellites, airplanes, birds) cross the disk of the Moon and are detected in the frames of both cameras (Fig.~\ref{fig:sat}). The event is characterized as false.
\\
\\
\textit{Case 4}: The second criterion is fulfilled (in cases of multi-frame events) or cannot be applied (in cases of single frame events) but the first is not, i.e. the event is detected in only one of the two cameras' frame(s). The latter produces two subcases, namely $Case~4R$ for the detection in the frame(s) of $R$~camera only and $Case~4I$ for the detection in $I$~camera only. This case is the most difficult but at the same time the most frequently met, since it is related to cosmic ray hits. In general, cosmic rays hit the sensors at random angles, producing, in most cases, elongated shapes, which can be easily discarded (Fig.~\ref{fig:Prof}b). However, there are cases that they hit almost perpendicularly the sensors providing round Point-Spread-Functions (PSF) like those of the stars and flashes (Fig.~\ref{fig:Prof}a).
\\
\\
\textit{Case 4R}: According to the first results given in Paper I as well as to the present results (Sections~\ref{sec:RES} and \ref{sec:Calc}) regarding the temperatures and the magnitudes of lunar impacts flashes, the apparent magnitude of a flash in the $I$ band is always brighter than in $R$, i.e. $R-I>0$. In addition, due to the Rayleigh scattering (i.e. the $R$ beam scatters more than the $I$ beam), the flashes in $I$ filter camera can be detected more easily, in a sense that they exceed the lowest software threshold. This information provides the means to directly discard events that have been detected only in the frame(s) of the $R$~camera and not in the respective one(s) of the $I$~camera. Therefore, the events of $Case-4R$ are characterized as false.
\\
\\
\textit{Case 4I}: Taking into account the second criterion regarding the non-movement of the event between successive frames, the $Case~4I$ has to be split into two subcases, which are addressed below.
\\
\\
\textit{Case 4I-1}: The first subcase concerns the satisfaction of the second criterion, which means that the event is detected in the same pixels of multiple successive frames and presents a round PSF. Cosmic ray hits last much less than the integration time of the image and they are detected in only one frame. The only exception is the particular case when the energy of the impacting particle is very large (i.e. produces saturation of the pixels) and its impact angle is almost perpendicular to the sensor. The latter produces a round PSF and also a residual signal that can be also detected in the next frame. However, this case is easy to be identified as a cosmic ray hit, because if it was a validated flash (i.e. following a Planck distribution) with such high luminosity level in this band, it would be certainly detected in the frames of the $R$~camera too. Therefore, $Case~4I-1$ events (i.e. multi-frame events detected in the $I$~camera only) are considered as `Suspected lunar impact flashes-Class 1'. Class 1 denotes that the events have high confidence to be considered as validated.
\\
\\
\textit{Case 4I-2}: In this subcase, the detection has been made only in one frame of the $I$~camera. For this case, the second criterion cannot be applied, since any possible movement cannot be verified. This case is the most difficult one and further verification is needed. The reason for this is that cosmic rays, with intensities well inside the dynamical range of the sCMOS (i.e. 0-65536~ADUs), hit almost perpendicularly to the sensor producing round PSFs and, therefore, cannot be easily distinguished from the fast validated impact flashes. As a diagnostic tool we use the comparison of the shapes (i.e. the Full Width at Half Maximum - FWHM) of a star and the event. In most of the cases, there are no field stars in the frame where the event is detected, so, the information comes from the standard stars observed between the lunar data chunks. On one hand, the observed standard stars have similar airmass values with that of the Moon, but, on the other hand, the Moon is observed typically at low altitude values where the seeing fluctuations are quite strong. Hence, we do not expect the event, if it is a validated impact flash, to have exactly the same FWHM value as that of the comparison star, but it is expected to vary within a certain range. For this, we plotted the FWHM values of the first 55 validated impact flashes with those of the standard stars used for their magnitude calibration and we found the differences in terms of FWHM (pixels) for each case. These differences are shown in Fig.~\ref{fig:FWHM} and produce a mean value of 0.56~pixels and a standard deviation of 0.71~pixels. An example of this comparison (i.e. large difference in terms of FWHM) can be seen in Fig.~\ref{fig:Prof}a and \ref{fig:Prof}c. Therefore, according to these results, the events that have difference $\pm1.3$~pixels in FWHM from that of the standard stars are considered as `Suspected lunar impact flashes-Class 2', with Class 2 to denote that the events have low confidence to be considered as validated.

\subsection{Cross check with orbits of man-made objects}
\label{sec:VAL2}
Most crossing man-made objects (i.e. (non) functioning satellites, known debris) can be easily recognized since they move during the exposure time and leave a trail (Fig.~\ref{fig:sat}-lower part). Normally, they also do not change their brightness quickly, so they are visible in several successive frames. On the other hand, another difficult case that has to be checked concerns the extremely slow moving satellites (i.e. geostationary), which produce round PSFs very similar to the true flashes. In cases of events similar to that shown in Fig.~\ref{fig:sat} (upper part), the satellite moves and spins around extremely slowly, so, it can be detected in only one frame every few seconds. Depending on its altitude and its speed, its crossing in front of the FoV of the NELIOTA setup (Fig.~\ref{fig:FoV}) may last from several seconds up to a few minutes. To exclude these events, we analyze whether any artificial objects cross the NELIOTA FoV.


We download orbit information about man-made objects in the so-called two-line element (TLE) format from two sources: (a) The database provided by the United States Strategic Command (USSTRATCOM\footnote{\url{http://www.space-track.org}}), and (b) TLE data provided by B. Gray on Github\footnote{\url{http://www.github.com/Bill-Gray/tles}}. Other data sources (CelesTrak\footnote{\url{https://celestrak.com/}}, Mike McCant\footnote{\url{https://www.prismnet.com/~mmccants/}}) were considered but not deemed useful.

Data files as close as possible to the date to be checked are downloaded from the web sites. A Python script using the Simplified General Perturbations (SGP4) orbit propagator is used to propagate the elements of all available objects to the epoch of the detected event. Using JPL’s SPICE\footnote{\url{https://naif.jpl.nasa.gov/naif/}} library, the apparent position in celestial coordinates (Right Ascension, Declination) of all objects, as seen from our telescope, is computed. The distance to the apparent position of the Moon is determined. If this value is smaller than a configurable threshold (set to match the size of the FoV), the object is flagged by the script. The script has been tested by checking several obvious satellite detections, where an object can be seen moving through the image.

For all potential impact flash events, this script can be used to check whether it could possibly be a man-made object. It should be noted, however, that not finding a match does not necessarily mean that an object can be excluded. It may as well be that this particular object is simply not in the database, or does not have TLEs with a good enough accuracy. E.g., \citet{KEL07} checked the accuracy of propagated TLEs compared to the measured position of GPS satellites and found that the typical deviation of the measured versus propagated in-track position is about 10~km after 5~days. For an orbit height of 800~km, this would correspond to an angular error of about 0.5~deg. This would already put the object outside the FoV of the instrument, thus not giving us a match. However, this check will provide more confidence about the event.

\begin{figure}
\centering
\includegraphics[width=8 cm]{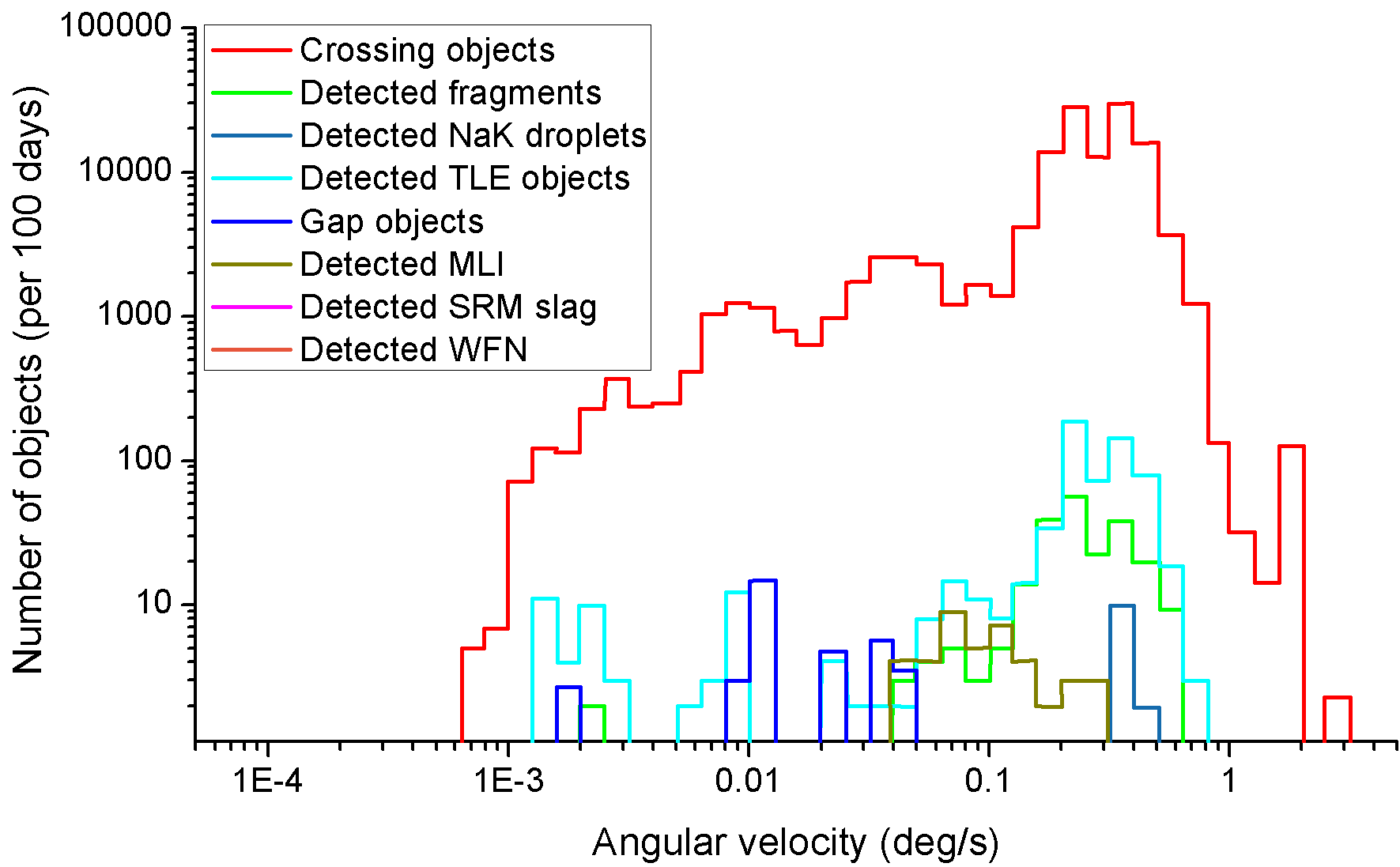}
\caption{Simulation of the Kryoneri telescope detections of artificial objects using the MASTER/PROOF tool of ESA. ‘Crossing objects’ are all objects crossing the field of view in a 100-day period. Out of those only a part are bright enough to be detected. These are shown in the other curves. NaK droplets and  solid rocket motors (SRM) slag are fuel remnants, MLI = Multi-layer Insulation, WFN stands for Westford Needles, small wires ejected by a space mission in 1963.}
\label{fig:MASTER}
\end{figure}

In addition, the MASTER/PROOF tool of ESA \citep{KRA00} was used to perform a statistical analysis. The tool was employed to find how many objects would cross the FoV of the Kryoneri telescope during a time period of 100 days, assuming a fixed pointing position. The resulting output is shown in Fig.~\ref{fig:MASTER}. PROOF distinguishes different object types. The top-most line is the total number of crossing objects. The other lines show the actually detected objects, taking their brightness into account.

It is assumed that we can recognize an object as moving if it moves at least two pixels during the exposure time of the images (23~ms), corresponding to an angular velocity of 1.3~arcsec~/~(23~ms~$\times$~3600~arcsec~deg$^{-1}$) = 0.016~deg~s$^{-1}$. Thus, only objects slower than this will not show a trail and are considered here. Integrating the number of objects with an apparent velocity below our threshold yields about 70 detectable objects. Assuming that a run of 100~days means that 1000~hours were dark, we get an average number of 0.07 crossing man-made objects per hour. For the current accumulated observing time of $\sim$110 hours this corresponds to about eight crossing man-made objects slow enough to not show a visible trail and thus potentially be mistaken with an impact flash. Note that they would still be detected in subsequent images and therefore can be excluded from the data set, unless they are just below the detection threshold and become visible due to specular reflections of sunlight. We consider this unlikely and conclude that we can safely assume that all the detected point sources are not satellites, except for four cases that are discussed in Section~\ref{sec:RES}. More details on these analyses can be found in \citet{EIS17}.

We argue that the possibility for a satellite flare, similar to an iridium flare, to be mistaken for an impact flash is very unlikely. Iridium flares are generated by three polished antennas, each about 2~m$^2$ in size. If they are oriented in the right way, sunlight can be specularly reflected and generate the flare. In principle, a rigid, fast rotating object would look similar to a lunar impact. At the distance of 15000~km, when an object in a circular orbit would move slow enough to not generate a streak in the image, a mirror of 1~cm$^2$ area would be enough to generate a 10~mag flare. We can, however, immediately exclude all multi-frame detections. A flare would show a symmetric light curve, whereas our observations have the brightness peak in the beginning, followed by a decay. For one-frame flashes, we use the following argument to show that a misidentification is unlikely. The apparent diameter of the Sun is 0.5~deg. For a flare to show up in only one frame (33~ms), the object would need to rotate once in 360~deg/0.5~deg $\times$~0.033~s = 23.8~s. To not show as a streak in a single exposure, the object has to move slower than 0.016~deg~s$^{-1}$. I.e. the next flash would should up in a distance of 23.8~s $\times$ 0.016~deg~s$^{-1}$ = 0.38~deg. Precisely, it would need to be a bit more, since the object has moved in space. However, this can be neglected at this height. This is indeed larger than our field of view and we would not see the object. Let's assume that we would see the object again if it is less than 0.1~deg away from the initial point, see e.g. Fig.~\ref{fig:sat} (upper part). For that, it has to move slower than 0.004~deg~s$^{-1}$, corresponding to about 29000~km altitude. Most man-made objects are either below 15000~km, or above 29000~km. We therefore argue that a misidentification is unlikely. Of course it could be that the flaring object only flies through a corner of our field of view. Or the object could be tumbling irregularly. Therefore we cannot fully exclude this to happen and would welcome other observatories in the same longitude range to take up parallel observations. With two different locations, one could fully exclude even these low-likelihood events.



\section{Photometry of the flashes}
\label{sec:PHOT}
\begin{figure}
\begin{tabular}{cc}
\includegraphics[width=4cm]{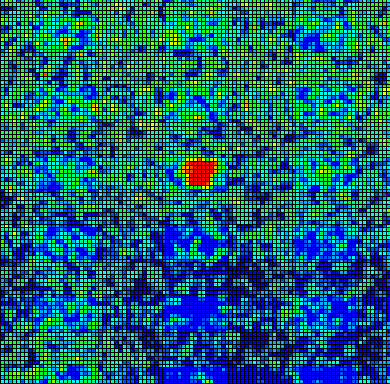}&\includegraphics[width=4cm]{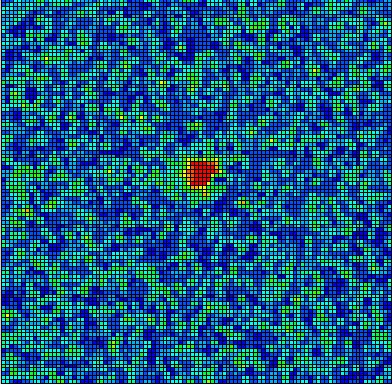}\\
\end{tabular}
\caption{Photometric profile example of a validated impact flash using pseudocolors before (left) and after (right) the lunar background subtraction. The images include an area of 100$\times$100~pixels around the flash with the colours to denote the intensity (red for flash). For the left image, it should be noticed the inhomogeneity of the background around the flash that makes the subtraction of the background determined in an annulus inaccurate. For the right image, the background around the flash is more smooth and more uniform and contains only the residual noise. For more details see text.}
\label{fig:BGD}
\end{figure}

The method followed to calculate the apparent magnitudes of the flashes was presented briefly in Paper I, but it was found useful to present it herein also in order to provide more details. The photometry on impact flashes cannot be considered a routine operation, since the inhomogeneity of the background due to the lunar features as well as due to the glare from the sunlit part of the Moon play a critical role. In addition, given that the Moon is usually observed at relatively high airmass values, the fluctuation of the background is not negligible. Therefore, very careful measurements have to be made. For measuring the intensity of the flashes, aperture photometry has been selected as the most appropriate method. However, the latter is strongly dependent on the background substraction around the area of interest, something that is complicated when measuring flashes on an inhomogeneous surface such that of the Moon. We initially measured the flashes on the frames containing the lunar background using standard aperture photometry techniques (i.e. use of star aperture and sky annulus). Various tests showed that the non-uniform background of the Moon plays an essential role to the calculation of the flux, In particular, in cases where the sky annulus included dark areas on one side and bright areas on the other side (e.g. craters and maria; see also Fig.~\ref{fig:BGD}), the mean background value, which is subtracted from the flux of the flash itself, was becoming unrealistic. The same situation was faced for flashes detected very close to the limb of the Moon. The one side of the sky annulus contained lunar background, while the other only sky background. In order to confront this situation, it was decided to perform aperture photometry on the lunar background subtracted frames (Fig.~\ref{fig:BGD}).

The NELIOTA-DET software provides a FITS data cube that contains seven frames before and seven after the frames in which the event is detected. It should be noted that the time difference between the first frame of this cube and the first frame of the event as well as between the last frame of the event and the last frame of the cube is 231~ms. The first step is to create a background image that will be subtracted from those that contain the flash. For this, the first five and the last five images of the data cube are combined producing a mean background image for each band. Hence, the background image is the mean image of a total of ten images taken 66~ms before and 66~ms after the frames of the event. The two frames before the first and after the last frames in which the event is detected are not used for the background calculation because the event may have begun in the previous frame with an intensity below the threshold of the software. The same applies for the frame after the last one in which the event is detected by the software. Using this method, the mean background image includes only the fluctuations of the seeing 231~ms before and after the event in contrast with the time-weighted background image automatically created by the software (see Paper II).


Subsequently, this mean background image is subtracted from the images containing the event producing the so-called `Difference' images in which the lunar background has been removed (Fig.~\ref{fig:BGD}) and only the event has been left. However, after the subtraction, the difference image has a non zero-level background, i.e. it contains a residual noise signal. The standard deviation of this noise depends on the glare of the sunlit part of the Moon (i.e. lunar phase) and the seeing conditions and plays an essential role to the error estimation of the magnitude of the flash. 

The photometric analysis is made with the software AIP4WIN \citep{BER00}. Optimal aperture values are used for both the flashes (in difference images) and the standard stars observed closest in time. The optimal aperture radius for the flash is selected according to its photometric profile (i.e. radius after that only background noise exists) and its curve of growth (i.e. radius that gives the maximum value in intensity). As instrumental flux of the flash ($f_{\rm f}$), we account only the ADUs measured in the first aperture. We do not subtract any sky value, since the background has been already removed as mentioned before. The optimal aperture radius for the standard stars is set as 4$\sigma$ of their FWHM. The flux value of the standard star ($f_{\rm s}$) is derived as the difference between the ADUs measured in the star aperture and the ADUs measured in the sky annulus. For each standard star observed, five images are acquired, hence, the final value of $f_{\rm s}$ is the average of these five measurements. The photometric errors of the fluxes are calculated based on the following equation (IRAF documentation\footnote{\url{http://stsdas.stsci.edu/cgi-bin/gethelp.cgi?phot.hlp}}):
\begin{equation}
\delta f_{\rm phot} = \sqrt{\frac{f}{G} + A\times \sigma_{\rm bgd}^2 + \frac{A^2\times \sigma_{\rm bgd}^2}{n_{\rm sky}}},
\label{eq:df}
\end{equation}
where $f$ is the instrumental flux in ADUs, $G$ the gain of the camera, $A$ the area that is covered by an aperture of a radius $r$ (i.e. $A=\pi r^2$) in pixels, $\sigma_{\rm bgd}$ is the standard deviation of the background, and $n_{\rm sky}$ is the number of pixels of the sky annulus. In the case of the flashes, the last term in Eq.~\ref{eq:df} is omitted, since we do not measure any sky background (i.e. no sky annulus used). The magnitude calculation of the flash is based on the Pogson law:
\begin{equation}
m_{\rm f} = m_{\rm s} + 2.5 \log \frac{f_{\rm s}}{f_{\rm f}},
\label{eq:pog}
\end{equation}
where $m_{\rm s}$ is the magnitude of the standard star as given in the catalogues and $m_{\rm f}$ is the magnitude of the flash. It should be noted that the $f_{\rm f}$ and $f_{\rm s}$ in Eq.~\ref{eq:pog} are normalized to the same integration time, since the standard stars are recorded with other exposure times than that of the flashes. The photometric magnitude error of the flash ($\delta m_{\rm f, phot}$) is derived according to the error propagation method and is based on the measured instrumental fluxes $f_{\rm s}$ and $f_{\rm f}$ and their respective errors ($\delta f_{\rm s}$ and $\delta f_{\rm f}$) as derived from Eq.~\ref{eq:df}. Every procedure described in this section is applied to the frames of each photometric band separately.

\subsection{Scintillation error}

\begin{figure}
\centering
\includegraphics[width=\columnwidth]{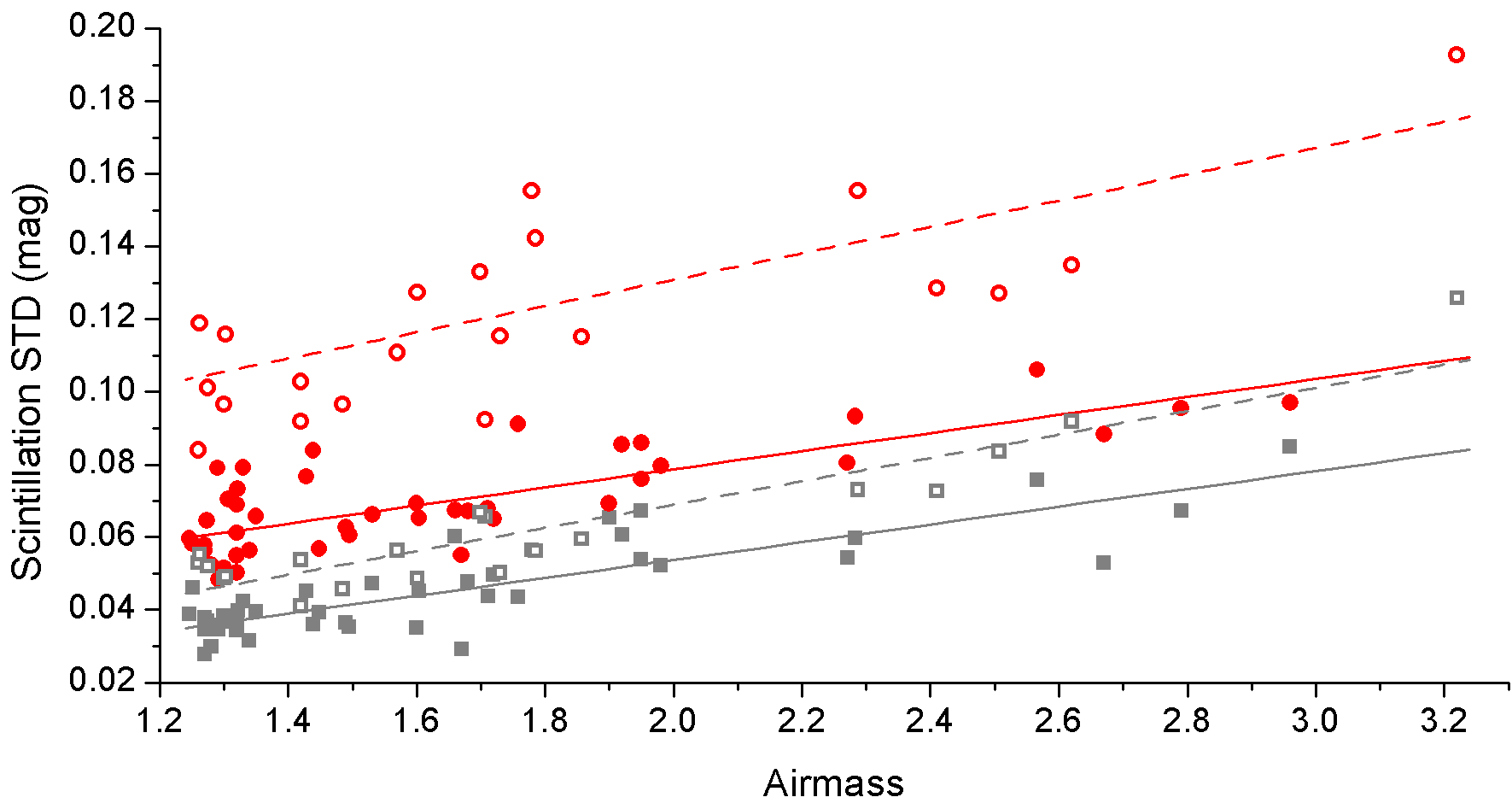}
\caption{Scintillation error against the airmass (X) for two magnitude ranges. Filled symbols and solid lines (red for $R$ band and grey for $I$ band) denote the observed values and the respective fittings in the magnitude range 9.5-10.5. Open symbols and dashed lines denote the same but for the magnitude range 10.5-11.5.}
\label{fig:SCINT}
\end{figure}

Scintillation has been proven as a significant contributor in the calculation of the magnitude errors values \citep[cf.][]{SUG14}, especially when observing in fast frame rates and with such a large aperture telescope. Therefore, in order to take into account this effect on our measurements, observations of standard stars with magnitudes between 9.5-11.5 in $R$ and $I$ passbands were obtained during several clear photometric nights between summer 2018-spring 2019. The stars were observed at various airmass ($X$) values during a given night in order to examine carefully the dependence of the scintillation effect on the altitude, hence the atmospheric transparency/turbulance, of the star. The standard stars for these observations are taken from the list of standards used also for the magnitude calibration of the flashes. In Fig.~\ref{fig:SCINT}, the standard deviations of the magnitudes of the stars in $R$ and $I$ bands (symbols) are plotted against the airmass. It was found that the scintillation effect has a different behaviour for the bright and the faint stars and depends also on the observed passband. Therefore, two magnitude ranges for each passband were selected for individual fits (lines in Fig.~\ref{fig:SCINT}), namely 9.5-10.5 and 10.5-11.5. The respective relations for the magnitude error due to scintillation $\delta m_{\rm f,~scin}$ according to these fits are:

\begin{equation}
  \delta m_{\rm f_{R},~scin}=
  \begin{cases}
    0.0289+0.0249X, & \text{for $9.5<R<10.5$}\\
    0.0585+0.0362X, & \text{for $10.5<R<11.5$}
  \end{cases}
\label{eq:scintR}
\end{equation}
and
\begin{equation}
  \delta m_{\rm f_{I},~scin}=
  \begin{cases}
    0.0047+0.0245X, & \text{for $9.5<I<10.5$}\\
    0.0048+0.0321X, & \text{for $10.5<I<11.5$}
  \end{cases}
  \label{eq:scintI}
\end{equation}

Therefore, the final magnitude error of the flashes based on both the photometry and the scintillation effect and their individual magnitudes and airmass values is calculated by the following formula:
\begin{equation}
\delta m_{\rm f} = \sqrt{(\delta m_{\rm f,~phot})^2 + (\delta m_{\rm f,~scin})^2}
\label{eq:finer}
\end{equation}

It should be noted that for flashes brighter than 9.5~mag or fainter than 11.5~mag the relation of the range that is closest to the observed magnitude value is used.


\section{Campaign statistics and results}
\label{sec:RES}

\begin{figure}
\centering
\begin{tabular}{c}
\includegraphics[width=\columnwidth]{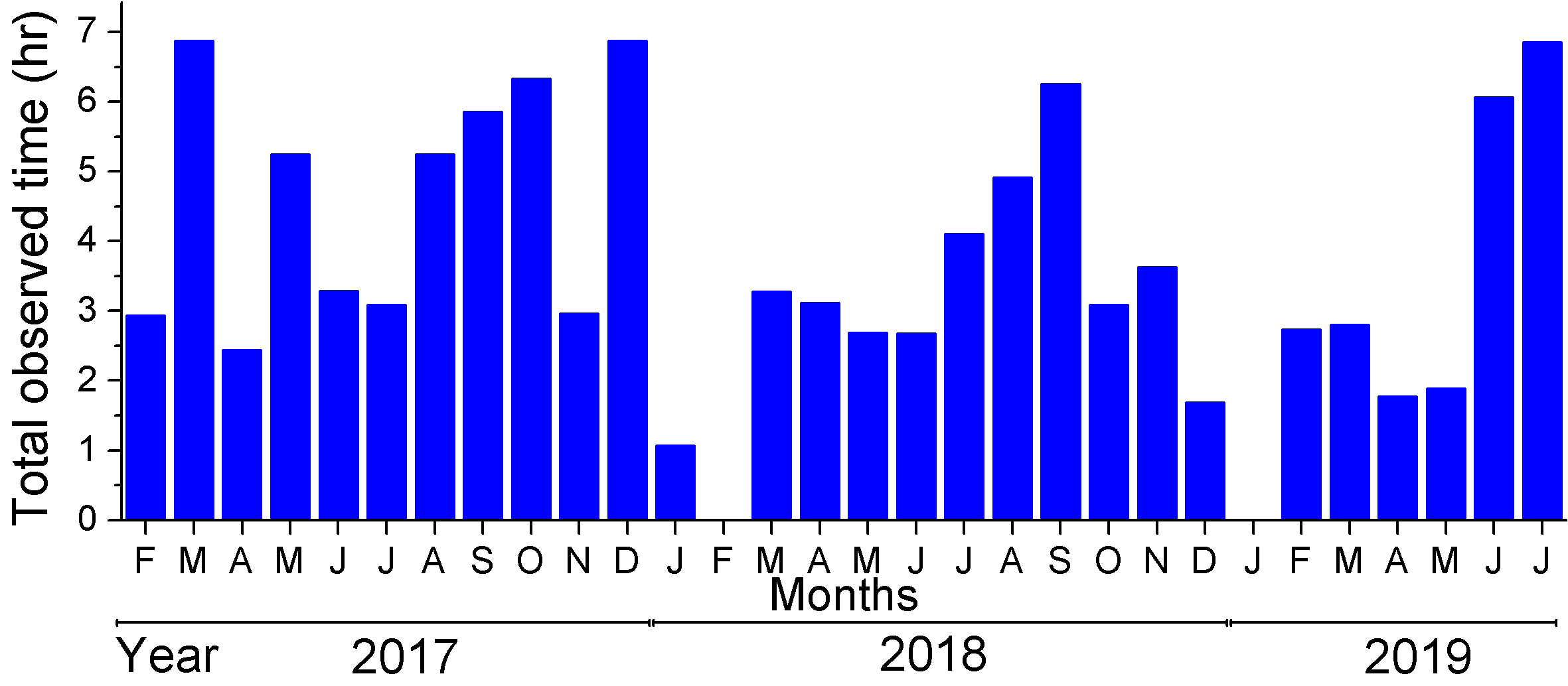}\\
\includegraphics[width=\columnwidth]{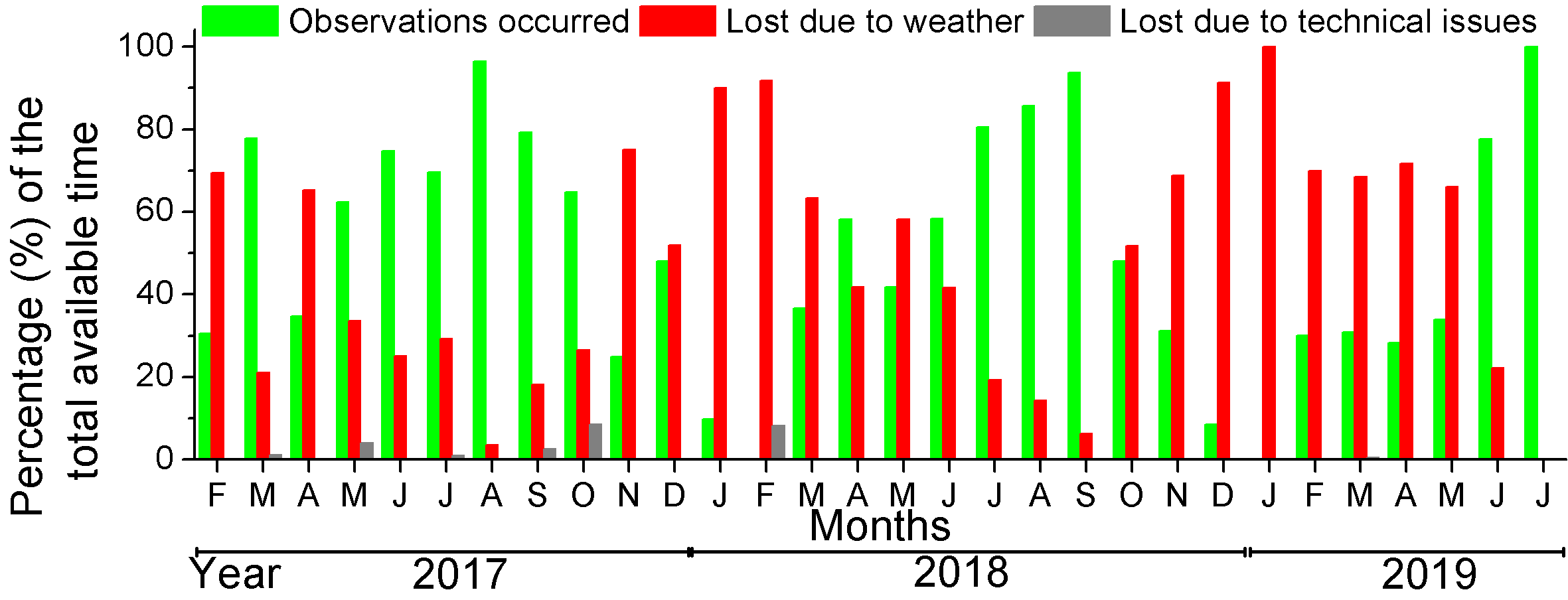}\\
\includegraphics[width=\columnwidth]{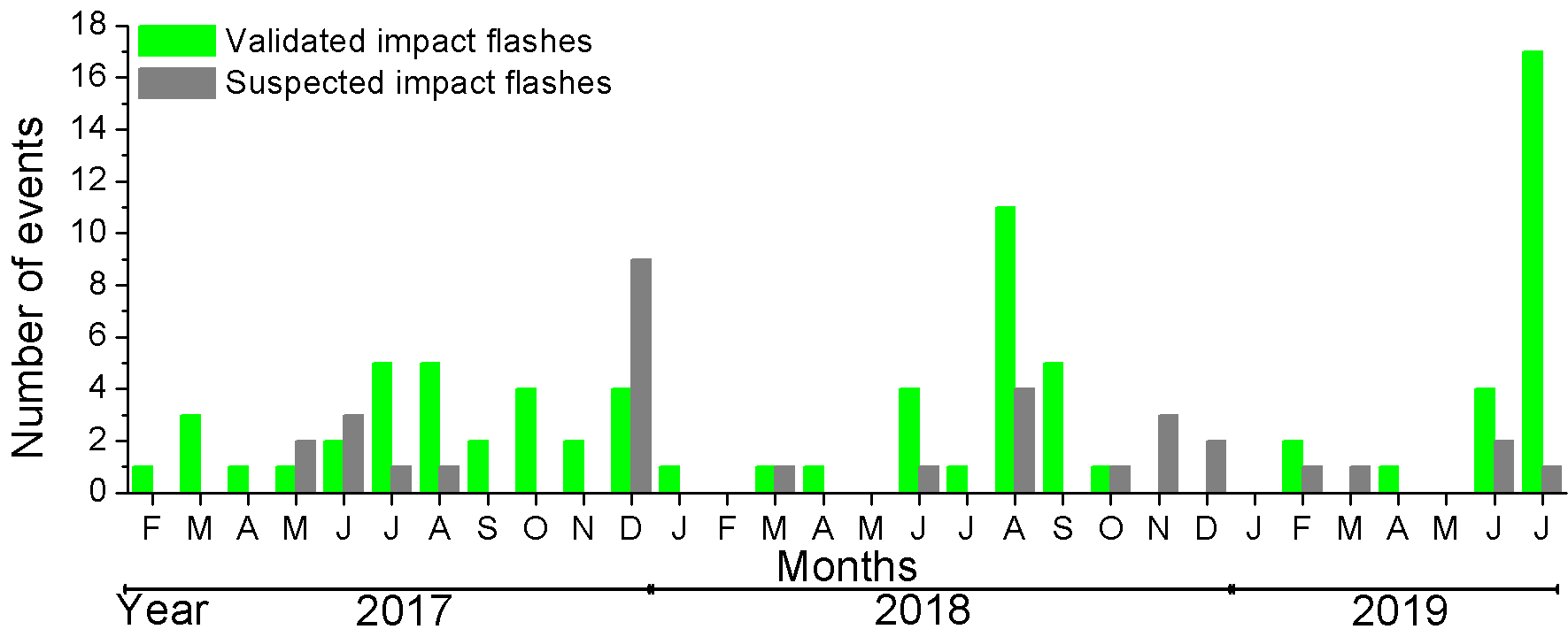}\\
\end{tabular}
\caption{Observational statistics log for the first 30 months of NELIOTA operations. Upper panel: Total observed hours on Moon. Middle panel: The NELIOTA time distribution. Lower panel: Distribution of the detected impact flashes.}
\label{fig:stat}
\end{figure}

In this section, we present the statistics of the campaign to date as well as the results for the events that are validated as flashes. The upper panel of Fig.~\ref{fig:stat} shows the histogram of the total observed hours on Moon. It should be noted that this plot includes only the real observed time of the Moon excluding the read-out time of the cameras and the time spent for the observations of the standard stars. The middle panel of Fig.~\ref{fig:stat} shows the distribution of the available time for lunar observations for each month of the campaign. In absolute numbers, the total available time on Moon for these 30 months of the campaign was 401.2~hrs. Excluding the time lost due to the read-out time of the cameras and the standard stars observations (i.e. $\sim40\%$ of the total available time; see Section~\ref{sec:Obs}), the total true available time becomes 248.2~hrs. Out of this, 110.48~hrs ($\sim44.5\%$) were spent for lunar recording, $\sim135.3$~hrs ($\sim54.5\%$) were lost due to bad weather conditions, and another $\sim2.4$~hrs ($\sim0.97\%$) were lost due to technical issues. The lower part of Fig.~\ref{fig:stat} shows the detection of flashes during the campaign.

\begin{figure}
\centering
\begin{tabular}{cc}
\includegraphics[width=4.3cm]{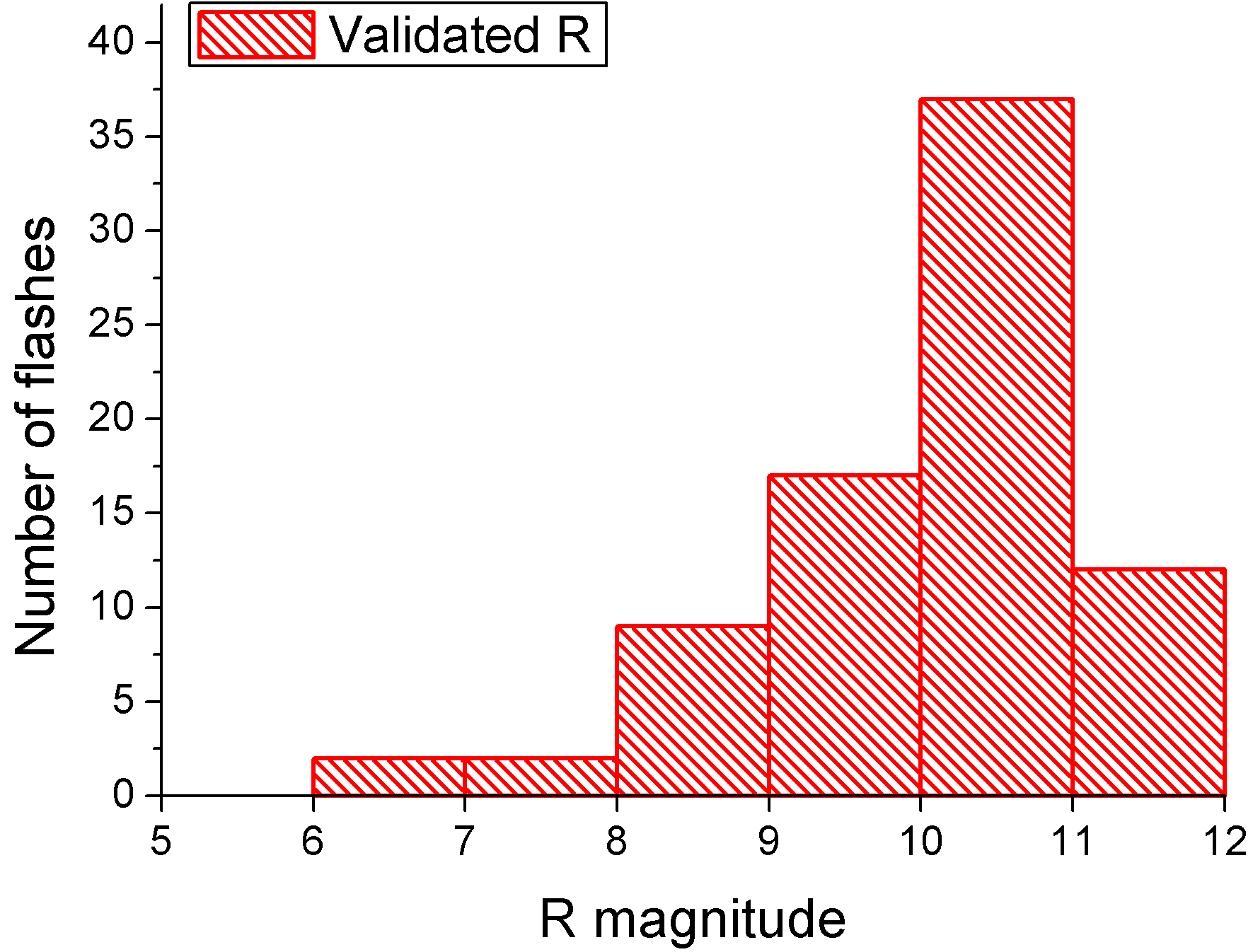}&\includegraphics[width=4.3cm]{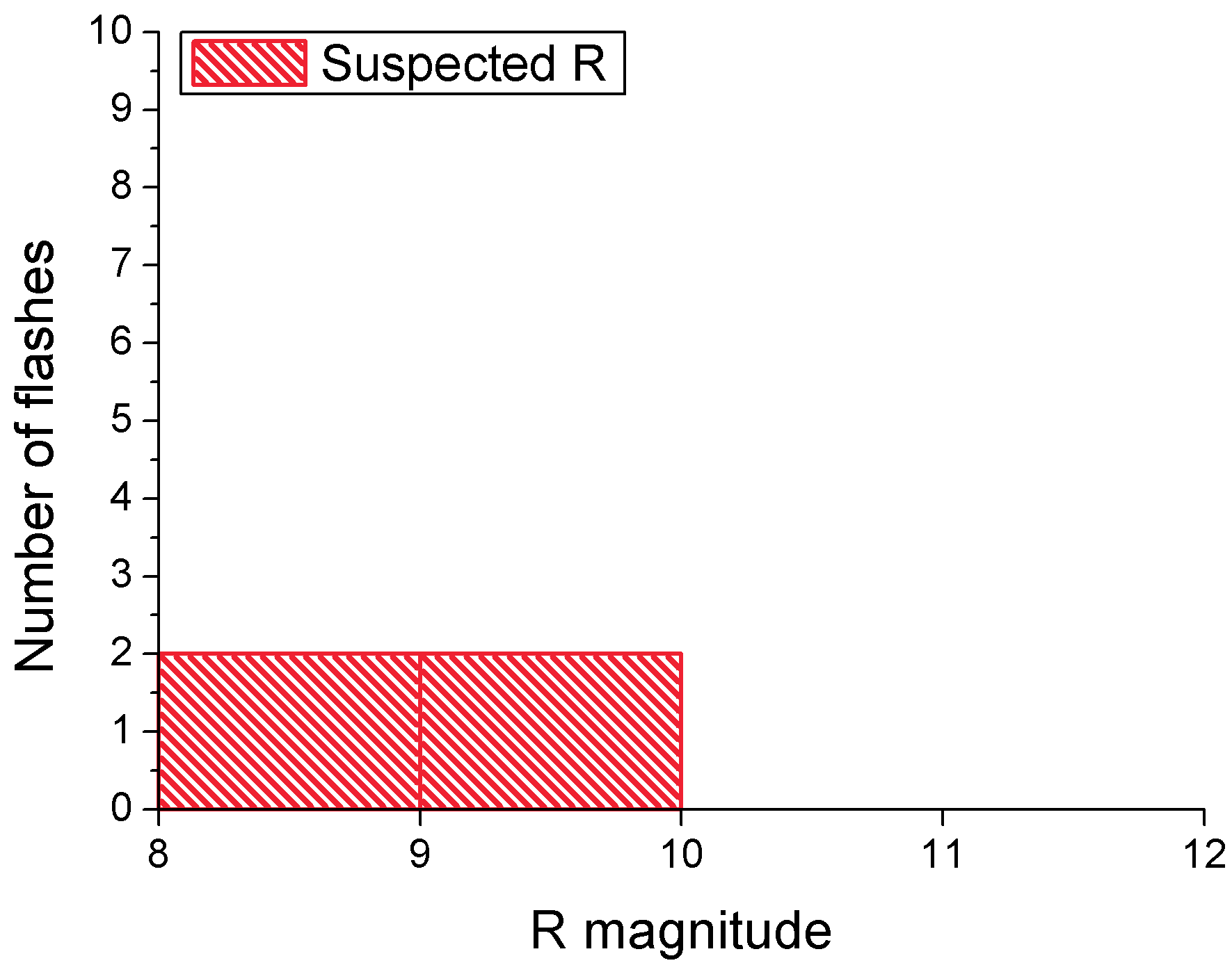}\\
\includegraphics[width=4.3cm]{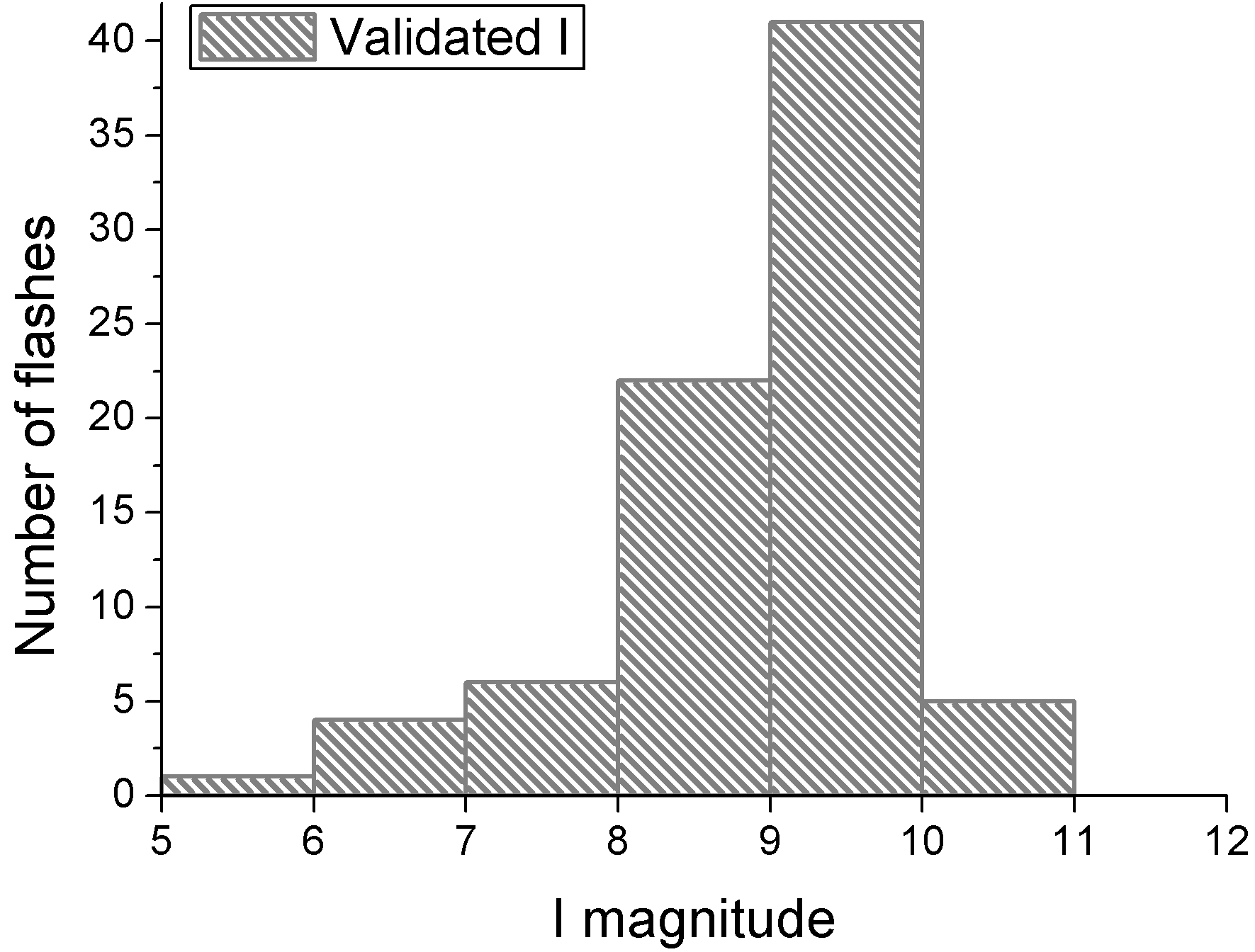}&\includegraphics[width=4.3cm]{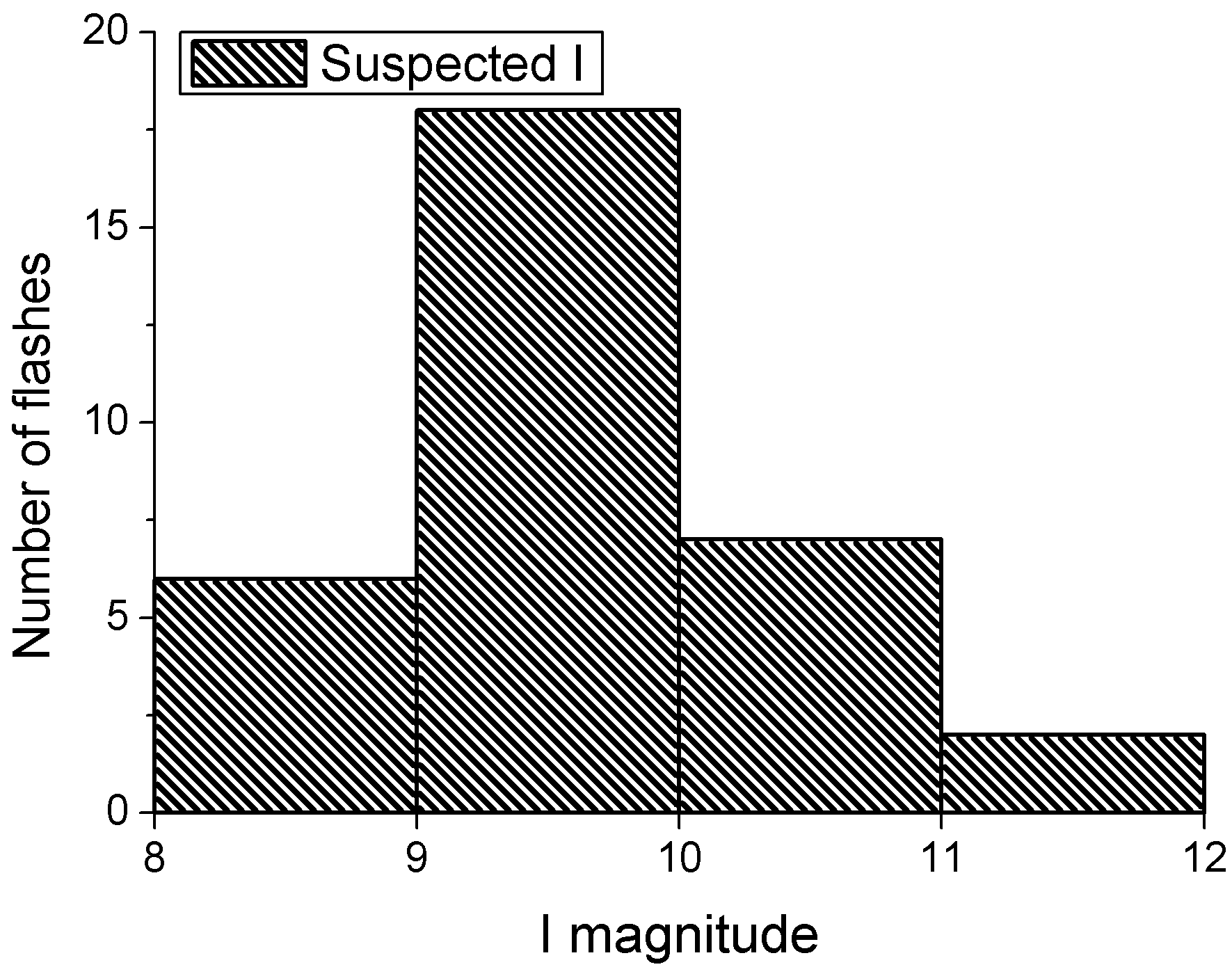}\\
\end{tabular}
\caption{Peak magnitude distributions in $R$ (upper panels) and $I$ (lower panels) bands for the validated (left) and the suspected (right) flashes.}
\label{fig:mag}
\end{figure}

\begin{figure}
\centering
\includegraphics[width=\columnwidth]{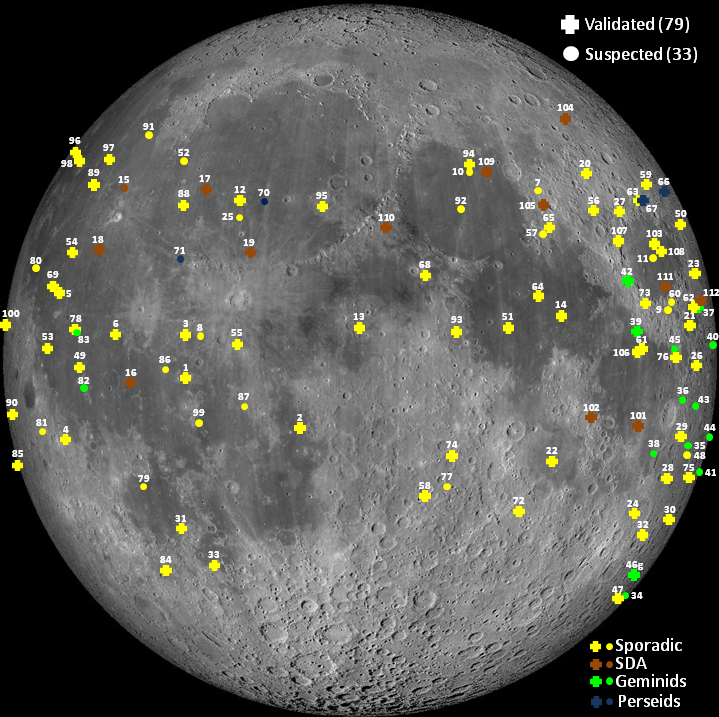}\\
\caption{Locations of the detected impact flashes on the lunar surface up to July 2019. Crosses and filled circles denote the validated and the suspected flashes, respectively. Different colours denote the most possible origin of the meteoroid.}
\label{fig:localization}
\end{figure}

In Table~\ref{tab:list}, the results for all the detected flashes are given in chronological order (increasing number). In particular, this table includes for each detected flash: the date and the UT timing of the frame containing the peak magnitude, the validation outcome according to the criteria set in Section~\ref{sec:VAL}, the maximum duration in ms (based on the number of frames detected) the peak magnitude(s) in $R$ and/or $I$ bands and the selenographic coordinates (latitude and longitude; for the method of localization see Appendix~\ref{sec:LOC}). The magnitude distributions of the validated and the suspected flashes are shown in Fig.~\ref{fig:mag}, while their locations on the lunar surface are shown in Fig.~\ref{fig:localization}.

Although impact flashes have common origin (i.e. meteoroids), their shape on the frames or their light curves differ from time to time due to various reasons. For example, the flash $\#19$ shows a peculiar PSF with two peaks. This may be caused either by the scintillation due to atmosphere or is, in fact, a double impact. However, it is not possible to be certain of its nature. Therefore, its magnitudes concern the total flux and in the following analysis is considered as a single impact. The $m_{\rm I}$ of flash $\#27$ is a rough estimation because it was detected at the edge of the frame. However, due to the cameras offset (see section~\ref{sec:VAL1}) the flash in the $R$ camera was completely inside the frame. The peak magnitudes of the flashes $\#28$ and $\#73$ were detected in the second set of frames, hence, the initial brightness increase (i.e. before the maximum) was also recorded. A few flashes, although detected in both bands, are characterized as suspected in Table~\ref{tab:list} for specific reasons. The flash $\#$52 shows an elongated shape, while its $R-I$ index has an extreme value in comparison with the rest validated flashes. Flash $\#$81 shows again an extreme $R-I$ value. The flashes $\#$70 and $\#$71 were detected with only 0.5~s difference but in different positions on the Moon and their $R-I$ indices were found to deviate slightly from the other validated ones. For all the aforementioned suspected flashes, except $\#$52, a possible cross match was found with slow moving satellites/space debris. In particular, the flashes $\#$70 and $\#$71 could be false positives of the NORAD~12406 (Kyokko 1, Japan) space debris, while flash $\#$81 could be the reflection of the rocket body NORAD~26738 (Breeze-M, Russia; Upper stage). No cross match was found for flash $\#$52 but its shape and $R-I$ index indicate that its nature is different than a meteoroid hit. Therefore, these four suspected flashes are excluded from the lists of flashes including calculations for the physical parameters (Tables~\ref{tab:ResultsReal}-\ref{tab:ResultsSusp}).

\begin{table*}
\centering
\caption{Photometric results and locations of the detected flashes during the first 30 months of NELIOTA operations. Errors are included in parentheses alongside magnitude values and correspond to the last digit(s). The error in the determination of the location is set as 0.5$\degr$.}
\label{tab:list}
\begin{tabular}{cccc cccc}
\hline\hline															
ID	&	Date \& UT	&	Val.	&	dt	&	$m_{\rm R}$	&	$m_{\rm I}$	&	Lat.	&	Long.	\\
	&		&		&	(ms)	&	(mag)	&	(mag)	&	($\degr$)	&	($\degr$)	\\
\hline															
1	&	2017 02 01    17:13:57.862	&	V	&	33	&	10.15(12)	&	9.05(5)	&	-1.5	&	-29.2	\\
2	&	2017 03 01    17:08:46.573	&	V	&	132	&	6.67(07)	&	6.07(6)	&	-10.3	&	-9.7	\\
3	&	2017 03 01    17:13:17.360	&	V	&	33	&	9.15(11)	&	8.23(7)	&	+4.5	&	-29.9	\\
4	&	2017 03 04    20:51:31.853	&	V	&	33	&	9.50(14)	&	8.79(6)	&	-12.7	&	-58.9	\\
5	&	2017 04 01    19:45:51.650	&	V	&	33	&	10.18(13)	&	8.61(3)	&	+11.6	&	-58.8	\\
6	&	2017 05 01    20:30:58.137	&	V	&	66	&	10.19(18)	&	8.84(5)	&	+4.7	&	-43.2	\\
7	&	2017 05 20    01:58:56.980	&	SC2	&	33	&		&	10.93(32)	&	29.5	&	38.5	\\
8	&	2017 05 29    19:00:05.083	&	SC2	&	33	&		&	9.78(12)	&	2.4	&	-25.8	\\
9	&	2017 06 19    01:50:34.560	&	SC1	&	66	&		&	9.60(9)	&	8.7	&	54.8	\\
10	&	2017 06 19    01:51:08.663	&	SC2	&	33	&		&	11.02(35)	&	33.9	&	18.8	\\
11	&	2017 06 19    02:39:13.590	&	SC2	&	33	&		&	10.99(40)	&	16.6	&	59.8	\\
12	&	2017 06 27    18:58:26.680	&	V	&	66	&	11.07(32)	&	9.27(6)	&	26.8	&	-22.5	\\
13	&	2017 06 28    18:45:25.568	&	V	&	66	&	10.56(38)	&	9.48(13)	&	5.6	&	0.0	\\
14	&	2017 07 19    02:00:36.453	&	V	&	66	&	11.23(40)	&	9.33(6)	&	7.8	&	35.0	\\
15	&	2017 07 27    18:31:06.720	&	SC1	&	66	&		&	9.34(10)	&	29.5	&	-46.7	\\
16	&	2017 07 28    18:21:44.850	&	V	&	66	&	11.24(34)	&	9.29(6)	&	-3.2	&	-40.0	\\
17	&	2017 07 28    18:42:58.027	&	V	&	33	&	10.72(24)	&	9.63(10)	&	28.5	&	-30.6	\\
18	&	2017 07 28    18:51:41.683	&	V	&	33	&	10.84(24)	&	9.81(9)	&	20.6	&	-50.7	\\
19	&	2017 07 28    19:17:18.307	&	V	&	165	&	8.27(04)	&	6.32(1)	&	18.1	&	-18.7	\\
20	&	2017 08 16    01:05:46.763	&	V	&	66	&	10.15(18)	&	9.54(10)	&	32.0	&	47.5	\\
21	&	2017 08 16    02:15:58.813	&	V	&	66	&	10.69(28)	&	9.11(6)	&	6.7	&	68.1	\\
22	&	2017 08 16    02:41:15.113	&	V	&	66	&	10.81(30)	&	9.08(6)	&	-15.6	&	34.6	\\
23	&	2017 08 18    02:02:21.417	&	V	&	66	&	10.92(18)	&	9.20(4)	&	-25.9	&	57.8	\\
24	&	2017 08 18    02:03:08.317	&	V	&	66	&	10.19(12)	&	8.83(4)	&	13.5	&	76.8	\\
25	&	2017 08 27    17:29:42.997	&	SC2	&	33	&		&	10.25(23)	&	24.6	&	-21.5	\\
26	&	2017 09 14    03:17:49.737	&	V	&	132	&	9.17(07)	&	8.07(3)	&	-1.1	&	70	\\
27	&	2017 09 16    02:26:24.933	&	V	&	231	&	8.52(03)	&	7.04(1)	&	24.7	&	52.5	\\
28	&	2017 10 13    01:54:21.482	&	V	&	132	&	9.28(11)	&	8.37(4)	&	-17.3	&	65.2	\\
29	&	2017 10 13    02:33:43.560	&	V	&	99	&	10.31(24)	&	9.89(12)	&	-12.5	&	66.5	\\
30	&	2017 10 16    02:46:45.613	&	V	&	99	&	10.72(16)	&	9.46(5)	&	-25.4	&	72.5	\\
31	&	2017 10 26    17:59:42.646	&	V	&	33	&	10.03(25)	&	9.42(12)	&	-27.9	&	-33.8	\\
32	&	2017 11 14    03:34:14.985	&	V	&	66	&	10.31(17)	&	9.31(6)	&	-29.5	&	64.4	\\
33	&	2017 11 23    16:17:33.000	&	V	&	66	&	10.45(23)	&	10.06(12)	&	-35.0	&	-30.5	\\
34	&	2017 12 11    03:46:22.300	&	SC2	&	33	&		&	9.65(10)	&	-41.0	&	84.5	\\
35	&	2017 12 12    01:49:26.480	&	SC1	&	66	&		&	8.91(8)	&	-14.0	&	70.7	\\
36	&	2017 12 12    02:06:11.777	&	SC2	&	33	&		&	9.63(8)	&	-7.2	&	64.7	\\
37	&	2017 12 12    02:48:08.178	&	V	&	66	&	10.50(24)	&	8.98(8)	&	9.0	&	74.0	\\
38	&	2017 12 12    03:33:05.912	&	SC2	&	33	&		&	9.61(14)	&	-15.4	&	58.4	\\
39	&	2017 12 12    04:30:00.398	&	V	&	33	&	10.58(28)	&	9.84(11)	&	5.4	&	51.2	\\
40	&	2017 12 12    04:58:00.343	&	SC2	&	33	&		&	10.13(20)	&	1.9	&	76.7	\\
41	&	2017 12 13    02:38:14.109	&	SC2	&	33	&		&	10.32(16)	&	-21.2	&	87.9	\\
42	&	2017 12 13    04:26:57.484	&	V	&	33	&	10.56(23)	&	9.95(11)	&	13.0	&	50.0	\\
43	&	2017 12 13    04:59:49.533	&	SC2	&	33	&		&	9.86(15)	&	-6.7	&	72.0	\\
44	&	2017 12 13    05:04:10.019	&	SC2	&	33	&		&	9.65(15)	&	-11.4	&	86.7	\\
45	&	2017 12 13    05:07:38.089	&	SC2	&	33	&		&	9.26(13)	&	1.7	&	62.1	\\
46	&	2017 12 14    04:35:09.737	&	V	&	132	&	7.94(5)	&	6.76(2)	&	-36.9	&	73.4	\\
47	&	2018 01 12    03:54:03.027	&	V	&	66	&	10.01(14)	&	9.31(7)	&	-40.7	&	79.2	\\
48	&	2018 03 10    03:30:05.884	&	SC2	&	33	&		&	9.65(11)	&	-13.0	&	71.0	\\
49	&	2018 03 23    17:24:19.012	&	V	&	33	&	9.93(26)	&	8.62(6)	&	-1.40	&	-52.0	\\
50	&	2018 04 10    03:36:57.535	&	V	&	33	&	8.84(13)	&	8.08(5)	&	21.7	&	74.5	\\
51	&	2018 06 09    02:29:18.467	&	V	&	33	&	9.92(23)	&	9.00(9)	&	4.3	&	24.6	\\
52	&	2018 06 18    19:16:44.473	&	SC2	&	33	&	8.85(9)	&	8.82(10)	&	33.9	&	-36.9	\\
53	&	2018 06 19    19:12:09.650	&	V	&	33	&	9.87(21)	&	9.03(9)	&	-59.0	&	3.6	\\
54	&	2018 06 19    20:00:48.490	&	V	&	33	&	9.92(28)	&	9.31(14)	&	-58.2	&	17.4	\\
55	&	2018 06 19    20:04:09.773	&	V	&	33	&	10.26(61)	&	8.63(11)	&	-20.0	&	2.5	\\
56	&	2018 07 09    01:44:19.410	&	V	&	33	&	11.16(28)	&	10.06(12)	&	24.9	&	46.0	\\
\hline																																														
\end{tabular}
\tablefoot{V=Validated flash, SC1/2=Suspected flash of Class 1/2 (Section~\ref{sec:VAL1})}
\end{table*}

\begin{table*}
\centering
\caption{Table~\ref{tab:list} cont.}
\label{tab:listCONT}
\begin{tabular}{cccc cccc}
\hline\hline																															
ID	&	Date \& UT	&	Val.	&	dt	&	$m_{\rm R}$	&	$m_{\rm I}$	&	Lat.	&	Long.	\\
	&		&		&	(ms)	&	(mag)	&	(mag)	&	($\degr$)	&	($\degr$)	\\
\hline																															
57	&	2018 08 06    01:12:10.939	&	SC2	&	33	&		&	10.47(30)	&	20.7	&	33.8	\\
58	&	2018 08 06    01:57:43.686	&	V	&	99	&	9.68(16)	&	8.14(4)	&	-22.1	&	10.6	\\
59	&	2018 08 06    02:38:14.302	&	V	&	99	&	9.16(9)	&	7.73(2)	&	28.8	&	67.2	\\
60	&	2018 08 06    03:15:10.684	&	SC2	&	33	&		&	9.80(25)	&	9.5	&	61.8	\\
61	&	2018 08 07    01:33:54.756	&	V	&	66	&	10.79(26)	&	9.31(7)	&	1.8	&	52.1	\\
62	&	2018 08 07    01:35:45.168	&	V	&	132	&	8.78(5)	&	7.74(2)	&	3.1	&	70.0	\\
63	&	2018 08 07    02:33:18.184	&	V	&	33	&	10.07(17)	&	9.46(7)	&	26.7	&	60.2	\\
64	&	2018 08 07    03:10:33.302	&	V	&	33	&	10.39(31)	&	9.80(14)	&	10.3	&	30.6	\\
65	&	2018 08 08    02:19:55.005	&	V	&	33	&	11.14(28)	&	9.90(7)	&	21.9	&	34.9	\\
66	&	2018 08 08    02:28:23.406	&	V	&	66	&	11.06(21)	&	10.40(13)	&	28.0	&	76.4	\\
67	&	2018 08 08    02:29:44.573	&	V	&	165	&	8.36(4)	&	7.30(2)	&	26.6	&	60.2	\\
68	&	2018 08 08    02:52:25.876	&	V	&	33	&	11.05(31)	&	9.74(10)	&	13.2	&	10.3	\\
69	&	2018 08 15    18:08:16.637	&	V	&	33	&	11.80(36)	&	9.56(9)	&	11.7	&	-62.4	\\
70	&	2018 08 17    19:00:54.395	&	SC2	&	33	&	8.92(14)	&	8.59(9)	&	26.8	&	-16.9	\\
71	&	2018 08 17    19:00:54.837	&	SC2	&	33	&	9.14(13)	&	8.63(9)	&	16.8	&	-32.0	\\
72	&	2018 09 04    01:33:52.975	&	V	&	33	&	9.87(30)	&	9.18(10)	&	-24.7	&	29.2	\\
73	&	2018 09 05    01:51:37.399	&	V	&	396	&	7.84(7)	&	6.60(2)	&	9.5	&	52.1	\\
74	&	2018 09 05    02:47:54.403	&	V	&	66	&	10.61(37)	&	9.09(9)	&	-15.5	&	15.2	\\
75	&	2018 09 06    02:00:33.053	&	V	&	33	&	10.95(30)	&	10.33(14)	&	-18.6	&	72.5	\\
76	&	2018 09 06    03:10:04.087	&	V	&	66	&	11.18(25)	&	9.86(9)	&	0	&	60.8	\\
77	&	2018 10 06    03:59:22.115	&	SC2	&	33	&		&	11.62(49)	&	-20.6	&	15.2	\\
78	&	2018 10 15    18:17:49.314	&	V	&	66	&	9.61(17)	&	8.84(8)	&	5.5	&	-53.3	\\
79	&	2018 11 12    16:09:13.209	&	SC2	&	33	&		&	9.72(11)	&	-20.3	&	-39.9	\\
80	&	2018 11 12    17:00:02.156	&	SC2	&	33	&		&	8.70(7)	&	14.6	&	-69.5	\\
81	&	2018 11 14    18:27:31.380	&	SC2	&	33	&	9.34(22)	&	9.26(17)	&	-11.2	&	-64.1	\\
82	&	2018 12 12    16:20:16.296	&	SC2	&	33	&		&	10.11(19)	&	-4.9	&	-50.4	\\
83	&	2018 12 12    17:45:58.713	&	SC2	&	33	&		&	8.94(10)	&	4.0	&	-51.4	\\
84	&	2019 02 09    17:29:38.338	&	V	&	33	&	10.32(28)	&	9.91(14)	&	-36	&	-43.1	\\
85	&	2019 02 09    18:17:00.009	&	V	&	66	&	10.39(25)	&	9.82(12)	&	-21.6	&	-93.3	\\
86	&	2019 02 10    19:10:05.599	&	SC2	&	33	&		&	9.02(14)	&	-1.2	&	-32.8	\\
87	&	2019 03 10    17:49:41.708	&	SC2	&	33	&		&	9.70(14)	&	-7.2	&	-19.1	\\
88	&	2019 04 10    19:53:21.200 	&	V	&	43	&	9.45(27)	&	8.55(12)	&	25.9	&	-33.3	\\
89	&	2019 06 08    19:14:58.325 	&	V	&	43	&	10.08(38)	&	8.64(10)	&	28.7	&	-57.4	\\
90	&	2019 06 08    19:26:58.103 	&	V	&	76	&	9.24(18)	&	8.04(7)	&	-7.6	&	-83.3	\\
91	&	2019 06 08    19:34:55.246 	&	SC2	&	43	&		&	9.58(24)	&	39.1	&	-50.2	\\
92	&	2019 06 26    02:24:58.028 	&	SC2	&	43	&		&	9.56(23)	&	25.4	&	18.49	\\
93	&	2019 06 28    01:56:47.678 	&	V	&	109	&	8.88(12)	&	7.59(7)	&	4.3	&	15.5	\\
94	&	2019 06 28    02:18:22.899 	&	V	&	109	&	10.12(20)	&	9.29(10)	&	33.2	&	21.2	\\
95	&	2019 07 06    19:12:55.225 	&	V	&	76	&	10.06(24)	&	9.08(10)	&	25.7	&	-6.8	\\
96	&	2019 07 07    18:32:55.695 	&	V	&	76	&	10.94(36)	&	9.63(11)	&	35.7	&	-77.5	\\
97	&	2019 07 07    18:40:20.874 	&	V	&	307	&	6.65(10)	&	5.49(06)	&	34.4	&	-57.5	\\
98	&	2019 07 07    18:48:48.082 	&	V	&	43	&	11.94(55)	&	9.86(12)	&	33.7	&	-70.9	\\
99	&	2019 07 08    18:31:17.676 	&	SC2	&	43	&		&	9.46(19)	&	-9.9	&	-27.4	\\
100	&	2019 07 08    19:11:44.449 	&	V	&	109	&	9.77(21)	&	8.19(10)	&	8.6	&	-83.0	\\
101	&	2019 07 26    00:18:27.627 	&	V	&	43	&	10.75(34)	&	9.65(15)	&	-11.0	&	53.3	\\
102	&	2019 07 26    00:41:35.185 	&	V	&	109	&	9.64(16)	&	8.21(7)	&	-9.3	&	40.8	\\
103	&	2019 07 27    01:13:12.236 	&	V	&	43	&	10.68(35)	&	9.46(10)	&	60.5	&	18.5	\\
104	&	2019 07 27    01:17:49.791 	&	V	&	142	&	8.95(13)	&	8.02(7)	&	50.4	&	42.3	\\
105	&	2019 07 27    02:12:25.049 	&	V	&	109	&	9.67(17)	&	8.67(7)	&	35.7	&	26.0	\\
106	&	2019 07 27    02:37:22.715 	&	V	&	76	&	10.16(21)	&	9.48(8)	&	49.4	&	0.5	\\
107	&	2019 07 27    02:59:56.458 	&	V	&	142	&	9.48(17)	&	8.25(7)	&	49.8	&	19.9	\\
108	&	2019 07 27    03:01:26.125 	&	V	&	109	&	8.90(14)	&	7.47(5)	&	60.8	&	17.4	\\
109	&	2019 07 28    01:33:40.121 	&	V	&	109	&	10.08(18)	&	8.93(10)	&	31.7	&	24.2	\\
110	&	2019 07 28    01:59:21.345 	&	V	&	76	&	10.80(31)	&	9.62(12)	&	21.2	&	4.2	\\
111	&	2019 07 28   02:00:53.885 	&	V	&	43	&	11.37(35)	&	151(16)	&	11.2	&	60.5	\\
112	&	2019 07 28   02:24:26.088 	&	V	&	76	&	11.04(29)	&	9.93(10)	&	2.9	&	74.2	\\
\hline
\end{tabular}
\tablefoot{V=Validated flash, SC1/2=Suspected flash of Class 1/2 (Section~\ref{sec:VAL1})}
\end{table*}
For all multi-frame flashes (i.e. those listed in Table~\ref{tab:list} with a duration longer than 43~ms), the full light curve magnitude values are given in Table~\ref{tab:multi}, while their light curves are shown in Figs.~\ref{fig:LCs1}-\ref{fig:LCs3}.


\section{Calculation of physical parameters}
\label{sec:Calc}

The origin of the flash is the impact, which is caused by the fall of a projectile on the lunar surface. In general, a projectile of mass $m_{\rm p}$, density $\rho_{\rm p}$, radius $r_{\rm p}$, velocity $V_{\rm p}$, and kinetic energy $E_{\rm kin}$ strikes the lunar surface (with density $\rho_{\rm Moon}$ and gravitational acceleration $g_{\rm Moon}$) and its $E_{\rm kin}$ is converted to: 1) luminous energy $E_{\rm lum}$ (flash generation; material melting and droplets heat) that increases rapidly the local temperature $T$, 2) kinetic energy of the ejected material, and 3) energy for the excavation of a crater of diameter $d_{\rm cr}$. The following sections describe in detail the methods followed to calculate the physical parameters of the projectiles through their observed $E_{\rm lum}$, the sudden local temperature increase and evolution, when possible, and the diameters of the resulted craters.

\subsection{Temperatures of flashes}
\label{sec:Temperature}

\begin{figure}
\centering
\includegraphics[width=\columnwidth]{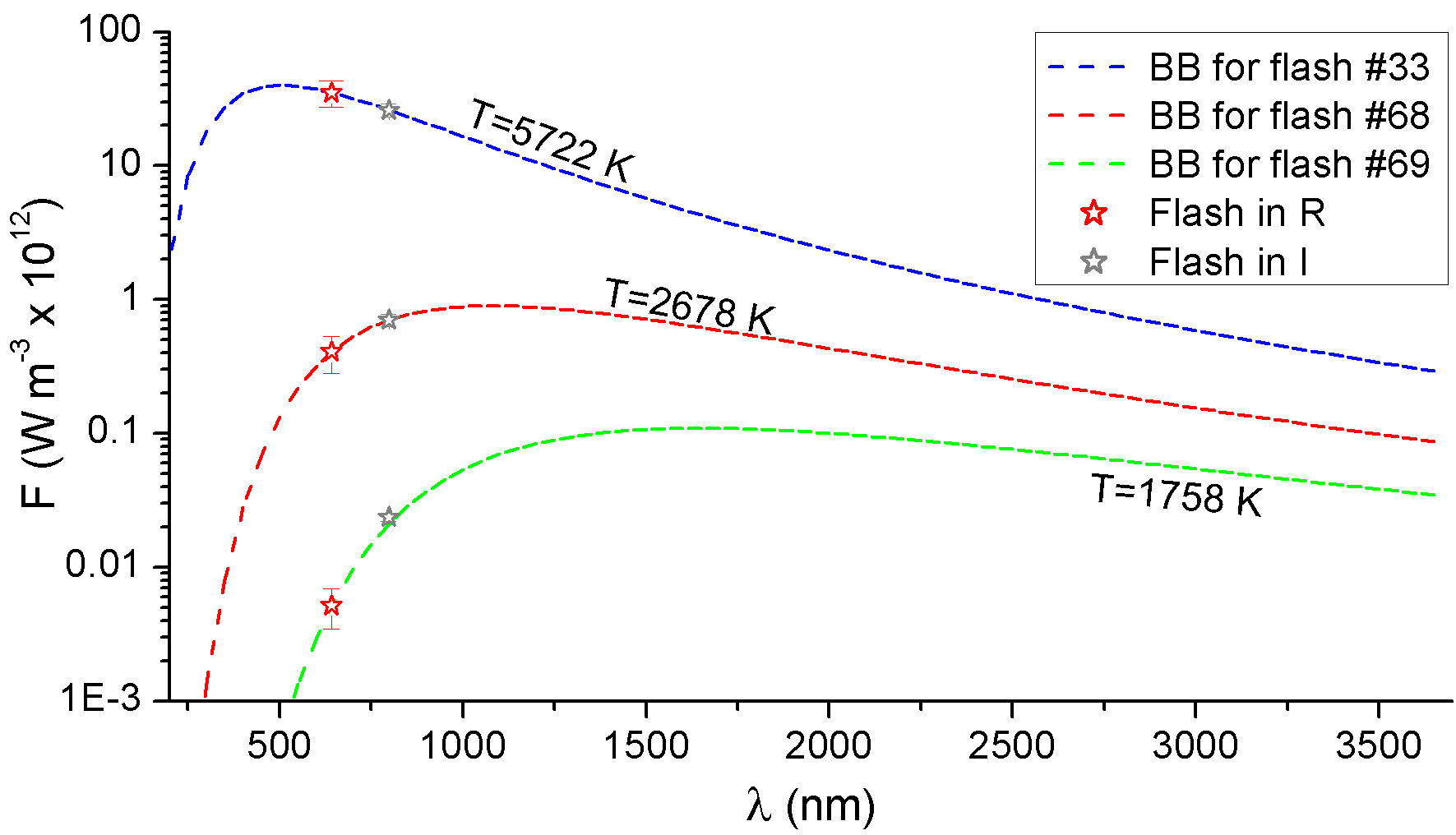}\\
\caption{Examples of BB curve fit (dashed colored lines) for two extreme flash temperatures (5722~K and 1758~K) and for a more typical one (2678~K). Stars (red=$R$ and grey=$I$ band) denote the energy flux per wavelength of the flashes in $R$ and $I$ passbands.}
\label{fig:Planck}
\includegraphics[width=\columnwidth]{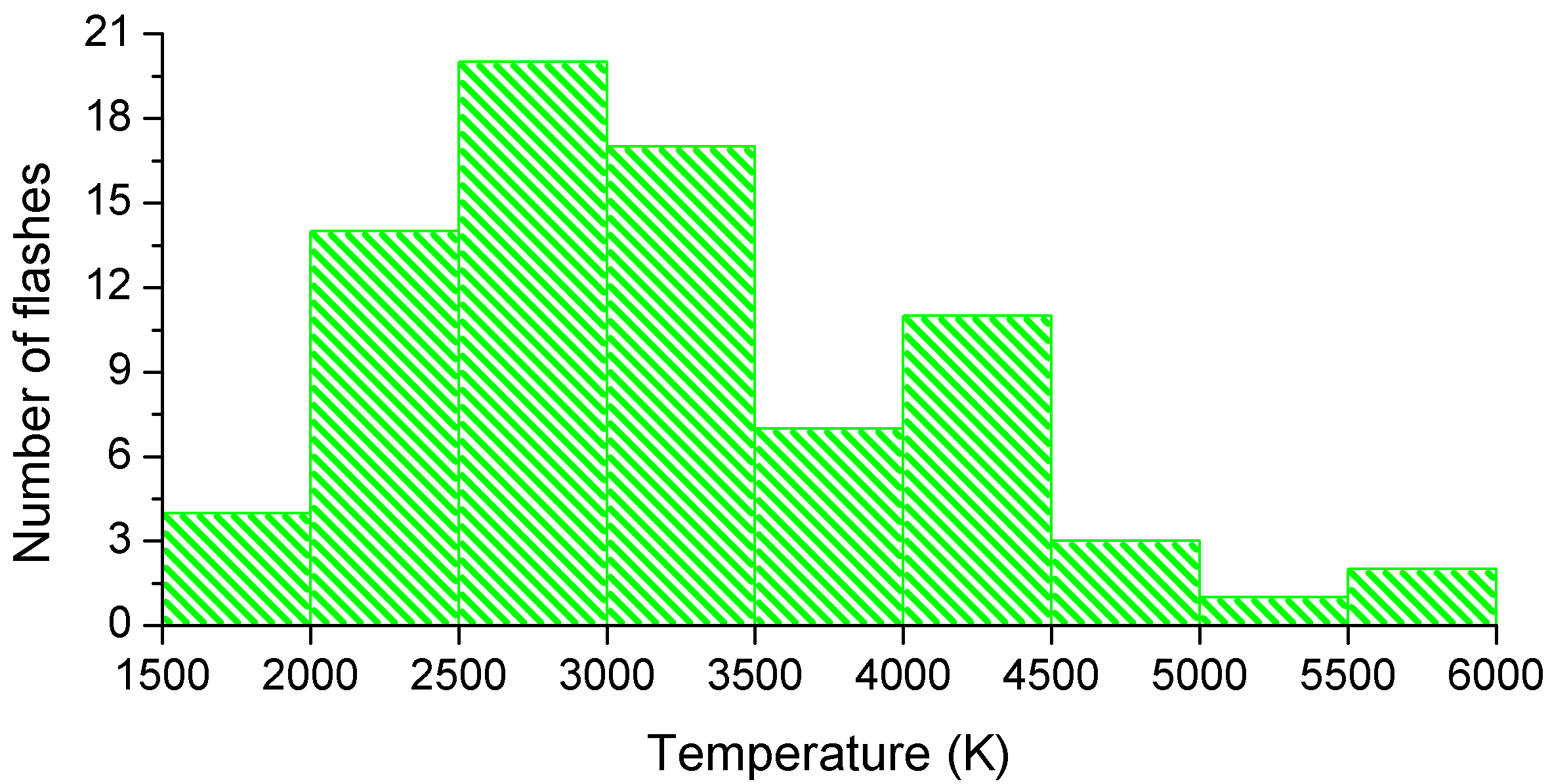}\\
\caption{Peak temperature distribution of validated flashes.}
\label{fig:Thist}
\end{figure}

According to \citet{BOU12}, the temperature of impact flashes is compatible with the formation of liquid silicate droplets, whereas volatile species may increase locally the gas pressure in the cloud of droplets. The following method has been revised in comparison with that presented in Paper I and is also followed by \citet{MAD18}. According to \citet[][Table A2\footnote{The values of $zp(f_{\lambda}$) and $zp(f_{\nu}$) should be interchanged}]{BES98b, BES98a}, the absolute flux $f_{\lambda}$ (i.e. energy per unit area per unit time per wavelength) of an emitting object (e.g. flash) can be calculated using its magnitude ($m_{\lambda}$) and a zeropoint that is based on the wavelength ($\lambda$) of the UBVRIJHKL Cousins-Glass-Johnson photometric system. NELIOTA observations use the $R$ and $I$ bands of this system (see Section~\ref{sec:Obs} and Paper II for details), therefore, the absolute fluxes of the flashes can be calculated using the $zp(f_{\rm R})=0.555$ and $zp(f_{\rm I})=1.271$ zeropoints for the $R$ and $I$ bands, respectively. Hence, solving the magnitude-absolute flux relation of \citet{BES98a} for $f_{\lambda}$, we get:
\begin{equation}
f_{\rm R}=2.18\times 10^{-8}~10^{-(m_{\rm R}/2.5)}~{\rm in~Wm^{-2}\mu m^{-1}},
\label{eq:fluxR}
\end{equation}
\begin{equation}
f_{\rm I}=1.13\times 10^{-8}~10^{-(m_{\rm I}/2.5)}~{\rm in~Wm^{-2}\mu m^{-1}}.
\label{eq:fluxI}
\end{equation}
Assuming that the light emission of the flashes follows the Planck's black body (BB) law, then its spectral density $B_{\rm \lambda}$ is:
\begin{equation}
B(\lambda, T)=\frac{2hc^2}{\lambda^5} \frac{1}{e^\frac{hc}{\lambda k_{\rm B} T}-1} ~{\rm in~Wm^{-3}sr^{-1}},
\label{eq:B}
\end{equation}
where $h=6.63\times10^{-34}$~J~s the Planck's constant, $c=2.998\times 10^8$~m~s$^{-1}$ the speed of light, $\lambda$ the wavelength in m, and $k_{\rm B}=1.38\times 10^{23}$~J~K$^{-1}$ the Boltzmann's constant. However, since the flashes are considered as BB that emit as half spheres, then the energy flux per wavelength to the line of sight of the observer is:
\begin{equation}
F_{\lambda}=\frac{1}{2}\pi~B(\lambda,~T) ~{\rm in~Wm^{-3}~or~in~\times10^{-6}~Wm^{-2} \mu m^{-1}}.
\label{eq:F}
\end{equation}
Using the inverse square law of the energy radiation transfer, we get:
\begin{equation}
f_{\rm \lambda}~f_{\rm a}~\pi~d^2=F_{\lambda}~2\pi R^2,
\label{eq:InvLaw}
\end{equation}
where $f_{\rm a}$ is the anisotropy factor ($f_{\rm a}=2$ for emissions from lunar surface, i.e we observe the emissions of a half sphere of radius $d$), $R$ is the radius of the BB emitting as a half sphere (i.e. area $2\pi R^2$) and $d$ is the Moon-Earth distance at the time of the flash in m. Therefore, combining Eqs.~\ref{eq:fluxR}-\ref{eq:InvLaw} yields:
 \begin{equation}
\frac{f_{\rm R}}{f_{\rm I}}=\bigg(\frac{\lambda_{\rm I}}{\lambda_{\rm R}}\bigg)^5 \frac{e^\frac{hc}{\lambda_{I} k_{\rm B} T}-1}{e^\frac{hc}{\lambda_{R} k_{\rm B} T}-1},
\label{eq:temp}
\end{equation}
where $\lambda_{\rm R}$ and $\lambda_{\rm I}$ are the effective wavelengths of the filters used (see Section~\ref{sec:Obs}) in units of m. The first part of Eq.~\ref{eq:temp} can be easily calculated using the absolute fluxes of Eqs.~\ref{eq:fluxR} and \ref{eq:fluxI} based on the magnitudes $m_{\rm R}$ and $m_{\rm I}$ of the flash. Then, Eq.~\ref{eq:temp} can be solved analytically for the temperature of the flash $T$. Examples of BB fit on flash data are shown in Fig.~\ref{fig:Planck}. The peak temperatures (i.e. the maximum temperature measured during a flash) of all validated flashes detected to date by the NELIOTA campaign are listed in Table~\ref{tab:ResultsReal} and their respective distribution is plotted in Fig.~\ref{fig:Thist}. From the latter figure, it can be plausibly concluded that the majority of the impacts ($\sim65\%$) produce temperatures between 2000~K and 3500~K. Moreover, the temperature values of the multi-frame flashes in both bands are also given in Table~\ref{tab:multi}.

\begin{figure}
\centering
\includegraphics[width=\columnwidth]{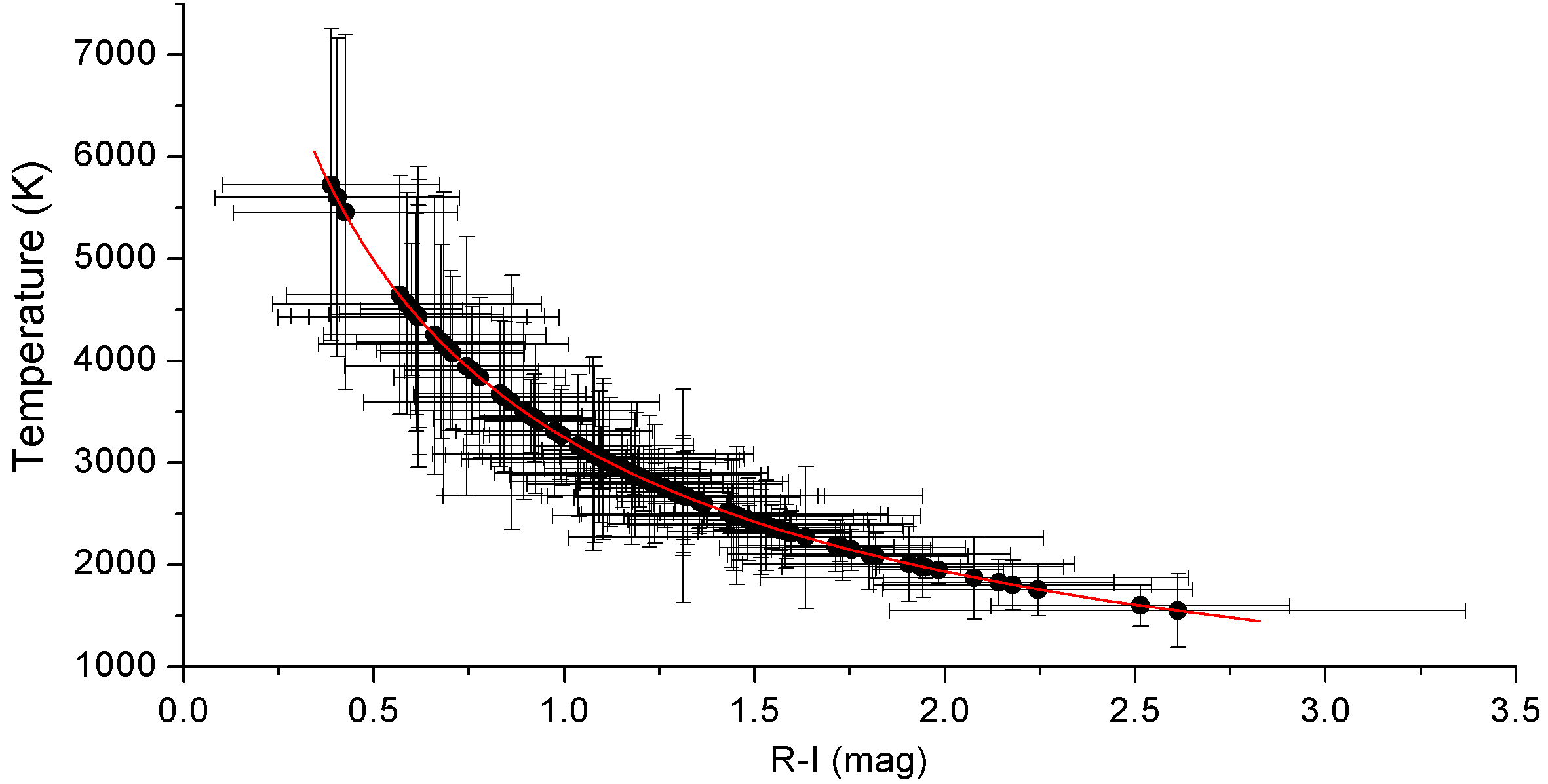}\\
\caption{Correlation between the temperature and $R-I$ indices of the flashes (symbols) fitted by a combination of Planck curves (line).}
\label{fig:R-I-T}
\end{figure}

In Fig.~\ref{fig:R-I-T}, we present the correlation between the temperature of the flashes $T$ and the color indices $R-I$. For this, all the values from Tables~\ref{tab:list}, \ref{tab:ResultsReal}, and \ref{tab:multi} were used. Considering the BB law, the data points were fit by the Planck curve as derived from the combination of Eqs.~\ref{eq:fluxR}, \ref{eq:fluxI}, and \ref{eq:temp}.

\begin{table}[b]
\centering
\caption{Thermal evolution rates for multi-frame flashes in both bands.}
\label{tab:tempevol}
\scalebox{0.82}{
\begin{tabular}{ccc ccc ccc}
\hline\hline																											
ID	&	t	&	rate	&	ID	&	t	&	rate	&	ID	&	t	&	rate	\\
	&	(ms)	&	(K~f$^{-1}$)	&		&	(ms)	&	(K~f$^{-1}$)	&		&	(ms)	&	(K~f$^{-1}$)	\\

\hline																	
2	&	33-66	&	-2357	&	62	&	33-66	&	-1553	&	\multirow{ 3}{*}{97}	&	33-66	&	-939	\\
19	&	33-66	&	-27	&	67	&	33-66	&	-1039	&		&	66-99	&	-157	\\
\multirow{ 2}{*}{27}	&	33-66	&	-267	&	\multirow{ 3}{*}{73}	&	33-66	&	+197	&		&	99-132	&	-280	\\
	&	66-99	&	-372	&		&	66-99	&	-82	&	102	&	33-66	&	-201	\\
28	&	33-66	&	-781	&		&	99-132	&	+130	&	104	&	33-66	&	-931	\\
30	&	33-66	&	-260	&	94	&	33-66	&	-1293	&	107	&	33-66	&	+89	\\
59	&	33-66	&	+565	&		&		&		&	108	&	33-66	&	-23	\\
\hline																									
\end{tabular}}
\end{table}

\subsubsection{Thermal evolution}
\label{sec:Tevol}

\begin{figure*}
\centering
\begin{tabular}{cccc}
\includegraphics[width=4.2cm]{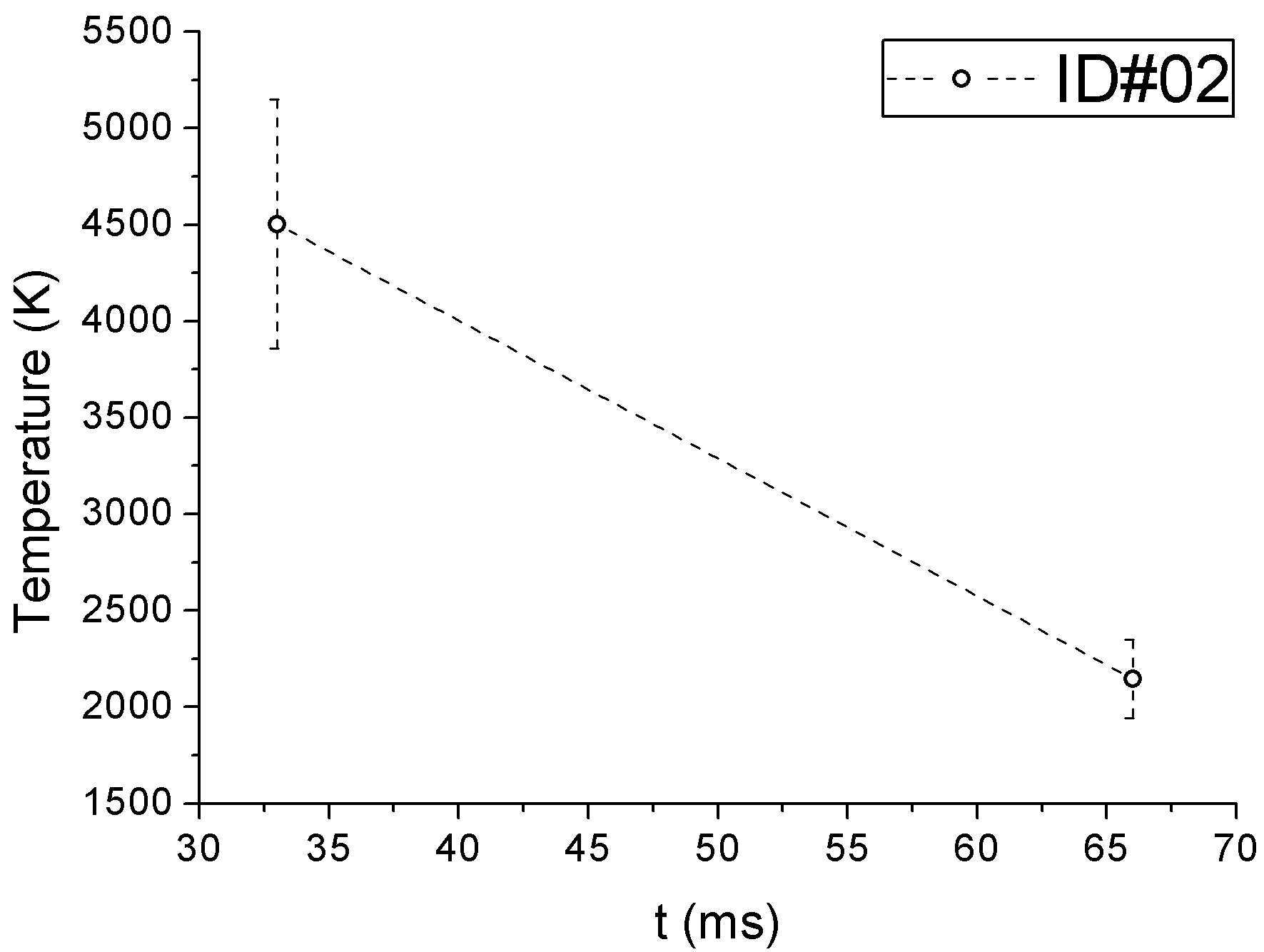}&\includegraphics[width=4.2cm]{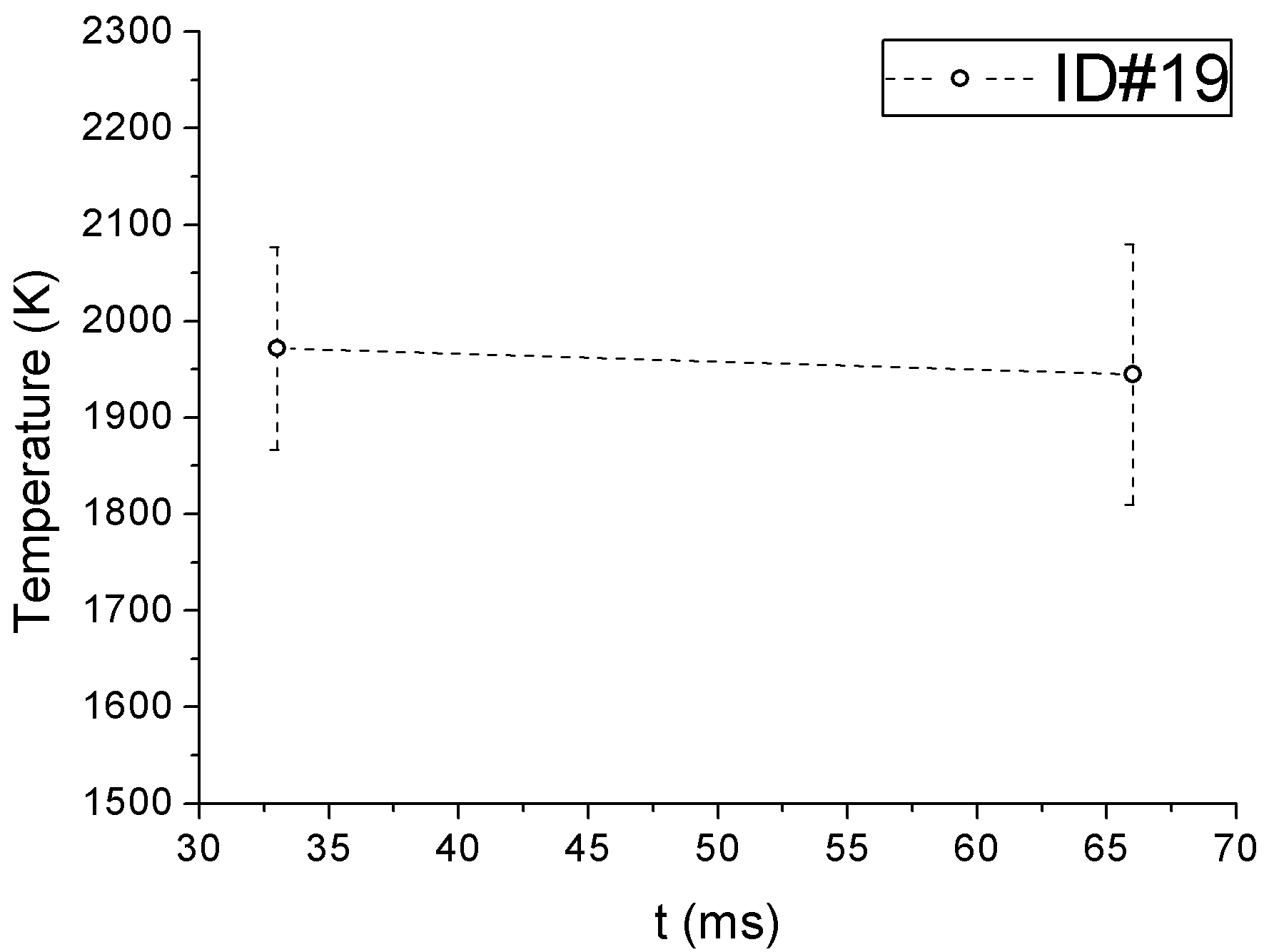}&\includegraphics[width=4.2cm]{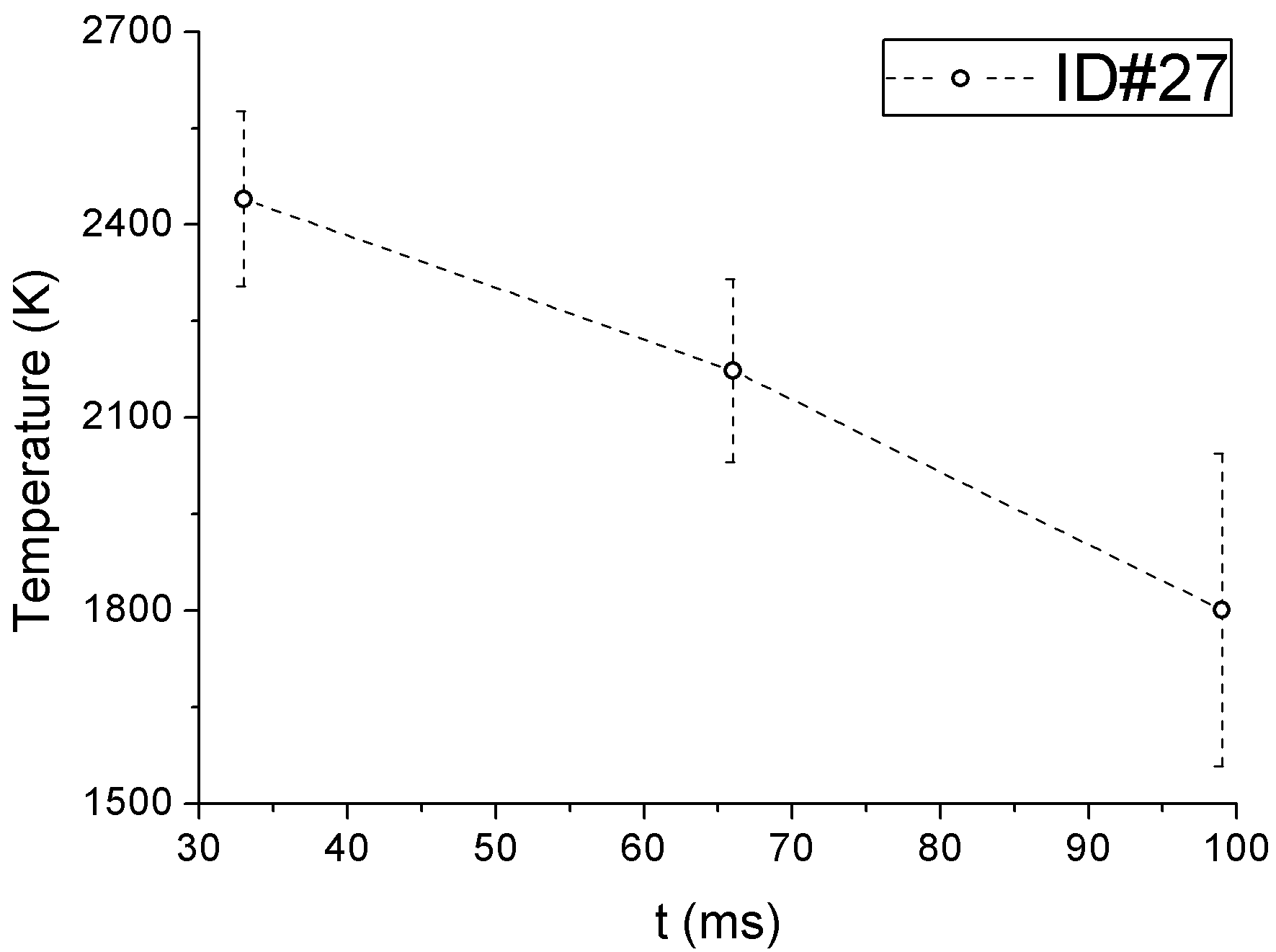}&\includegraphics[width=4.2cm]{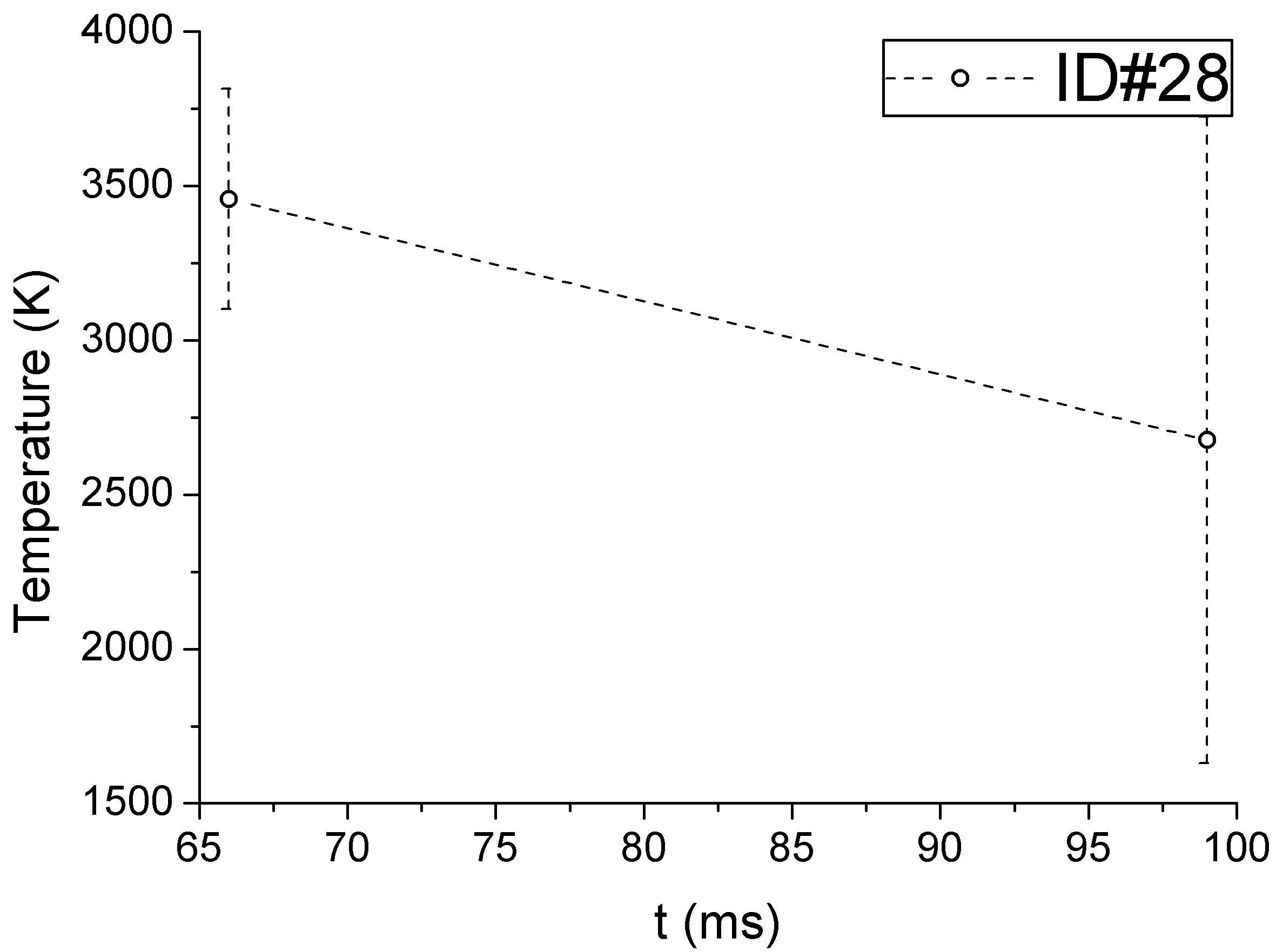}\\
\includegraphics[width=4.2cm]{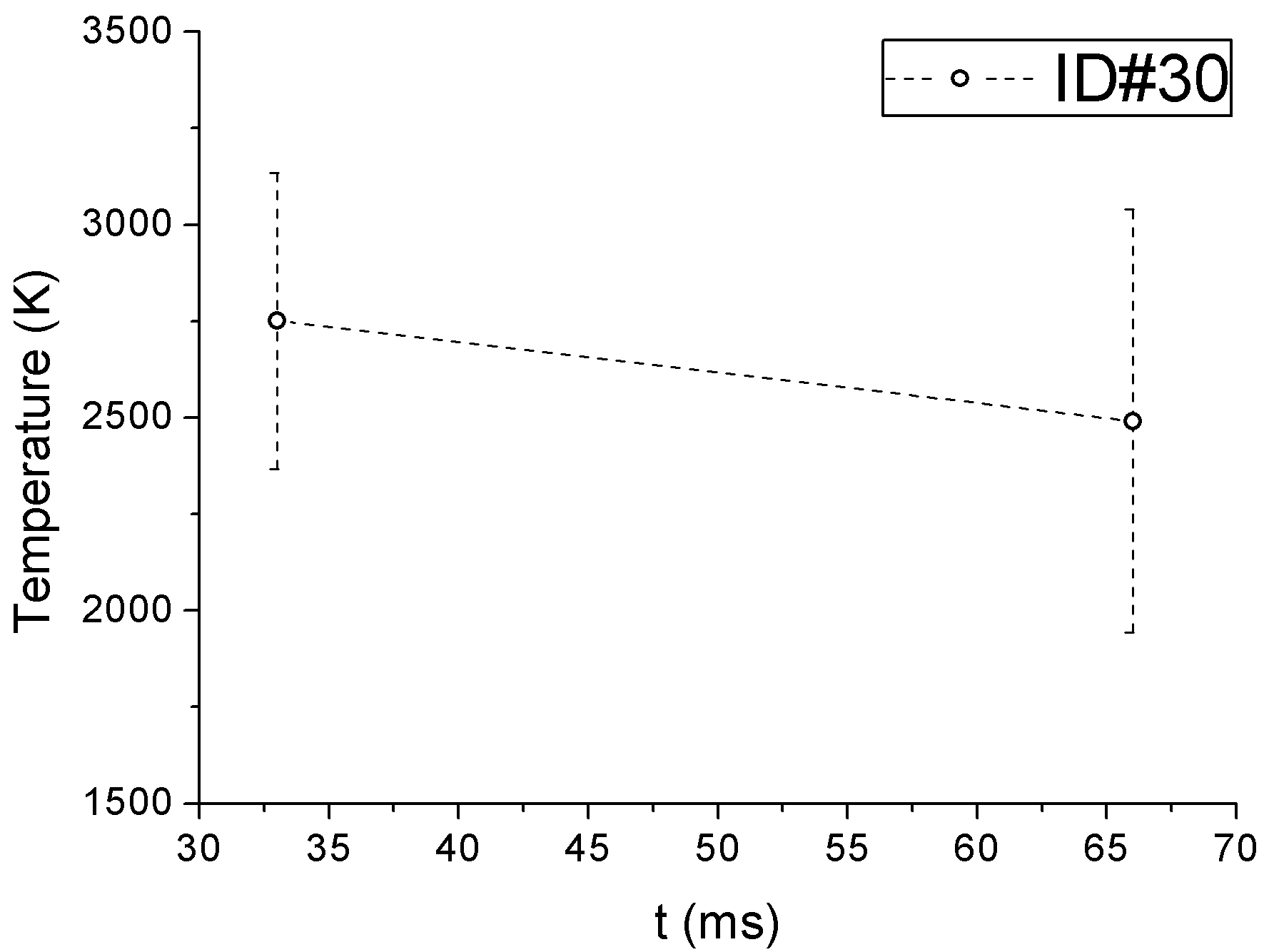}&\includegraphics[width=4.2cm]{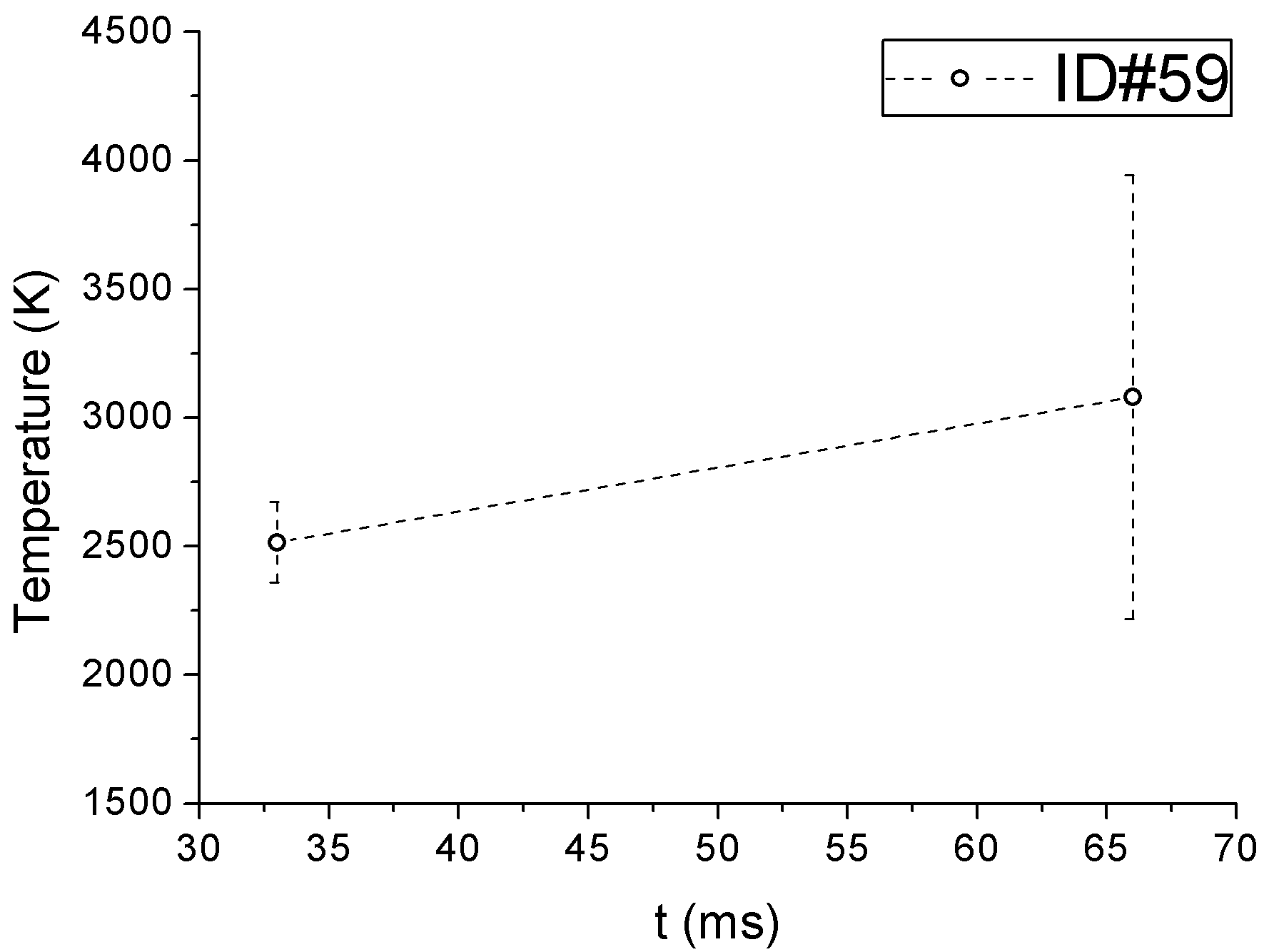}&\includegraphics[width=4.2cm]{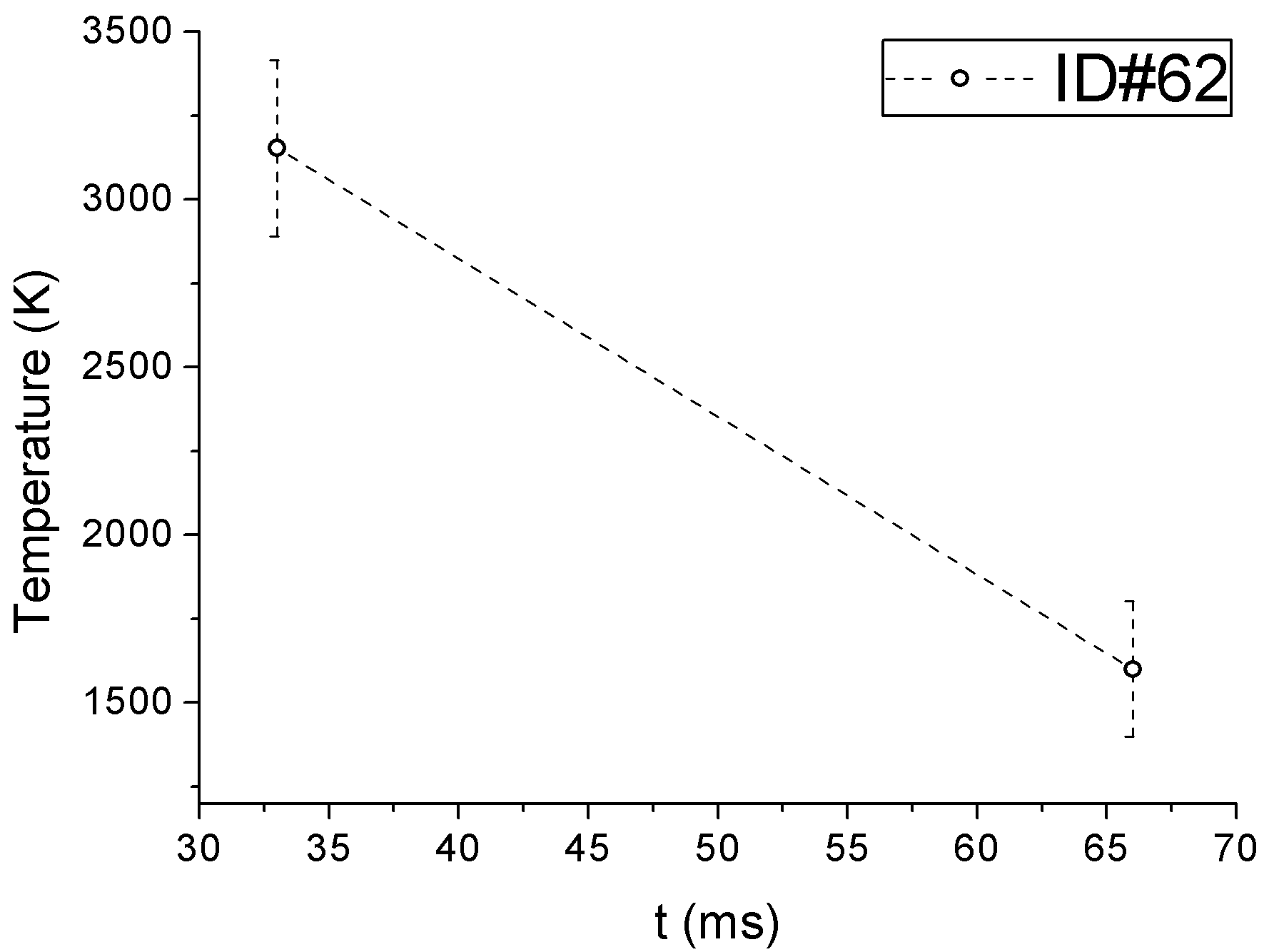}&\includegraphics[width=4.2cm]{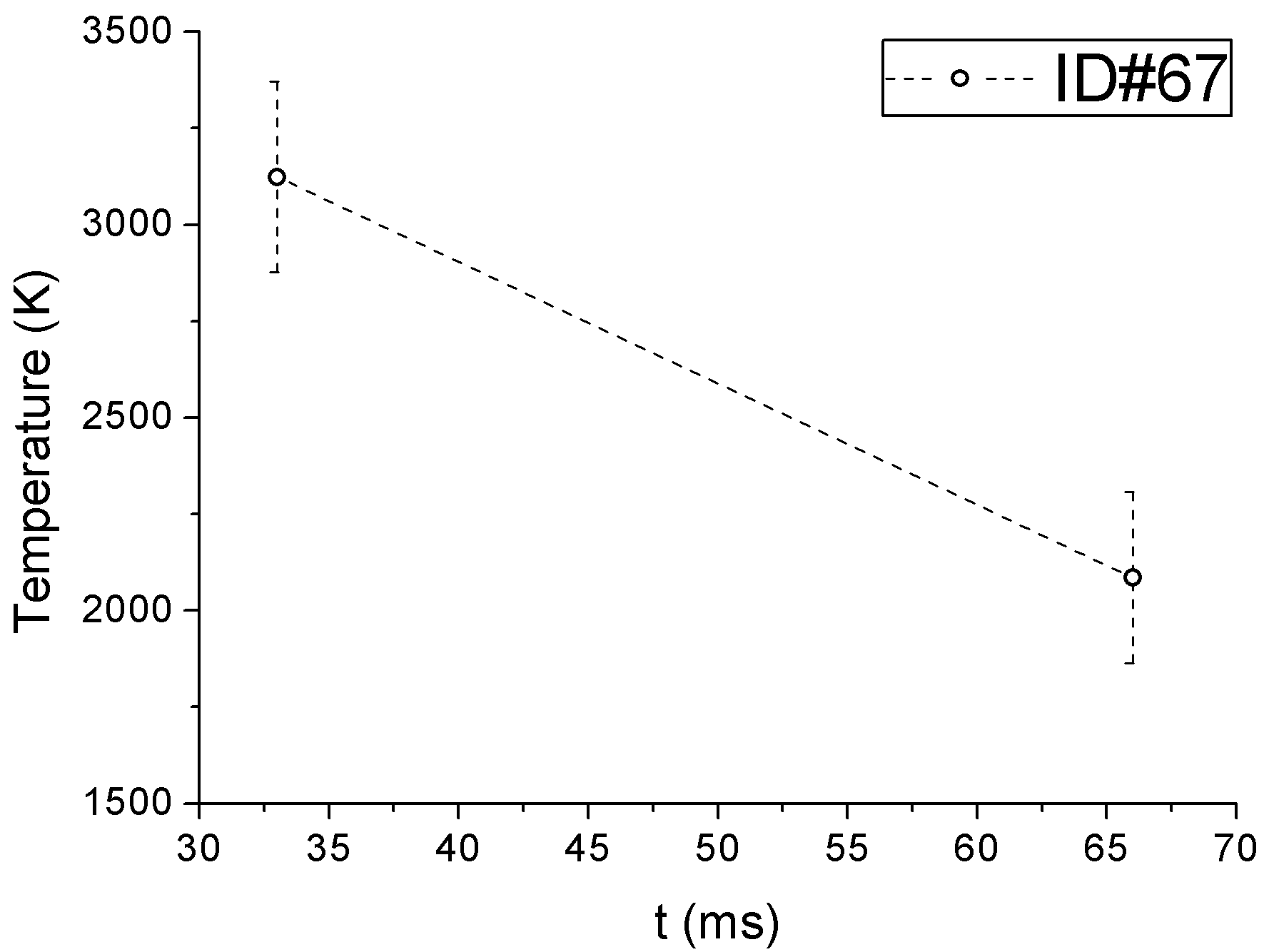}\\
\includegraphics[width=4.2cm]{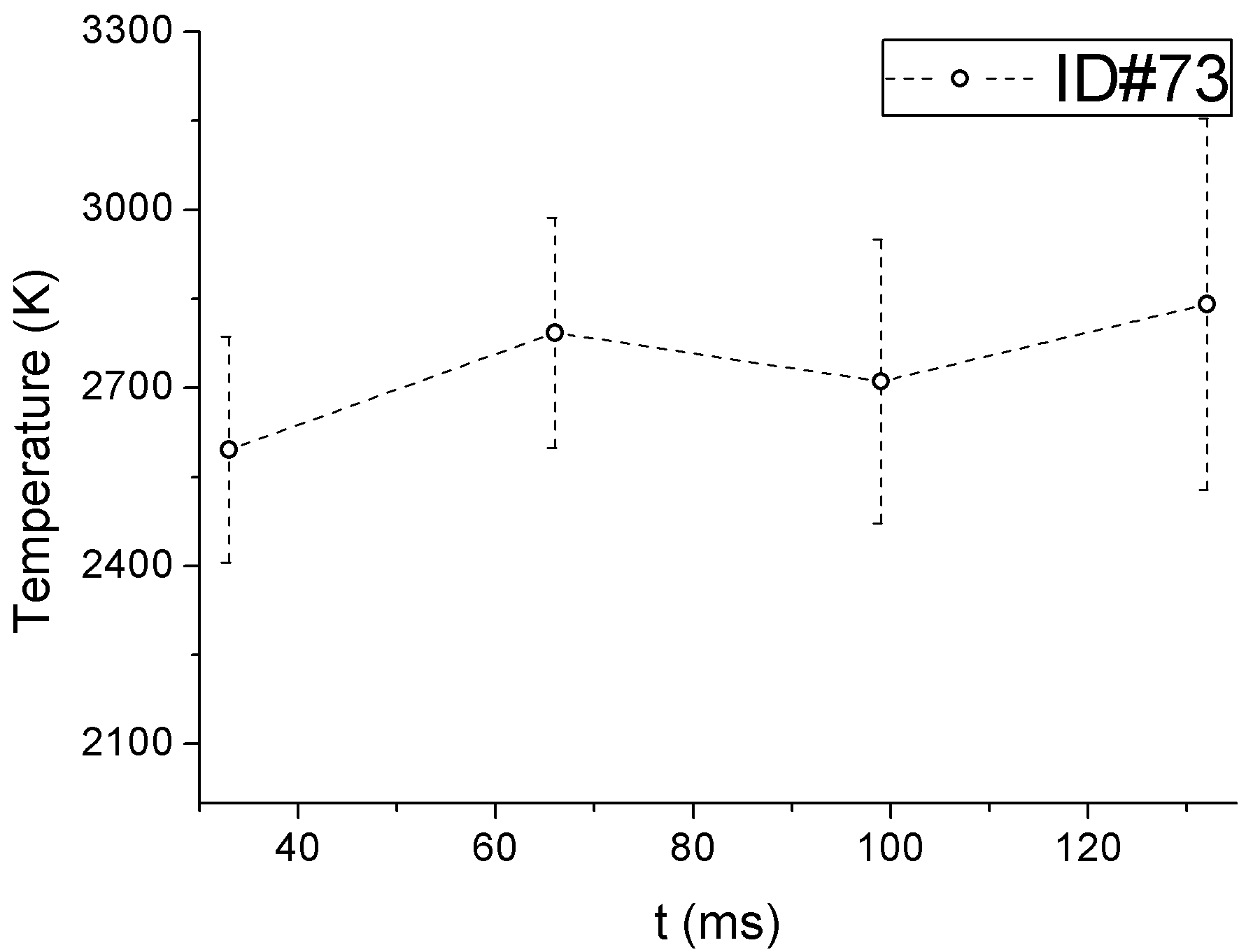}&\includegraphics[width=4.2cm]{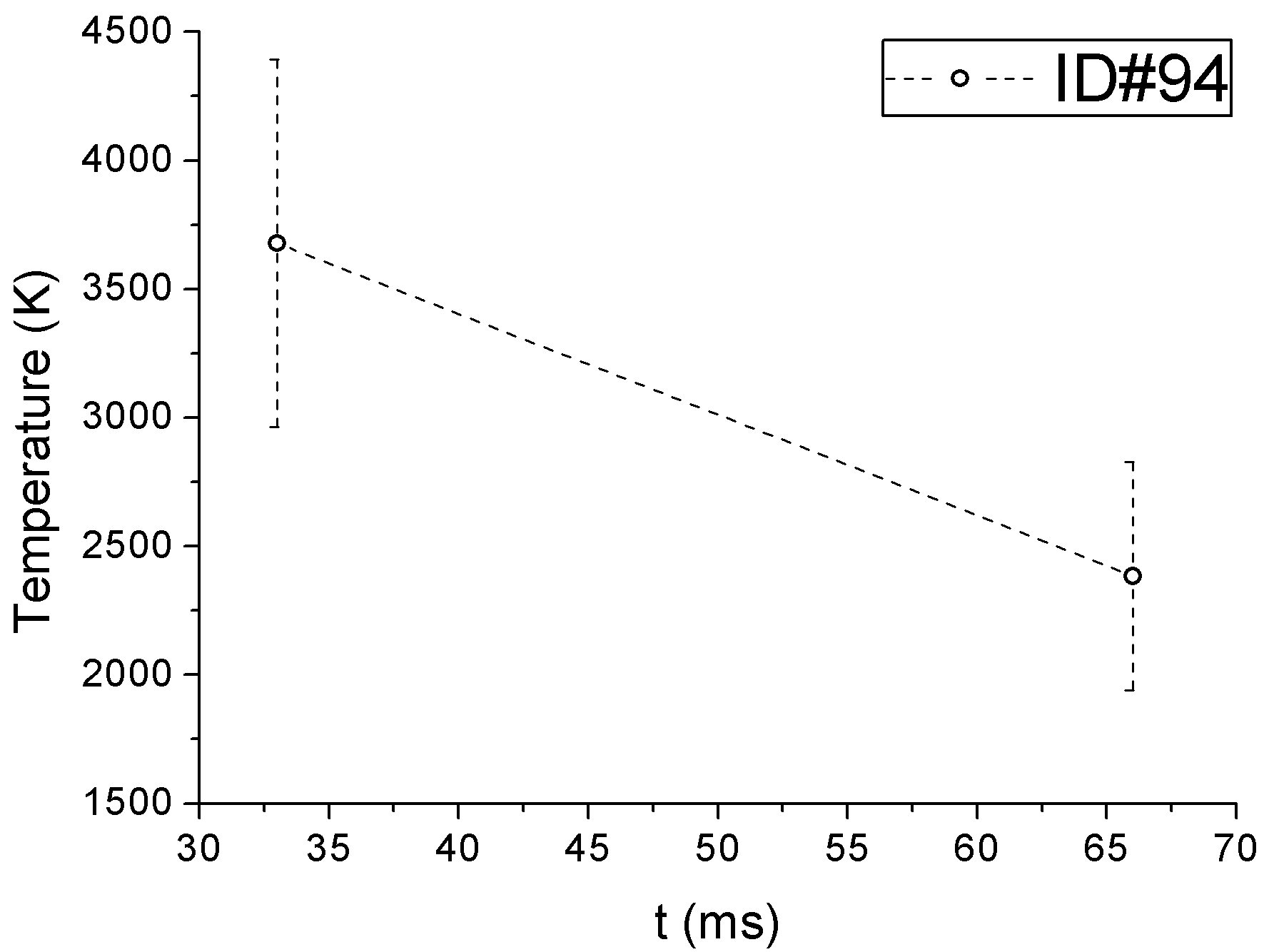}&\includegraphics[width=4.2cm]{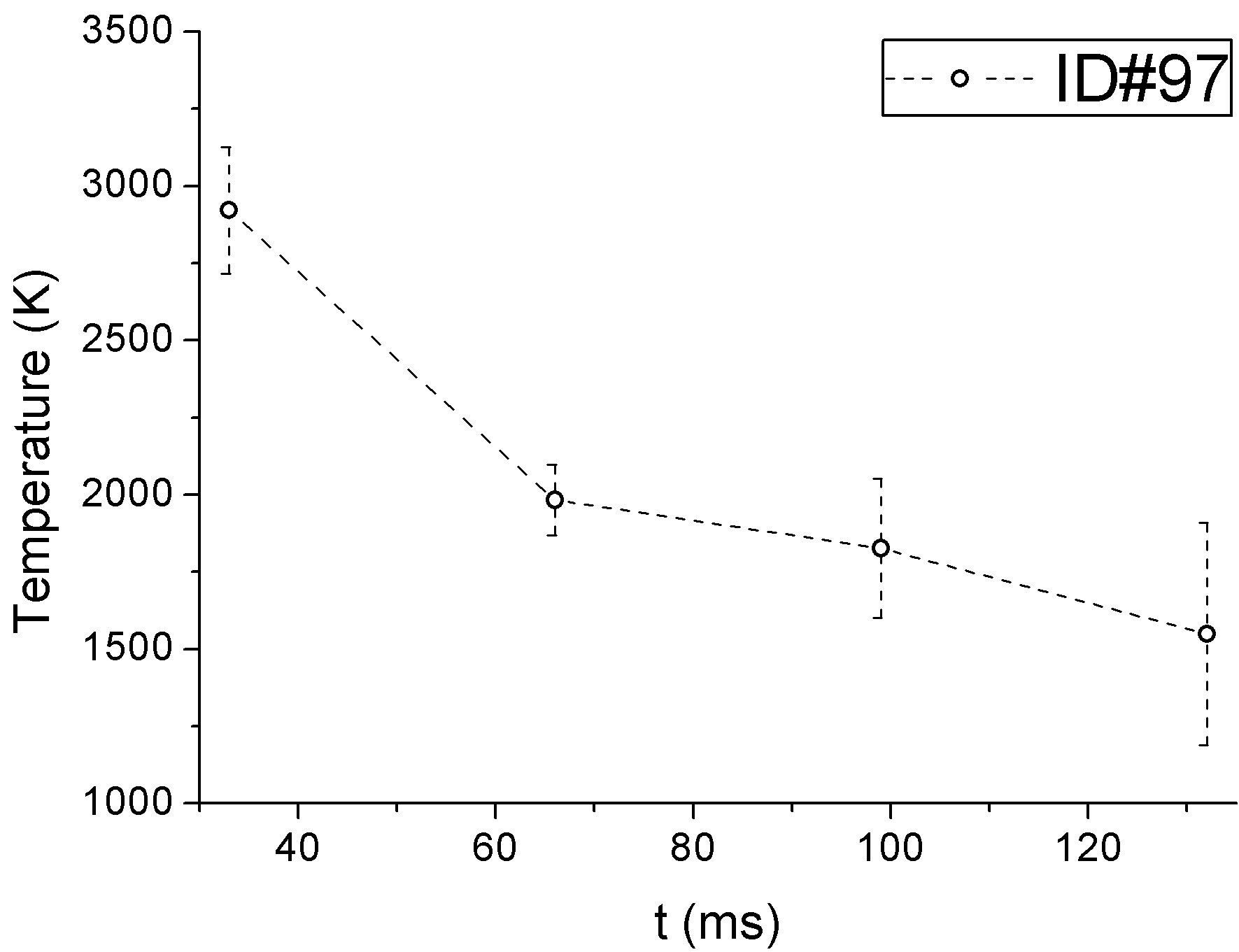}&\includegraphics[width=4.2cm]{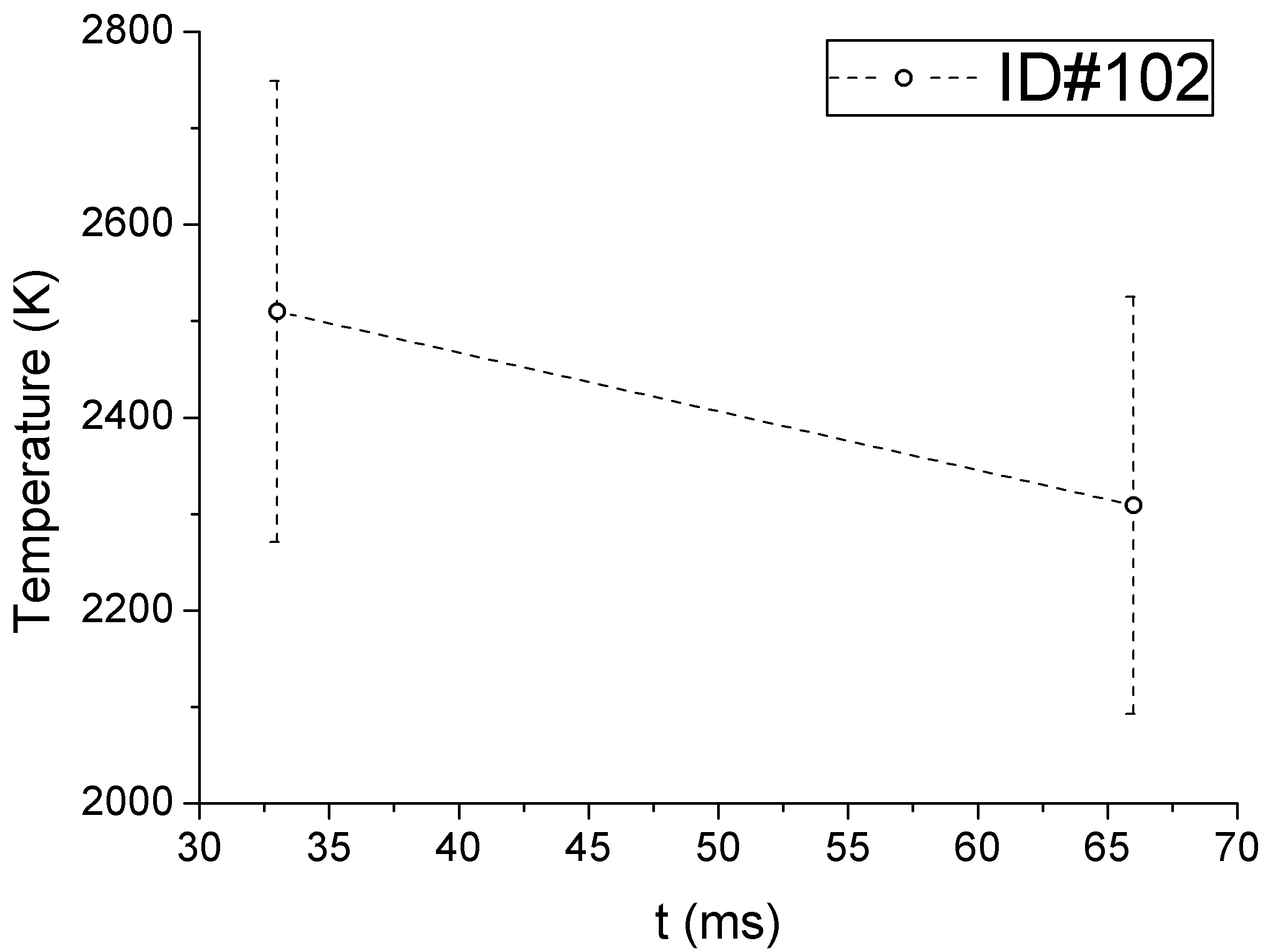}\\
\includegraphics[width=4.2cm]{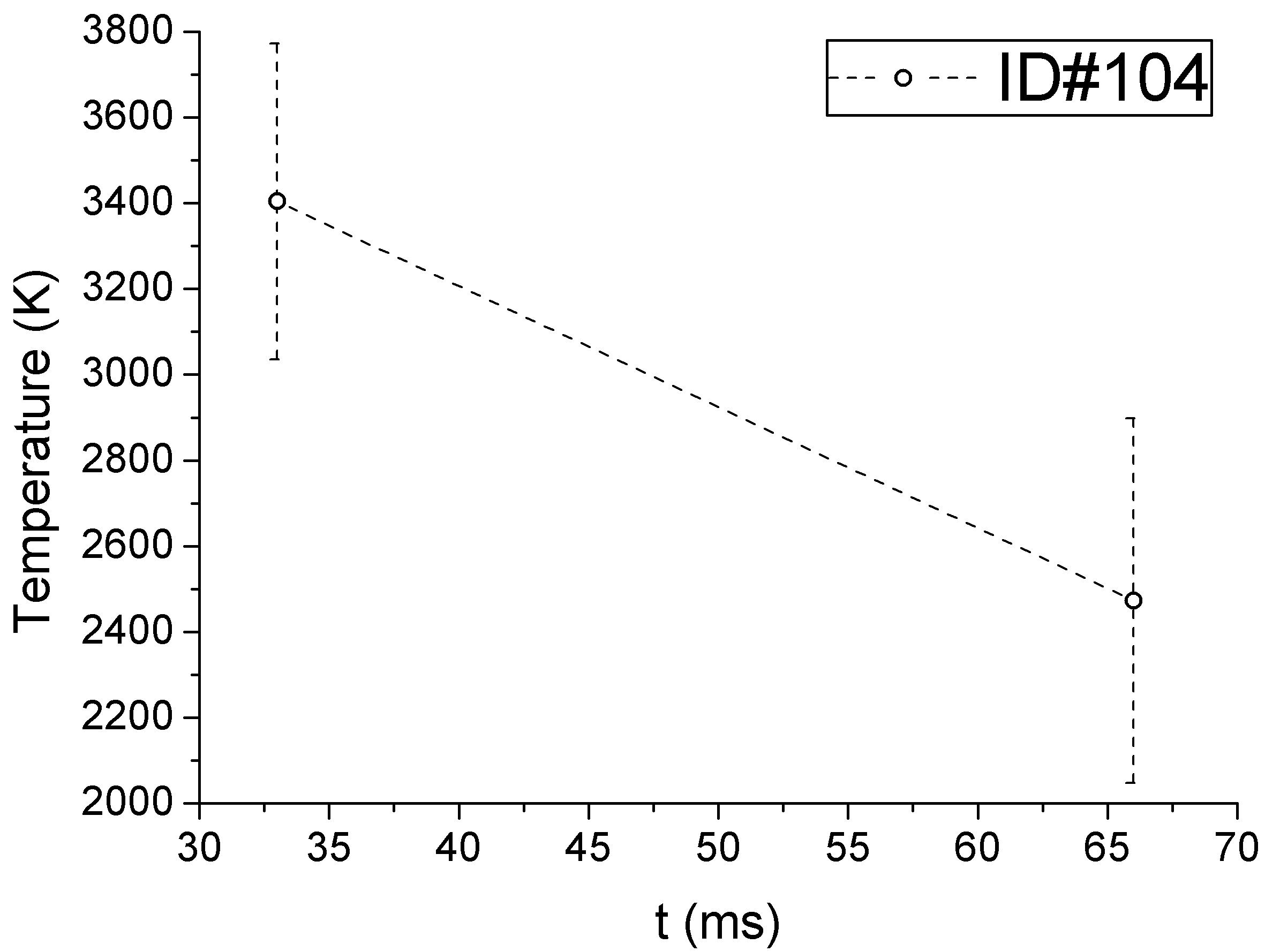}&\includegraphics[width=4.2cm]{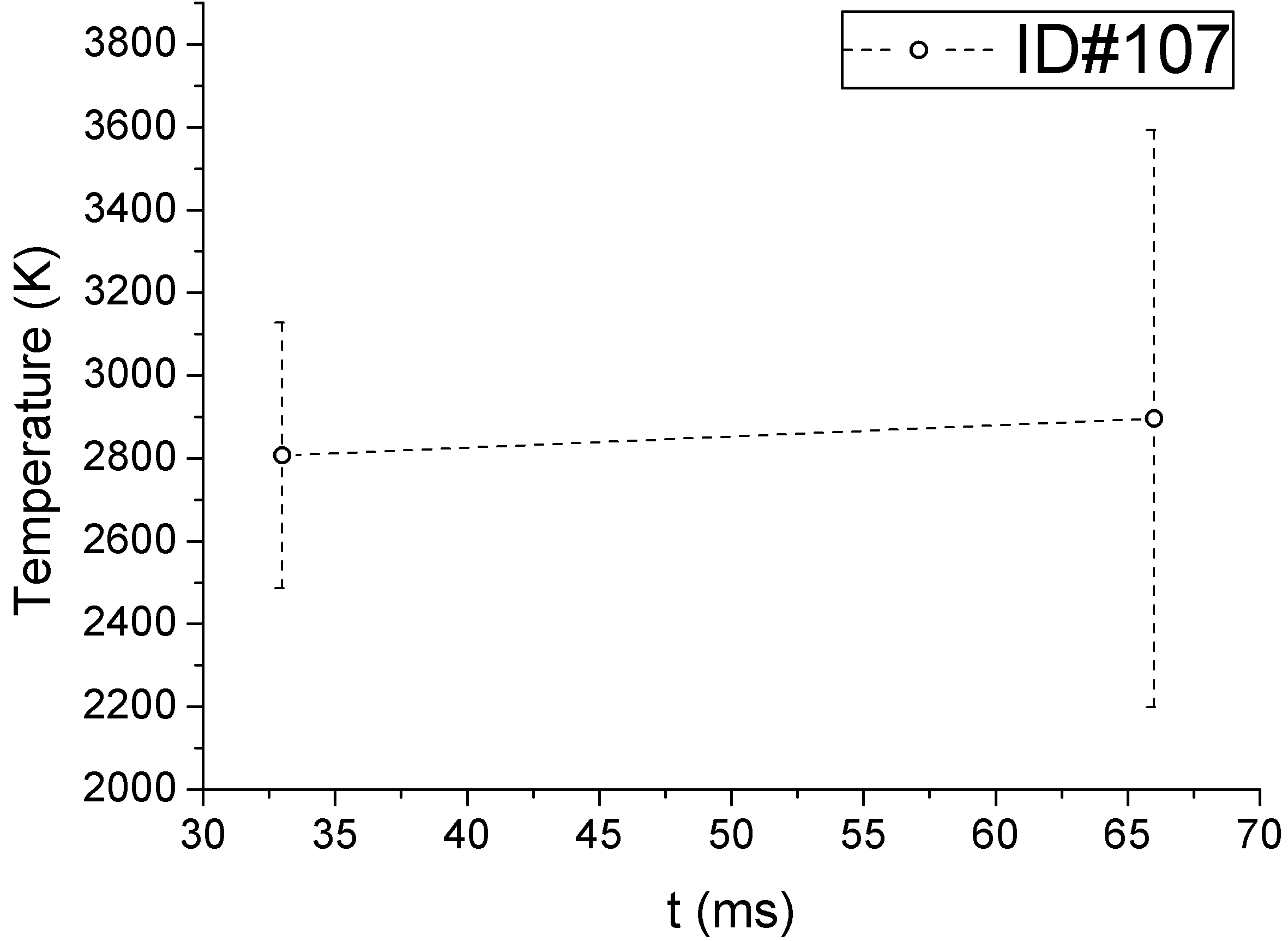}&\includegraphics[width=4.2cm]{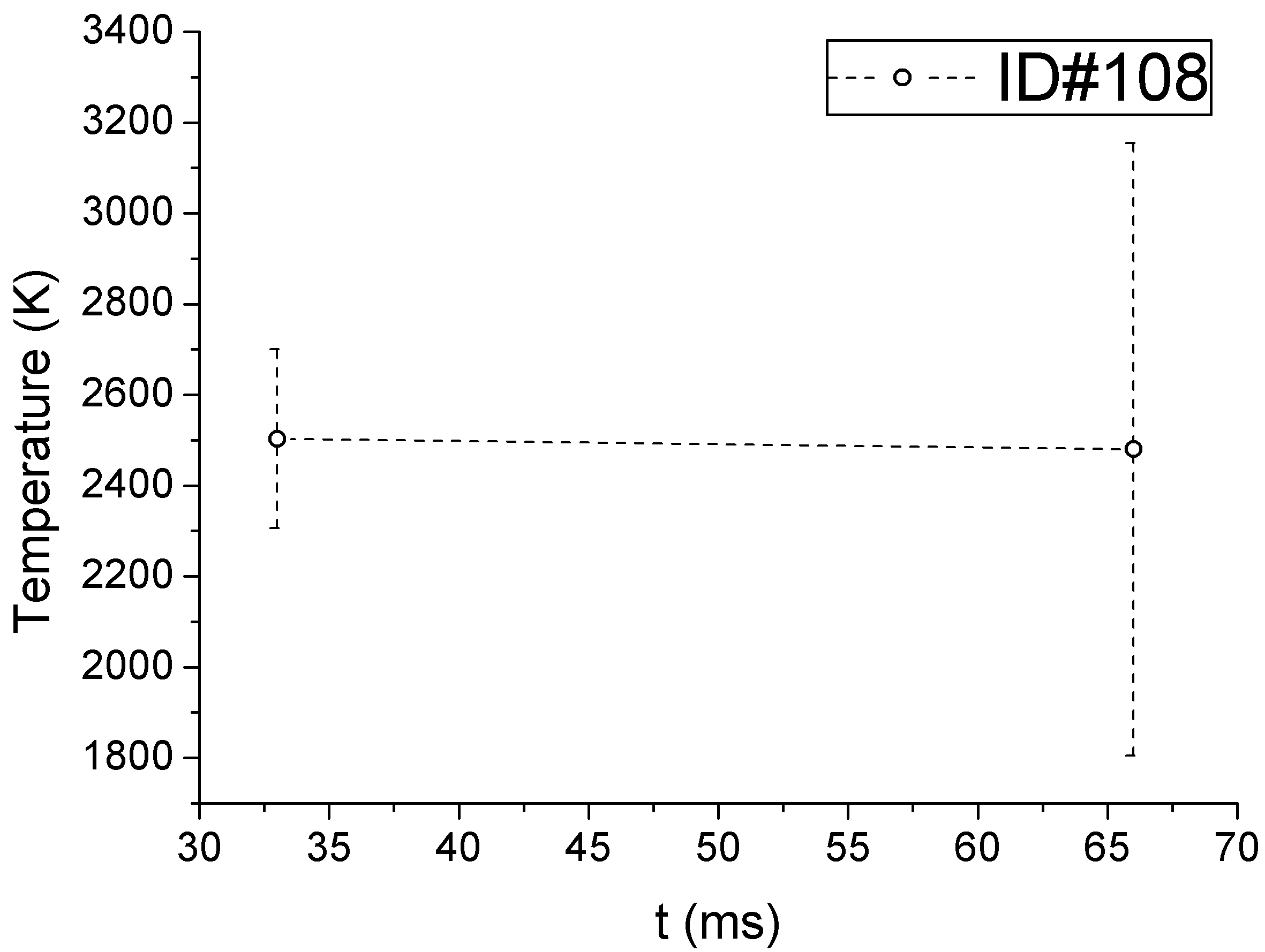}& \\
\end{tabular}
\caption{Thermal evolution for validated flashes detected in more than one set of frames (multi-frame flashes in both bands).}
\label{fig:Tevol}
\end{figure*}

For the flashes detected in more than one set of frames in $R$ and $I$ bands (i.e. multi-frame flashes in both bands), it is feasible to calculate their thermal evolution. Using the temperature values of Table~\ref{tab:multi}, the thermal evolution graphs for 15 validated flashes are plotted in Fig.~\ref{fig:Tevol}. All flashes except $\#59$, $\#73$, and $\#107$ are cooling in time, which is what is expected from an impact. However, it should be noted that the cooling rate is different for each flash. The thermal evolution rates (in units of Kelvin degrees per frame, i.e. K~f$^{-1}$) for all the multi-frame flashes in both bands are given in Table~\ref{tab:tempevol}. Intentionally, the term `frame' is used rather than `time', because, given the integration time of the images (23~ms exposures and 10~ms read-out time), we cannot be certain when exactly the peak temperatures occurred. Different cooling rates probably are connected to the ejecta of the heated material. Flashes $\#19$ and $\#108$ show almost no cooling between the successive frames, while flashes $\#59$ and $\#107$ show a slight temperature increase. Finally, flash $\#73$ shows again a temperature increase between the first and the second frames, and then it remains almost constant. For the cases of flashes $\#59$ and $\#107$ it seems that the peak temperature occurred during the read-out time of the cameras or the cooling rate after the second frame was that rapid so the flash did not emit to the $R$ band any more. Flash $\#73$ can be considered as a special event. It is the longest in duration flash ever observed during the NELIOTA campaign and was recorded before the peak magnitude. Again, very probably, after the fourth set of frames, the cooling rate was so rapid, that it could not be observed in the $R$ band. Obviously, all the above regarding the unexpected behaviour of some flashes (i.e. not cooling in time) can be considered as rough estimations, given the large errors bars in temperature.

\subsection{Energy and efficiency of the flashes}
\label{sec:LE}

In this section we describe in detail the definitions of luminous energy and luminous efficiency of the flashes. For the latter, two possible ways for its calculation following different approximations are shown.

\subsubsection{Luminous efficiency of the flashes}
\label{sec:eta}
Luminous energy ($E_{\rm lum}$) of a flash is defined as the energy emitted through irradiation during an impact and is directly proportional to the kinetic energy ($E_{\rm kin}$) of the projectile:
\begin{equation}
E_{\rm kin}=\frac{E_{\rm lum}}{\eta},
\label{eq:eta}
\end{equation}
where $\eta$ is a factor called luminous efficiency. It should be noticed that in Eq.~\ref{eq:eta} the $E_{\rm lum}$ refers to the total radiation energy emitted in all wavelengths. It seems that in the literature there is a confusion regarding the definition of $E_{\rm lum}$ and $\eta$ \citep[cf.][]{MAD19c}. The factor $\eta$ has been found to vary between $5\times10^{-4}$ and $5\times10^{-3}$ \citep[c.f.][]{ORT06, BOU12, MAD15a, MAD18}, while \citet{MOS11} and \citet{SUG14} found an empirical relation between $\eta$ and the velocity of the projectiles $V_{\rm p}$. However, $\eta=1.5\times10^{-3}$ can be considered as a typical value \citep[][and references therein]{BOU12, SWI11}. All the above are based on observations in the visual and the near-infrared bands (i.e. unfiltered observations). Observing without filters simply means that the derived magnitude depends on the camera's wavelength sensitivity range. In addition, all these studies used the method of \citet{BES98b, BES98a} for the calculation of the absolute flux (i.e. spectral energy density) from the observed magnitudes of the flashes and the radiation transfer inverse square law (Eqs. \ref{eq:InvLaw} and \ref{eq:L2}) to calculate the so-called $E_{\rm lum}$. In fact, they calculated the energy emitted ($E_{\rm emitted}$) through the observed passband and given that the $\eta$ range is quite large and has been calculated for the 400-900~nm band (i.e. this range covers their observing passband), then they assumed that the $E_{\rm emitted}$ was actually the $E_{\rm lum}$. Before NELIOTA, all the observations for lunar impact flashes, except those made by \citet{MAD18}, were performed without filters. Moreover, the same bands were used for the observations of the meteor showers on Earth on which the $\eta$ range is based on. As can be seen from the temperature values of the flashes ({Table~\ref{tab:ResultsReal}, see also Fig.~\ref{fig:Thist}) and their peak wavelengths (based on the Wien's law, see Fig.~\ref{fig:Wien}), the majority of their energy is emitted in the near-infrared band. Hence, the approximation that the $E_{\rm emitted}$ is in fact the $E_{\rm lum}$ can be considered reasonable. Therefore, it can plausibly concluded that different $\eta$ should be used for wavelengths other than the optical, hence, $\eta$ should be wavelength depended i.e. $\eta_{\lambda}$ \citep[cf.][]{MAD18, MAD19c}. Solving Eq.~\ref{eq:eta} for $E_{\rm lum}$, for which we assume that is the sum of the emitted energies in the 400-900~nm wavelength range, i.e. $E_{\rm lum}=E_{\rm R} + E_{\rm I}$ for NELIOTA case we get:
\begin{equation}
E_{\rm lum}=\eta~E_{\rm kin} \Rightarrow E_{\rm R} + E_{\rm I}=\eta~E_{\rm kin}=\eta_{\rm R}E_{\rm kin} + \eta_{\rm I}E_{\rm kin},
\label{eq:ELEREI}
\end{equation}
where $\eta_{\rm R}$ and $\eta_{\rm I}$ are the luminous efficiencies for the specific observing bands and they are related to $\eta$ (for 400-900~nm) with the simple equation:
\begin{equation}
\eta=\eta_{\rm R}+\eta_{\rm I}
\label{eq:ETA3}
\end{equation}
Therefore, combining the above equations, we get \citep[see also][]{MAD18}:
\begin{equation}
E_{\rm kin}=\frac{E_{\rm R}}{\eta_{\rm R}}= \frac{E_{\rm I}}{\eta_{\rm I}} \Rightarrow \frac{\eta_{\rm R}}{\eta_{\rm I}} = \frac{E_{\rm R}}{E_{\rm I}}
\label{eq:eta2}
\end{equation}

\begin{figure}
\centering
\includegraphics[width=\columnwidth]{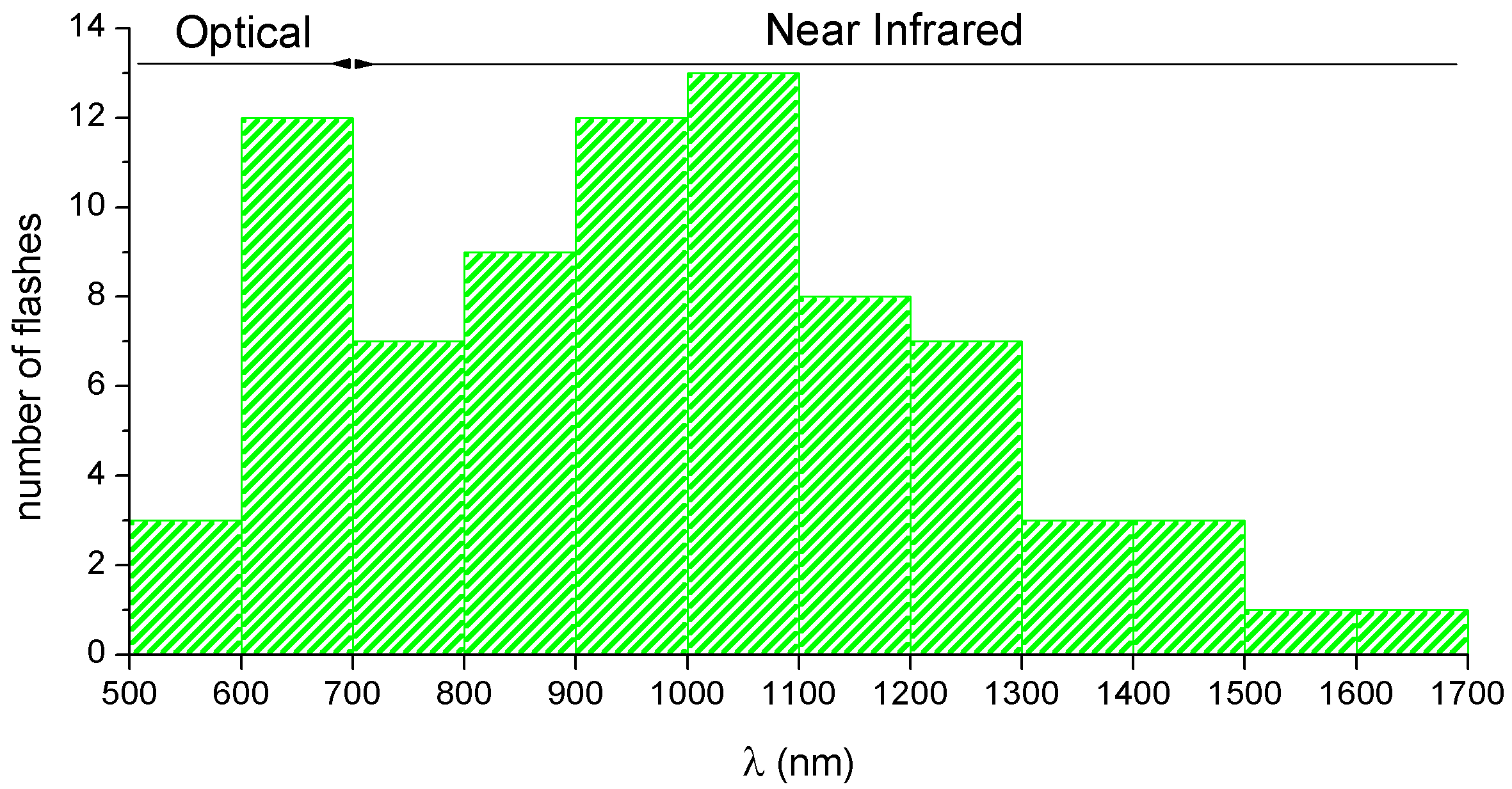}\\
\caption{Histogram of the peak wavelengths of the lunar impacts flashes based on their calculated temperatures and the Wien's law.}
\label{fig:Wien}
\end{figure}

\subsubsection{Luminous energy calculation based on the emitted energies from each band}
\label{sec:LE1}

The first method for the calculation of $E_{\rm lum}$ is the standard method used in various works \citep[e.g.][]{SUG14, MAD18}. Using the definition of the stellar luminosity $L_{\lambda}$ adjusted for flashes for a specific wavelength range $\Delta\lambda$ around a central wavelength $\lambda$:
\begin{equation}
L_{\lambda}=F_{\lambda}\Delta\lambda~2\pi R^2~{\rm in~W}
\label{eq:L}
\end{equation}
and the observed energy flux within the same specific wavelength range $\Delta\lambda$ around the same central wavelength $\lambda$:
\begin{equation}
\int f_{\lambda}d\lambda~{\rm in~W~m^{-2}},
\label{eq:f}
\end{equation}
and combining Eq.~\ref{eq:f} with Eq.~\ref{eq:InvLaw}, we get:
\begin{equation}
L_{\lambda}=f_{\lambda}~\Delta\lambda~f_{\rm a}~\pi d^2~{\rm in~W},
\label{eq:L2}
\end{equation}
hence,
\begin{equation}
E_{\lambda}=L_{\lambda}~t~{\rm in~J},
\label{eq:E}
\end{equation}
where $t$ is the duration of the flash \citep[for Eq.~\ref{eq:L2} see also e.g.][]{SUG14, REM15, MAD15b, MAD15a}. Therefore, using the magnitudes of a flash and assuming that its observed energy flux can be approximated by the method of \citet{BES98b, BES98a}, and that $\Delta\lambda_R=0.158~\mu$m $\Delta\lambda_I=0.154~\mu$m for $R$ and $I$ bands, respectively, we can calculate the luminosity and the energy for each band. The problem with this method concerns the marginal assumption that the flashes are BBs. By definition, the BB are objects in thermal equilibrium with a temperature $T$  emitting at all wavelengths. This assumption can stand for flashes only for a very short amount of time, i.e. the exposure time of the camera. So, for single frame flashes in both bands, the BB assumption is plausible. Therefore, Eq.~\ref{eq:E}, using the exposure time of the camera, can derive the emitted $E$ of the band. However, for the multi-frame flashes, the total emitted energy of a band has to take into account all the individual energies that are calculated from the magnitude of all frames in which the flash was detected. Hence, the total emitted energy of a band $E_{\lambda}$ becomes:
\begin{equation}
E_{\lambda}=\int_{0}^{t}L_{\lambda}(t)dt~{\rm in~J},
\label{eq:Emf}
\end{equation}
where $L_{\lambda}(t)$ the luminosity of the flash as measured in one frame with an exposure time $t$ during the time range $dt$. It is noticed that the duration of the multi-frame flashes is different for the $R$ and $I$ bands (Table~\ref{tab:multi}). Results are given for each flash in Table~\ref{tab:ResultsReal}. In this table are listed the maximum durations of the flashes $t_{\rm max}$, where $t_{\rm max}=n\times t_{\rm exp}+(n+1)\times t_{\rm ro}$, where $n$ the number of frames in which the flash is detected in $I$ band, $t_{\rm exp}=23$~ms the exposure time of one frame and $t_{\rm ro}=$10~ms the read out time of the camera. For the single frame flashes there is uncertainty about their true duration. They may last less than 23~ms but not more than 43~ms, i.e. by taking into account the sum of $t_{\rm ro}$ before and after the frame. Hence, we cannot be certain if the total emitted energy was actually captured in the frame. For the multi-frame flashes, again, we cannot be sure about their total duration but it ranges between [$(n-1)\times t_{\rm exp}+n\times t_{\rm ro}]<t<[n\times t_{\rm exp}+(n+1)\times t_{\rm ro}]$~ms and undoubtedly a portion of the total emitted energy has not been recorded (i.e. during $t_{\rm ro}$) .

In order to calculate the energy lost during the $t_{\rm ro}$ of the multi-frame flashes, an energy interpolation for these specific time ranges has been made. In particular, the energies from the recorded light during the exposures assigned time values that correspond to the half of their exposure times ranges. For example, the energy calculated in the second frame of a multi-frame flash corresponds to the time range $33<t<56$~ms and, therefore, a corresponding timing of $t=44.5$~ms is assigned to it. Hence, using this way, the energy curves of the flashes can be derived in an similar way their light curves are produced (Figs.~\ref{fig:LCs1}- \ref{fig:LCs3}). Assuming an exponential energy drop (i.e. form of $E(t)=E_0 + Ae^{-Bt}$), we can interpolate the values of energies lost during the read out time of the cameras. These values, then, are summed to the total observed emitted energy of the flash for each passband. In Fig.~\ref{fig:Interpol} is shown an example of calculation of the lost energy during the read out time ranges of a multi-frame flash. Therefore, the total emitted energy for a band of the multi-frame flash is given by the following formula:
\begin{equation}
E_{\lambda}~(\rm total)=E_{\lambda} (t)\mid_{0}^{t_1} + E_{\lambda} (t)\mid_{t_1}^{t_1+t_{\rm ro}} +...+ E_{\lambda} (t)\mid_{t_{n-1}+t_{\rm ro}}^{t_n}~{\rm in~J},
\label{eq:Emf2}
\end{equation}

\begin{figure}
\centering
\includegraphics[width=8.6cm]{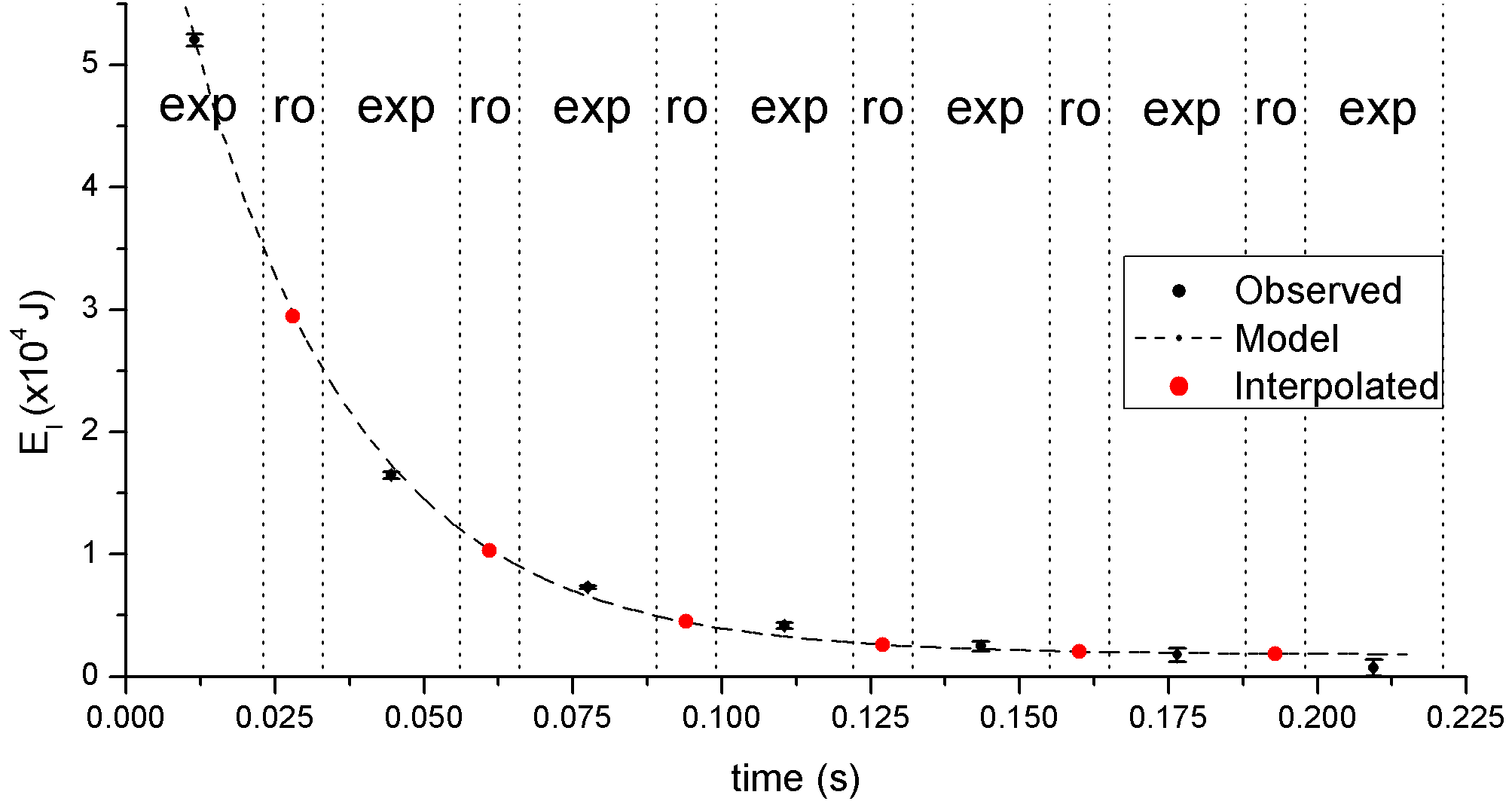}\\
\caption{Example of exponential fit on the $E_{\rm I}$ curve of the flash $\#27$. Black symbols denote the observed $E_{\rm I}$ values, while red symbols the interpolated values based on the model (dashed line). Vertical dotted lines distinguish the exposures (exp) and the read out (ro) time ranges.}
\label{fig:Interpol}
\end{figure}


One useful correlation is between the emitted $E$ of both bands. Using Eq.~\ref{eq:Emf} for the single frame or Eq.~\ref{eq:Emf2} for the multi-frame flashes, the total $E$ for each flash per band was calculated. The correlation diagram is given in the upper panel of Fig.~\ref{fig:MagEREI}. A linear fit to these data points was made and produced the following empirical relation:
\begin{equation}
\log E_{\rm R} = 0.3(2) + 0.85(5)\log E_{\rm I}~{\rm with}~r=0.9,
\label{eq:LELR}
\end{equation}
where $r$ is the correlation coefficient. This relation can be used to estimate an average value of $E_{\rm R}$ for the suspected flashes in order to be able to estimate the $E_{\rm lum}$ and subsequently the rest of the parameters of the projectiles (see Section~\ref{sec:MRC}). In Fig.~\ref{fig:MagEREI} is shown the $E$ distribution of all flashes (validated and suspected) detected in both bands against their peak magnitude values. In this plot, it is clearly seen that the majority of the suspected flashes are systematically fainter, hence, not feasible to be detected in the $R$ filter too. At this point, it should be noted, that the detection of a flash is not only matter of energy, but also matter of seeing conditions of the observing site and of the lunar phase. The detection limits in $R$ band for NELIOTA equipment and site have been estimated as 11.4~mag during the less bright lunar phases and as 10.5~mag during the more bright ones (see Paper II and Fig.~\ref{fig:Phamag}).

Using Eqs.~\ref{eq:LELR} and \ref{eq:ELEREI} we are able to perform a rough estimation of the impactors parameters detected only in $I$ band, assuming again only the $\eta$ for the 400-900~nm range. The $E_{\rm R}$ and $E_{\rm I}$ values for the validated flashes are given in Table~\ref{tab:ResultsReal}, while the $E_{\rm I}$ values of the suspected ones along with the estimation for the $E_{\rm R,~est}$ based on the correlation found between $E_{\rm I}-E_{\rm R}$ for the validated flashes (Eq.~\ref{eq:LELR}) are listed in Table~\ref{tab:ResultsSusp}. In Table~\ref{tab:ResultsSusp}, we present also the estimated $m_{\rm R,~est}$ values calculated by solving backwards the Eqs.~\ref{eq:LELR}, \ref{eq:E}, \ref{eq:L2}, and \ref{eq:fluxR}. These values are plotted along with those of the validated flashes in Fig.~\ref{fig:Phamag}. At this point it should be noticed, that the majority of the calculated $m_{\rm R,~est}$ are fainter than 11.4~mag (i.e. NELIOTA detection limit in low brightness lunar phases). Contrary to these flashes, the $\#35$, $\#80$, $\#83$ and $\#86$ have $m_{\rm R,~est}$ between 10.5-11.4~mag, which is above the detection limit for the lunar phase during which they were detected (i.e. phase $>115\degr$ and $<250\degr$; see Fig.~\ref{fig:Phamag}). However, their non detection indicates that either their $m_{\rm R,~est}$ are underestimated or the seeing conditions were not that good. Finally, using the $m_{\rm R,~est}$ and $m_{\rm I}$ of the suspected flashes, their temperatures are also roughly estimated and listed in Table~\ref{tab:ResultsSusp}.

\begin{figure}
\begin{tabular}{c}
\includegraphics[width=8.6cm]{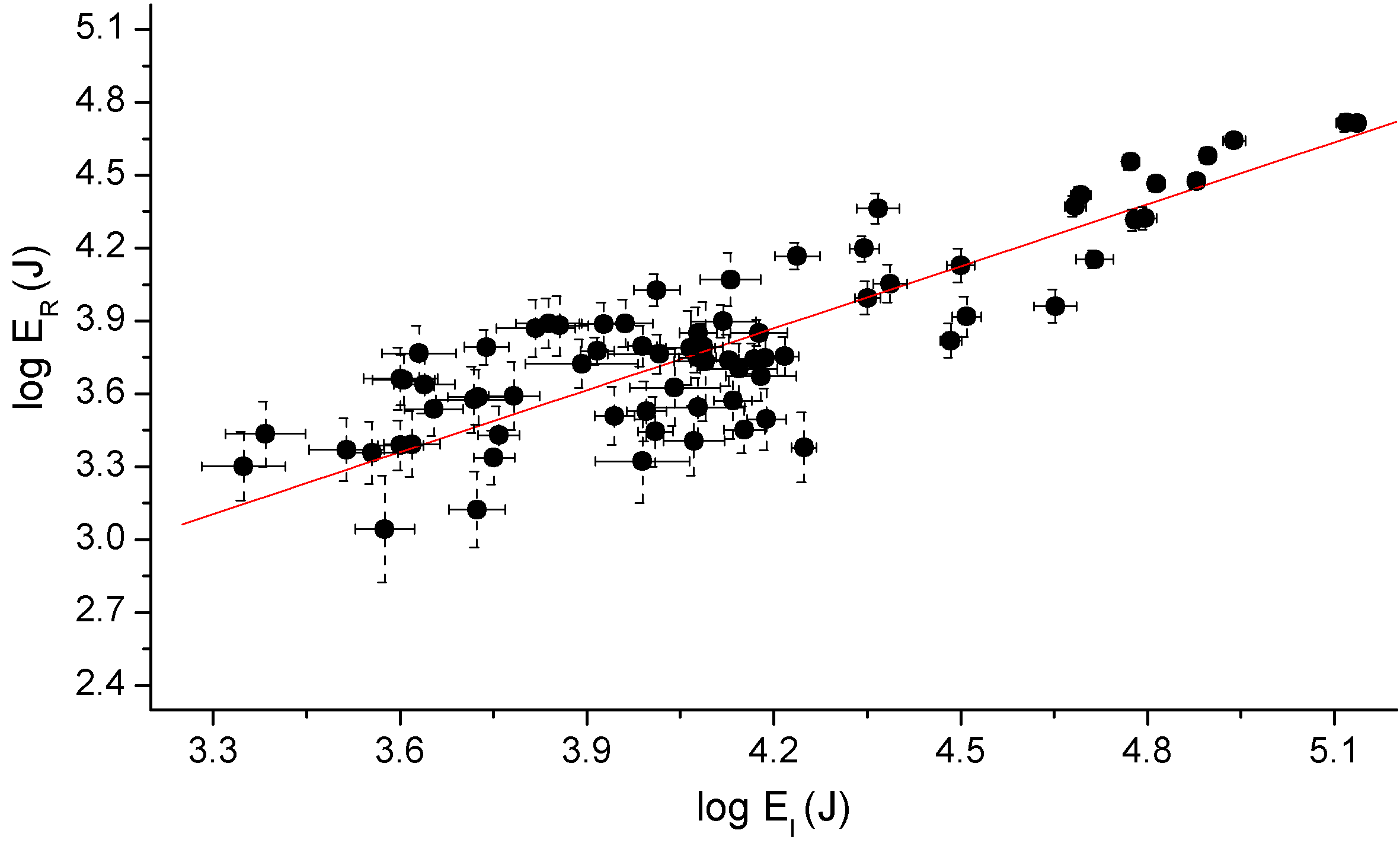}\\
\includegraphics[width=8.6cm]{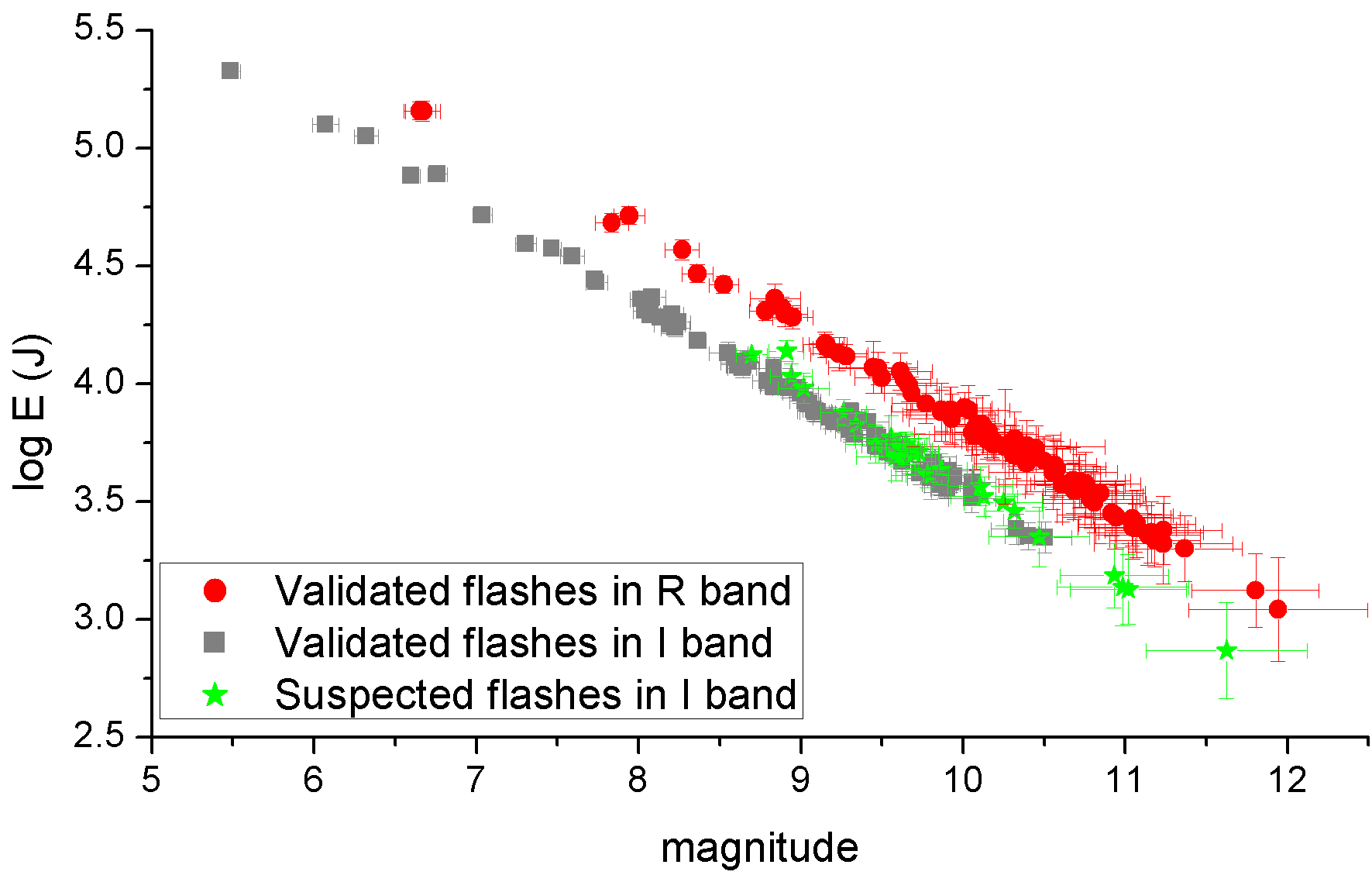}\\
\end{tabular}
\caption{Upper panel: Linear correlation (solid line) between the emitted energies $E_{\rm R}$ and $E_{\rm I}$ for the validated flashes (symbols). Lower panel: Distribution of the emitted $E$ and their respective peak magnitudes for the validated and the suspected flashes for both bands.}
\label{fig:MagEREI}
\end{figure}
\begin{figure}
\includegraphics[width=8.7cm]{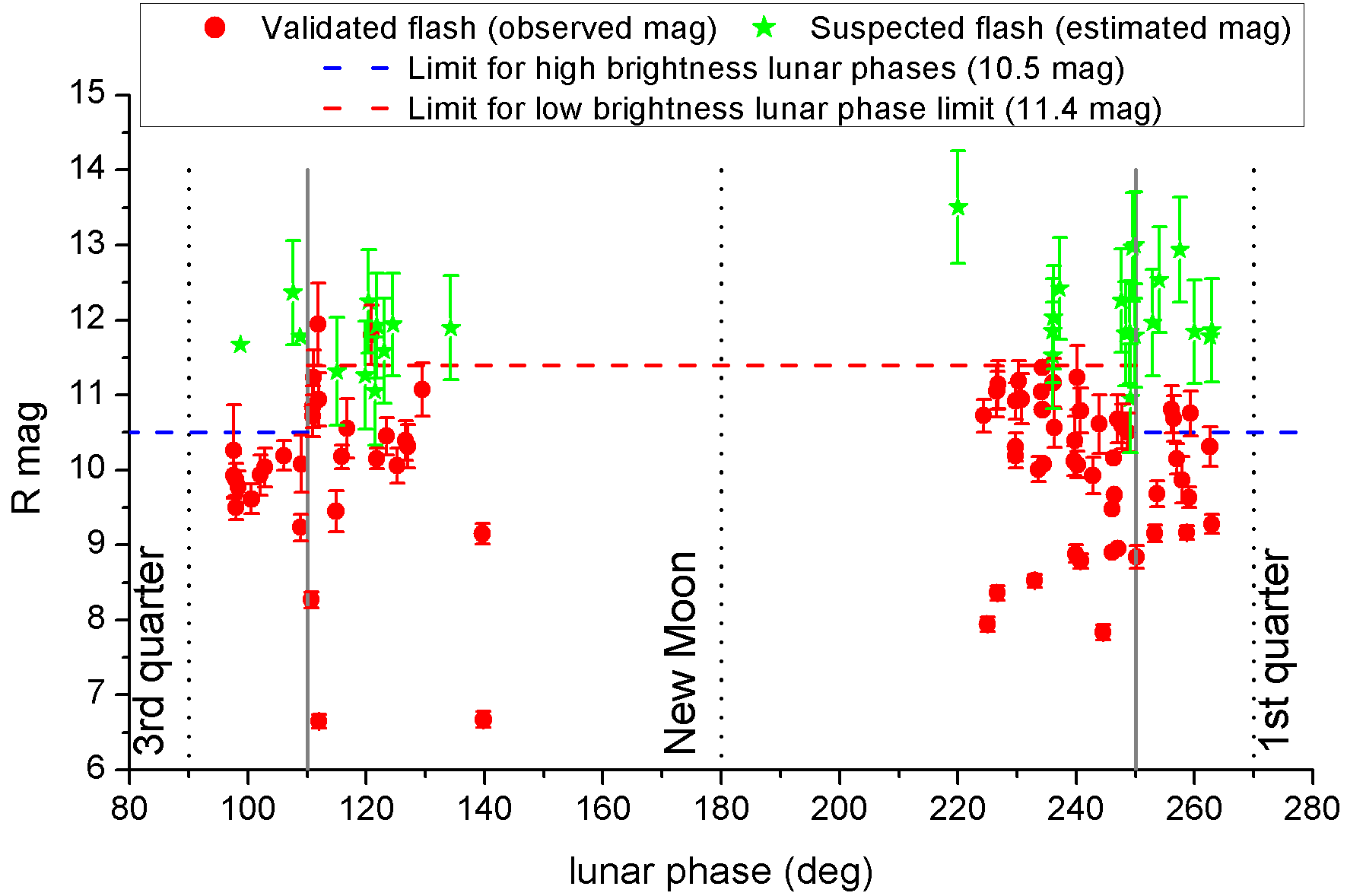}
\caption{Validated flash detections and suspected flashes estimated magnitudes in $R$ band against lunar phases. Black dotted vertical lines denote the quarters (90$\degr$ and 270$\degr$) and the New Moon (180$\degr$) phases. Gray vertical solid lines correspond to the boundaries of lunar phases (110$\degr$ and 250$\degr$) empirically set to separate the lunar phases with different $m_{\rm R,~lim}$. Blue and red dashed horizontal lines represent the two upper limits of $m_{\rm R}$ of flash detections according to the lunar phase.}
\label{fig:Phamag}
\end{figure}

The $E$ for each band can be calculated from Eqs.~\ref{eq:E} and \ref{eq:Emf} and can be connected to the temperature found for each set of frames. It should be noted that the temperature assigned to each data set is not connected to the total energy of the flash but to the energy recorded in the particular set of frames. E.g. for a flash detected in two frames in $R$ and in four frames in $I$, we calculate the $E$ for the first two set of frames for which the temperature could be calculated. The correlation between the peak temperature of the flashes against their $\eta$ ratio in $R$ and $I$ bands is plotted in Fig.~\ref{fig:Teta}. As expected, this plot shows that hotter flashes have larger efficiency in $R$ than in $I$ band and that approximately at 4000~K the efficiencies become equal. For a $T_{\rm f}\sim6200$~K the $\eta$ ratio becomes $\sim1.5$, while for a $T_{\rm f}\sim1600$~K becomes $\sim0.2$. Therefore, based on the lunar impact flashes observed during the NELIOTA campaign, it can be concluded that $\eta_{\rm I}$ ranges between $0.66\eta_{\rm R}<\eta_{\rm I}<5\eta_{\rm R}$. Given that the majority of the $T_{\rm f}$ lie in the range 2000-3500~K (Fig.~\ref{fig:Thist}), a typical value for $\eta_{\rm I}$ can be considered between $1.1\eta_{\rm R}<\eta_{\rm I}<3.1\eta_{\rm R}$.

\begin{figure}
\includegraphics[width=8.7cm]{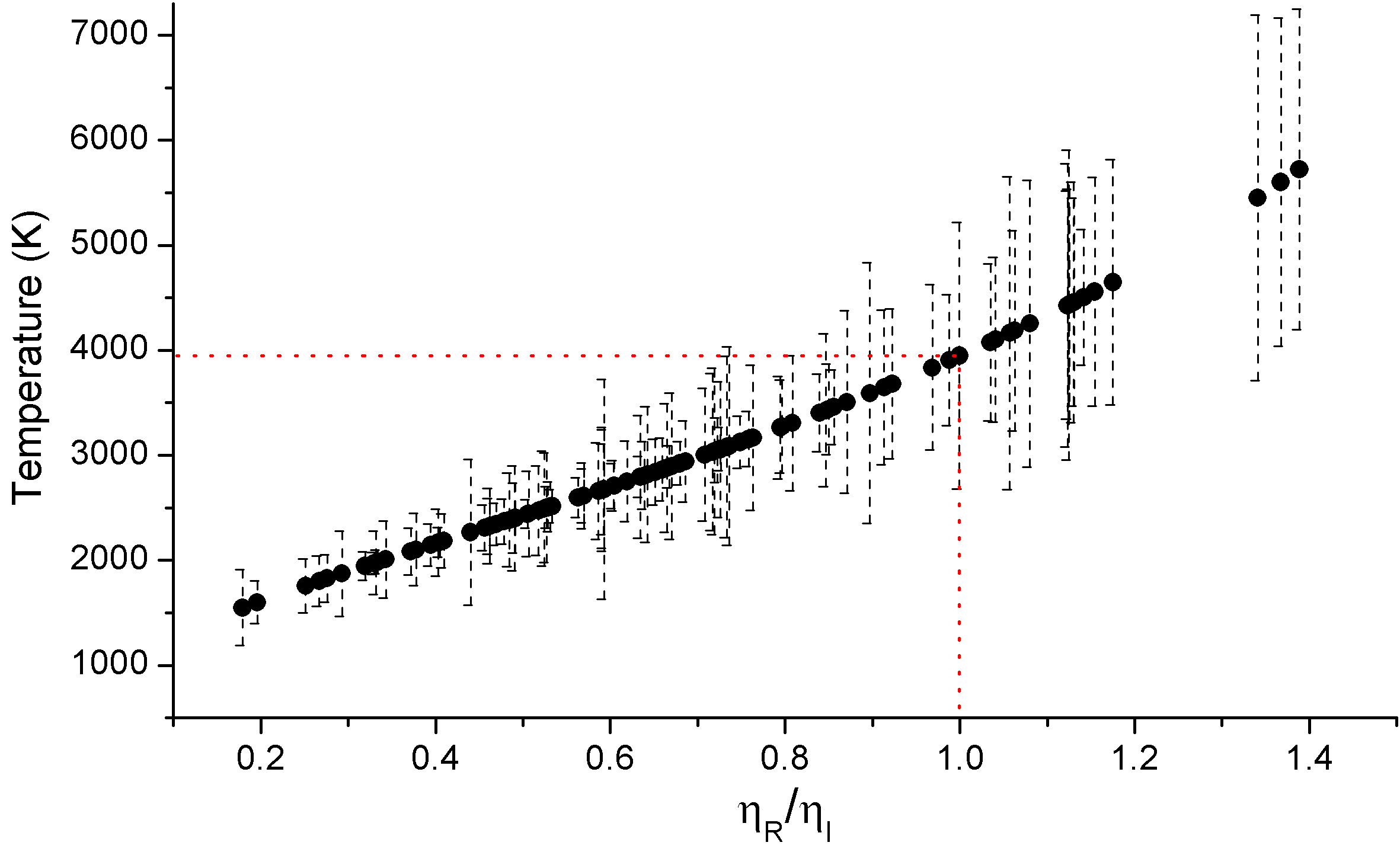}
\caption{Correlation between peak temperature and luminous efficiency ratio for the validated flashes. The crossing of the red dashed lines indicates the temperature value of the flash for which the luminous efficiencies of $R$ and $I$ bands are equal.}
\label{fig:Teta}
\end{figure}

\subsubsection{Luminous energy calculation based on the BB curve}
\label{sec:LE2}

Another way to calculate the $E_{\rm lum}$ of a flash is to use its temperature $T$ \citep[c.f.][]{AVD19}. This method has the advantage that takes into account the spectral density based on the BB curve, hence, for a given $T$ it can be integrated over a chosen wavelength range to produce the energy flux. The greatest disadvantage, which the reason for not using it in our study, is that takes into account only the set of frames from which the $T$ can be calculated. This means, that the emitted energies from the $I$ band frames for which there is no respective detection in the $R$ band are totaly neglected (notice that these flashes are the majority). This method can be applied successfully only for single frame flashes, while for the multi-frame ones can be considered only as an approximation. In general, flashes fade and cool over time, therefore the $L$ has to be calculated for each set of frames in which flash is detected. Combining Eqs.~\ref{eq:fluxR}, \ref{eq:fluxI}, and \ref{eq:InvLaw} a mean radius of the emitted area can be derived:
\begin{equation}
R=\sqrt{\frac{f_{\lambda}\Delta\lambda~f_{\rm a}~d^2}{2\sigma T^4}}~{\rm in~m},
\label{eq:Remit}
\end{equation}
where $\sigma=5.7\times10^{-8}$~W~m$^{-2}$~K$^4$ the Stefan-Boltzmann's constant, and $f_{\rm a}=2$, $f_{\lambda}$, $\Delta\lambda$, and $d$ as defined above. Then, by simply using the definition of the luminosity of the BB, taking into account that emits as half sphere (area $2\pi R^2$) we get:
\begin{equation}
L_{\rm T}=2\pi R^2~\sigma T^4~{\rm in~W},
\label{eq:LT}
\end{equation}
thus, $E_{\rm lum}$ can be calculated using Eq.~\ref{eq:E}. However, for multi-frame flashes Eq.~\ref{eq:Emf} can be rewritten as:
\begin{equation}
E_{\rm lum}=\int_{0}^{t}L_{\rm T}(t)dt~{\rm in~J},
\label{eq:ET}
\end{equation}
where $L_{\rm T}(t)$ is the luminosity of the flash for a specific time range $dt$ during which the flash emitted as a BB with temperature $T$. For a specific wavelength range $d\lambda$ the following equation \citep[see also][]{AVD19} can be also used:
\begin{equation}
E_{\rm lum}=\pi^2 R^2\int_{0}^{t} \int_{\lambda_1}^{\lambda_2} B(\lambda,~T,~t)d\lambda dt~{\rm in~J}
\label{eq:Eb}
\end{equation}

\subsection{Origin of the impactors}
\label{sec:streamassociation}

The main difficulty in the estimation of the mass of a projectile is the assumptions of its velocity and the $\eta$ of the flash. For meteoroid streams there are great certainties about the velocities of the meteoroids based on calculations from meteor showers. Contrary to that, for the sporadic impacts the uncertainty in velocity is quite large ($17<V_{\rm p}<24$~km~s$^{-1}$). Therefore, it is very useful to associate any impacts with active meteoroid streams, since a significant constraint on the velocity is provided, hence, its parameters can be considered more accurate (see Section~\ref{sec:MRC}). In order to check whether any of the detected flashes from the NELIOTA campaign can be potentially associated to active meteoroid streams, the method of \citet{ORT15, MAD15b, MAD15a} was applied. This method calculates the probability of a meteoroid to be member of a stream ($p^{\rm ST}$) and is given by the following relation \citep{MAD15b, MAD15a}:
\begin{equation}
\begin{split}
&p^{\rm ST}= \\
&\frac{\nu^{\rm ST}~\gamma^{\rm ST}~\cos\theta~\sigma~ZHR^{\rm ST}_{\rm E}(\rm max) 10^{-B|\lambda_{\sun} - \lambda_{\sun, \rm max}|}} {\nu^{\rm SPO}\gamma^{\rm SPO} HR^{\rm SPO}_{\rm E} + \nu^{\rm ST} \gamma^{\rm ST}\cos\theta~\sigma~ZHR^{\rm ST}_{\rm E}(\rm max) 10^{-B|\lambda_{\sun} - \lambda_{\sun, \rm max}|}},
\end{split}
\label{eq:POS}
\end{equation}
where $\nu$ is a multiplicative factor which is expressed by \citep{BEL00b}:
\begin{equation}
\nu=\Bigg(\frac{m_0~V_{\rm p}^2}{2}\Bigg)^{s-1} E_{\rm kin,~min}^{1-s},
\label{eq:ni}
\end{equation}
where $V_{\rm p}$ is the impact velocity in km~s$^{-1}$, $E_{\rm kin,~min}$ is the minimum $E_{\rm kin}$ that a projectile must have to be detected. $s$ is the mass index of the shower, which is connected to the respective population index $r$ with the following relation \citep{MAD15a}:
\begin{equation}
s=1+2.5\log r.
\label{eq:s}
\end{equation}
$m_0$ is the mass of a meteoroid producing on Earth a meteor of 6.5~mag. According to the empirical relations of \citet[][Eqs.~1 and 2]{HUG87}, $m_0$ can be calculated using the formula:
\begin{equation}
m_0=1.66 V_{\rm p}^{-4.25}~\rm in~kg.
\label{eq:m0}
\end{equation}
Returning to the description of Eq.~\ref{eq:POS}, the $\gamma$ factor is the ratio of the gravitational focusing factors $\Phi$ of Moon and Earth and it is given by the following relation \citep{MAD15a}:
\begin{equation}
\gamma = \frac{\Phi_{\rm Moon}}{\Phi_{\rm Earth}}=\frac{1+\frac{V^2_{\rm esc, ~Moon}}{V^2_{\rm p}}}{1+\frac{V^2_{\rm esc, ~Earth}}{V^2_{\rm p}}},
\label{eq:gamma}
\end{equation}
where $V_{\rm esc}$ is the escape velocity from a central body (e.g. Moon, Earth). The $\theta$ parameter in Eq.~\ref{eq:POS} is the angle between the position of the impact and the subradiant point of the stream on Moon and can be calculated by the method given by \citet{BEL00b}. The $\sigma$ factor is the ratio of the distances of Earth and Moon from the center of the stream, which is assumed to have the shape of a tube (density decreases from the center to the edges). $ZHR_{\rm E}^{ST}$~(max) is the maximum zenithal hourly rate of a stream as observed from Earth on a specific date represented by the solar longitude $\lambda_{\sun,\rm max}$. The $HR^{\rm SPO}_{\rm E}$ is the hourly rate of sporadic meteors as observed from Earth, $B$ (expressed in meteors per solar longitude degrees) is the slope of the $ZHR_{\rm E}^{ST}$ of a stream, i.e. the $ZHR$ around its maximum value against the solar longitude, and $\lambda_{\sun}$ the corresponding solar longitude of the time of the impact.

The first step to calculate the $p^{\rm ST}$ of the association of a meteoroid with a stream using NELIOTA setup is to estimate the $E_{\rm kin,~min}$ that should have to be detected. The latter estimation has difficulties since it is connected to the $E_{\rm lum}$, hence, the magnitude of the flash, which depends on the lunar phase. For this, it was preferred to add another factor for the $E_{\rm kin,~min}$ based on the detections of NELIOTA in $R$ band. The detection limits of the NELIOTA setup according to the lunar phase are shown in Fig.~\ref{fig:Phamag}. Therefore, for lunar phases between $110\degr-250\degr$ (i.e. before and after New Moon) the limiting magnitude is $m_{\rm R,~lim}=11.4$~mag (see also Paper II), while for lunar phases less than 110$\degr$ and greater 250$\degr$ (i.e. after the third and before the first quarters) is estimated as $m_{\rm R,~lim}=10.5$~mag.

Hence, using these $m_{\rm R,~lim}$ we are able to calculate the $E_{\rm kin,~min}$ using Eqs.~\ref{eq:fluxR}, \ref{eq:E}, and \ref{eq:eta} for a specific value of $\eta$. Eq.~\ref{eq:POS} is very sensitive to the $\nu$, $B$, and $\lambda_{\sun}$ values. In addition, $\nu$ is extremely sensitive to the factors $r$ and $s$ (Eqs.~\ref{eq:ni} and \ref{eq:s}). In order to calculate the $p^{\rm ST}$ for the flashes, we gathered data from the literature \citep{HUG87, JEN94, BEL00b, BRO02, BRO10, MAD15a} and updated online databases (American Meteor Society\footnote{\url{https://www.amsmeteors.org/}}, International Meteor Organization\footnote{\url{https://www.imo.net/}}) for the factors $r$, $s$, $B$, $V_{\rm p}$, and $ZHR^{\rm ST}_{\rm E, ~max}$ of ten strong meteoroid streams (i.e. those with the greatest $ZHR^{\rm ST}_{\rm E,~max}$) and for each of them the $m_0$ was calculated and listed in Table~\ref{tab:streams}. The same table includes also for each stream: the beginning ($\lambda_{\sun, \rm beg}$) and ending ($\lambda_{\sun, \rm end}$) solar longitudes, the $\lambda_{\sun, \rm max}$, and the equatorial ($RA$, $Dec$) and the ecliptic ($l$, $b$) coordinates of the radiant point. Fig.~\ref{fig:streams} illustrates an example of the calculation of the probabilities of the meteoroids detected from impact flashes during the maximum of Geminids stream in December 2017 to be associated with the stream. It should be noted that the stream association probability decreases as moving away from the sub-radiant point of the stream.

\begin{figure}
\includegraphics[width=\columnwidth]{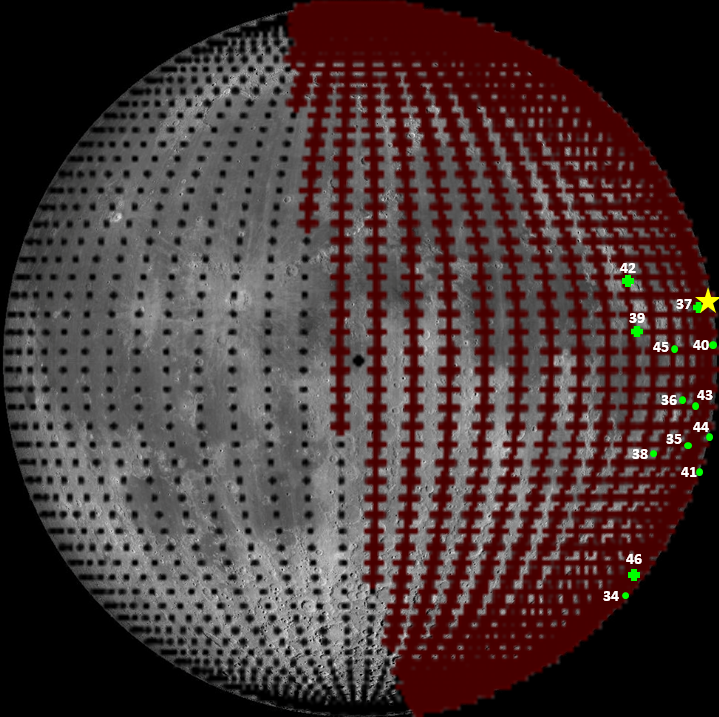}\\
\caption{Example of the calculation of meteoroid probabilities to be associated with an active stream. Red thick crosses denote the lunar area where the stream meteoroids are expected to impact Moon during the night of 14 December 2017 (Geminids maximum). Yellow star is the sub-radiant point of the Geminids stream on the Moon, while green crosses and circles represent the detected validated and suspected flashes, respectively, during 11-14 December 2017.}
\label{fig:streams}
\end{figure}

For each detected flash, the $p^{\rm ST}$ was calculated based on the aforementioned method. We associate impact flashes to an active meteoroid stream only if $p^{\rm ST}>0.5$. Otherwise, they are considered as sporadics. In Eqs.~\ref{eq:POS} and \ref{eq:gamma}, except for the stream depended parameters (e.g. $V_{\rm p}$, $\nu^{\rm ST}$, $\gamma^{\rm ST}$, $ZHR^{\rm ST}_{\rm E}(\rm max)$), the following values were assumed: $\gamma^{\rm SPO}=0.77$ \citep{ORT06, MAD15b}, $V_{\rm esc,~Earth}=11.19$~km~s$^{-1}$, $V_{\rm esc,~Moon}=2.38$~km~s$^{-1}$, $\sigma=1$ \citep{MAD15b, MAD15a}, and $HR^{\rm SPO}_{\rm E}=10$~meteors~hr$^{-1}$ \citep{DUB10}. The various values of $\nu^{\rm SPO}$ are given in Table~\ref{tab:nu} according to the assumed $\eta$ and the $m_{\rm R,~lim}$.  The possible associations of the detected flashes with active meteoroid streams (using their code names, see Table~\ref{tab:streams}) and their assumed $V_{\rm p}$ are listed in Tables~\ref{tab:ResultsReal}-\ref{tab:ResultsSusp}. The calculations show that 18 out of 79 validated flashes can be associated with meteoroid streams ($\sim23\%$), while, if the suspected flashes are taken also into account, then their percentage can be extended to $\sim28\%$ (30 out of 108 total flashes).

\subsection{Physical parameters of the impactors and the crater}
\label{sec:MRC}

In this section we describe in detail the methods and the assumptions for the calculation of the physical parameters of the impacting bodies, namely the mass $m_{\rm p}$ and the radius $r_{\rm p}$ as well as that of the generated crater, namely its diameter $d_{\rm c}$. As mentioned already in Section~\ref{sec:LE}, one of the most important factors to calculate the $m_{\rm p}$ of a projectile is the assumption for the $\eta$ value. Then, $E_{\rm kin}$ is derived using Eq.~\ref{eq:eta} based on the calculated $E_{\rm lum}$ of the flash. According to the definition of $E_{\rm kin}$ and the assumed $V_{\rm p}$ (see Section~\ref{sec:streamassociation}), the $m_{\rm p}$ is calculated using the formula:
\begin{equation}
m_{\rm p}=\frac{2E_{\rm kin}}{V^2_{\rm p}}~{\rm in~kg}
\label{eq:mass}
\end{equation}
The calculation of the $r_{\rm p}$ is strongly depended on the density $\rho_{\rm p}$ of the projectile. In the work of \citet{BAD09} are listed the bulk densities of the meteoroids of ten strong streams according to their parent body. These values along with the bulk density of the sporadics are included in Table~\ref{tab:streams}. Therefore, based on these $\rho_{\rm p}$ values and the fundamental formula:
\begin{equation}
r_{\rm p}=\Bigg(\frac{3m_{\rm p}}{4 \pi \rho_{\rm p}}\Bigg)^{1/3}~{\rm in~cm,}
\label{eq:radius}
\end{equation}
the radius of a projectile can be calculated.

\begin{sidewaystable*}
\centering
\caption{Results for the validated flashes, projectiles and craters for various $\eta$ values. Errors are included in parentheses alongside values and correspond to the last digit(s).}
\label{tab:ResultsReal}
\scalebox{0.9}{
\begin{tabular}{ccccc|cccc|cccc|cccc}
\hline\hline																																	
	&		&		&		&		&	\multicolumn{4}{c}{$\eta=5\times10^{-3}$}					&	\multicolumn{4}{c}{$\eta=1.5\times10^{-3}$}					&	\multicolumn{4}{c}{$\eta=5\times10^{-4}$}											\\
\hline																																	
ID	&	Stream	&	$E_{\rm R}$	&	$E_{\rm I}$	&	$T$	&	$E_{\rm kin,~p}$	&	$m_{\rm p}$	&	$r_{\rm p}$	&	$d_{\rm c}$	&	$E_{\rm kin,~p}$	&	$m_{\rm p}$	&	$r_{\rm p}$	&	$d_{\rm c}$	&	$E_{\rm kin,~p}$	&	$m_{\rm p}$	&	$r_{\rm p}$	&	$d_{\rm c}$	\\
	&		&	($\times 10^4$~J)	&	($\times 10^4$~J)	&	(K)	&	($\times 10^6$~J)	&	(g)	&	(cm)	&	(m)	&	($\times 10^6$~J)	&	(g)	&	(cm)	&	(m)	&	($\times 10^7$~J)	&	(g)	&	(cm)	&	(m)	\\
\hline																																	
1	&	spo	&	0.60(8)	&	0.83(5)	&	3046(307)	&	2.8(2)	&	20(1)	&	1.4(4)	&	1.33(4)	&	9(1)	&	66(4)	&	2.0(7)	&	1.89(6)	&	2.8(2)	&	197(13)	&	3(1)	&	2.6(1)	\\
2	&	spo	&	18(1)	&	21(1)	&	4503(646)	&	78(4)	&	541(24)	&	4(1)	&	3.48(11)	&	260(12)	&	1802(80)	&	6(2)	&	4.93(15)	&	78(4)	&	5407(240)	&	9(3)	&	6.8(2)	\\
3	&	spo	&	1.47(19)	&	1.73(14)	&	3438(431)	&	6.4(5)	&	44(3)	&	1.8(6)	&	1.68(6)	&	21(2)	&	147(11)	&	2.7(9)	&	2.39(8)	&	6.4(5)	&	442(33)	&	4(1)	&	3.3(1)	\\
4	&	spo	&	1.06(16)	&	1.03(9)	&	4076(748)	&	4.2(4)	&	29(3)	&	1.6(5)	&	1.49(6)	&	14(1)	&	96(9)	&	2.3(8)	&	2.11(8)	&	4.2(4)	&	289(26)	&	3(1)	&	2.9(1)	\\
5	&	spo	&	0.56(8)	&	1.20(8)	&	2343(200)	&	3.5(2)	&	24(2)	&	1.5(5)	&	1.42(5)	&	12(1)	&	81(5)	&	2.2(7)	&	2.01(7)	&	3.5(2)	&	243(16)	&	3(1)	&	2.8(1)	\\
6	&	spo	&	0.57(10)	&	1.65(8)	&	2615(315)	&	4.4(3)	&	31(2)	&	1.6(5)	&	1.51(5)	&	15(1)	&	102(6)	&	2.4(8)	&	2.15(7)	&	4.4(3)	&	307(18)	&	3(1)	&	3.0(1)	\\
12	&	spo	&	0.25(8)	&	1.18(13)	&	2101(343)	&	2.9(3)	&	20(2)	&	1.4(5)	&	1.33(6)	&	10(1)	&	66(7)	&	2.0(7)	&	1.89(8)	&	2.9(3)	&	199(22)	&	3(1)	&	2.6(1)	\\
13	&	spo	&	0.42(15)	&	1.10(18)	&	3088(945)	&	3.0(5)	&	21(3)	&	1.4(5)	&	1.36(7)	&	10(2)	&	70(11)	&	2.1(7)	&	1.92(10)	&	3.0(5)	&	211(33)	&	3(1)	&	2.6(1)	\\
14	&	spo	&	0.21(8)	&	0.98(17)	&	2008(367)	&	2.4(4)	&	16(3)	&	1.3(4)	&	1.26(7)	&	8(1)	&	55(9)	&	1.9(6)	&	1.79(10)	&	2.4(4)	&	164(26)	&	3(1)	&	2.5(1)	\\
16	&	SDA	&	0.24(8)	&	1.77(8)	&	1978(300)	&	4.0(2)	&	4.8(3)	&	0.8(9)	&	1.54(7)	&	13(1)	&	16(1)	&	1.2(8)	&	2.19(10)	&	4.0(2)	&	48(3)	&	2(2)	&	3.0(1)	\\
17	&	SDA	&	0.39(10)	&	0.53(6)	&	3056(647)	&	1.8(2)	&	2.2(3)	&	0.6(7)	&	1.23(7)	&	6(1)	&	7(1)	&	0.9(5)	&	1.74(10)	&	1.8(2)	&	22(3)	&	1(1)	&	2.4(1)	\\
18	&	SDA	&	0.34(9)	&	0.45(5)	&	3168(692)	&	1.6(2)	&	1.9(2)	&	0.6(7)	&	1.18(7)	&	5(1)	&	6(1)	&	0.9(5)	&	1.67(9)	&	1.6(2)	&	19(2)	&	1(1)	&	2.3(1)	\\
19	&	SDA	&	7.11(40)	&	27.2(8)	&	1972(105)	&	69(2)	&	82(2)	&	2(1)	&	3.51(15)	&	228(6)	&	272(7)	&	3(2)	&	4.98(21)	&	69(2)	&	815(22)	&	4(5)	&	6.9(3)	\\
20	&	spo	&	0.58(11)	&	1.04(20)	&	4455(1146)	&	3.2(5)	&	22(3)	&	1.4(5)	&	1.38(7)	&	11(2)	&	75(10)	&	2.1(7)	&	1.96(10)	&	3.2(5)	&	224(31)	&	3(1)	&	2.7(1)	\\
21	&	spo	&	0.35(10)	&	1.20(21)	&	2326(356)	&	3.1(5)	&	21(3)	&	1.4(5)	&	1.36(7)	&	10(2)	&	71(11)	&	2.1(7)	&	1.93(10)	&	3.1(5)	&	214(32)	&	3(1)	&	2.7(1)	\\
22	&	spo	&	0.31(9)	&	1.54(11)	&	2167(319)	&	3.7(3)	&	26(2)	&	1.5(5)	&	1.44(5)	&	12(1)	&	86(7)	&	2.2(7)	&	2.04(7)	&	3.7(3)	&	257(20)	&	3(1)	&	2.8(1)	\\
23	&	spo	&	0.28(6)	&	1.42(11)	&	2185(255)	&	3.4(2)	&	24(2)	&	1.5(5)	&	1.40(5)	&	11(1)	&	79(6)	&	2.2(7)	&	1.99(7)	&	3.4(2)	&	236(17)	&	3(1)	&	2.7(1)	\\
24	&	spo	&	0.55(8)	&	1.47(9)	&	2615(256)	&	4.1(2)	&	28(2)	&	1.5(5)	&	1.48(5)	&	14(1)	&	94(5)	&	2.3(8)	&	2.09(7)	&	4.1(2)	&	281(16)	&	3(1)	&	2.9(1)	\\
26	&	spo	&	1.42(12)	&	5.19(36)	&	3058(207)	&	13.2(8)	&	92(5)	&	2.3(7)	&	2.08(7)	&	44(3)	&	305(18)	&	3(1)	&	2.95(10)	&	13.2(8)	&	915(53)	&	5(2)	&	4.1(1)	\\
27	&	spo	&	5.16(24)	&	13.7(3)	&	2440(136)	&	37.7(8)	&	261(6)	&	3(1)	&	2.82(8)	&	126(3)	&	870(19)	&	5(2)	&	3.99(11)	&	37.7(8)	&	2609(57)	&	7(2)	&	5.5(2)	\\
28	&	spo	&	2.06(21)	&	6.02(14)	&	3458(357)	&	16.2(5)	&	112(3)	&	2.4(8)	&	2.20(6)	&	54(2)	&	373(12)	&	4(1)	&	3.12(9)	&	16.2(5)	&	1119(35)	&	5(2)	&	4.3(1)	\\
29	&	spo	&	0.50(12)	&	1.39(20)	&	5453(1740)	&	3.8(5)	&	26(3)	&	1.5(5)	&	1.45(6)	&	13(2)	&	87(11)	&	2.2(7)	&	2.05(9)	&	3.8(5)	&	262(32)	&	3(1)	&	2.8(1)	\\
30	&	spo	&	0.71(9)	&	1.50(16)	&	2751(384)	&	4.4(4)	&	31(3)	&	1.6(5)	&	1.51(6)	&	15(1)	&	102(8)	&	2.4(8)	&	2.14(8)	&	4.4(4)	&	306(25)	&	3(1)	&	2.9(1)	\\
31	&	spo	&	0.78(18)	&	0.69(8)	&	4431(1088)	&	2.9(4)	&	20(3)	&	1.4(5)	&	1.34(7)	&	10(1)	&	68(9)	&	2.1(7)	&	1.90(9)	&	2.9(4)	&	203(28)	&	3(1)	&	2.6(1)	\\
32	&	spo	&	0.56(10)	&	1.54(7)	&	3264(488)	&	4.2(2)	&	29(2)	&	1.6(5)	&	1.49(5)	&	14(1)	&	97(5)	&	2.3(8)	&	2.11(7)	&	4.2(2)	&	290(16)	&	3(1)	&	2.9(1)	\\
33	&	spo	&	0.53(12)	&	0.78(16)	&	5722(1528)	&	2.6(4)	&	18(3)	&	1.3(4)	&	1.30(7)	&	9(1)	&	60(9)	&	2.0(7)	&	1.84(10)	&	2.6(4)	&	181(28)	&	3(1)	&	2.5(1)	\\
37	&	GEM	&	0.47(11)	&	1.51(20)	&	2402(334)	&	4.0(5)	&	6(1)	&	0.8(3)	&	1.59(8)	&	13(2)	&	20(2)	&	1.0(5)	&	2.25(11)	&	4.0(5)	&	61(7)	&	2(3)	&	3.1(1)	\\
39	&	GEM	&	0.44(12)	&	0.44(5)	&	3948(1267)	&	1.7(3)	&	2.7(4)	&	0.6(1)	&	1.25(7)	&	6(1)	&	9(1)	&	0.9(5)	&	1.77(10)	&	1.7(3)	&	27(4)	&	1(2)	&	2.4(1)	\\
42	&	GEM	&	0.45(11)	&	0.40(5)	&	4432(1475)	&	1.7(2)	&	2.6(4)	&	0.6(1)	&	1.24(7)	&	6(1)	&	9(1)	&	0.9(5)	&	1.76(9)	&	1.7(2)	&	26(4)	&	1(2)	&	2.4(1)	\\
46	&	GEM	&	5.18(45)	&	13.2(5)	&	2889(200)	&	37(1)	&	57(2)	&	1.7(8)	&	3.03(11)	&	122(5)	&	189(7)	&	2.5(9)	&	4.29(15)	&	37(1)	&	567(21)	&	4(6)	&	5.9(2)	\\
47	&	spo	&	0.79(12)	&	1.32(16)	&	4101(782)	&	4.2(4)	&	29(3)	&	1.6(5)	&	1.49(6)	&	14(1)	&	97(9)	&	2.3(8)	&	2.11(8)	&	4.2(4)	&	291(28)	&	3(1)	&	2.9(1)	\\
49	&	spo	&	0.71(17)	&	1.20(8)	&	2675(431)	&	3.8(4)	&	26(3)	&	1.5(5)	&	1.45(6)	&	13(1)	&	88(9)	&	2.2(7)	&	2.05(8)	&	3.8(4)	&	264(27)	&	3(1)	&	2.8(1)	\\
50	&	spo	&	2.3(3)	&	2.33(19)	&	3905(626)	&	9.3(8)	&	64(5)	&	2.0(7)	&	1.87(7)	&	31(3)	&	214(17)	&	3(1)	&	2.66(10)	&	9.3(8)	&	641(52)	&	4(1)	&	3.7(1)	\\
51	&	spo	&	0.77(17)	&	0.91(9)	&	3428(727)	&	3.4(4)	&	23(3)	&	1.5(5)	&	1.40(6)	&	11(1)	&	78(9)	&	2.2(7)	&	1.98(9)	&	3.4(4)	&	234(27)	&	3(1)	&	2.7(1)	\\
53	&	spo	&	0.77(15)	&	0.85(8)	&	3646(736)	&	3.2(3)	&	22(2)	&	1.4(5)	&	1.38(6)	&	11(1)	&	75(8)	&	2.1(7)	&	1.96(8)	&	3.2(3)	&	224(24)	&	3(1)	&	2.7(1)	\\
54	&	spo	&	0.74(20)	&	0.66(10)	&	4436(1094)	&	2.8(4)	&	19(3)	&	1.4(4)	&	1.32(7)	&	9(1)	&	64(10)	&	2.0(7)	&	1.88(10)	&	2.8(4)	&	193(31)	&	3(1)	&	2.6(1)	\\
55	&	spo	&	0.54(31)	&	1.23(14)	&	2267(695)	&	3.5(7)	&	25(5)	&	1.5(5)	&	1.42(9)	&	12(2)	&	82(16)	&	2.2(7)	&	2.01(12)	&	3.5(7)	&	245(47)	&	3(1)	&	2.8(2)	\\
56	&	spo	&	0.23(7)	&	0.33(5)	&	3032(748)	&	1.1(2)	&	8(1)	&	1.0(3)	&	1.02(5)	&	4(1)	&	26(4)	&	1.5(5)	&	1.44(7)	&	1.1(2)	&	78(12)	&	2(1)	&	2.0(1)	\\
58	&	spo	&	0.91(14)	&	4.5(4)	&	2369(215)	&	10.8(8)	&	75(5)	&	2.1(7)	&	1.96(7)	&	36(3)	&	249(18)	&	3(1)	&	2.78(10)	&	10.8(8)	&	747(53)	&	5(2)	&	3.8(1)	\\
59	&	spo	&	2.6(2)	&	4.9(2)	&	2515(156)	&	15.1(5)	&	104(4)	&	2.4(8)	&	2.16(6)	&	50(2)	&	348(12)	&	3(1)	&	3.06(9)	&	15.1(5)	&	1045(37)	&	5(2)	&	4.2(1)	\\
\hline														
\end{tabular}}
\end{sidewaystable*}

\begin{sidewaystable*}
\centering
\caption{Table~\ref{tab:ResultsReal} cont.}
\label{tab:ResultsReal2}
\scalebox{0.9}{
\begin{tabular}{ccccc|cccc|cccc|cccc}
\hline\hline																																	
	&		&		&		&		&	\multicolumn{4}{c}{$\eta=5\times10^{-3}$}					&	\multicolumn{4}{c}{$\eta=1.5\times10^{-3}$}					&	\multicolumn{4}{c}{$\eta=5\times10^{-4}$}											\\
\hline																																	
ID	&	Stream	&	$E_{\rm R}$	&	$E_{\rm I}$	&	$T$	&	$E_{\rm kin,~p}$	&	$m_{\rm p}$	&	$r_{\rm p}$	&	$d_{\rm c}$	&	$E_{\rm kin,~p}$	&	$m_{\rm p}$	&	$r_{\rm p}$	&	$d_{\rm c}$	&	$E_{\rm kin,~p}$	&	$m_{\rm p}$	&	$r_{\rm p}$	&	$d_{\rm c}$	\\
	&		&	($\times 10^4$~J)	&	($\times 10^4$~J)	&	(K)	&	($\times 10^6$~J)	&	(g)	&	(cm)	&	(m)	&	($\times 10^6$~J)	&	(g)	&	(cm)	&	(m)	&	($\times 10^7$~J)	&	(g)	&	(cm)	&	(m)	\\
\hline																																	
61	&	spo	&	0.32(9)	&	0.88(6)	&	2444(405)	&	2.4(2)	&	17(2)	&	1.3(4)	&	1.27(5)	&	8(1)	&	55(5)	&	1.9(6)	&	1.80(7)	&	2.4(2)	&	166(15)	&	3(1)	&	2.5(1)	\\
62	&	spo	&	2.91(20)	&	6.5(2)	&	3153(262)	&	18.8(6)	&	130(4)	&	2.6(8)	&	2.30(7)	&	63(2)	&	435(13)	&	4(1)	&	3.27(10)	&	18.8(6)	&	1304(40)	&	5(2)	&	4.5(1)	\\
63	&	spo	&	0.62(10)	&	0.55(5)	&	4460(988)	&	2.3(2)	&	16(2)	&	1.3(4)	&	1.26(5)	&	8(1)	&	54(5)	&	1.9(6)	&	1.78(7)	&	2.3(2)	&	162(16)	&	3(1)	&	2.5(1)	\\
64	&	spo	&	0.46(14)	&	0.40(5)	&	4558(1087)	&	1.7(3)	&	12(2)	&	1.2(4)	&	1.15(7)	&	6(1)	&	40(7)	&	1.7(6)	&	1.63(9)	&	1.7(3)	&	119(20)	&	2(1)	&	2.2(1)	\\
65	&	spo	&	0.23(7)	&	0.36(3)	&	2793(583)	&	1.2(2)	&	8(1)	&	1.0(3)	&	1.03(5)	&	4(1)	&	27(3)	&	1.5(5)	&	1.46(7)	&	1.2(2)	&	81(10)	&	2(1)	&	2.0(1)	\\
66	&	PER	&	0.24(6)	&	0.40(3)	&	4253(1364)	&	1.3(1)	&	0.7(1)	&	0.5(1)	&	0.99(4)	&	4(1)	&	2(1)	&	0.8(1)	&	1.40(6)	&	1.3(1)	&	7(1)	&	1.1(1)	&	1.9(1)	\\
67	&	PER	&	4.38(27)	&	8.7(4)	&	3124(247)	&	26.1(9)	&	14.0(5)	&	1.4(1)	&	2.37(7)	&	87(3)	&	47(2)	&	2.1(2)	&	3.36(10)	&	26.1(9)	&	140(5)	&	3.1(3)	&	4.6(1)	\\
68	&	spo	&	0.25(8)	&	0.42(4)	&	2678(563)	&	1.3(2)	&	9(1)	&	1.1(3)	&	1.07(5)	&	4(1)	&	31(4)	&	1.6(5)	&	1.51(7)	&	1.3(2)	&	92(12)	&	2(1)	&	2.1(1)	\\
69	&	spo	&	0.13(5)	&	0.53(5)	&	1758(257)	&	1.3(1)	&	9(1)	&	1.1(3)	&	1.07(4)	&	4(1)	&	31(3)	&	1.6(5)	&	1.51(6)	&	1.3(1)	&	92(10)	&	2(1)	&	2.1(1)	\\
72	&	spo	&	0.76(22)	&	0.72(8)	&	4163(1490)	&	3.0(5)	&	20(3)	&	1.4(5)	&	1.35(7)	&	10(2)	&	68(11)	&	2.1(7)	&	1.91(10)	&	3.0(5)	&	204(32)	&	3(1)	&	2.6(1)	\\
73	&	spo	&	19.3(6)	&	38.5(6)	&	2793(194)	&	115(2)	&	799(12)	&	5(2)	&	3.90(11)	&	385(6)	&	2665(40)	&	7(2)	&	5.53(15)	&	116(2)	&	7994(120)	&	10(3)	&	7.6(2)	\\
74	&	spo	&	0.37(13)	&	1.36(9)	&	2401(501)	&	3.5(3)	&	24(2)	&	1.5(5)	&	1.41(5)	&	12(1)	&	80(8)	&	2.2(7)	&	2.00(8)	&	3.5(3)	&	240(23)	&	3(1)	&	2.7(1)	\\
75	&	spo	&	0.27(8)	&	0.24(4)	&	4428(1348)	&	1.0(2)	&	7(1)	&	1.0(3)	&	0.99(6)	&	3(1)	&	24(4)	&	1.5(5)	&	1.41(8)	&	1.0(2)	&	71(13)	&	2(1)	&	1.9(1)	\\
76	&	spo	&	0.22(6)	&	0.56(4)	&	2660(459)	&	1.6(1)	&	11(1)	&	1.1(4)	&	1.12(4)	&	5(1)	&	36(3)	&	1.7(5)	&	1.59(6)	&	1.6(1)	&	108(10)	&	2(1)	&	2.2(1)	\\
78	&	spo	&	1.13(21)	&	2.4(2)	&	3836(787)	&	7.1(5)	&	49(4)	&	1.9(6)	&	1.74(6)	&	24(2)	&	165(12)	&	2.8(9)	&	2.46(9)	&	7.1(5)	&	494(35)	&	4(1)	&	3.4(1)	\\
84	&	spo	&	0.58(15)	&	0.43(6)	&	5601(1560)	&	2.0(3)	&	14(2)	&	1.2(4)	&	1.21(7)	&	7(1)	&	47(8)	&	1.8(6)	&	1.71(9)	&	2.0(3)	&	140(23)	&	3(1)	&	2.4(1)	\\
85	&	spo	&	0.55(13)	&	1.34(8)	&	4647(1169)	&	3.8(3)	&	26(2)	&	1.5(5)	&	1.45(5)	&	13(1)	&	87(7)	&	2.2(7)	&	2.05(8)	&	3.8(3)	&	262(22)	&	3(1)	&	2.8(1)	\\
88	&	spo	&	1.2(3)	&	1.4(2)	&	3506(868)	&	5.1(7)	&	35(5)	&	1.7(5)	&	1.57(7)	&	17(2)	&	117(15)	&	2.5(8)	&	2.23(11)	&	5.1(7)	&	350(46)	&	4(1)	&	3.1(1)	\\
89	&	spo	&	0.61(22)	&	1.2(1)	&	2499(522)	&	3.6(5)	&	25(3)	&	1.5(5)	&	1.42(7)	&	12(2)	&	82(11)	&	2.2(7)	&	2.01(10)	&	3.6(5)	&	246(33)	&	3(1)	&	2.8(1)	\\
90	&	spo	&	1.34(22)	&	3.2(2)	&	2864(298)	&	9.0(5)	&	62(4)	&	2.0(7)	&	1.86(6)	&	30(2)	&	208(13)	&	3(1)	&	2.64(09)	&	9.0(5)	&	623(38)	&	4(1)	&	3.6(1)	\\
93	&	spo	&	2.10(22)	&	6.3(3)	&	2709(215)	&	16.7(7)	&	116(5)	&	2.5(8)	&	2.22(7)	&	56(2)	&	386(16)	&	4(1)	&	3.15(10)	&	16.7(7)	&	1157(48)	&	5(2)	&	4.3(1)	\\
94	&	spo	&	1.76(19)	&	2.2(1)	&	3678(715)	&	7.9(5)	&	55(3)	&	1.9(6)	&	1.79(6)	&	26(2)	&	183(10)	&	2.9(9)	&	2.54(8)	&	7.9(5)	&	550(31)	&	4(1)	&	3.5(1)	\\
95	&	spo	&	0.62(13)	&	1.2(1)	&	3307(644)	&	3.7(3)	&	26(2)	&	1.5(5)	&	1.44(5)	&	12(1)	&	85(7)	&	2.2(7)	&	2.03(8)	&	3.7(3)	&	255(22)	&	3(1)	&	2.8(1)	\\
96	&	spo	&	0.28(09)	&	1.0(1)	&	2678(590)	&	2.6(2)	&	18(2)	&	1.3(4)	&	1.30(5)	&	9(1)	&	60(5)	&	2.0(6)	&	1.84(7)	&	2.6(2)	&	180(16)	&	2.8(9)	&	2.5(1)	\\
97	&	spo	&	20(1)	&	38(1)	&	2922(205)	&	116(4)	&	806(24)	&	5(2)	&	3.91(11)	&	388(12)	&	2686(80)	&	7(2)	&	5.54(16)	&	116.4(3.5)	&	8058(239)	&	10(3)	&	7.6(2)	\\
98	&	spo	&	0.11(06)	&	0.38(4)	&	1873(405)	&	1.0(1)	&	7(1)	&	1.0(3)	&	0.98(5)	&	3(1)	&	22(3)	&	1.4(5)	&	1.38(7)	&	1.0(1)	&	67(10)	&	2.1(7)	&	1.9(1)	\\
100	&	spo	&	0.83(16)	&	3.2(2)	&	2325(266)	&	8.1(5)	&	56(3)	&	1.9(6)	&	1.80(6)	&	27(2)	&	187(11)	&	2.9(9)	&	2.56(8)	&	8.1(5)	&	562(33)	&	4(1)	&	3.5(1)	\\
101	&	SDA	&	0.38(12)	&	0.52(7)	&	3036(790)	&	1.8(3)	&	2(1)	&	0.6(3)	&	1.22(7)	&	6(1)	&	7(1)	&	0.9(6)	&	1.73(11)	&	1.8(3)	&	21(3)	&	1.3(9)	&	2.4(1)	\\
102	&	SDA	&	2.98(22)	&	7.56(21)	&	2510(239)	&	21.1(6)	&	25(1)	&	1.4(9)	&	2.50(11)	&	70(2)	&	84(2)	&	2.0(1)	&	3.54(15)	&	21.1(6)	&	251(7)	&	3(2)	&	4.9(2)	\\
103	&	spo	&	0.39(13)	&	0.61(6)	&	2817(646)	&	2.0(3)	&	14(2)	&	1.2(4)	&	1.20(6)	&	7(1)	&	46(6)	&	1.8(6)	&	1.70(8)	&	2.0(3)	&	138(19)	&	3(1)	&	2.3(1)	\\
104	&	SDA	&	3.58(26)	&	5.93(19)	&	3404(368)	&	19.0(6)	&	23(1)	&	1.3(9)	&	2.42(10)	&	63(2)	&	75(3)	&	2(1)	&	3.44(15)	&	19.0(6)	&	226(8)	&	3(2)	&	4.7(2)	\\
105	&	SDA	&	0.99(16)	&	2.24(11)	&	3273(441)	&	6.5(4)	&	8(1)	&	0.9(6)	&	1.77(8)	&	22(1)	&	26(1)	&	1.4(9)	&	2.51(11)	&	6.5(4)	&	77(4)	&	2(1)	&	3.5(2)	\\
106	&	spo	&	0.63(12)	&	0.98(5)	&	4186(953)	&	3.2(3)	&	22(2)	&	1.4(5)	&	1.38(5)	&	11(1)	&	74(6)	&	2.1(7)	&	1.95(7)	&	3.2(3)	&	222(18)	&	3(1)	&	2.7(1)	\\
107	&	spo	&	2.35(23)	&	4.83(19)	&	2896(697)	&	14.4(6)	&	99(4)	&	2.3(8)	&	2.13(6)	&	48(2)	&	331(14)	&	4(1)	&	3.02(9)	&	14.4(6)	&	994(42)	&	5(2)	&	4.2(1)	\\
108	&	spo	&	3.80(27)	&	7.90(22)	&	2503(197)	&	23.4(7)	&	162(5)	&	2.8(9)	&	2.45(7)	&	78(2)	&	540(16)	&	4(1)	&	3.48(10)	&	23.4(7)	&	1620(49)	&	6(2)	&	4.8(1)	\\
109	&	SDA	&	0.66(11)	&	3.05(12)	&	2941(386)	&	7.4(3)	&	9(1)	&	1.0(8)	&	1.84(8)	&	25(1)	&	29(1)	&	1.4(6)	&	2.61(11)	&	7.4(3)	&	88(4)	&	2(2)	&	3.6(2)	\\
110	&	SDA	&	0.34(10)	&	0.99(7)	&	2879(611)	&	2.7(2)	&	3(1)	&	0.7(5)	&	1.37(7)	&	9(1)	&	11(1)	&	1.0(8)	&	1.94(10)	&	2.7(2)	&	32(3)	&	2(2)	&	2.7(1)	\\
111	&	SDA	&	0.20(06)	&	0.22(3)	&	3592(1243)	&	0.8(1)	&	1(1)	&	0.5(4)	&	0.98(6)	&	3(1)	&	3(1)	&	0.7(6)	&	1.39(9)	&	0.8(1)	&	10(2)	&	1.0(8)	&	1.9(1)	\\
112	&	SDA	&	0.27(07)	&	0.57(4)	&	3004(633)	&	1.7(2)	&	2(1)	&	0.6(5)	&	1.20(6)	&	6(1)	&	7(1)	&	0.9(8)	&	1.70(9)	&	1.7(2)	&	20(2)	&	1.3(9)	&	2.3(1)	\\
\hline														
\end{tabular}}
\end{sidewaystable*}

\begin{table*}
\centering
\caption{Rough estimations of the parameters of the suspected flashes and their corresponding projectiles and impact craters for $\eta=1.5\times10^{-3}$. Errors are included in parentheses alongside values and correspond to the last digit(s).}
\label{tab:ResultsSusp}
\begin{tabular}{ccccccccccc}
\hline\hline																					
ID	&	Stream	&	$m_{\rm I}$	&	$m_{\rm R, ~est}$	&	$E_{\rm I}$	&	$E_{\rm R, ~est}$	&	$T_{\rm est}$	&	$E_{\rm kin,~p,~est}$	&	$m_{\rm p, ~est}$	&	$r_{\rm p, ~est}$	&	$d_{\rm c, ~est}$	\\
	&		&	(mag)	&	(mag)	&	($\times 10^3$~J)	&	($\times 10^3$~J)	&	(K)	&	($\times 10^6$~J)	&	(g)	&	(cm)	&	(m)	\\
\hline																					
7	&	spo	&	10.9(3)	&	12.9(7)	&	1.5(5)	&	1.0(7)	&	1928(599)	&	1.7(4)	&	12(3)	&	1.2(4)	&	1.1(1)	\\
8	&	spo	&	9.8(1)	&	11.9(7)	&	4.1(5)	&	2(2)	&	1811(465)	&	4(1)	&	30(7)	&	1.6(5)	&	1.5(1)	\\
9	&	spo	&	9.6(1)	&	11.8(7)	&	5.0(5)	&	3(2)	&	1791(453)	&	5(1)	&	36(8)	&	1.7(6)	&	1.6(1)	\\
10	&	spo	&	11.0(4)	&	13.0(7)	&	1.3(5)	&	0.9(6)	&	1946(624)	&	1.5(4)	&	10(3)	&	1.1(4)	&	1.1(1)	\\
11	&	spo	&	11.0(4)	&	13.0(7)	&	1.4(5)	&	0.9(6)	&	1942(654)	&	1.5(4)	&	11(3)	&	1.1(4)	&	1.1(1)	\\
15	&	SDA	&	9.3(1)	&	11.6(7)	&	6.7(7)	&	4(2)	&	1758(442)	&	7(2)	&	8(2)	&	0.9(1.1)	&	1.8(1)	\\
25	&	spo	&	10.2(2)	&	12.4(7)	&	3.1(7)	&	2(1)	&	1843(510)	&	3.3(8)	&	23(6)	&	1.4(5)	&	1.4(1)	\\
34	&	GEM	&	9.6(1)	&	11.8(7)	&	5.0(5)	&	3(2)	&	1789(450)	&	5(1)	&	8(2)	&	0.9(1.5)	&	1.7(1)	\\
35	&	GEM	&	8.9(1)	&	11.0(7)	&	14(1)	&	7(4)	&	1898(544)	&	14(3)	&	21(5)	&	1.2(2.0)	&	2.3(2)	\\
36	&	GEM	&	9.6(1)	&	11.8(7)	&	5.3(5)	&	3(2)	&	1784(449)	&	5(1)	&	8(2)	&	0.9(1.5)	&	1.7(1)	\\
38	&	GEM	&	9.6(1)	&	11.8(7)	&	5.4(8)	&	3.(2)	&	1782(458)	&	5(1)	&	9(2)	&	0.9(1.5)	&	1.7(1)	\\
40	&	GEM	&	10.1(2)	&	12.3(7)	&	3.3(6)	&	2(1)	&	1836(394)	&	3.5(8)	&	5(1)	&	0.8(1.3)	&	1.5(1)	\\
41	&	GEM	&	10.3(2)	&	12.4(7)	&	2.9(5)	&	2(1)	&	1852(492)	&	3.1(7)	&	5(1)	&	0.7(1.2)	&	1.5(1)	\\
43	&	GEM	&	9.9(2)	&	12.0(7)	&	4.3(7)	&	3(2)	&	1805(469)	&	4(1)	&	7(2)	&	0.8(1.4)	&	1.7(1)	\\
44	&	GEM	&	9.7(2)	&	11.9(7)	&	5.3(8)	&	3(2)	&	1784(459)	&	6(1)	&	8(2)	&	0.9(1.5)	&	1.7(1)	\\
45	&	GEM	&	9.3(1)	&	11.5(7)	&	8(1)	&	4(3)	&	1741(440)	&	8(2)	&	12(3)	&	1.0(1.7)	&	1.9(1)	\\
48	&	spo	&	9.6(1)	&	11.9(7)	&	5.6(7)	&	3(2)	&	1777(455)	&	6(1)	&	40(9)	&	1.7(6)	&	1.6(1)	\\
57	&	spo	&	10.5(3)	&	12.5(7)	&	2.2(7)	&	1.4(9)	&	1884(501)	&	2.4(6)	&	17(4)	&	1.3(4)	&	1.3(1)	\\
60	&	spo	&	9.8(2)	&	12.0(7)	&	4(1)	&	2(2)	&	1813(500)	&	4(1)	&	30(7)	&	1.6(5)	&	1.5(1)	\\
77	&	spo	&	11.6(5)	&	13.5(8)	&	0.7(3)	&	0.5(4)	&	2013(551)	&	0.9(3)	&	6(2)	&	0.9(3)	&	0.9(1)	\\
79	&	spo	&	9.7(1)	&	11.9(7)	&	5.1(9)	&	3(2)	&	1786(447)	&	5(1)	&	37(9)	&	1.7(6)	&	1.6(1)	\\
80	&	spo	&	8.7(1)	&	11.1(7)	&	13.3(3)	&	6(4)	&	1675(433)	&	13(3)	&	91(19)	&	2.3(8)	&	2.1(1)	\\
82	&	GEM	&	10.1(2)	&	12.3(7)	&	3.7(7)	&	2(1)	&	1826(466)	&	3.9(9)	&	6(1)	&	0.8(1.3)	&	1.6(1)	\\
83	&	GEM	&	8.9(1)	&	11.3(7)	&	11(1)	&	5(4)	&	1700(433)	&	11(2)	&	17(4)	&	1.1(1.9)	&	2.1(2)	\\
86	&	spo	&	9.0(2)	&	11.3(7)	&	10(1)	&	5(3)	&	1714(400)	&	10(2)	&	67(15)	&	2.1(7)	&	1.9(1)	\\
87	&	spo	&	9.7(2)	&	11.9(7)	&	5.1(7)	&	3(2)	&	1788(480)	&	5(1)	&	36(8)	&	1.7(6)	&	1.6(1)	\\
91	&	spo	&	9.6(2)	&	11.8(7)	&	5(1)	&	3(2)	&	1792(479)	&	5(1)	&	35(8)	&	1.7(6)	&	1.6(1)	\\
92	&	spo	&	9.6(2)	&	11.8(7)	&	6(1)	&	3(2)	&	1770(458)	&	6(1)	&	42(10)	&	1.8(6)	&	1.7(1)	\\
99	&	spo	&	9.5(2)	&	11.7(7)	&	6(1)	&	3(2)	&	1778(464)	&	6(1)	&	39(9)	&	1.7(6)	&	1.6(1)	\\
\hline																					
\end{tabular}
\tablefoot{All values, except for $m_{\rm I}$ and $E_{\rm I}$, are estimations based on Eqs.~\ref{eq:fluxR}, \ref{eq:temp}, \ref{eq:E}, \ref{eq:eta2}, \ref{eq:LELR}, \ref{eq:mass}, \ref{eq:radius}, \ref{eq:crater}.}
\end{table*}

The crater size (i.e. diameter $d_{\rm c}$) can be calculated using the scaling law of \citet{GAU74} \citep[c.f.][]{MEL89}:
\begin{equation}
d_{\rm c}=0.25\rho_{\rm p}^{1/6}~\rho_{\rm M}^{-1/2} g_{\rm M}^{-0.165} E_{\rm kin,~p}^{0.29} \sin^{1/3}\vartheta~{\rm in~m},
\label{eq:crater}
\end{equation}
where $\rho_{\rm M}=1.6$~g~cm$^{-3}$ the density of the lunar regolith \citep{MIT73, HAY17}, $g_{\rm M}=1.67$~m~s$^{-2}$ the gravity acceleration value of the Moon, and $\vartheta=45\degr$ the assumed average impacting angle \citep[c.f.][]{BOU12, MAD15b}.

For the consistency of the units, in Eq.~\ref{eq:mass} the $V_{\rm p}$ should be given in m~s$^{-1}$, in Eq.~\ref{eq:radius} the $\rho_{\rm p}$ should be given in g~cm$^{-3}$ and the $m_{\rm p}$ in g, and in Eq.~\ref{eq:crater} the $\rho_{\rm M}$ and $\rho_{\rm p}$ should be given in kg~cm$^{-3}$, $E_{\rm kin}$ in J, and $g_{\rm M}$ in m~s$^{-2}$.

The results for the physical parameters of the validated flashes are given in Tables~\ref{tab:ResultsReal} and \ref{tab:ResultsReal2}. In addition, rough estimations for the parameters (maximum values for a given $\eta$) of the suspected flashes are listed in Table~\ref{tab:ResultsSusp}. The latter parameters are based in the calculation of $E_{\rm R,~max}$ from the correlation between $E_{\rm R}-E_{\rm I}$ for the validated flashes (Eq.~\ref{eq:LELR}). It is noticed that the assumed $V_{\rm p}$ and $\rho_{\rm p}$ values according to the origin of the projectiles are gathered in Table~\ref{tab:streams}. The large discrepancies between the values of the first ten validated flashes of this study with those of Paper I are due to the different methods followed for the derivation of the physical parameters and assumptions for the $\eta$.


\section{Correlations and distributions}
\label{sec:CorDib}
This section is dedicated in exploring possible correlations between the parameters of the meteoroids as well as to their frequency distributions. We investigated the correlation between the total $E_{\rm lum}$ with the temperature. For this, we used the $E_{\rm lum}$ ($=E_{\rm R}+E_{\rm I}$) values that correspond to the set of frames from which the respective peak temperature was calculated, and not the total $E_{\rm lum}$ of the flash. Moreover, in order to check any possible dependence of the peak temperature on the composition of the soil, where the impact occurred, the area of each flash was characterized as bright, grey or dark based on visual inspection on high detailed lunar images (see Appendix~\ref{sec:LOC}). In addition, flashes are further characterized based on their frame multiplicity (i.e. total duration), because, as mentioned in Section~\ref{sec:LE1}, for the single frame flashes there is uncertainty regarding the total measured flux, hence, the temperature, but for the multi-frame flashes is certain that a part of the energy has not been recorded. The second correlation is between the mass of the meteoroid, as derived using the total $E_{\rm lum}$ of the flash (see Section~\ref{sec:MRC}), and the temperature of the flash. The diagrams of temperature against $E_{\rm lum}$ (upper panel) and $m_{\rm p}$ (lower panel) for $\eta=1.5\times10^{-3}$ are shown in Fig.~\ref{fig:LEMT}. Both diagrams show no obvious correlation between these quantities no matter their frame multiplicity or the composition of the soil. In matter of fact, they seem uncorrelated and that for any $E_{\rm lum}$ or mass value of the meteoroid, a flash temperature between 1700-6000~K may be developed. At a first glance, this result seems to be counterintuitive. However, it should be taken into account that the materials have specific heat capacity and thermal conductivity, hence, after a specific temperature threshold their temperature cannot be increased any more. Therefore, it seems that high energy meteoroids produce heat that melts the lunar material as well themselves but the temperature cannot exceed the threshold that depends strongly on both materials \citep[c.f. section 3 in][]{BOU12}.

\begin{figure}
\begin{tabular}{c}
\includegraphics[width=\columnwidth]{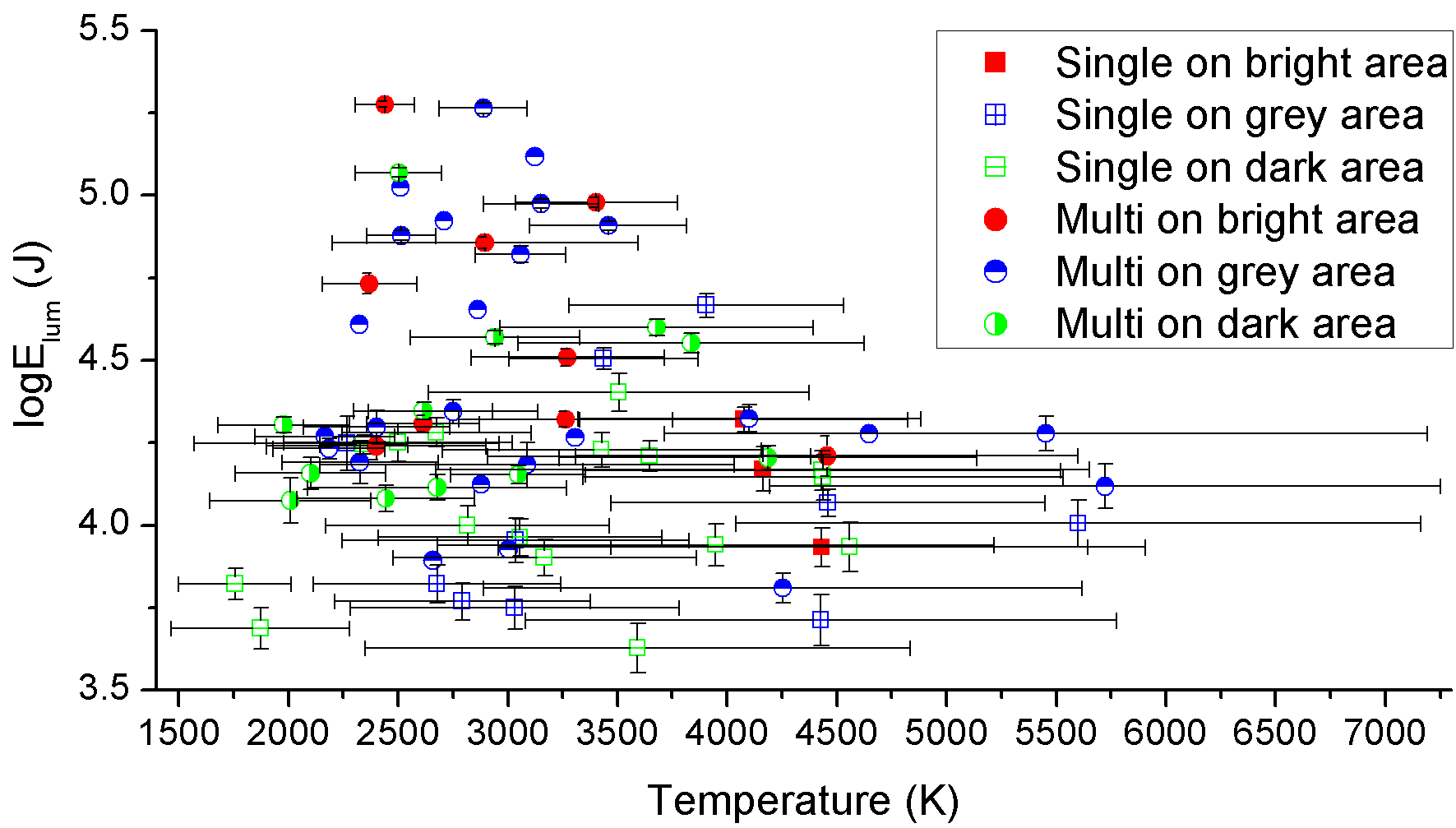}\\
\includegraphics[width=\columnwidth]{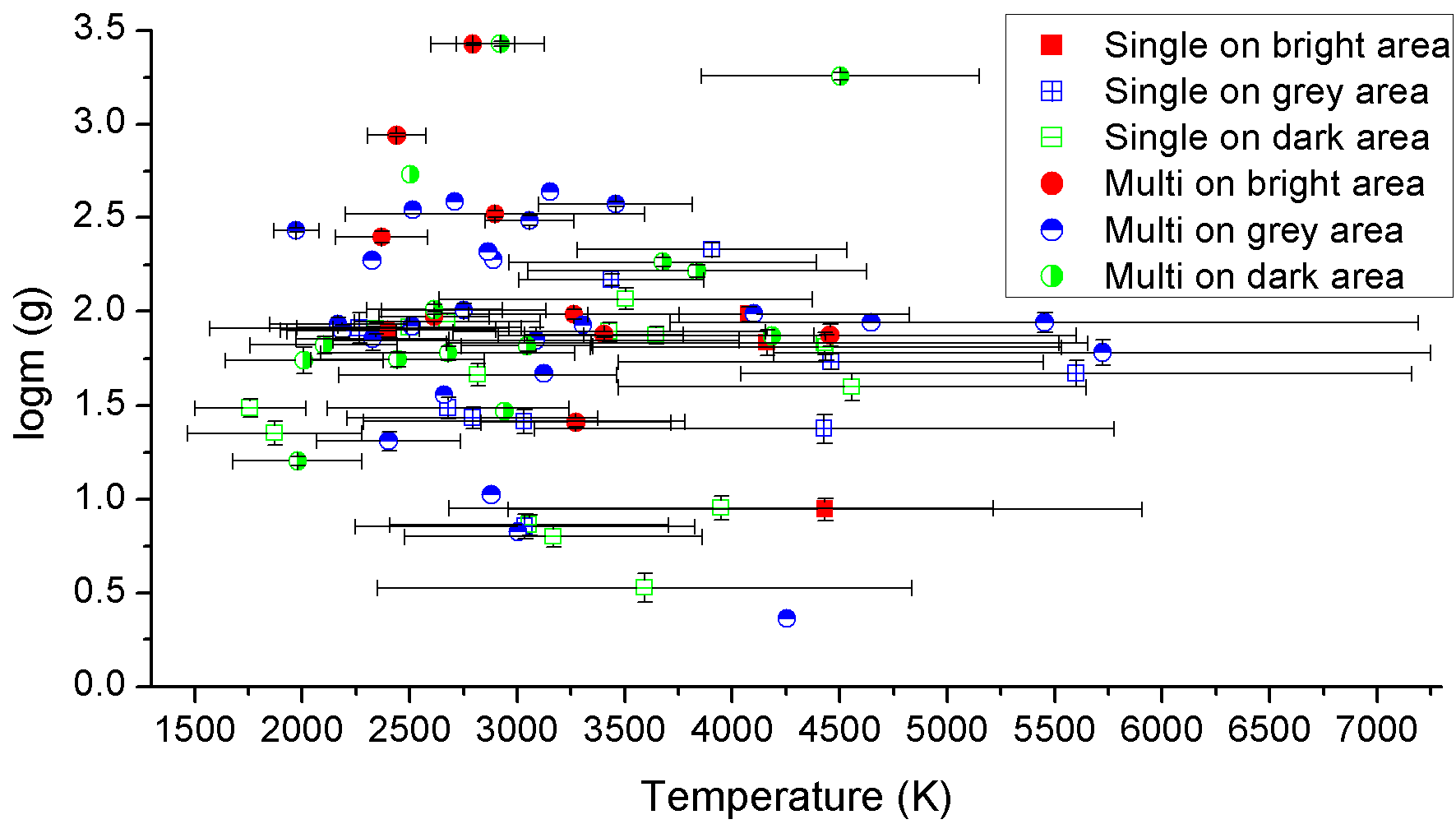}\\
\end{tabular}
\caption{Correlation between the peak temperatures of the validated flashes against their respective $E_{\rm lum}$ (upper panel) and against the $m_{\rm p}$ of the respective meteoroids (lower panel) for $\eta=1.5\times10^{-3}$. Squares denote the single frame flashes, while circles the multi-frames. Red, blue, and green colours denote that the flashes were detected in bright, grey, and dark lunar areas, respectively.}
\label{fig:LEMT}
\includegraphics[width=\columnwidth]{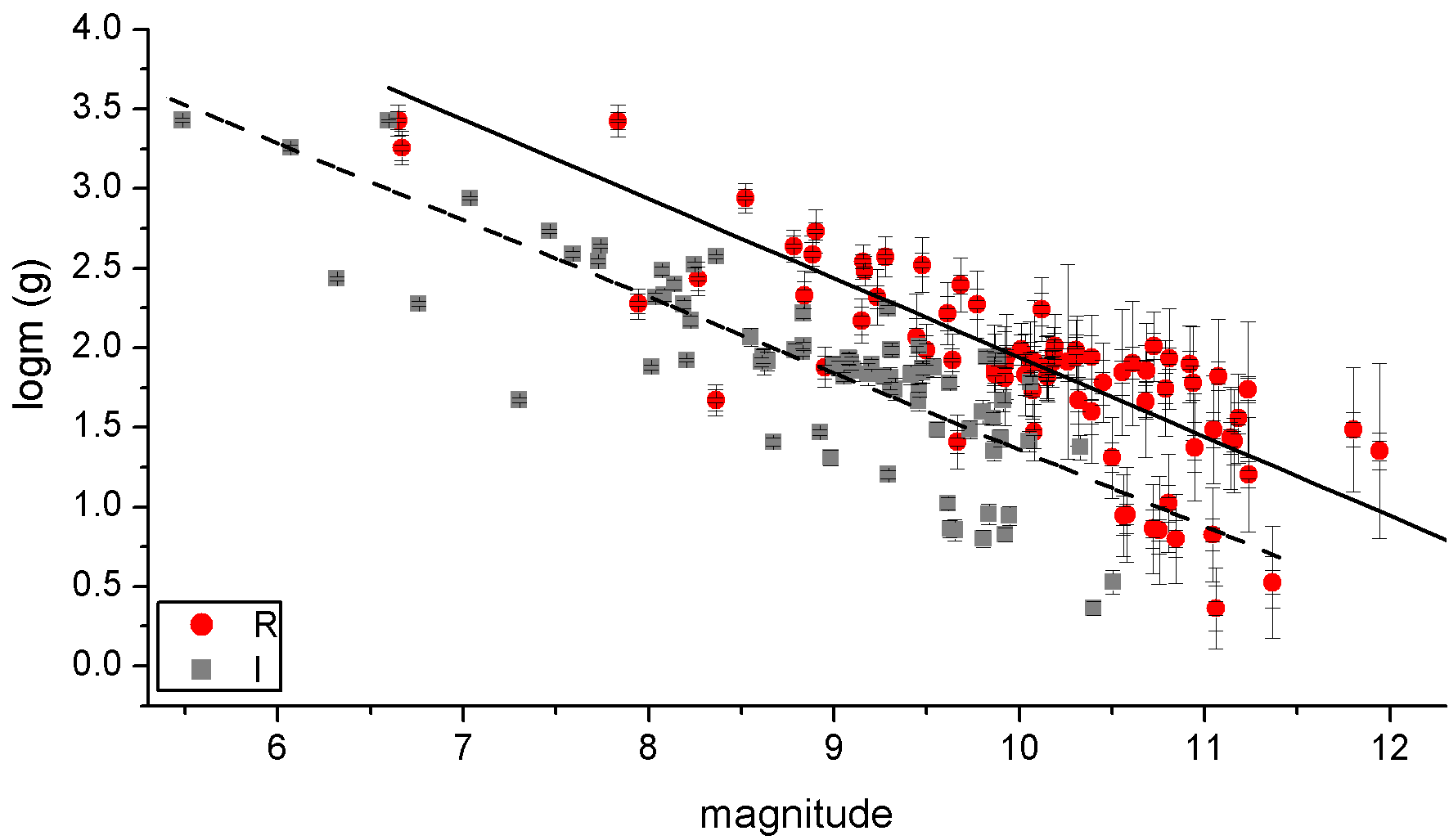}\\
\caption{Correlation between the observed magnitudes (red circles=$R$ band and grey squares=$I$ band) of the validated flashes against the meteoroid masses for $\eta=1.5\times10^{-3}$. Empirical linear fits on the individual data sets for $R$ (solid line) and $I$ (dashed line) are also presented.}
\label{fig:magmass}
\end{figure}
\begin{figure}
\includegraphics[width=\columnwidth]{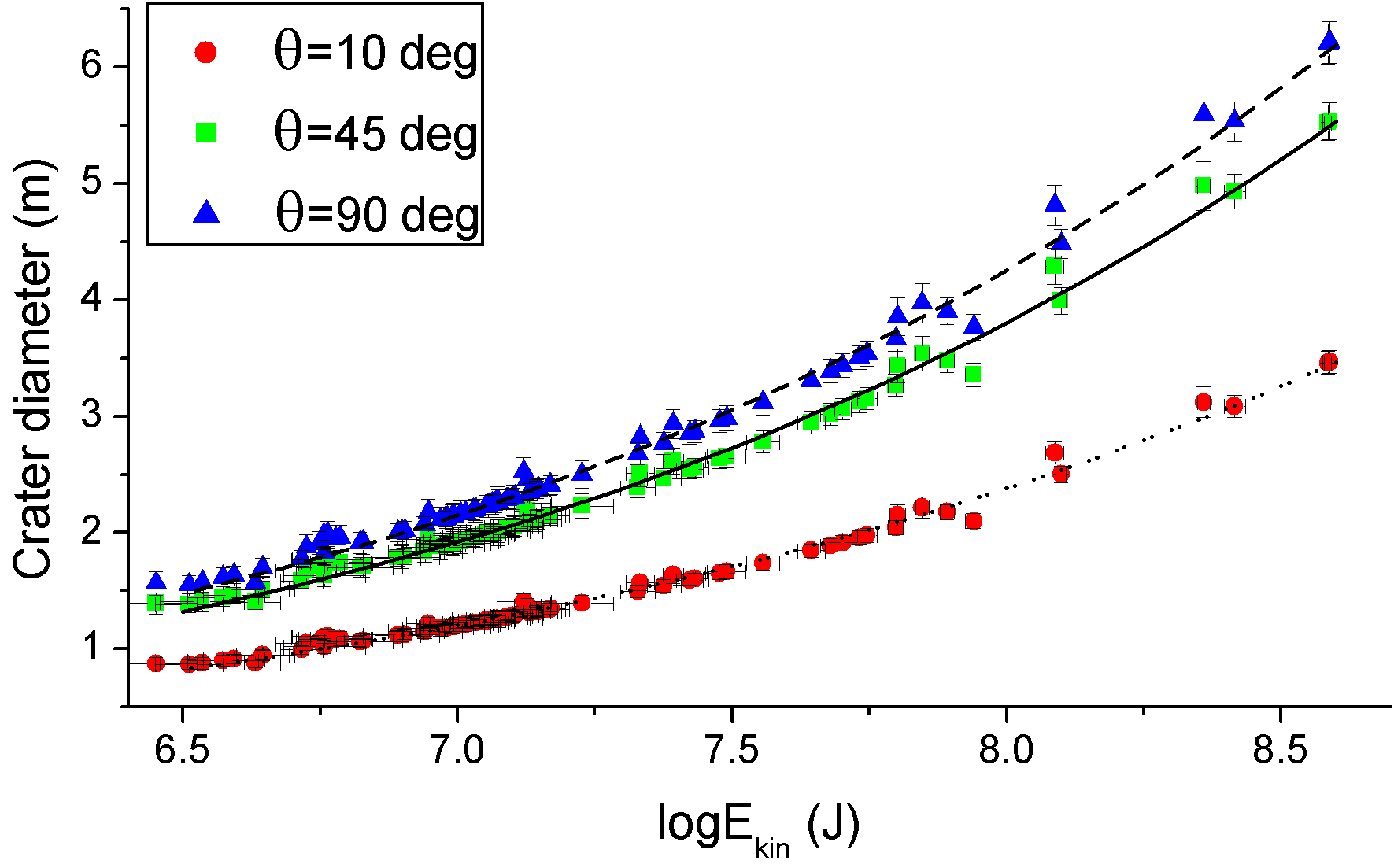}\\
\caption{Crater sizes against the $E_{\rm kin}$ of the meteoroids for various impact angles for $\eta=1.5\times10^{-3}$. The data points are based on the validated impact flashes. Lines correspond to the respective power laws based on Eq.~\ref{eq:crater}.}
\label{fig:crater}
\end{figure}

The observed magnitudes of the flashes against the corresponding calculated masses of the meteoroids for $\eta=1.5\times10^{-3}$ are plotted in Fig.~\ref{fig:magmass}. Likely Fig.~\ref{fig:MagEREI}, this plot clearly shows the dependence of the mass on the emitted energy (magnitude), i.e. the more massive the meteoroid the brightest the flash. This is very plausible, since the $E_{\rm lum}$ is directly connected to both the emitted energies from each band (see Sections~\ref{sec:Temperature} and \ref{sec:LE1}) and the mass through the relation of $E_{\rm kin}$ and $\eta$. In order to make a first correlation between the magnitude of the flash $m_{\lambda}$ and the mass $m$ of the projectile, we used linear functions to fit the data sets of $R$ and $I$ bands, respectively, and they are shown in Fig.~\ref{fig:magmass}. These fittings resulted in the following correlations:
\begin{equation}
\log m=6.92(\pm0.02)-0.498(\pm0.002) m_{\rm R}~{\rm with}~r=0.82
\label{eq:m-R}
\end{equation}
and
\begin{equation}
\log m=6.17(\pm0.02)-0.481(\pm0.002)m_{\rm I}~{\rm with}~r=0.81,
\label{eq:m-I}
\end{equation}
where $r$ is the correlation coefficient.

The distributions of the impact crater sizes against the $E_{\rm kin}$ of the projectiles are plotted in Fig.~\ref{fig:crater}. The diameters of the craters range between 0.87-8.5~m (depending on angle) and can be larger by a factor of $\sim1.4$ for the extreme value of $\eta=5\times10^{-4}$. Fig.~\ref{fig:DistMR} shows the distributions of masses and diameters of the meteoroids associated with lunar impact flashes found during the NELIOTA campaign. For $\eta=5\times10^{-3}$ the majority of the meteoroids have masses less than 100~g and sizes less than 5~cm. For $\eta=5\times10^{-4}$ the respective masses are less than 1~kg and the sizes less than 12~cm. For all $\eta$ values the masses and sizes of the meteoroids of the suspected flashes have small values. Taking into account only the validated flashes, for the extreme value of $\eta=5\times10^{-3}$ the least massive meteoroid has a mass value of 0.7~g and a size of 0.93~cm, while for the other extreme value of $\eta=5\times10^{-4}$, the most massive meteoroid has a mass of 8.06~kg and a size of 19.98~cm.

\begin{figure*}
\begin{tabular}{cccc}
\includegraphics[width=4.2cm]{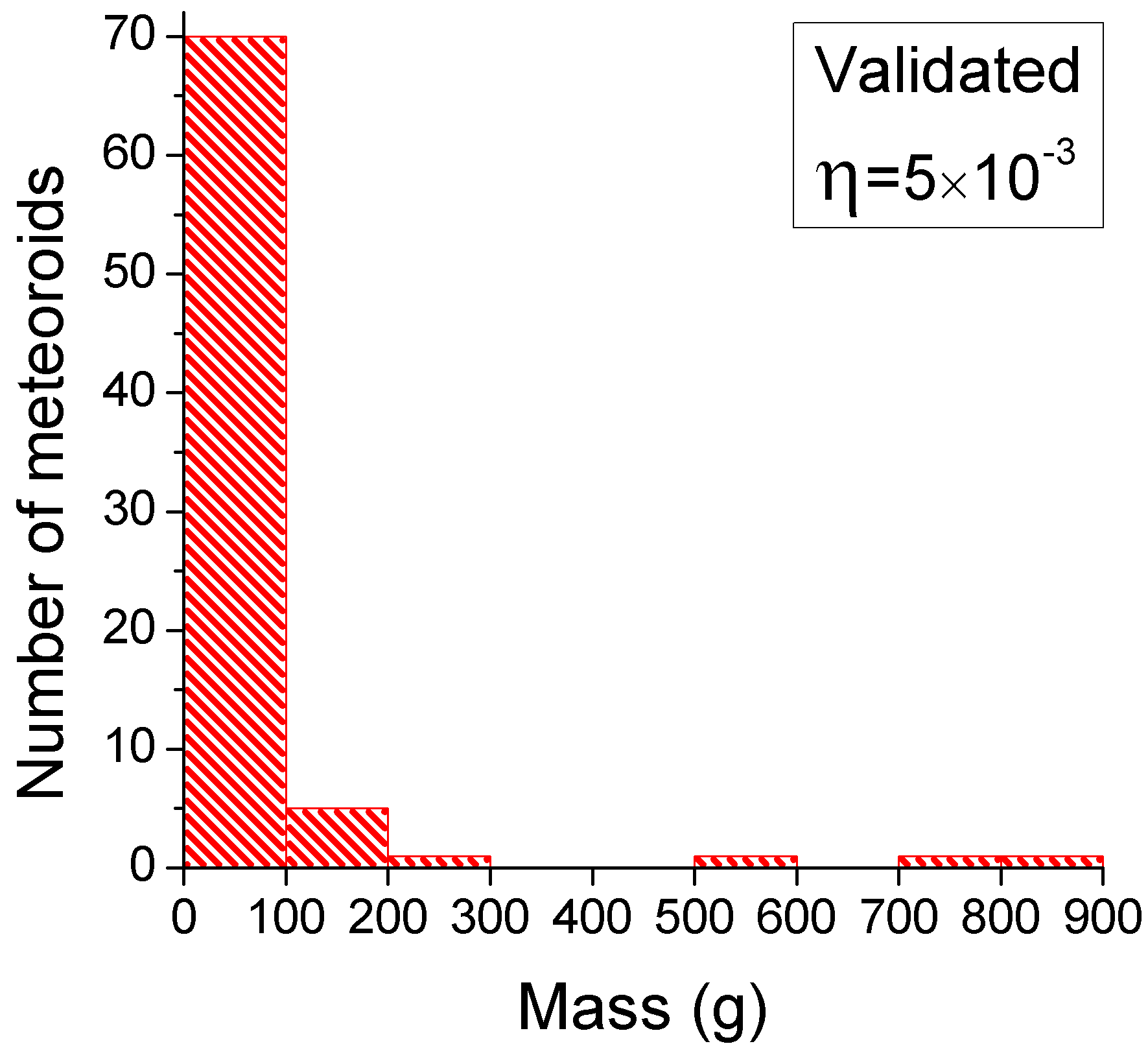}&\includegraphics[width=4.2cm]{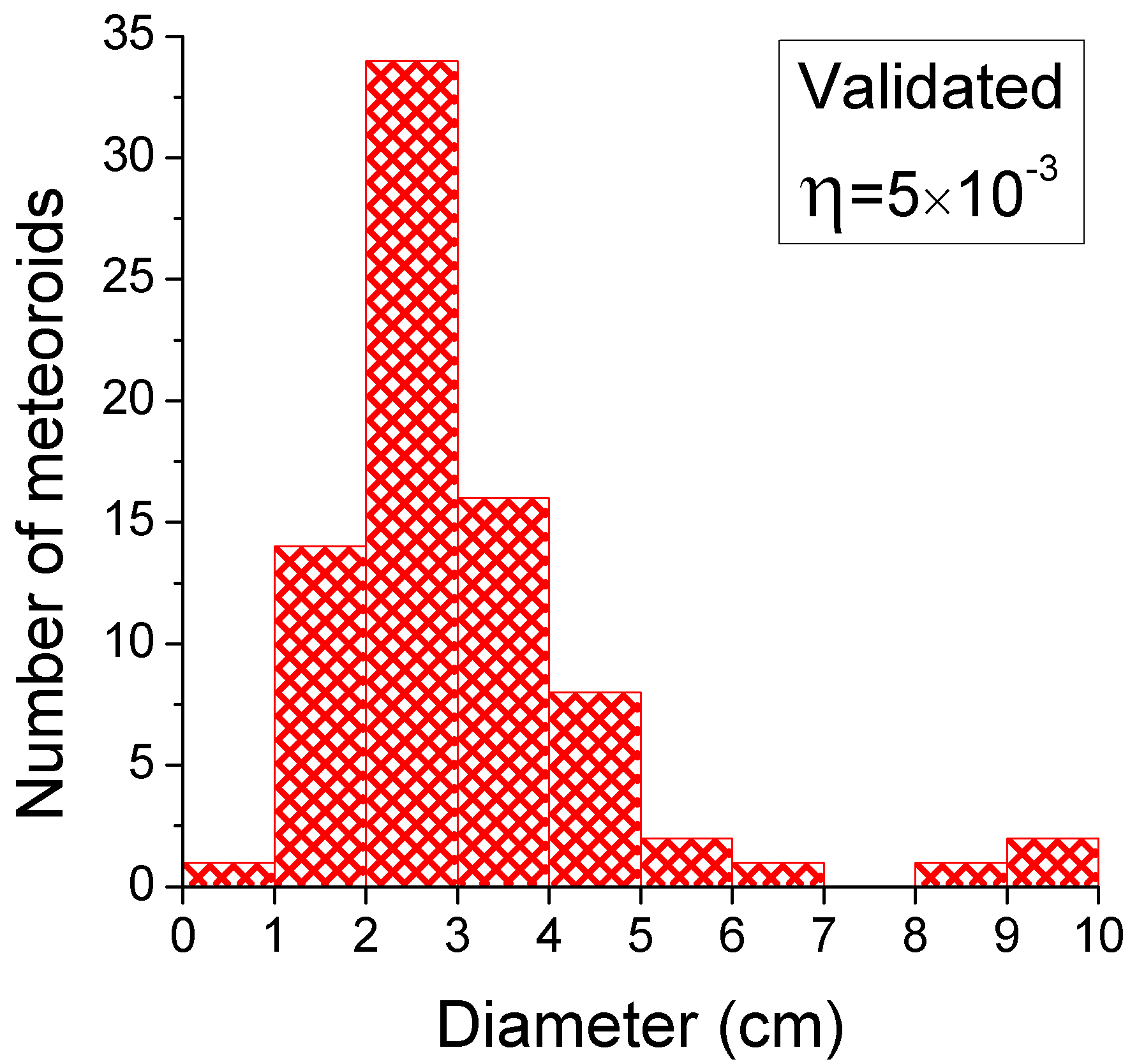}&\includegraphics[width=4.2cm]{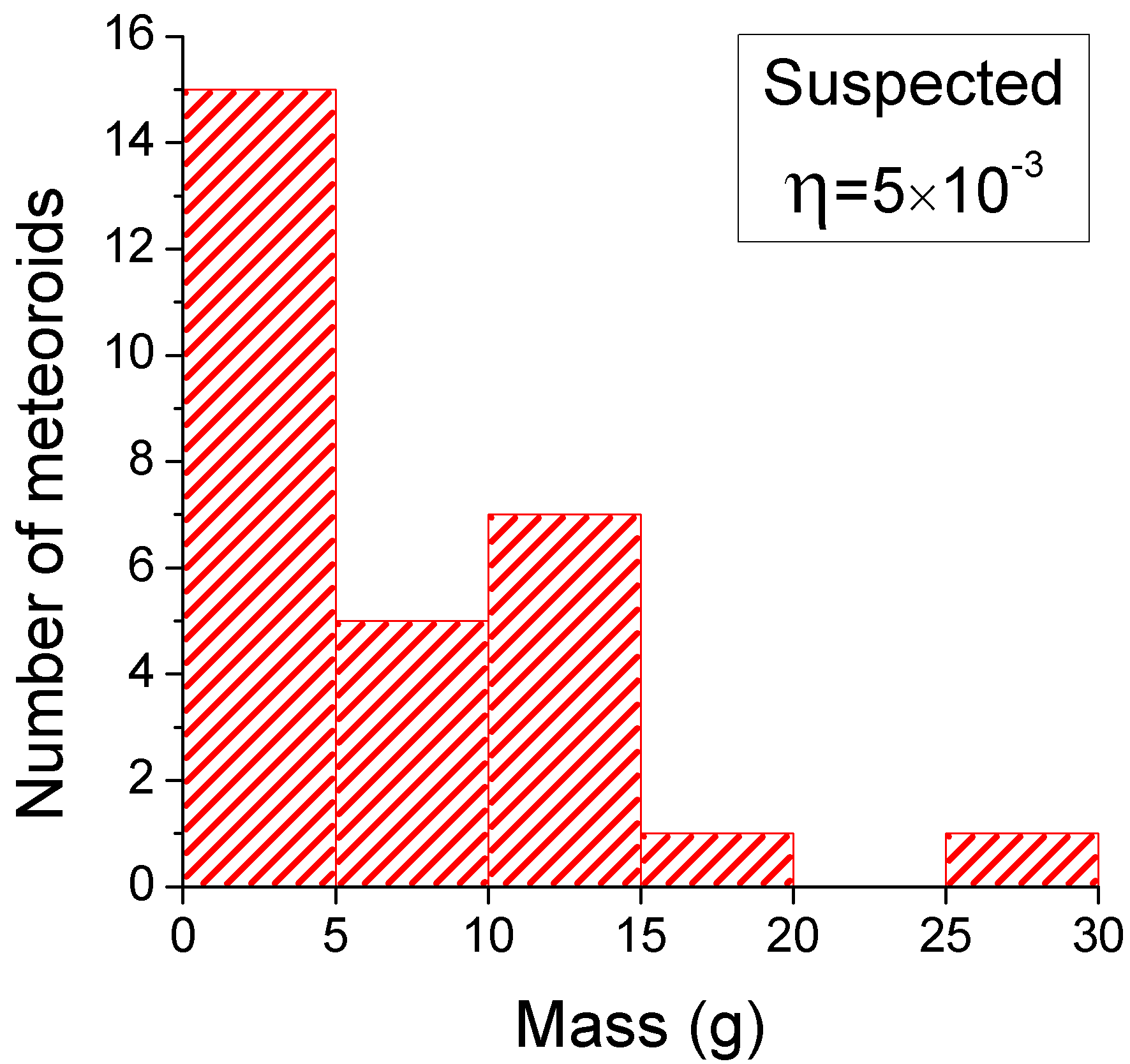}&\includegraphics[width=4.2cm]{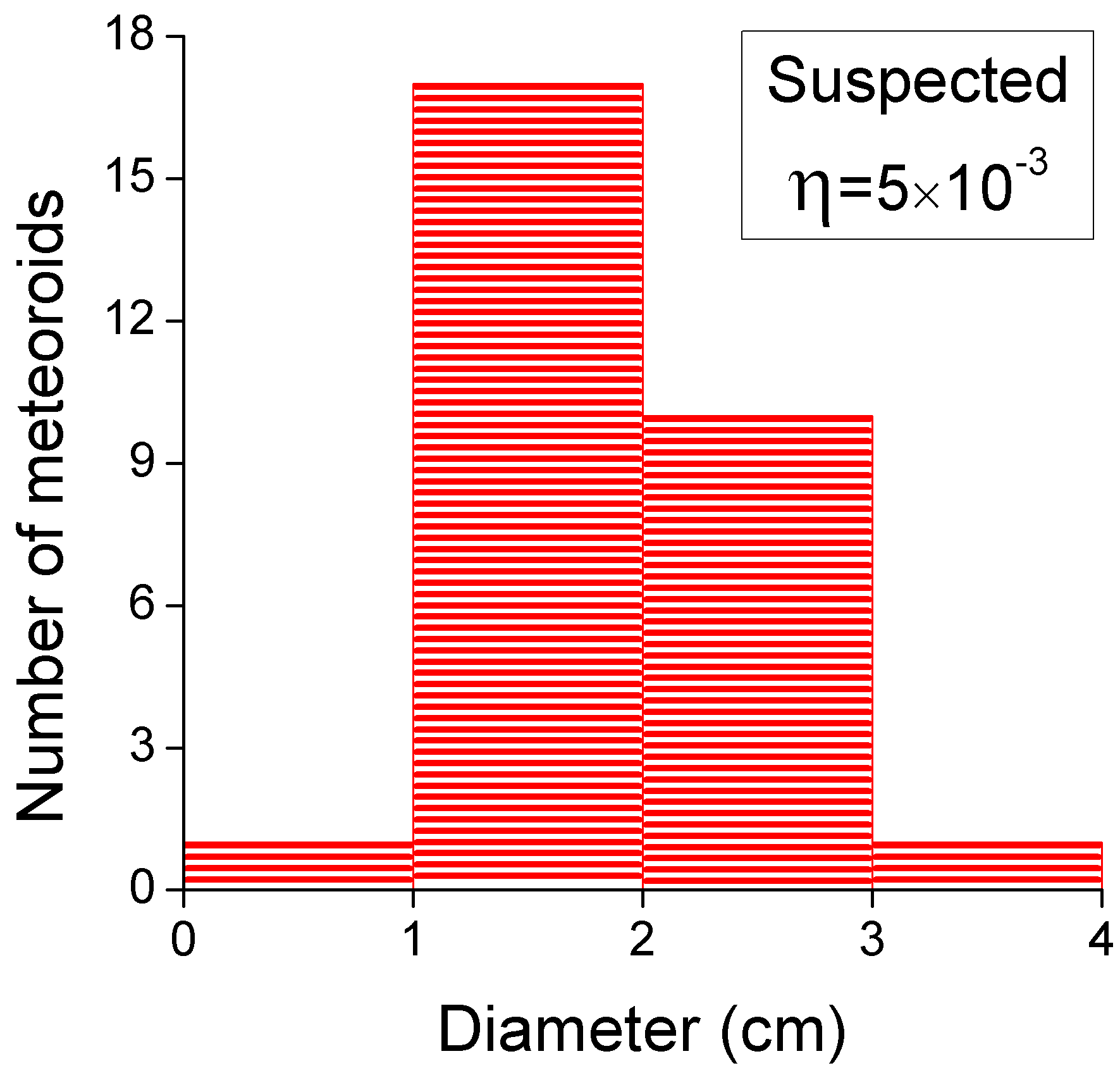}\\
\includegraphics[width=4.2cm]{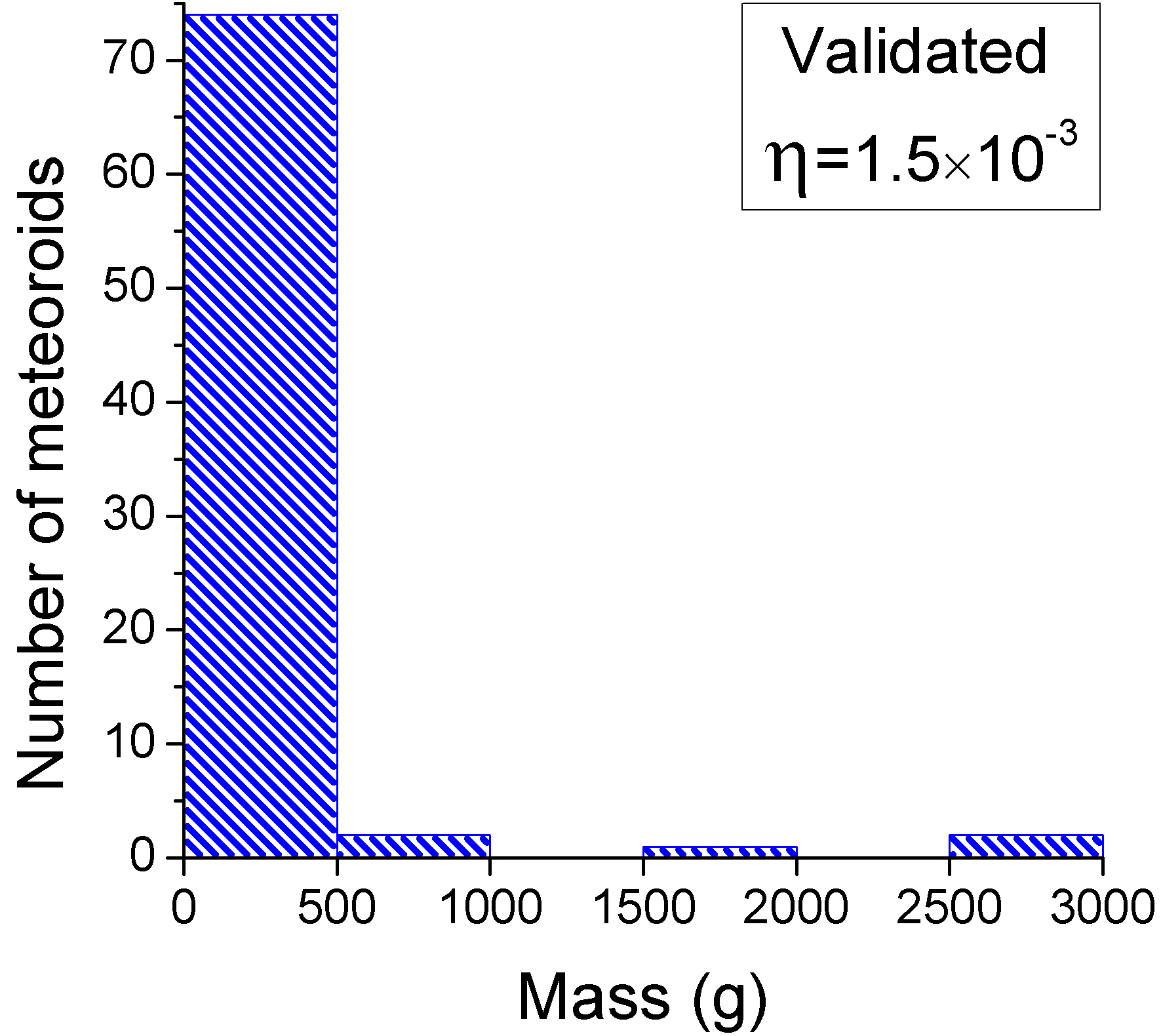}&\includegraphics[width=4.2cm]{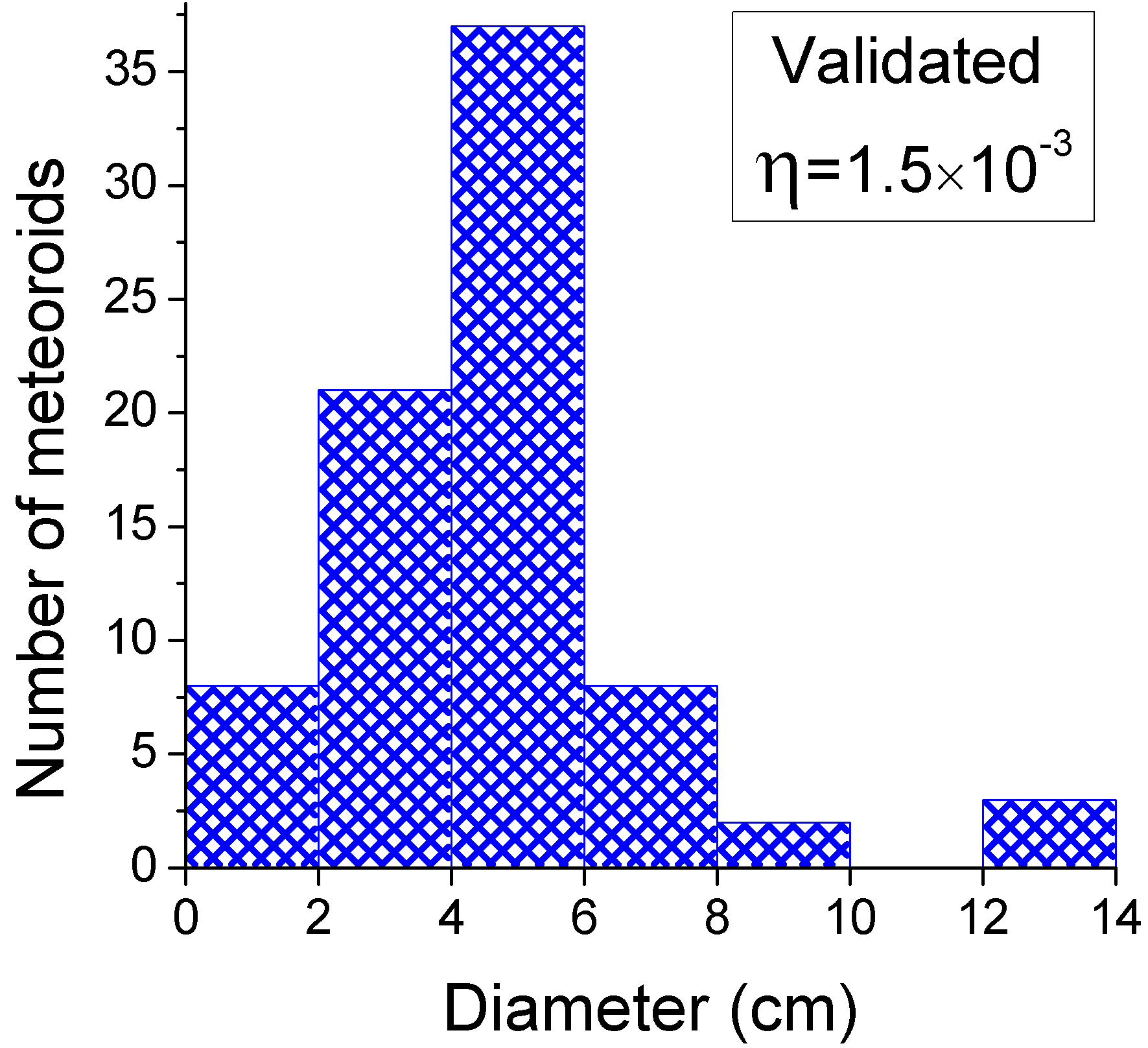}&\includegraphics[width=4.2cm]{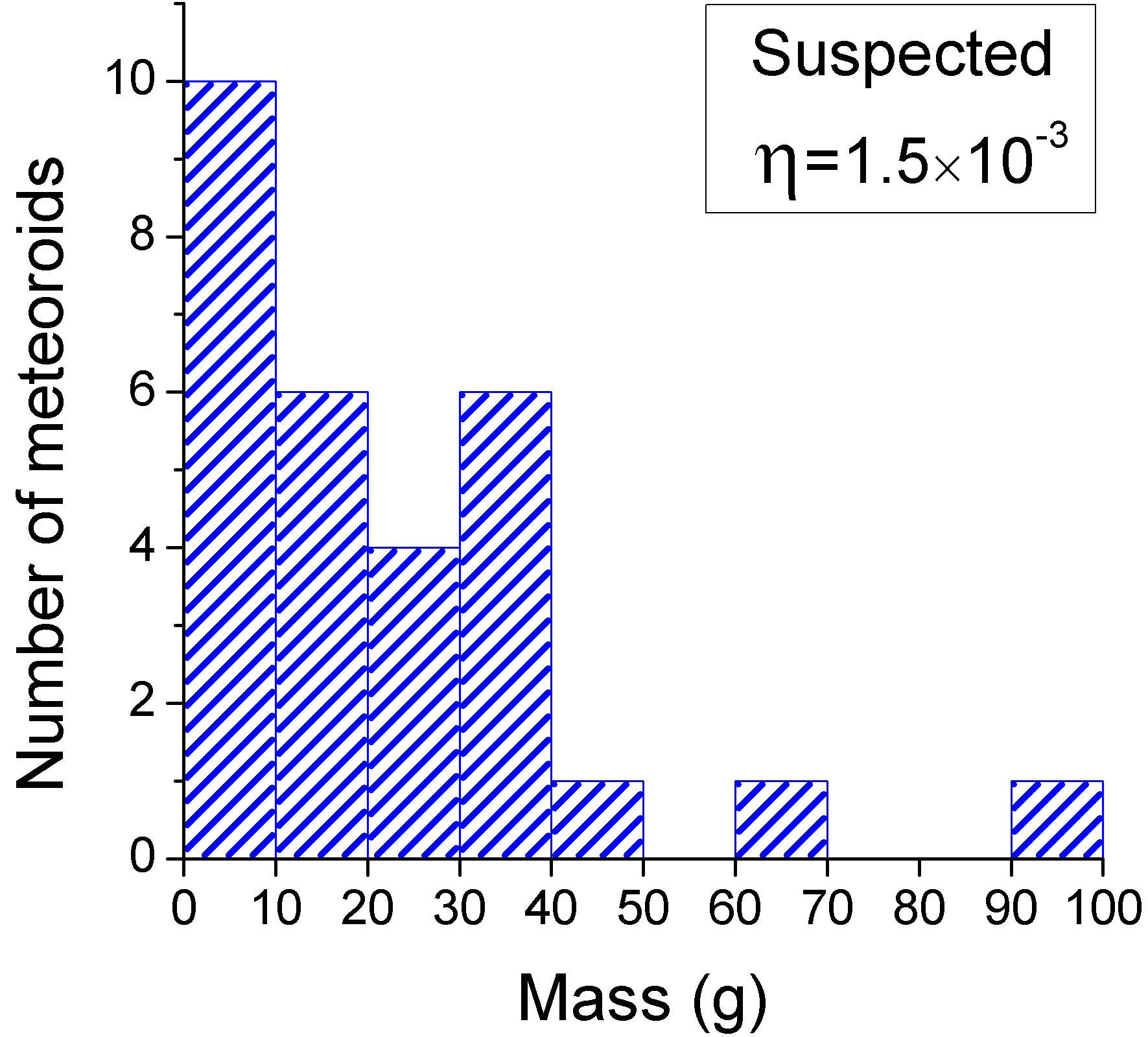}&\includegraphics[width=4.2cm]{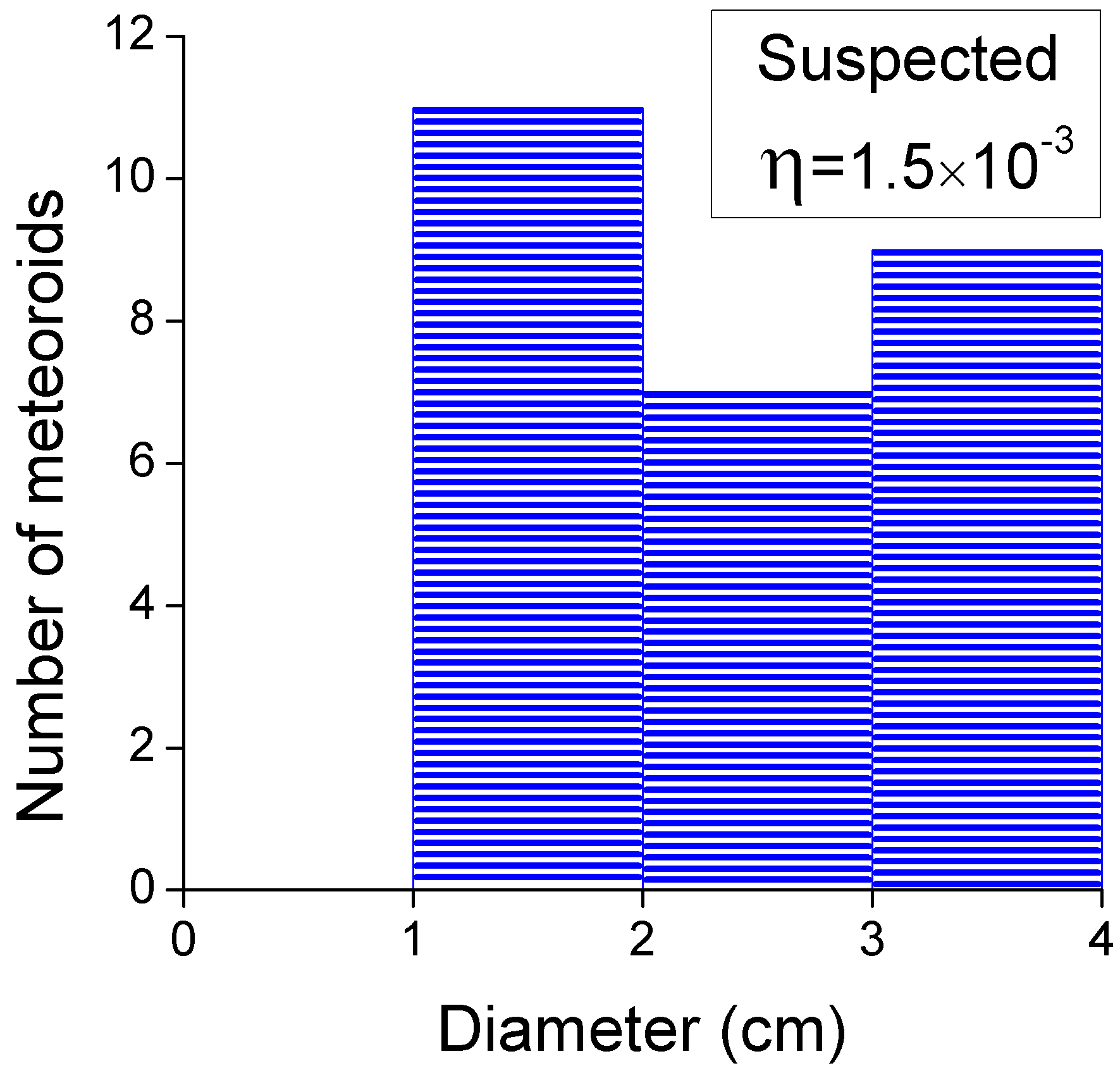}\\
\includegraphics[width=4.2cm]{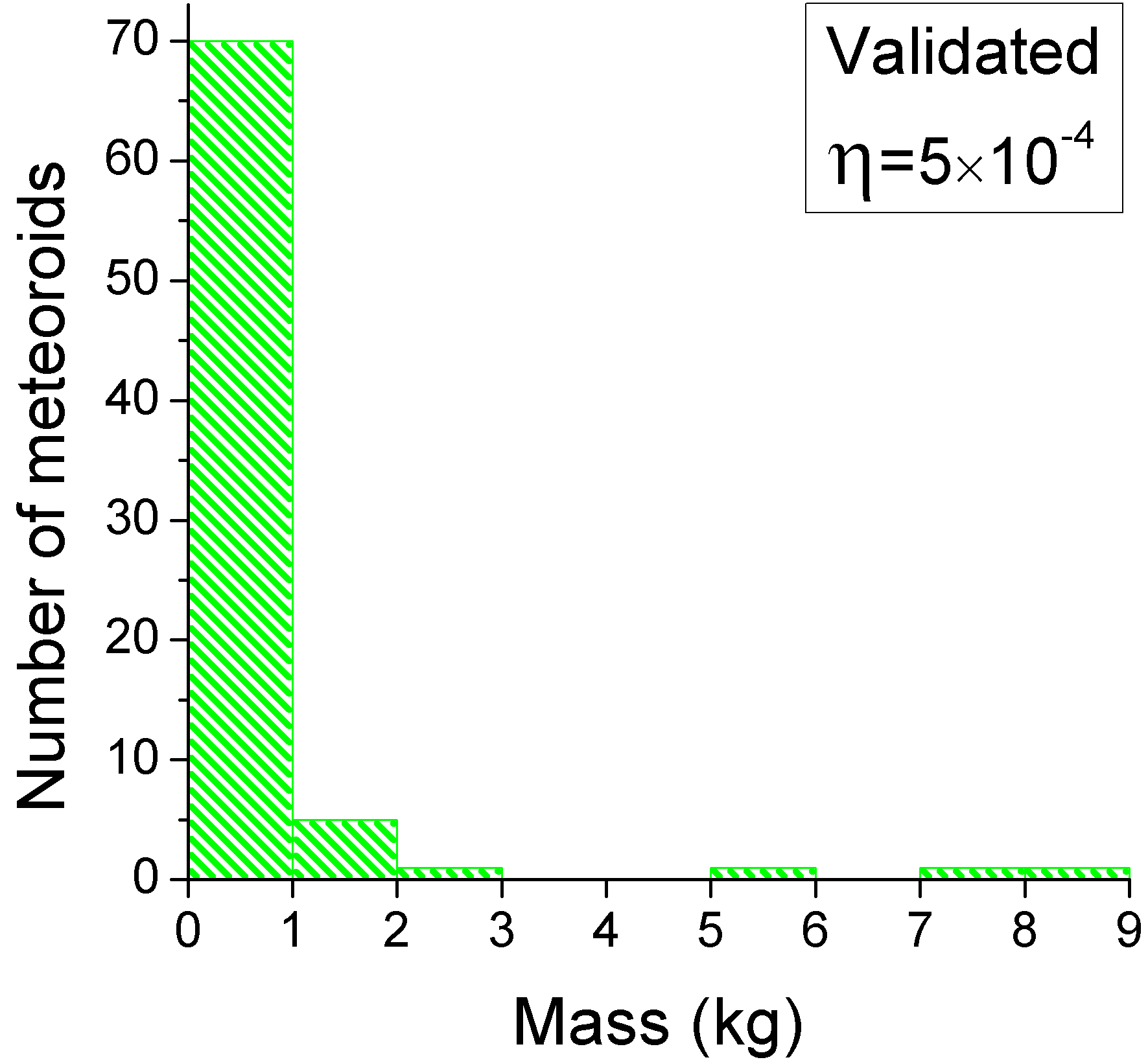}&\includegraphics[width=4.2cm]{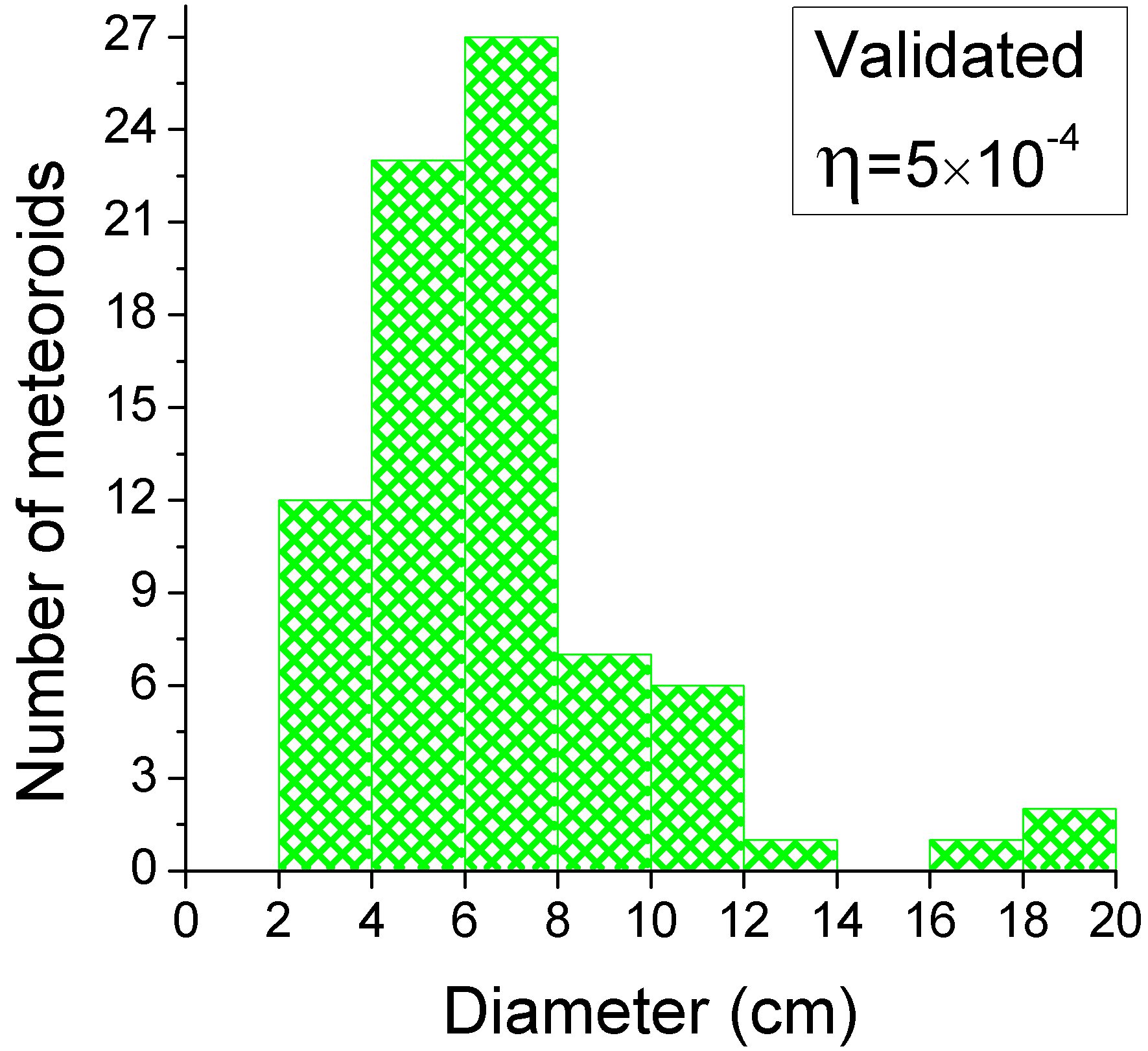}&\includegraphics[width=4.2cm]{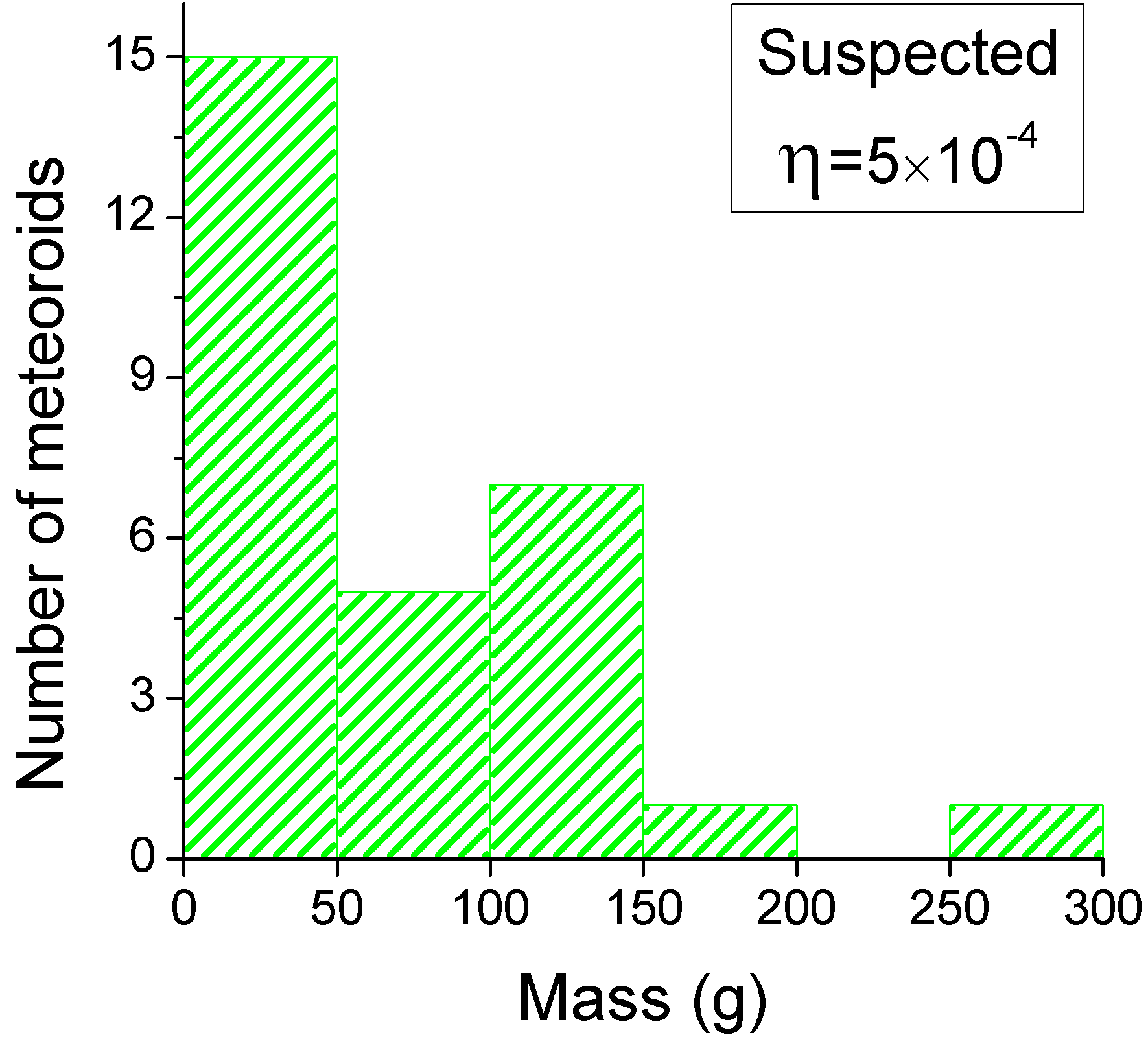}&\includegraphics[width=4.2cm]{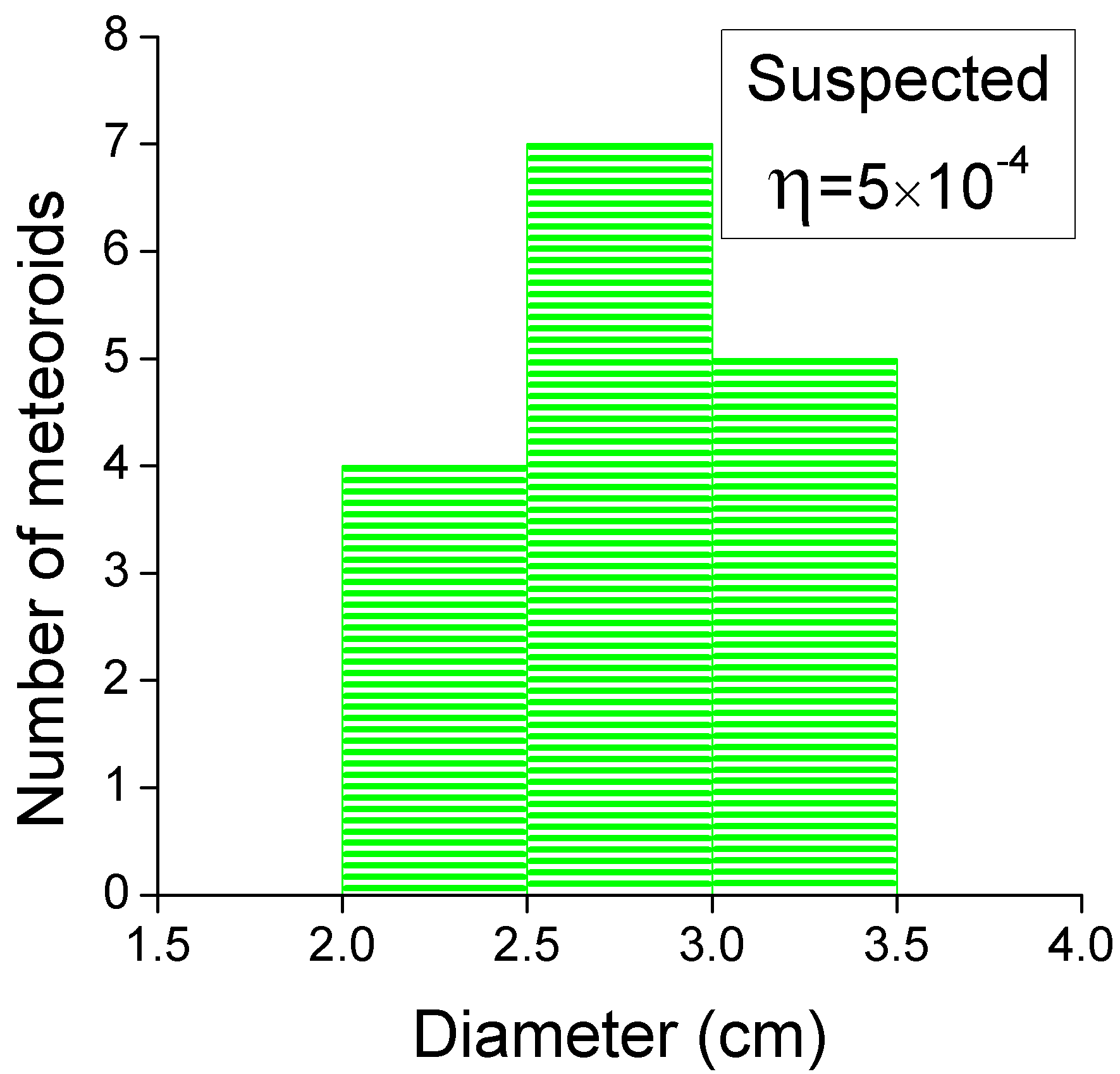}\\
\end{tabular}
\caption{Distributions of masses and diameters of meteoroids for various values of $\eta$ associated with the detected lunar impact flashes during the NELIOTA campaign. Left panels correspond to validated and right panels to suspected flashes, respectively.}
\label{fig:DistMR}
\end{figure*}


\section{Meteoroid flux rates and extrapolation to Earth}
\label{sec:Rates}

Using the total observed hours on the Moon and the total covered area (i.e. FoV; see Section~\ref{sec:Obs}) it is feasible to calculate the detection rate of NELIOTA setup. It should be noted that the observed area varies between the observing runs because: 1) it depends on the Earth-Moon distance and 2) during relatively bright lunar phases (i.e. $>0.35$) the covered area was less up to $50\%$ in order to avoid the glare (these nights are the longest in duration). The latter reason results in a slight underestimation of the derived rates, which can be considered as the minimum ones. Therefore, we approximate the mean covered lunar area from our equipment as $\sim3.11\times10^6$~km$^2$ at an Earth-Moon distance of $3.84\times10^5$~km. Another important issue that has to be taken into account concerns the meteoroid origin. Obviously, the frequency of the meteoroids varies as the Earth orbits Sun because of the passage of our planet through the cometary/asteroidal debris. For this, it was preferred to calculate various rates according to the possible association of the meteoroids with active streams. The flashes whose corresponding meteoroids are possibly associated with streams are listed in Tables~\ref{tab:ResultsReal}-\ref{tab:ResultsSusp}. So, the observed hours during the nights when streams found to be active (i.e. detection of flashes from stream meteoroids; see Section~\ref{sec:streamassociation}) were subtracted from the total observed time, and a different rate that corresponds only to the streams is calculated. In addition, rates taking into account the suspected flashes too are also calculated and can be considered as the upper values of the respective ones of the validated flashes. The observed hours, the number of validated and suspected flashes and the detection rates for each case are given in Table~\ref{tab:rates}.

\begin{table}
\centering
\caption{Flash detection rates of NELIOTA according to the origin of the meteoroids.}
\label{tab:rates}
\scalebox{0.93}{
\begin{tabular}{c ccc |ccc}
\hline													
\hline													
	&	\multicolumn{ 3}{c}{Validated}					&	\multicolumn{ 3}{c}{Validated and suspected}					\\
\hline													
	&	Spo	&	Str	&	Sum	&	Spo	&	Str	&	Sum	\\
\hline													
Obs. Hours	&	95.66	&	14.82	&	110.48	&	95.66	&	14.82	&	110.48	\\
Detections	&	61	&	18	&	79	&	78	&	30	&	108	\\
Rate$^a$ 	&	2.05	&	3.90	&	2.30	&	2.62	&	6.51	&	3.14	\\
\hline																					
\end{tabular}}
\tablefoot{$^a$ in units of 10$^{-7}$~meteoroids~hr$^{-1}$~km$^{-2}$, Spo=sporadic, Str=stream}
\end{table}

\begin{table}
\centering
\caption{Appearance frequencies of meteoroids on and around Moon and Earth.}
\label{tab:freqs}
\scalebox{0.9}{
\begin{tabular}{c| ccc |ccc}
\hline													
\hline													
	&		&	\multicolumn{ 2}{c}{Moon}			&	\multicolumn{3}{c}{Earth}					\\
\hline													
	&		&	Sur	&	Orbit	&	MS	&	LEO	&	GEO	\\
Distance (km)	&		&		&	70	&	90	&	2000	&	36000	\\
\hline													
validated 	&	Spo	&	7.8	&	8.4	&	108	&	181	&	3338	\\
(meteoroids~hr$^{-1}$)         &	Str	&	14.8 &	16.0&	205	&	344	&	6356	\\
	                  &	Total	&	8.7	&	9.4	&	121	&	202	&	3742	\\
\hline													
validated and	&	Spo	    &	9.9	    &	10.8	&	138	&	231	&	4268	\\
suspected     &	Str	    &	24.7	&	26.7	&	341	&	573	&	10594	\\
 (meteoroids~hr$^{-1}$)&	Total	&	11.9	&	12.9	&	165	&	277	&	5117	\\	
\hline													
\end{tabular}}
\tablefoot{Spo=sporadic, Str=stream, Sur=surface, MS=Mesosphere, LEO=Low Earth Orbit, GEO=Geosynchronous Earth Orbit.}
\end{table}	

Based on the various detection rates (i.e. taking into account only the validated flashes) it is feasible to calculate the meteoroid appearance frequency distributions for various distances from Earth or on/from the Moon (i.e. for specific orbit distances). All the rates are based on the assumption of an isotropic distribution of the meteoroids in space i.e. all areas on Earth or Moon have the same probability to be hit. Appearance frequency values for specific distances from the Moon and the Earth are listed in Table~\ref{tab:freqs}. Fig.~\ref{fig:FreqD} illustrates the distributions for various distances from the surface of the Earth according to the respective rates found for the cases of: 1) validated sporadic, 2) validated stream, 3) validated sporadic and stream, 4) validated and suspected sporadic, 5) validated and suspected stream, and 6) validated and suspected sporadic and stream flashes. In the same figure, some characteristic distance ranges from the Earth such as the mesosphere (i.e. the atmospheric layer where meteoroids turn to meteors), the orbit used by some of the space telescopes (e.g. Hubble), the Low Earth Orbit (LEO), where the majority of the satellites orbit Earth, the Medium Earth Orbit (MEO), and the Geosynchronous orbit (GEO) are also indicated.

\begin{figure}
\begin{tabular}{c}
\includegraphics[width=8.4cm]{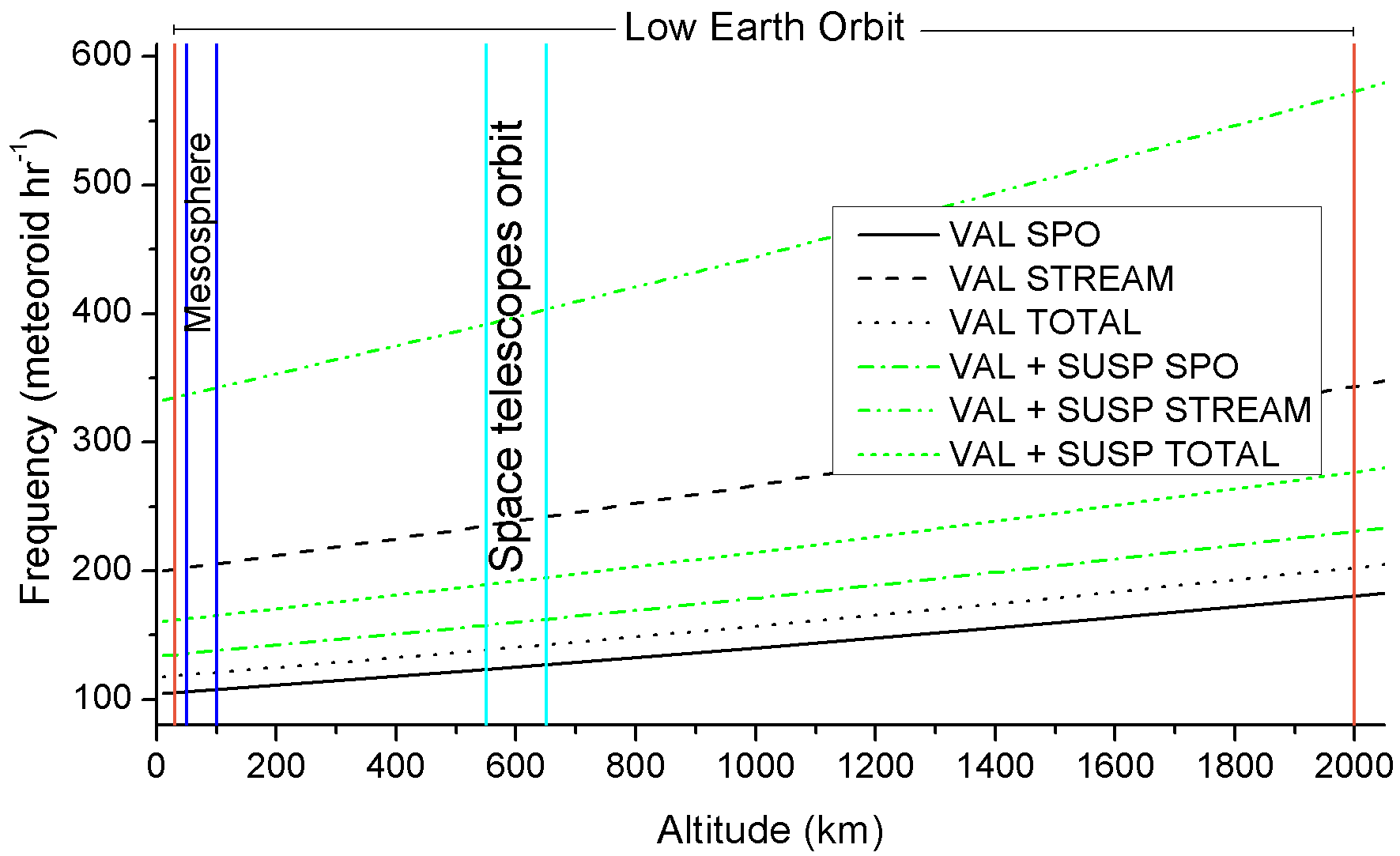}\\
\includegraphics[width=8.4cm]{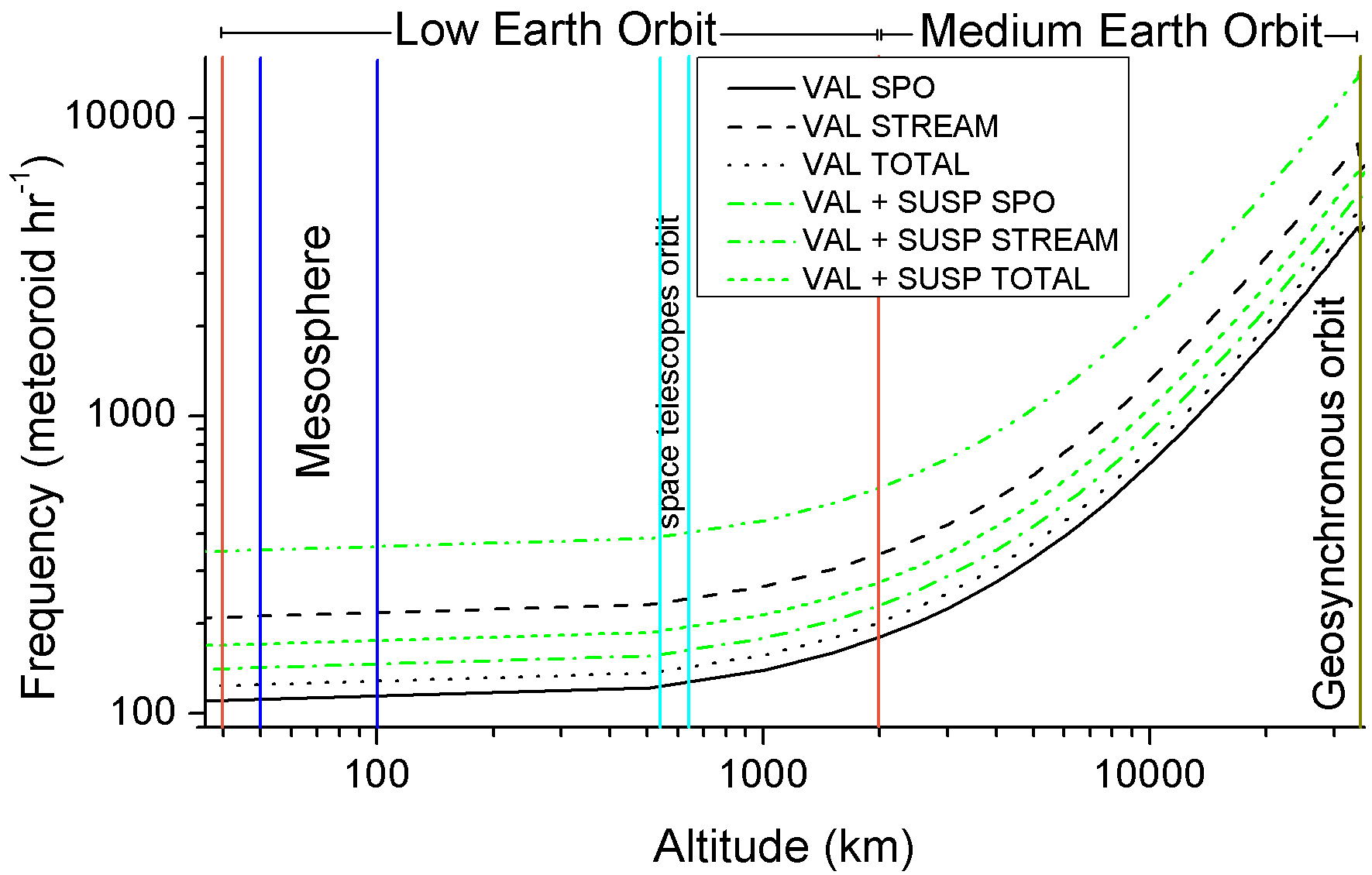}
\end{tabular}
\caption{Frequency distributions of meteoroids in the vicinity of Earth based on the NELIOTA detection rate of lunar impact flashes. Upper panel shows the frequency distributions for the first 2000~km above the surface of Earth, while the lower panel extends up to 36,000~km. Black lines denote the calculated frequencies taking into account only the validated flashes, while green lines the respective frequencies based on the sum of validated and suspected flashes. Different styles of lines (e.g. solid, dashed, doted) represent the frequencies of sporadic, stream, and total meteoroids according to the sample used (validated or validated and suspected flashes). Vertical solid lines indicate the boundaries of the low, medium and geosynchronous Earth orbits, space telescope orbits, and mesosphere.}
\label{fig:FreqD}
\end{figure}


\section{Summary and discussion}
\label{sec:Disc}
This study includes all the methods for the detection, validation, and photometry of lunar impact flashes as implemented by the NELIOTA team. Results for 79 validated and 29 suspected flashes during the first 30 months of NELIOTA operations were presented. Statistics of observations of the campaign showed that NELIOTA operates on a quite efficient site and with the current infrastructure is able to provide the most detailed results for such events. The aperture of the telescope, which is the largest worldwide dedicated to this kind of research, allows the detection of faint flashes up to $\sim11.5$~mag in $R$ band with a time resolution of 30~fps. Simultaneous observations in two different wavelength bands permit the validation of the lunar impact flashes using a single telescope, while they also allow the systematic derivation of their temperatures.

The temperatures of the flashes were calculated using the approximation of BB, while the thermal evolution of the multi-frame ones (in both bands) were also presented. It was found that the thermal evolution rates of the flash emitting areas are not the same for all impacts. The latter is probably connected to the material of the projectiles and of the lunar soil and it is a subject of another study. Using the emitted energies per band, it was found that the luminous efficiency is wavelength, hence temperature, dependent. In particular, for $R$ and $I$ bands in the range $1600<T<6200$~K the respective luminous efficiencies range between $5\eta_{\rm R}>\eta_{\rm I}>0.2\eta_{\rm R}$.

We presented a detailed description of the $E_{\rm lum}$ calculation and showed that the energy lost during the read-out time of the cameras plays a significant role in the total energy amount. Moreover, given that the majority of the flashes emit mostly in the near-infrared passband, our approximation that the sum of the emitted energies in $R$ and $I$ bands is in fact the total $E_{\rm lum}$ is very plausible. However, it should be noted that the calculation of $E_{\rm lum}$ using the temperature of the flash is possible but it can be considered valid only for the single frame flashes i.e. for very short durations that do not exceed the exposure times of the frames. An empirical relation between the emitted energies $E_{\rm R}$ and $E_{\rm I}$ of each band was derived and provides the means for a rough estimation of the physical parameters of the meteoroids (masses and sizes) of the suspected flashes and their respective developed impact temperatures. Based on these calculations, it was found that the majority of the suspected flashes should be fainter than the detection limit of NELIOTA in $R$ passband, but a few of them probably should have been detected. For the latter, either their magnitudes are underestimated or the seeing conditions and/or the glare from the dayside part of the Moon did not allow them to exceed the background threshold.

Using reasonable assumptions for $V_{\rm p}$ and $\rho_{\rm p}$ of the meteoroids according to their origin, their masses and sizes were calculated for various values of $\eta$. Considering a typical value of $\eta=1.5\times10^{-3}$ and taking into account only the validated flashes, we found that their masses range between 2.3~g and 2.7~kg with their majority ($\sim71\%$) to be less than 100~g. Their respective sizes (diameters) lie in the range 1.4-14~cm with their majority ($\sim76\%$) to be less than 5~cm. The produced $E_{\rm lum}$ were found to be in the range $4.2\times10^3-5.8\times10^5$~J corresponding to the range $2.8\times10^6-3.9\times10^8$~J of $E_{\rm kin}$, while their majority ($\sim68\%$) have values less than $2.53\times10^4$~J and $1.68\times10^7$~J, respectively. The crater sizes (diameters) for an impact angle of $45\degr$ range between 1.4-5.5~m, with their majority ($\sim68\%$) to be less than 2.25~m. Correlations between the emitted energy per band and the masses have been found, showing that the more massive the projectile the more luminous the flash. Contrary to this, a similar correlation between the masses of the projectiles and the temperatures of the flashes is not verified. The latter is probably connected to the heat capacity and the thermal conductivity of the materials of both meteoroids and lunar soil, factors that play a crucial role to the development and the evolution of the temperature.

In order to validate the potential of NELIOTA in the determination of low mass meteoroids, it was found useful to compare the present results with those of other campaigns. The most systematic campaign so far has been made by \citet{SUG14}, who detected 126 lunar impact flashes in approximately seven years of systematic observations. The $R$ magnitude distributions of flashes of that and our studies are shown in Fig.~\ref{fig:magS}. The comparison shows the significant contribution of NELIOTA in faint impact flashes. Fig.~\ref{fig:KEM} illustrates the comparison between the results of NELIOTA and those of \citet{SUG14} for the correlation between the $m_{\rm p}$ and $E_{\rm kin}$. In order the NELIOTA results to be directly comparable with those of \citet{SUG14}, $\eta=5\times10^{-3}$ and $V_{\rm p}^{\rm SPO}=24$~km~s$^{-1}$ were assumed. The comparison shows very good agreement between the two studies and the significant contribution of NELIOTA in the detection of low mass meteoroids.

\begin{figure}
\includegraphics[width=\columnwidth]{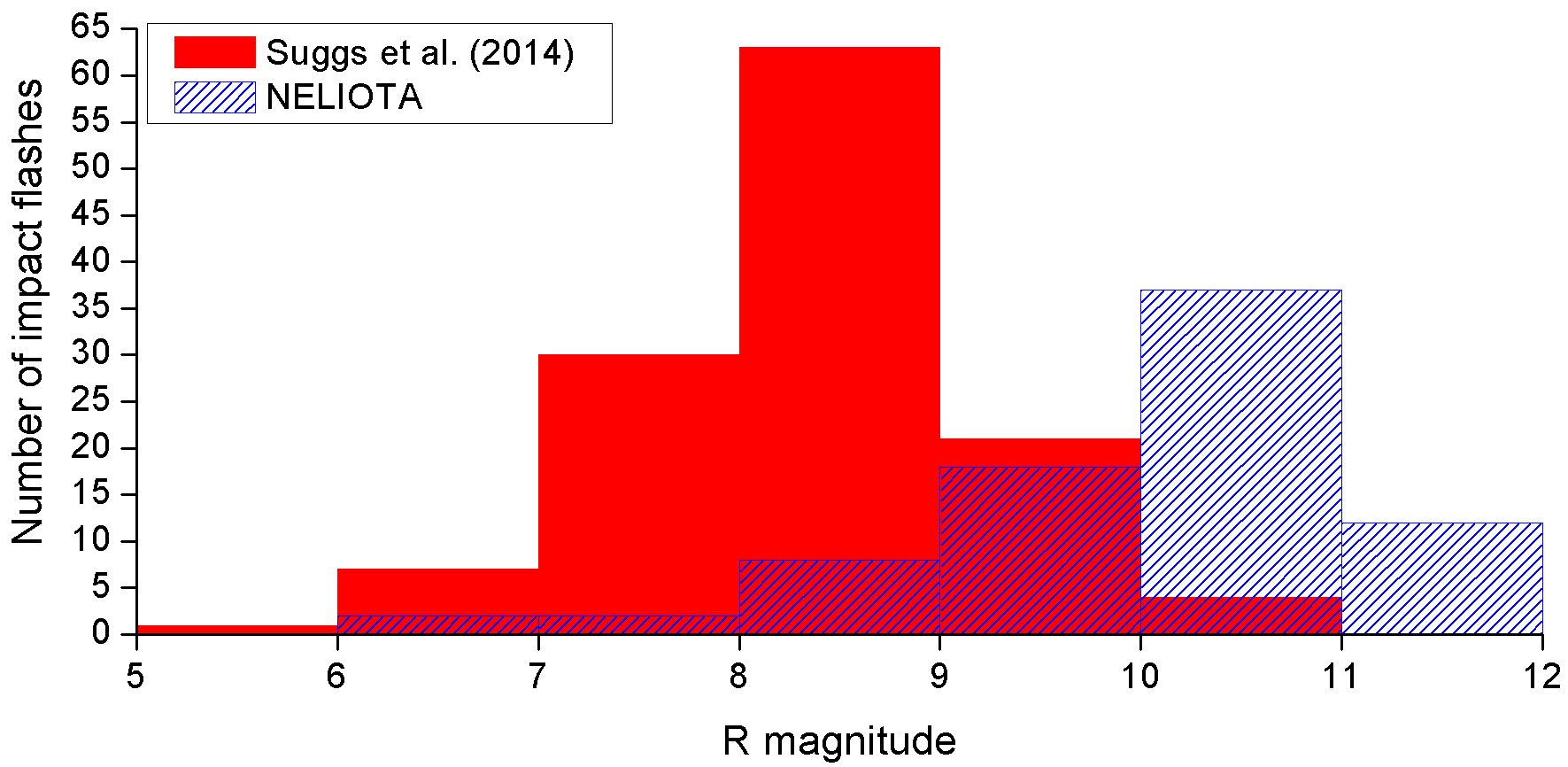}\\
\caption{Comparison histograms between NELIOTA (blue) and \citet{SUG14} (red) for the flash detection in $R$ band.}
\label{fig:magS}
\end{figure}

\begin{figure}
\includegraphics[width=\columnwidth]{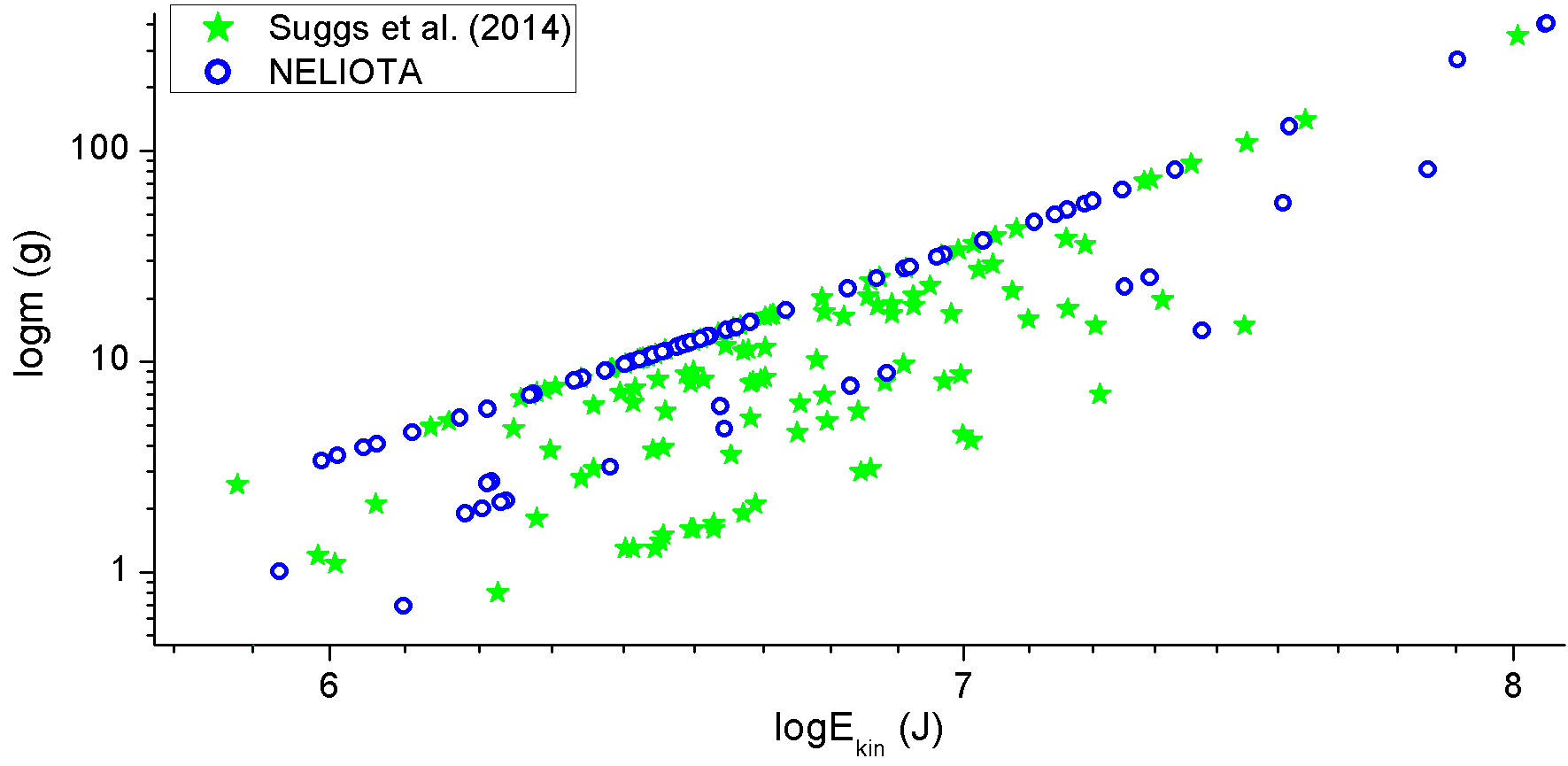}\\
\caption{Comparison diagram for the $m$ and $E_{\rm kin}$ of meteoroids (based only on the validated flashes) between NELIOTA (blue empty circles) and \citet{SUG14} (green stars).}
\label{fig:KEM}
\end{figure}

Recently, \citet{AVD19} published a similar work with the present study regarding the temperature and the calculation of meteoroids parameters. Although these two studies follow different approximations for the temperature calculation, the agreement between these values is quite good, well inside the error ranges. For the rest of the parameters (masses and sizes of the meteoroids), there are discrepancies due to the different method followed for the $E_{\rm lum}$ calculation and the different assumptions on the $V_{\rm p}$, which are connected to the methods followed to determine the origin of the meteoroids.

A detection rate of $2.3\times10^{-7}$~meteoroids~hr$^{-1}$~km$^{-2}$ based on the validated flash detections (from sporadic and stream meteoroids), which can be extended to $3.14\times10^{-7}$~meteoroids~hr$^{-1}$~km$^{-2}$ if we take into account the suspected flashes too, has been derived for the NELIOTA campaign so far. If we compare this rate with the respective ones of \citet{SUG14} ($1.03\times10^{-7}$~meteoroids~hr$^{-1}$~km$^{-2}$) and \citet{REM15} ($1.09\times10^{-7}$~meteoroids~hr$^{-1}$~km$^{-2}$), it can be clearly seen that NELIOTA has over doubled the detection rate of lunar impact flashes, and obviously this reflects the use of a telescope with large diameter. Using the detection rate of NELIOTA campaign and assuming an isotropic distribution of the meteoroids in space, the frequency distributions of meteoroids near Earth and Moon were calculated. For example, an appearance frequency of sporadic meteoroids between 181-231~meteoroids~hr$^{-1}$ was calculated for the upper boundary of LEO orbit, below which the majority of the satellites orbit Earth, while the respective frequency for the total surface of the Moon is $\sim7.8-9.9$~meteoroids~hr$^{-1}$.

These frequency distributions along with the respective distributions for masses and sizes of the meteoroids can be considered as a very powerful knowledge for the space industry i.e. for the spacecraft/satellite engineers or for structure engineers for lunar bases. They provide the probability for an orbiting satellite or for an infrastructure on Moon to be hit by a meteoroid of specific $E_{\rm kin}$ and size. Therefore, using the present results, the thickness of armor or the material that should be used for surviving from a possible impact with such objects can be better estimated. Thinking reversely, considering the cost for a vehicle launch, which is directly connected to the weight of the payload and the distance that should cover (i.e. fuel), a thinner armor may be used, if the probability of a meteoroid impact is extremely small. Therefore, NELIOTA results can be considered extremely valuable for estimating the risk of a collision with meteoroids of a space mission according to its total duration and covered distance (e.g. to the Moon) or to its orbital distance from Earth or Moon.

The greatest challenge of the ground based observations is the atmosphere of the Earth. Bad weather or bad seeing conditions affect negatively the detections of faint fast flashes and the longer recording of the brighter ones. Another preventing reason is the glare from the dayside part of the Moon (its effect is proportional to the telescope diameter), that does not allow to observe it when its phase is brighter than $\sim45\%$, i.e. in nights when the duration of the observations is longer. Unfortunately, all the above can be overcome only with space telescopes in orbit around Earth or, even better, in Moon-stationary orbit above its night side part. NELIOTA will continue to operate at least until January 2021 and is expected to approximately double the flashes detections providing the most accurate statistics that will have ever been made.

\begin{acknowledgements}
The authors gratefully acknowledge financial support by the European Space Agency under the NELIOTA program, contract No. 4000112943. The NELIOTA team express its gratitude to Mr. G. Dimou for his excellent technical support. A.L. thanks Dr. Chiotellis for the fruitful discussion on the luminous energy calculation. This work has made use of data from the European Space Agency NELIOTA project, obtained with the 1.2~m Kryoneri telescope, which is operated by IAASARS, National Observatory of Athens, Greece.
\end{acknowledgements}

\bibliography{NELIOTA}

\begin{thebibliography}{44}
\expandafter\ifx\csname natexlab\endcsname\relax\def\natexlab#1{#1}\fi

\bibitem[{{Ait Moulay Larbi} {et~al.}(2015{\natexlab{a}}){Ait Moulay Larbi},
  {Benkhaldoun}, {Baratoux}, {Daassou}, \& {Bouley}}]{AIT15a}
{Ait Moulay Larbi}, M., {Benkhaldoun}, Z., {Baratoux}, D., {Daassou}, A., \&
  {Bouley}, S. 2015{\natexlab{a}}, European Planetary Science Congress, 10,
  EPSC2015

\bibitem[{{Ait Moulay Larbi} {et~al.}(2015{\natexlab{b}}){Ait Moulay Larbi},
  {Daassou}, {Baratoux}, {Bouley}, {Benkhaldoun}, {Lazrek}, {Garcia}, \&
  {Colas}}]{AIT15b}
{Ait Moulay Larbi}, M., {Daassou}, A., {Baratoux}, D., {et~al.}
  2015{\natexlab{b}}, Earth Moon and Planets, 115, 1

\bibitem[{{Avdellidou} \& {Vaubaillon}(2019)}]{AVD19}
{Avdellidou}, C. \& {Vaubaillon}, J. 2019, \mnras, 484, 5212

\bibitem[{{Babadzhanov} \& {Kokhirova}(2009)}]{BAD09}
{Babadzhanov}, P.~B. \& {Kokhirova}, G.~I. 2009, \aap, 495, 353

\bibitem[{{Bellot Rubio} {et~al.}(2000){Bellot Rubio}, {Ortiz}, \&
  {Sada}}]{BEL00b}
{Bellot Rubio}, L.~R., {Ortiz}, J.~L., \& {Sada}, P.~V. 2000, Earth Moon and
  Planets, 82, 575

\bibitem[{{Berry} \& {Burnell}(2000)}]{BER00}
{Berry}, R. \& {Burnell}, J. 2000, {The handbook of astronomical image
  processing} (Willmann-Bell, Inc.)

\bibitem[{{Bessell} {et~al.}(1998{\natexlab{a}}){Bessell}, {Castelli}, \&
  {Plez}}]{BES98b}
{Bessell}, M.~S., {Castelli}, F., \& {Plez}, B. 1998{\natexlab{a}}, \aap, 337,
  321

\bibitem[{{Bessell} {et~al.}(1998{\natexlab{b}}){Bessell}, {Castelli}, \&
  {Plez}}]{BES98a}
{Bessell}, M.~S., {Castelli}, F., \& {Plez}, B. 1998{\natexlab{b}}, \aap, 333,
  231

\bibitem[{{Bonanos} {et~al.}(2016{\natexlab{a}}){Bonanos}, {Liakos},
  {Xilouris}, {Boumis}, {Bellas-Velidis}, {Marousis}, {Dapergolas}, {Fytsilis},
  {Noutsopoulos}, {Charmandaris}, {Tsiganis}, {Tsinganos}, {Els}, {Koschny},
  {Lock}, \& {Navarro}}]{BON16b}
{Bonanos}, A., {Liakos}, A., {Xilouris}, M., {et~al.} 2016{\natexlab{a}}, in
  \procspie, Vol. 9911, Modeling, Systems Engineering, and Project Management
  for Astronomy VI, 991122

\bibitem[{{Bonanos} {et~al.}(2015){Bonanos}, {Xilouris}, {Boumis},
  {Bellas-Velidis}, {Maroussis}, {Dapergolas}, {Fytsilis}, {Charmandaris},
  {Tsiganis}, \& {Tsinganos}}]{BON15}
{Bonanos}, A., {Xilouris}, M., {Boumis}, P., {et~al.} 2015, IAU General
  Assembly, 22, 2251039

\bibitem[{{Bonanos} {et~al.}(2018){Bonanos}, {Avdellidou}, {Liakos},
  {Xilouris}, {Dapergolas}, {Koschny}, {Bellas-Velidis}, {Boumis},
  {Charmandaris}, {Fytsilis}, \& {Maroussis}}]{BON18}
{Bonanos}, A.~Z., {Avdellidou}, C., {Liakos}, A., {et~al.} 2018, \aap, 612, A76

\bibitem[{{Bonanos} {et~al.}(2016{\natexlab{b}}){Bonanos}, {Xilouris},
  {Boumis}, {Bellas-Velidis}, {Maroussis}, {Dapergolas}, {Fytsilis},
  {Charmandaris}, {Tsiganis}, \& {Tsinganos}}]{BON16a}
{Bonanos}, A.~Z., {Xilouris}, M., {Boumis}, P., {et~al.} 2016{\natexlab{b}}, in
  IAU Symposium, Vol. 318, Asteroids: New Observations, New Models, ed. S.~R.
  {Chesley}, A.~{Morbidelli}, R.~{Jedicke}, \& D.~{Farnocchia}, 327--329

\bibitem[{{Bouley} {et~al.}(2012){Bouley}, {Baratoux}, {Vaubaillon}, {Mocquet},
  {Le Feuvre}, {Colas}, {Benkhaldoun}, {Daassou}, {Sabil}, \&
  {Lognonn{\'e}}}]{BOU12}
{Bouley}, S., {Baratoux}, D., {Vaubaillon}, J., {et~al.} 2012, \icarus, 218,
  115

\bibitem[{{Brown} {et~al.}(2002){Brown}, {Spalding}, {ReVelle}, {Tagliaferri},
  \& {Worden}}]{BRO02}
{Brown}, P., {Spalding}, R.~E., {ReVelle}, D.~O., {Tagliaferri}, E., \&
  {Worden}, S.~P. 2002, \nat, 420, 294

\bibitem[{{Brown} {et~al.}(2010){Brown}, {Wong}, {Weryk}, \& {Wiegert}}]{BRO10}
{Brown}, P., {Wong}, D.~K., {Weryk}, R.~J., \& {Wiegert}, P. 2010, \icarus,
  207, 66

\bibitem[{{Dubietis} \& {Arlt}(2010)}]{DUB10}
{Dubietis}, A. \& {Arlt}, R. 2010, Earth Moon and Planets, 106, 105

\bibitem[{{Eisele}(2017)}]{EIS17}
{Eisele}, J. 2017, Master's thesis, Lehrstuhl für Raumfahrttechnik, TU Munich

\bibitem[{{Gault}(1974)}]{GAU74}
{Gault}, D.~E. 1974, in A Primer in Lunar Geology, ed. R.~{Greeley} \& P.~H.
  {Schultz}

\bibitem[{{Hayne} {et~al.}(2017){Hayne}, {Bandfield}, {Siegler}, {Vasavada},
  {Ghent}, {Williams}, {Greenhagen}, {Aharonson}, {Elder}, {Lucey}, \&
  {Paige}}]{HAY17}
{Hayne}, P.~O., {Bandfield}, J.~L., {Siegler}, M.~A., {et~al.} 2017, Journal of
  Geophysical Research (Planets), 122, 2371

\bibitem[{{Hughes}(1987)}]{HUG87}
{Hughes}, D.~W. 1987, \aap, 187, 879

\bibitem[{{Hurley} {et~al.}(2017){Hurley}, {Cook}, {Retherford}, {Greathouse},
  {Gladstone}, {Mandt}, {Grava}, {Kaufmann}, {Hendrix}, {Feldman}, {Pryor},
  {Stickle}, {Killen}, \& {Stern}}]{HUR17}
{Hurley}, D.~M., {Cook}, J.~C., {Retherford}, K.~D., {et~al.} 2017, \icarus,
  283, 31

\bibitem[{{Jenniskens}(1994)}]{JEN94}
{Jenniskens}, P. 1994, \aap, 287, 990

\bibitem[{{Kelso}(2007)}]{KEL07}
{Kelso}, T. 2007, in 17th AAS/AIAA Space Flight Mechanics Meeting, AAS/AIAA
  Space Flight Mechanics Meeting No. AAS 07-127 (AAS Publications Office), 14

\bibitem[{{Koschny} {et~al.}(2019){Koschny}, {Soja}, {Engrand}, {Flynn},
  {Lasue}, {Levasseur-Regourd}, {Malaspina}, {Nakamura}, {Poppe}, {Sterken}, \&
  {Trigo-Rodr{\'\i}guez}}]{KOS19}
{Koschny}, D., {Soja}, R.~H., {Engrand}, C., {et~al.} 2019, \ssr, 215, 34

\bibitem[{{Krag} {et~al.}(2000){Krag}, {Beltrami-Karlezi}, {Bendisch},
  {Klinkrad}, {Rex}, {Rosebrock}, \& {Schildknecht}}]{KRA00}
{Krag}, H., {Beltrami-Karlezi}, P., {Bendisch}, J., {et~al.} 2000, Acta
  Astronautica, 47, 687

\bibitem[{{Liakos} {et~al.}(2019){Liakos}, {Bonanos}, {Xilouris},
  {Bellas-Velidis}, {Boumis}, {Charmandaris}, {Dapergolas}, {Fytsilis},
  {Maroussis}, {Koschny}, {Moissl}, \& {Navarro}}]{LIA19}
{Liakos}, A., {Bonanos}, A., {Xilouris}, E., {et~al.} 2019, arXiv e-prints
  [\eprint[arXiv]{1901.11414}]

\bibitem[{{Madiedo} {et~al.}(2017){Madiedo}, {Ortiz}, \& {Morales}}]{MAD17}
{Madiedo}, J.~M., {Ortiz}, J.~L., \& {Morales}, N. 2017, in Lunar and Planetary
  Science Conference, 1319

\bibitem[{{Madiedo} {et~al.}(2018){Madiedo}, {Ortiz}, \& {Morales}}]{MAD18}
{Madiedo}, J.~M., {Ortiz}, J.~L., \& {Morales}, N. 2018, \mnras, 480, 5010

\bibitem[{{Madiedo} {et~al.}(2014){Madiedo}, {Ortiz}, {Morales}, \&
  {Cabrera-Ca{\~n}o}}]{MAD14}
{Madiedo}, J.~M., {Ortiz}, J.~L., {Morales}, N., \& {Cabrera-Ca{\~n}o}, J.
  2014, \mnras, 439, 2364

\bibitem[{{Madiedo} {et~al.}(2015{\natexlab{a}}){Madiedo}, {Ortiz}, {Morales},
  \& {Cabrera-Ca{\~n}o}}]{MAD15b}
{Madiedo}, J.~M., {Ortiz}, J.~L., {Morales}, N., \& {Cabrera-Ca{\~n}o}, J.
  2015{\natexlab{a}}, \planss, 111, 105

\bibitem[{{Madiedo} {et~al.}(2019{\natexlab{a}}){Madiedo}, {Ortiz}, {Morales},
  {Roman}, \& {Alonso}}]{MAD19a}
{Madiedo}, J.~M., {Ortiz}, J.~L., {Morales}, N., {Roman}, A., \& {Alonso}, S.
  2019{\natexlab{a}}, in Lunar and Planetary Science Conference, Lunar and
  Planetary Science Conference, 1406

\bibitem[{{Madiedo} {et~al.}(2019{\natexlab{b}}){Madiedo}, {Ortiz}, {Morales},
  \& {Santos-Sanz}}]{MAD19b}
{Madiedo}, J.~M., {Ortiz}, J.~L., {Morales}, N., \& {Santos-Sanz}, P.
  2019{\natexlab{b}}, \mnras, 486, 3380

\bibitem[{{Madiedo} {et~al.}(2015{\natexlab{b}}){Madiedo}, {Ortiz}, {Organero},
  {Ana-Hern{\'a}ndez}, {Fonseca}, {Morales}, \& {Cabrera-Ca{\~n}o}}]{MAD15a}
{Madiedo}, J.~M., {Ortiz}, J.~L., {Organero}, F., {et~al.} 2015{\natexlab{b}},
  \aap, 577, A118

\bibitem[{{Madiedo} {et~al.}(2019{\natexlab{c}}){Madiedo}, {Ortiz},
  {Yanagisawa}, {Aceituno}, \& {Aceituno}}]{MAD19c}
{Madiedo}, J.~M., {Ortiz}, J.~L., {Yanagisawa}, M., {Aceituno}, J., \&
  {Aceituno}, F. 2019{\natexlab{c}}, {Impact Flashes of Meteoroids on the
  Moon}, ed. G.~O. {Ryabova}, D.~J. {Asher}, \& M.~J. {Campbell-Brown}
  (Cambridge University Press), 136--158

\bibitem[{{Melosh}(1989)}]{MEL89}
{Melosh}, H.~J. 1989, {Impact cratering: A geologic process} (Oxford University
  Press)

\bibitem[{{Mitchell} {et~al.}(1973){Mitchell}, {Carrier}, {Costes}, {Houston},
  \& {Scott}}]{MIT73}
{Mitchell}, J.~K., {Carrier}, III, W.~D., {Costes}, N.~C., {Houston}, W.~N., \&
  {Scott}, R.~F. 1973, in Lunar and Planetary Science Conference Proceedings,
  Vol.~4, Lunar and Planetary Science Conference Proceedings, 2437

\bibitem[{{Moser} {et~al.}(2011){Moser}, {Suggs}, {Swift}, {Suggs}, {Cooke},
  {Diekmann}, \& {Koehler}}]{MOS11}
{Moser}, D.~E., {Suggs}, R.~M., {Swift}, W.~R., {et~al.} 2011, in Meteoroids:
  The Smallest Solar System Bodies, ed. W.~J. {Cooke}, D.~E. {Moser}, B.~F.
  {Hardin}, \& D.~{Janches}, 142

\bibitem[{{Ortiz} {et~al.}(2006){Ortiz}, {Aceituno}, {Quesada}, {Aceituno},
  {Fern{\'a}ndez}, {Santos-Sanz}, {Trigo-Rodr{\'{\i}}guez}, {Llorca},
  {Mart{\'{\i}}n-Torres}, {Monta{\~n}{\'e}s-Rodr{\'{\i}}guez}, \&
  {Pall{\'e}}}]{ORT06}
{Ortiz}, J.~L., {Aceituno}, F.~J., {Quesada}, J.~A., {et~al.} 2006, \icarus,
  184, 319

\bibitem[{{Ortiz} {et~al.}(2015){Ortiz}, {Madiedo}, {Morales}, {Santos-Sanz},
  \& {Aceituno}}]{ORT15}
{Ortiz}, J.~L., {Madiedo}, J.~M., {Morales}, N., {Santos-Sanz}, P., \&
  {Aceituno}, F.~J. 2015, \mnras, 454, 344

\bibitem[{{Rembold} \& {Ryan}(2015)}]{REM15}
{Rembold}, J.~J. \& {Ryan}, E.~V. 2015, \planss, 117, 119

\bibitem[{{Suggs} {et~al.}(2014){Suggs}, {Moser}, {Cooke}, \& {Suggs}}]{SUG14}
{Suggs}, R.~M., {Moser}, D.~E., {Cooke}, W.~J., \& {Suggs}, R.~J. 2014,
  \icarus, 238, 23

\bibitem[{{Swift} {et~al.}(2011){Swift}, {Moser}, {Suggs}, \& {Cooke}}]{SWI11}
{Swift}, W.~R., {Moser}, D.~E., {Suggs}, R.~M., \& {Cooke}, W.~J. 2011, in
  Meteoroids: The Smallest Solar System Bodies, ed. W.~J. {Cooke}, D.~E.
  {Moser}, B.~F. {Hardin}, \& D.~{Janches}, 125

\bibitem[{{Tucker} {et~al.}(2019){Tucker}, {Farrell}, {Killen}, \&
  {Hurley}}]{TUC19}
{Tucker}, O.~J., {Farrell}, W.~M., {Killen}, R.~M., \& {Hurley}, D.~M. 2019,
  Journal of Geophysical Research (Planets), 124, 278

\bibitem[{{Xilouris} {et~al.}(2018){Xilouris}, {Bonanos}, {Bellas-Velidis},
  {Boumis}, {Dapergolas}, {Maroussis}, {Liakos}, {Alikakos}, {Charmandaris},
  {Dimou}, {Fytsilis}, {Kelley}, {Koschny}, {Navarro}, {Tsiganis}, \&
  {Tsinganos}}]{XIL18}
{Xilouris}, E.~M., {Bonanos}, A.~Z., {Bellas-Velidis}, I., {et~al.} 2018, \aap,
  619, A141

\end{thebibliography}

\begin{appendix}

\section{Localization of the impact flashes}
\label{sec:LOC}
The positions of the validated and the suspected flashes on the lunar surface is performed with a semi-automatic method. After the positive validation of an event, another tool of the NELIOTA-DET software, dedicated to cross match images, is used. In particular, this `localization' tool requires as input a very detailed lunar map. For the latter, the software Virtual Moon Atlas 6.0\footnote{\url{https://www.ap-i.net/avl/en/start}} (VMA6), that provides such images, is used. VMA6 can produce the image of the Moon based on the coordinates of the observing site, and the exact timing of the detected event. Moreover, the lunar libration is also taken into account. The `localization' tool of the NELIOTA-DET sums all the images that are included in the data cube (see Section~\ref{sec:PHOT}) of the flash producing a high-contrast image, in which lunar features can be identified more easily. The NELIOTA-DET displays in a single screen the high-contrast (data cube) and the high detailed (VMA6) images side by side. At this point, the user has to pair at least three lunar features in these images. It should be noticed that the pixel scale is not the same for these two images. Therefore, the tool, via the cross match of the features, calculates the pixel images transformation matrix. This matrix is then used to translate the position of the flash (included in the header) in the high-contract image to a position (in pixels) in the high detailed image. The output result is the high detailed image in which the location of the flash is marked. Finally, the user has to cross match this image with the interactive screen in the VMA6 software to identify its lunar coordinates. This method provides an error of $\sim0.5$~degrees in the selenographic coordinates system (i.e. area with a diameter of $\sim$20~km).

\begin{figure}
\centering
\includegraphics[width=\columnwidth]{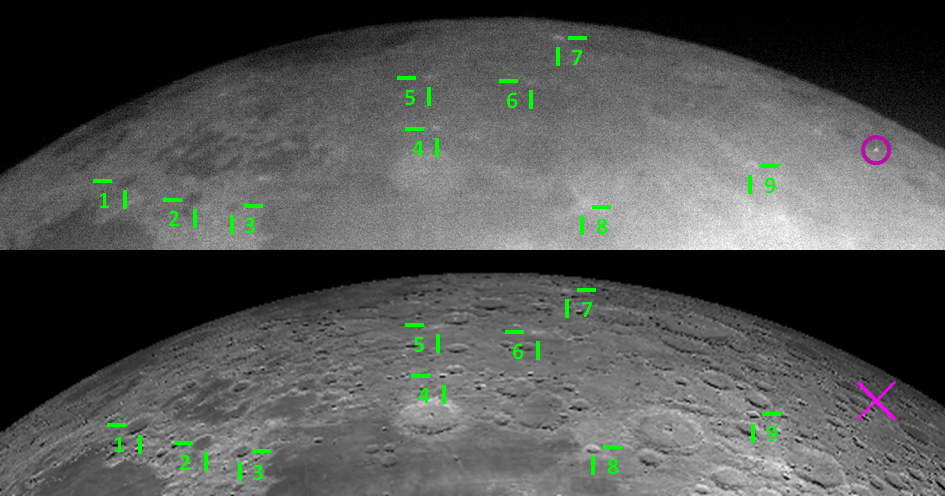}
\caption{Example of the implementation of the localization tool of NELIOTA-DET. The upper figure is the high-contrast image (NELIOTA data), while the lower figure is a high resolution lunar map obtained from VMA6. Both figures are portion of the original ones. The paired lunar features are indicated as numbers. The detected flash is shown in the purple circle in the upper figure, while the purple `X' in the lower figure denotes its position. Subsequently, the lower image can be cross matched with the interactive image of VMA6 to obtain the selenographic coordinates of the flash.}
\label{fig:loc}
\end{figure}

\section{Archiving and public availability}
\label{sec:WEB}
After the validation, the photometry and the localization of the flash, the data are sent to the NELIOTA-ARC system in another virtual server in NOA HQ for archiving. Later, the expert user has to operate the NELIOTA-WEB software in order to upload the results to the official NELIOTA website within 24 hrs after the end of the observations. An unregistered user of the website is able to see online the images of only the validated flashes as well as some essential information about them (e.g. time, magnitude). Registered users are permitted to see and also download the data of both the validated and the suspected flashes. The data downloading can be done using the `download cart' tool that is available in the website under the data access menu. The data package contains for each photometric band: a) The data cube that includes the frames of the flash and the seven previous and the seven successive images (see Section~\ref{sec:PHOT}), b) the mean image of the lunar background produced by the pipeline (see Paper II-Section~5.2), and c) the data cube with the `difference' images derived from the subtraction of the background image (produced by the pipeline) and the images containing the flash and its successive frames. At this point, it should be noticed that for the present results we did not use the difference images produced by the pipeline, but we followed the method described in Section~\ref{sec:PHOT}.

\section{Multi frame flashes}
\label{sec:App1}
In this appendix are given the magnitude values for all the multi-frame flashes (Table~\ref{tab:multi}). In the case of flashes for which more than two set of frames in $R$ and $I$ bands obtained, the corresponding temperature of each set of frames is also given. Moreover, the light curves of these flashes are plotted in Figs.~\ref{fig:LCs1}-\ref{fig:LCs3}.

\begin{table*}
\centering
\caption{List of multi-frame flashes. Errors are included in parentheses alongside values and correspond to the last digit(s).}
\label{tab:multi}
\scalebox{0.85}{
\begin{tabular}{ccccc | ccccc }
\hline\hline	
ID	&	Date \& time	&	$m_{\rm R}$	&	$m_{\rm I}$	&	T	&	ID	&	Date \& time	&	$m_{\rm R}$	&	$m_{\rm I}$	&	T	\\
	& (UT)		&	(mag)	&	(mag)	&	(K)	&		&	(UT)	&	(mag)	&	(mag)	&	(K)	\\
\hline	
2a	&	2017 03 01 17:08:46.573	&	6.67(11)	&	6.07(8)	&	4503(646)	&	59a	&	2018 08 06 02:38:14.302	&	9.16(11)	&	7.73(5)	&	2515(156)	\\
2b	&		&	10.01(19)	&	8.26(9)	&	2146(202)	&	59b	&		&	10.63(37)	&	9.55(11)	&	3080(863)	\\
2c	&		&		&	9.27(11)	&		&	59c	&		&		&	10.53(21)	&		\\
2d	&		&		&	10.57(16)	&		&	61a	&	2018 08 07 01:33:54.756	&	10.79(30)	&	9.31(10)	&	2444(405)	\\
6a	&	2017 05 01 20:30:58.137	&	10.19(20)	&	8.84(8)	&	2615(315)	&	61b	&		&		&	12.03(57)	&		\\
6b	&		&		&	10.44(22)	&		&	62a	&	2018 08 07 01:35:45.168	&	8.78(10)	&	7.74(7)	&	3153(262)	\\
9a	&	2017 06 19 01:50:34.560	&		&	9.60(12)	&		&	62b	&		&	11.30(39)	&	8.79(7)	&	1600(202)	\\
9b	&		&		&	10.24(18)	&		&	62c	&		&		&	10.43(15)	&		\\
12a	&	2017 06 27 18:58:26.680	&	11.07(36)	&	9.27(9)	&	2101(343)	&	62d	&		&		&	11.25(28)	&		\\
12b	&		&		&	10.80(23)	&		&	66a	&	2018 08 08 02:28:23.406	&	11.06(25)	&	10.40(14)	&	4253(1364)	\\
13a	&	2017 06 28 18:45:25.568	&	10.56(40)	&	9.48(14)	&	3088(945)	&	66b	&		&		&	11.90(27)	&		\\
13b	&		&		&	10.65(27)	&		&	67a	&	2018 08 08 02:29:44.573	&	8.36(10)	&	7.30(7)	&	3124(247)	\\
14a	&	2017 07 19 02:00:36.453	&	11.23(43)	&	9.33(8)	&	2008(367)	&	67b	&		&	10.63(23)	&	8.81(7)	&	2085(222)	\\
14b	&		&		&	11.28(29)	&		&	67c	&		&		&	9.69(10)	&		\\
15a	&	2017 07 27 18:31:06.720	&		&	9.34(12)	&		&	67d	&		&		&	10.46(14)	&		\\
15b	&		&		&	11.25(30)	&		&	67e	&		&		&	10.84(19)	&		\\
16a	&	2017 07 28 18:21:44.850	&	11.24(36)	&	9.29(8)	&	1978(300)	&	73a	&	2018 09 05 01:51:37.399	&	8.53(12)	&	7.17(6)	&	2596(190)	\\
16b	&		&		&	9.67(12)	&		&	73b	&		&	7.84(10)	&	6.60(5)	&	2793(194)	\\
19a	&	2017 07 28 19:17:18.307	&	8.27(11)	&	6.32(7)	&	1972(105)	&	73c	&		&	8.72(14)	&	7.43(6)	&	2711(239)	\\
19b	&		&	9.43(15)	&	7.44(8)	&	1945(135)	&	73d	&		&	9.39(16)	&	8.18(7)	&	2841(313)	\\
19c	&		&		&	8.89(10)	&		&	73e	&		&		&	8.63(9)	&		\\
19d	&		&		&	9.38(13)	&		&	73f	&		&		&	9.09(11)	&		\\
19e	&		&		&	10.29(24)	&		&	73g	&		&		&	9.80(15)	&		\\
20a	&	2017 08 16 01:05:46.763	&	10.15(20)	&	9.54(11)	&	4455(1146)	&	73h	&		&		&	10.01(13)	&		\\
20b	&		&		&	10.50(27)	&		&	73i	&		&		&	10.30(18)	&		\\
21a	&	2017 08 16 02:15:58.813	&	10.69(30)	&	9.11(7)	&	2326(356)	&	73j	&		&		&	10.89(20)	&		\\
21b	&		&		&	11.09(27)	&		&	73k	&		&		&	10.80(28)	&		\\
22a	&	2017 08 16 02:41:15.113	&	10.81(32)	&	9.08(7)	&	2167(319)	&	73l	&		&		&	11.41(42)	&		\\
22b	&		&		&	10.11(15)	&		&	74a	&	2018 09 05 02:47:54.403	&	10.61(39)	&	9.09(10)	&	2401(501)	\\
23a	&	2017 08 18 02:02:21.417	&	10.92(24)	&	9.20(8)	&	2185(255)	&	74b	&		&		&	10.52(27)	&		\\
23b	&		&		&	10.12(10)	&		&	76a	&	2018 09 06 03:10:04.087	&	11.18(28)	&	9.86(10)	&	2660(459)	\\
24a	&	2017 08 18 02:03:08.317	&	10.19(16)	&	8.83(8)	&	2615(256)	&	76b	&		&		&	12.03(37)	&		\\
24b	&		&		&	11.07(17)	&		&	78a	&	2018 10 15 18:17:49.314	&	9.61(20)	&	8.84(11)	&	3836(787)	\\
26a	&	2017 09 14 03:17:49.737	&	9.17(9)	&	8.07(4)	&	3058(207)	&	78b	&		&		&	9.71(19)	&		\\
26b	&		&		&	8.93(6)	&		&	80a	&	2018 11 12    17:00:02.156	&		&	8.70(10)	&		\\
26c	&		&		&	10.28(13)	&		&	80b	&		&		&	9.42(14)	&		\\
26d	&		&		&	11.33(27)	&		&	85a	&	2019 02 09 18:17:00.009	&	10.39(27)	&	9.82(13)	&	4647(1169)	\\
27a	&	2017 09 16 02:26:24.933	&	8.52(9)	&	7.04(6)	&	2440(136)	&	85b	&		&		&	9.91(15)	&		\\
27b	&		&	10.01(13)	&	8.29(6)	&	2173(1423)	&	90a	&	2019 06 08 19:26:58.103	&	9.24(18)	&	8.04(7)	&	2864(298)	\\
27c	&		&	11.35(36)	&	9.17(7)	&	1801(243)	&	90b	&		&		&	10.1(3)	&		\\
27d	&		&		&	9.78(7)	&		&	93a	&	2019 06 28 01:56:47.678	&	8.88(12)	&	7.59(7)	&	2709(215)	\\
27e	&		&		&	10.35(9)	&		&	93b	&		&		&	9.43(13)	&		\\
27f	&		&		&	10.70(12)	&		&	93c	&		&		&	10.22(23)	&		\\
27g	&		&		&	11.70(23)	&		&	94a	&	2019 06 28 02:18:22.899	&	10.12(20)	&	9.29(10)	&	3678(715)	\\
28a	&	2017 10 13 01:54:21.482	&		&	8.49(5)	&		&	94b	&		&	10.63(38)	&	9.10(10)	&	2385(444)	\\
28b	&		&	9.28(13)	&	8.37(12)	&	3458(357)	&	94c	&		&		&	10.95(33)	&		\\
28c	&		&	11.09(62)	&	9.77(21)	&	2677(1047)	&	95a	&	2019 07 06 19:12:55.225	&	10.06(24)	&	9.08(10)	&	3307(644)	\\
28d	&		&		&	10.35(21)	&		&	95b	&		&		&	11.05(36)	&		\\
29a	&	2017 10 13 02:33:43.560	&	10.31(26)	&	9.89(14)	&	5453(1740)	&	96a	&	2019 07 07 18:32:55.695	&	10.94(36)	&	9.63(11)	&	2678(590)	\\
29b	&		&		&	10.05(14)	&		&	96b	&		&		&	10.36(21)	&		\\
29c	&		&		&	11.36(32)	&		&	97a	&	2019 07 07 18:40:21.170	&	6.65(10)	&	5.49(6)	&	2922(205)	\\
30a	&	2017 10 16 02:46:45.613	&	10.72(22)	&	9.46(8)	&	2751(384)	&	97b	&		&	9.53(12)	&	7.59(6)	&	1983(114)	\\
30b	&		&	11.86(39)	&	10.41(11)	&	2491(548)	&	97c	&		&	10.86(32)	&	8.72(9)	&	1827(227)	\\
30c	&		&		&	11.20(21)	&		&	97d	&		&	12.07(76)	&	9.46(9)	&	1549(360)	\\
32a	&	2017 11 14 03:34:14.985	&	10.31(19)	&	9.31(8)	&	3264(488)	&	97e	&		&		&	10.03(11)	&		\\
32b	&		&		&	10.03(12)	&		&	97f	&		&		&	10.17(16)	&		\\
33a	&	2017 11 23 16:17:33.000	&	10.45(25)	&	10.06(14)	&	5722(1528)	&	97g	&		&		&	10.94(31)	&		\\
33b	&		&		&	11.00(29)	&		&	97h	&		&		&	11.57(38)	&		\\
35a	&	2017 12 12 01:49:26.480	&		&	8.91(11)	&		&	97i	&		&		&	11.36(36)	&		\\
35b	&		&		&	10.08(19)	&		&	100a	&	2019 07 08 19:11:44.449	&	9.77(21)	&	8.19(10)	&	2325(266)	\\
37a	&	2017 12 12 02:48:08.178	&	10.50(26)	&	8.98(9)	&	2402(334)	&	100b	&		&		&	9.98(15)	&		\\
37b	&		&		&	10.98(35)	&		&	100c	&		&		&	10.94(33)	&		\\
46a	&	2017 12 14 04:35:09.737	&	7.94(9)	&	6.76(6)	&	2889(200)	&	102a	&	2019 07 26 00:41:35.185 	&	9.64(14)	&	8.21(7)	&	2510(239)	\\
46b	&		&		&	8.88(9)	&		&	102b	&		&	9.76(18)	&	8.17(7)	&	2309(216)	\\
46c	&		&		&	10.02(15)	&		&	102c	&		&		&	9.72(13)	&		\\
46d	&		&		&	10.34(16)	&		&	104a	&	2019 07 27 01:17:49.791	&	8.95(13)	&	8.02(7)	&	3404(368)	\\
47a	&	2018 01 12 03:54:03.027	&	10.01(17)	&	9.31(10)	&	4101(782)	&	104b	&		&	10.48(31)	&	9.02(09)	&	2473(425)	\\
47b	&		&		&	10.88(27)	&		&	104c	&		&		&	10.07(17)	&		\\
58a	&	2018 08 06 01:57:43.686	&	9.68(17)	&	8.14(6)	&	2369(215)	&	104d	&		&		&	10.89(28)	&		\\
58b	&		&		&	9.27(12)	&		&	105a	&	2019 07 27 02:12:25.049	&	9.67(17)	&	8.67(7)	&	3273(441)	\\
58c	&		&		&	10.21(25)	&		&	105b	&		&		&	10.53(26)	&		\\
	&		&		&		&		&	105c	&		&		&	11.31(35)	&		\\
\hline
\end{tabular}}
\end{table*}

\begin{table*}
\centering
\caption{Table~\ref{tab:multi} cont.}
\label{tab:multiCONT}
\scalebox{0.85}{
\begin{tabular}{ccccc | ccccc }
\hline\hline	
ID	&	Date \& time	&	$m_{\rm R}$	&	$m_{\rm I}$	&	T	&	ID	&	Date \& time	&	$m_{\rm R}$	&	$m_{\rm I}$	&	T	\\
	&	(UT)	&	(mag)	&	(mag)	&	(K)	&		&	(UT)	&	(mag)	&	(mag)	&	(K)	\\
\hline
106a	&	2019 07 27 02:37:22.715	&	10.16(21)	&	9.48(8)	&	4186(953)	&	109a	&	2019 07 28 01:33:40.121	&	10.08(18)	&	8.93(10)	&	2941(386)	\\
106b	&		&		&	11.26(30)	&		&	109b	&		&		&	9.26(10)	&		\\
107a	&	2019 07 27 02:59:56.458	&	9.48(17)	&	8.25(7)	&	2807(321)	&	109c	&		&		&	10.57(19)	&		\\
107b	&		&	10.48(33)	&	9.30(16)	&	2896(697)	&	110a	&	2019 07 28 01:59:21.345	&	10.80(31)	&	9.62(12)	&	2879(611)	\\
107c	&		&		&	10.12(24)	&		&	110b	&		&		&	10.73(27)	&		\\
107d	&		&		&	10.78(41)	&		&	112a	&	2019 07 28 02:24:26.088 	&	11.04(29)	&	9.93(10)	&	3004(633)	\\
108a	&	2019 07 27 03:01:26.125	&	8.90(14)	&	7.47(5)	&	2503(197)	&	112b	&		&		&	12.15(54)	&		\\
108b	&		&	10.58(48)	&	9.13(15)	&	2480(675)	&		&		&		&		&		\\
108c	&		&		&	9.80(23)	&		&		&		&		&		&		\\
\hline
\end{tabular}}
\end{table*}

\begin{figure*}[h]
\begin{tabular}{ccc}
\includegraphics[width=5.6cm]{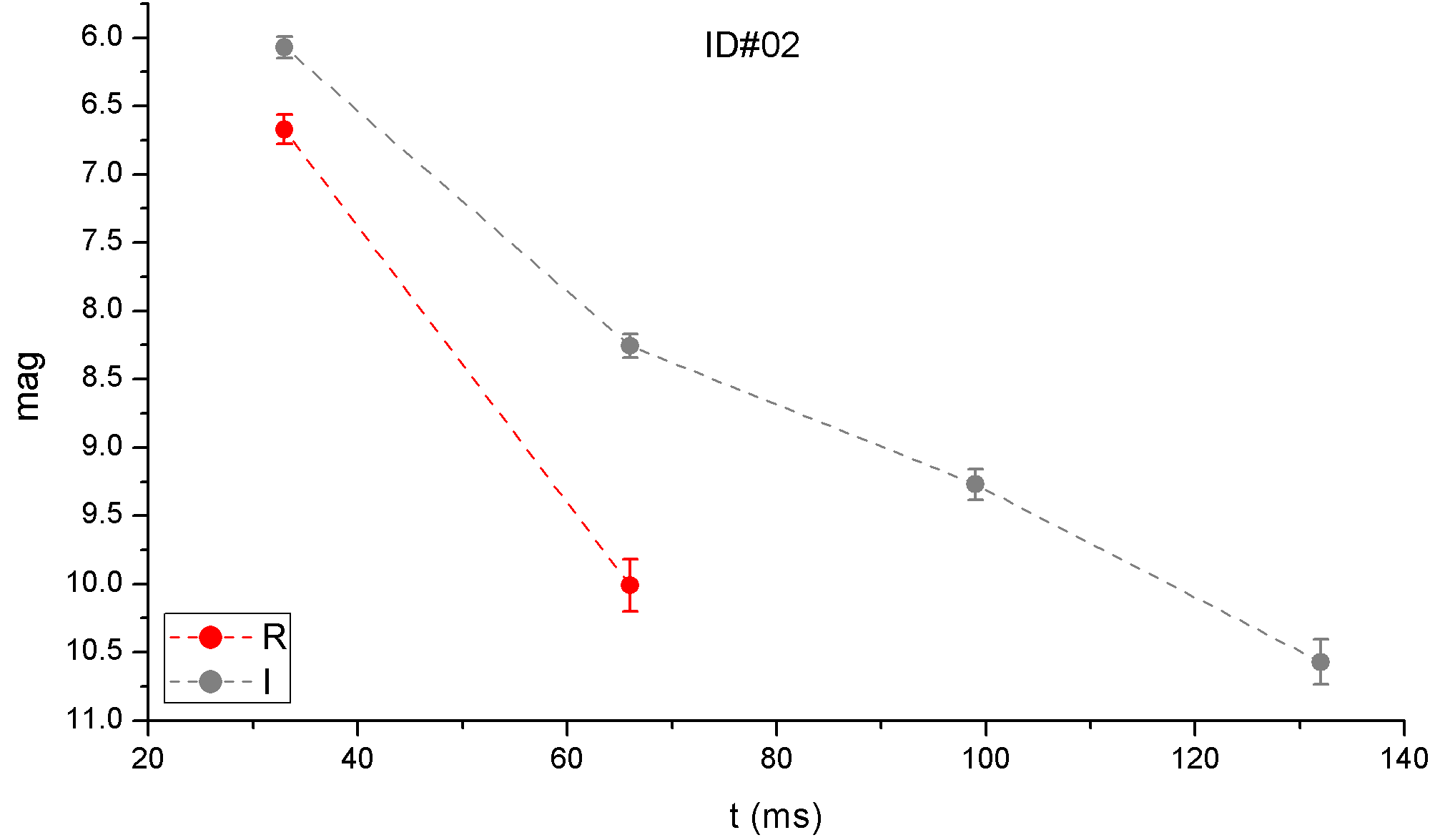}&\includegraphics[width=5.6cm]{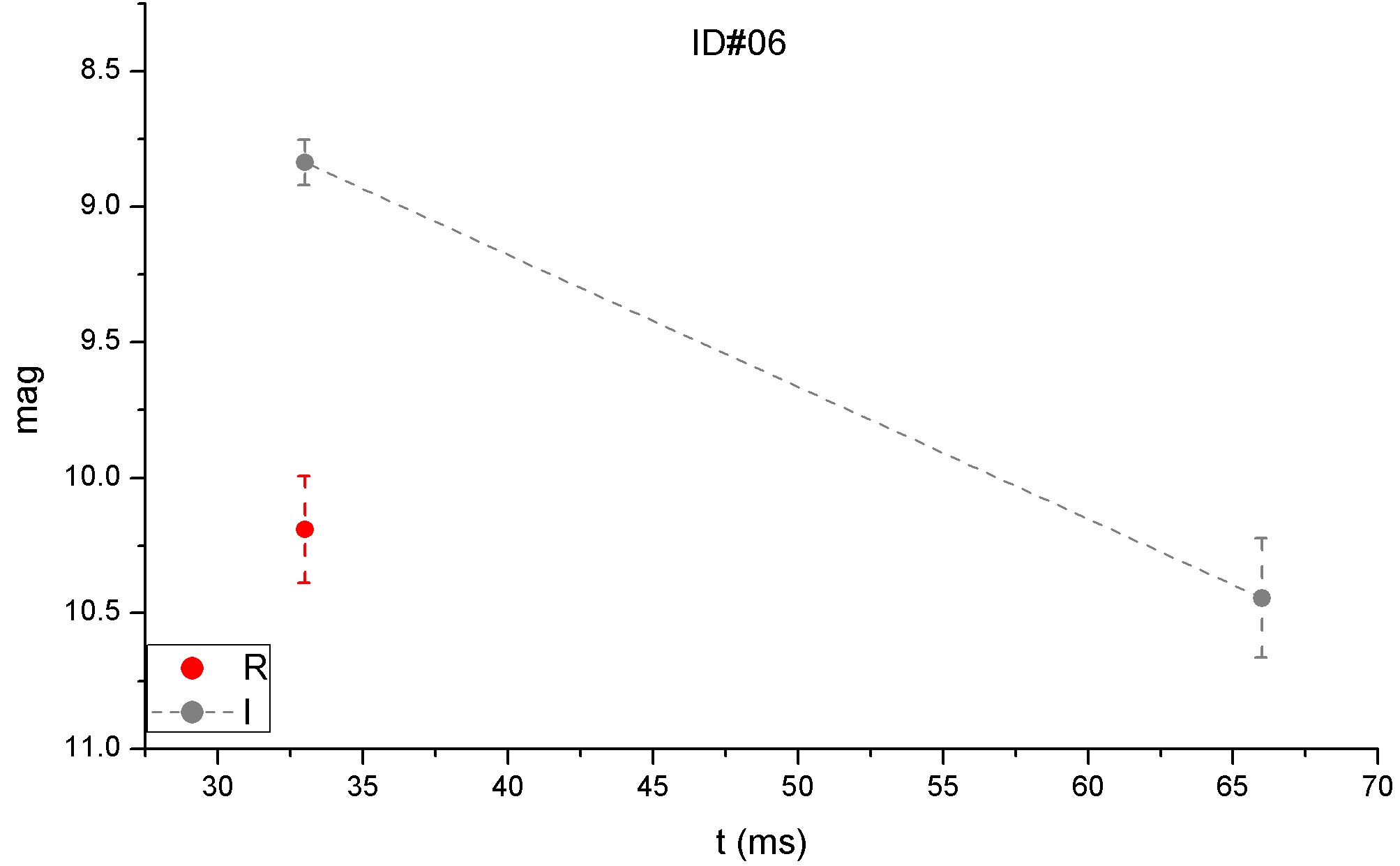}&\includegraphics[width=5.6cm]{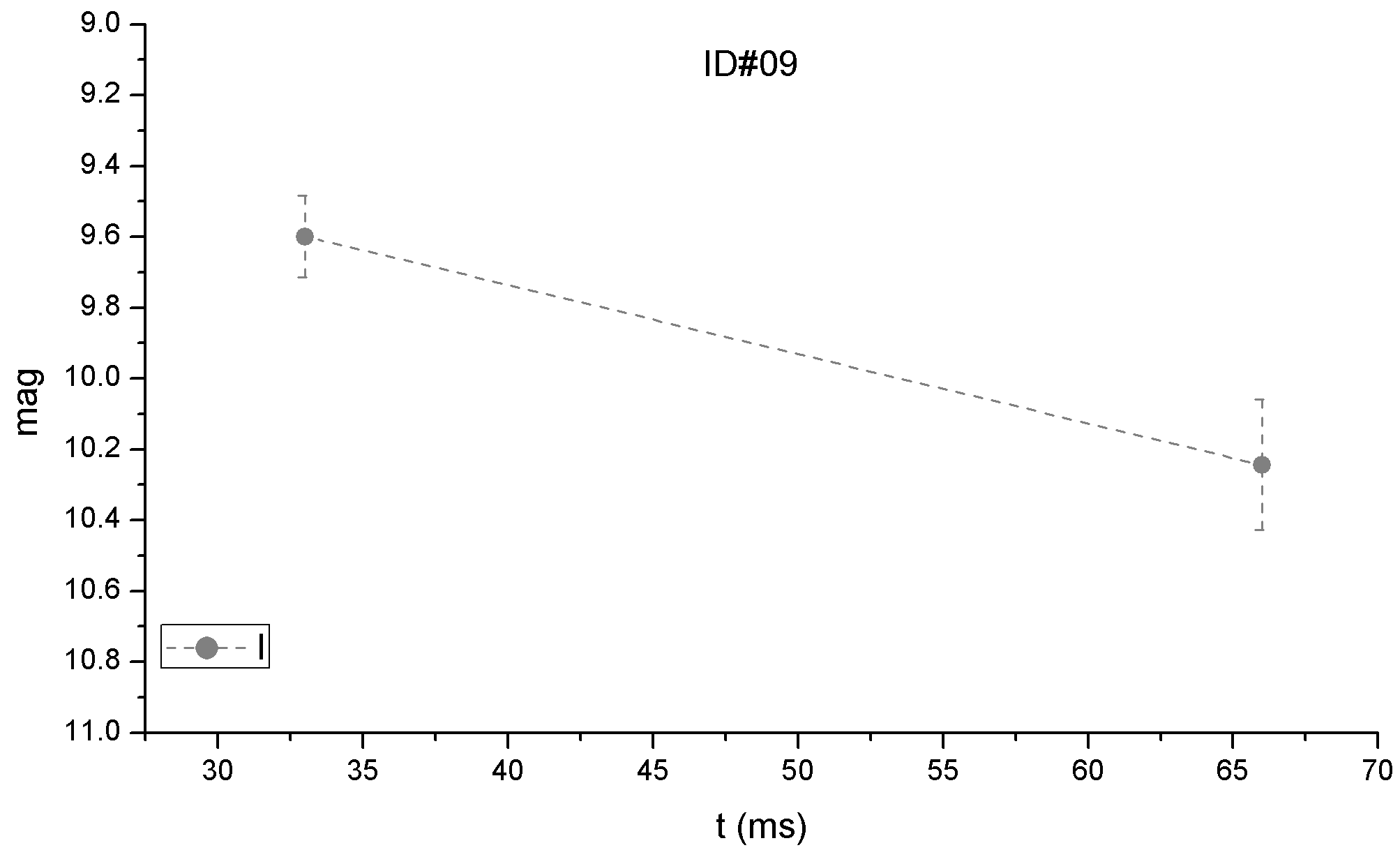}\\
\includegraphics[width=5.6cm]{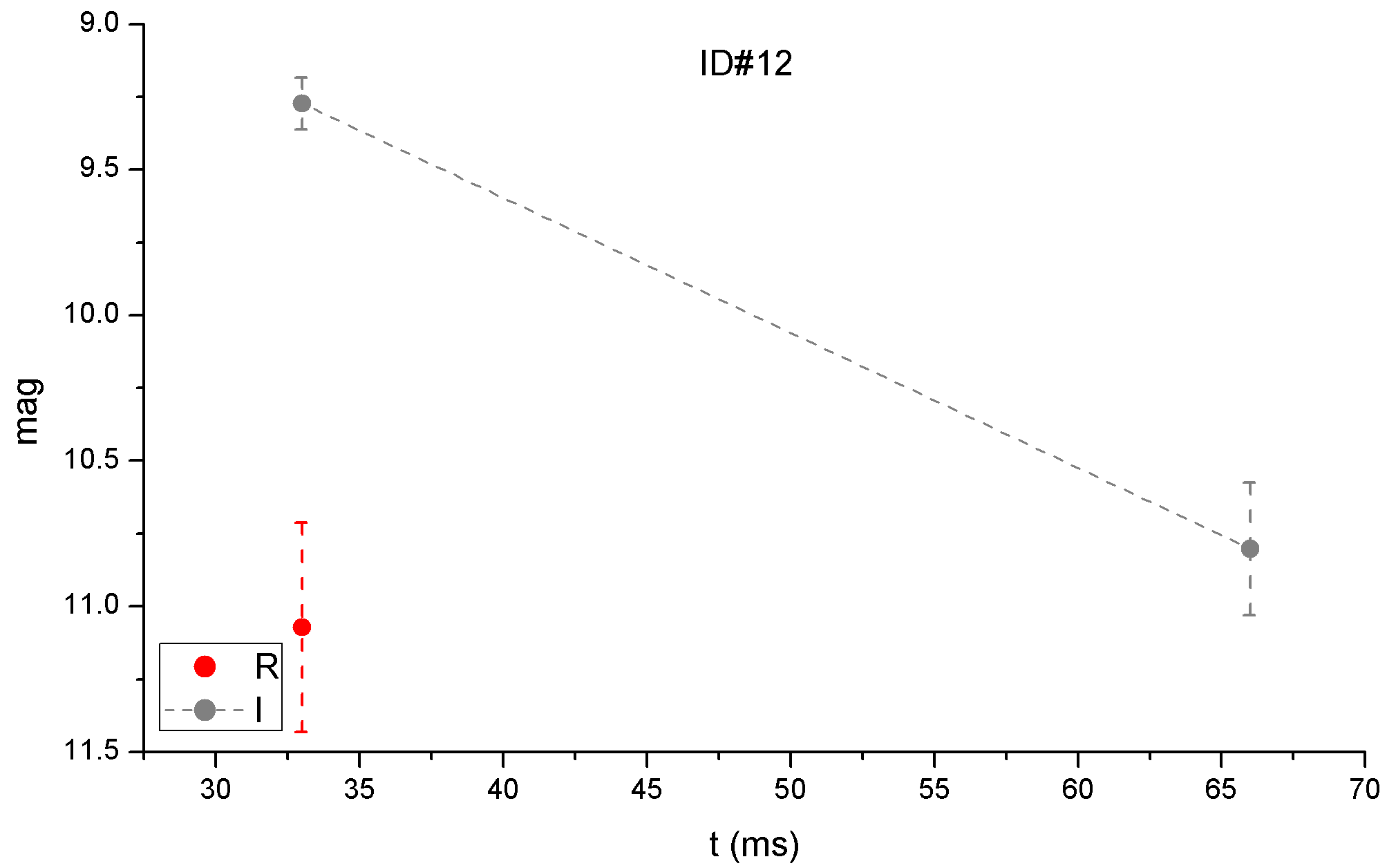}&\includegraphics[width=5.6cm]{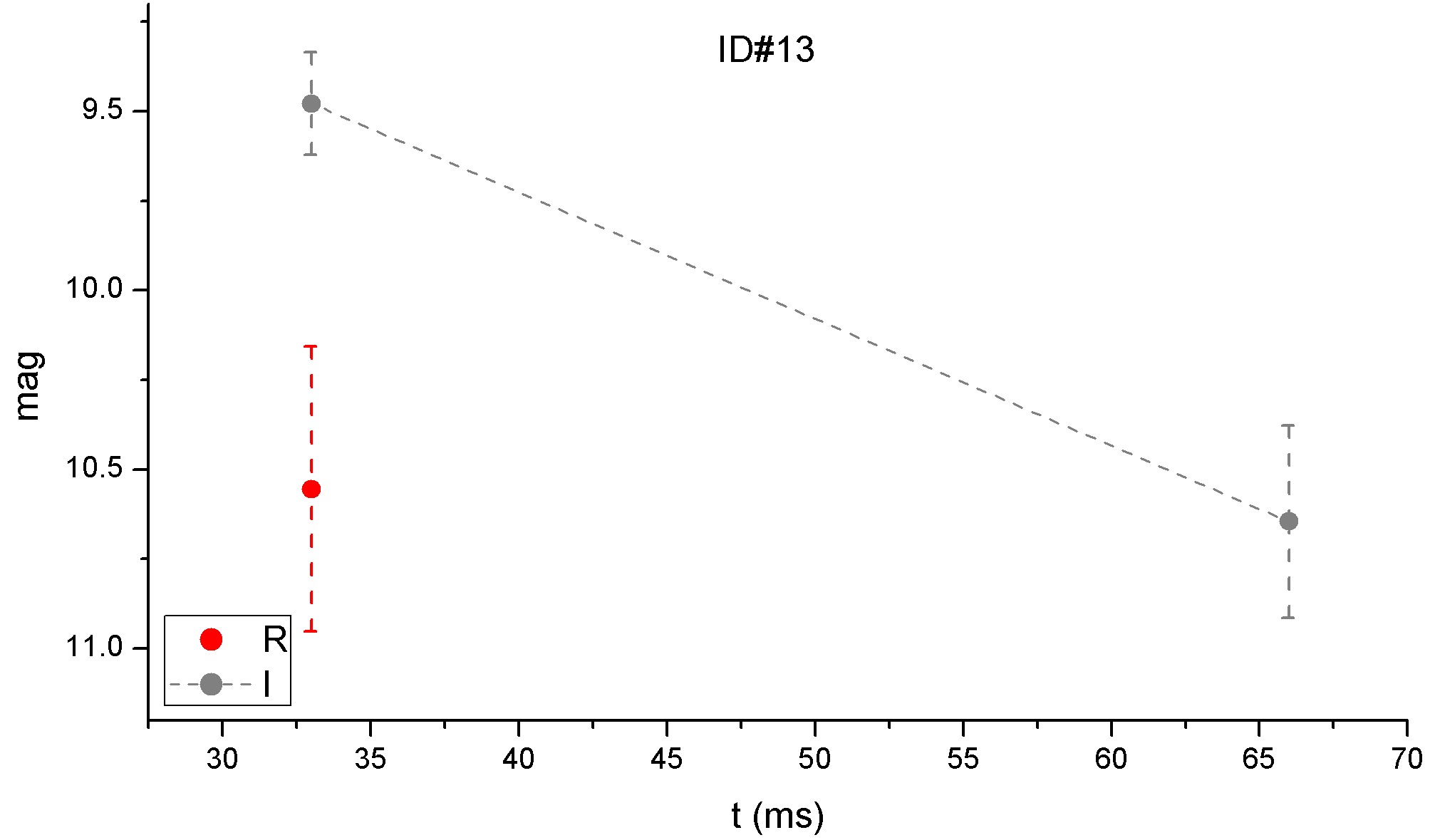}&\includegraphics[width=5.6cm]{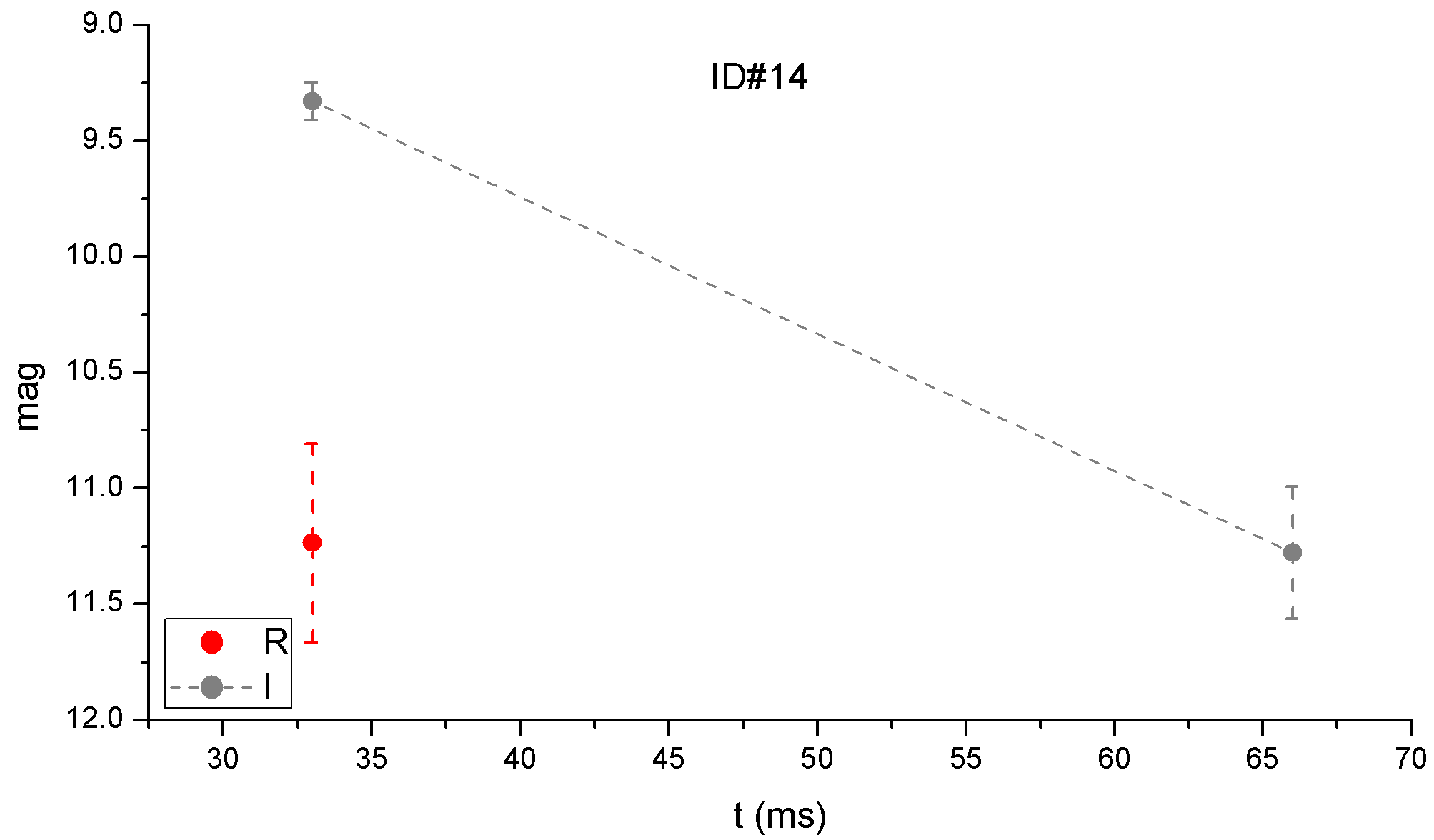}\\
\includegraphics[width=5.6cm]{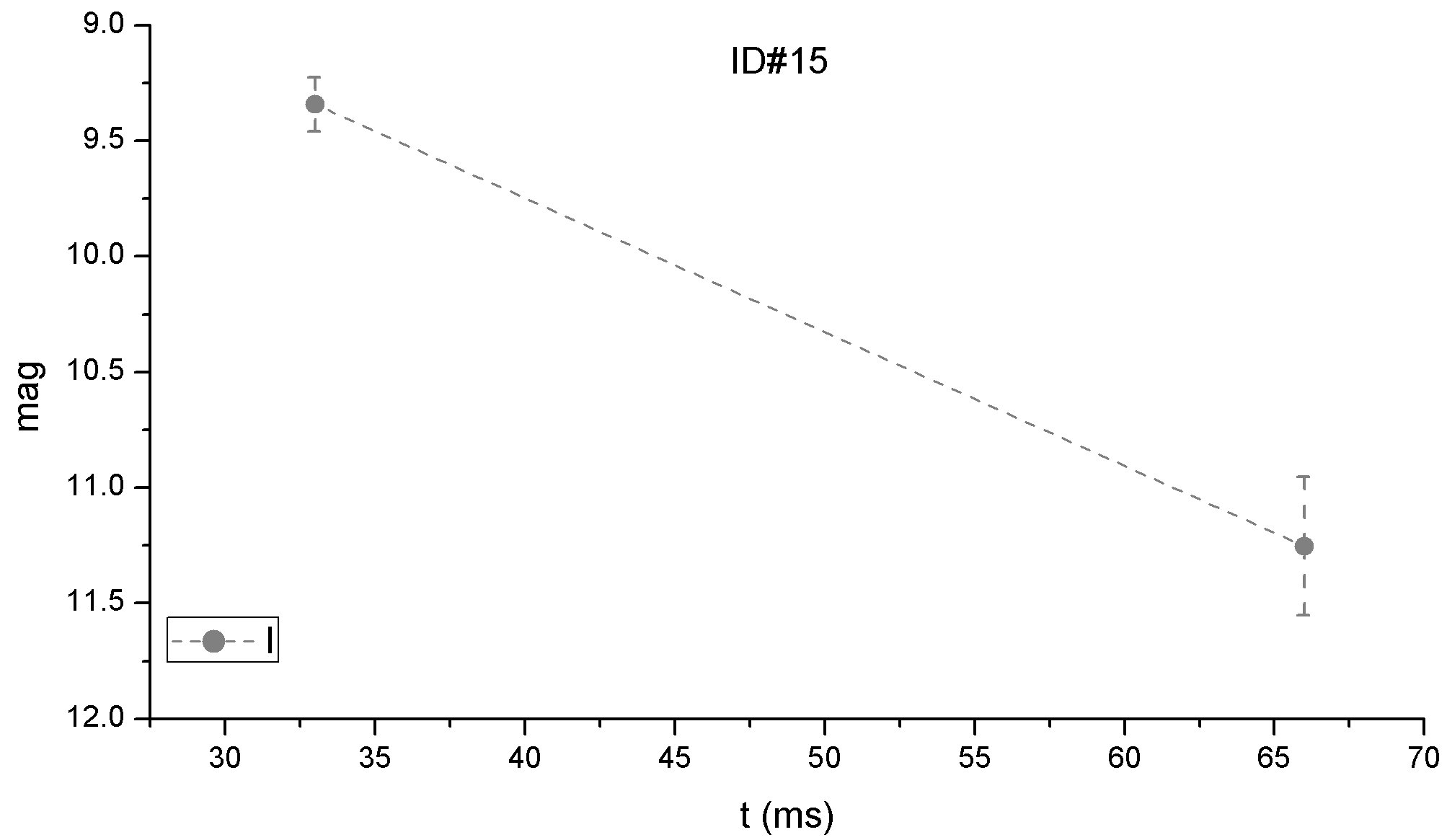}&\includegraphics[width=5.6cm]{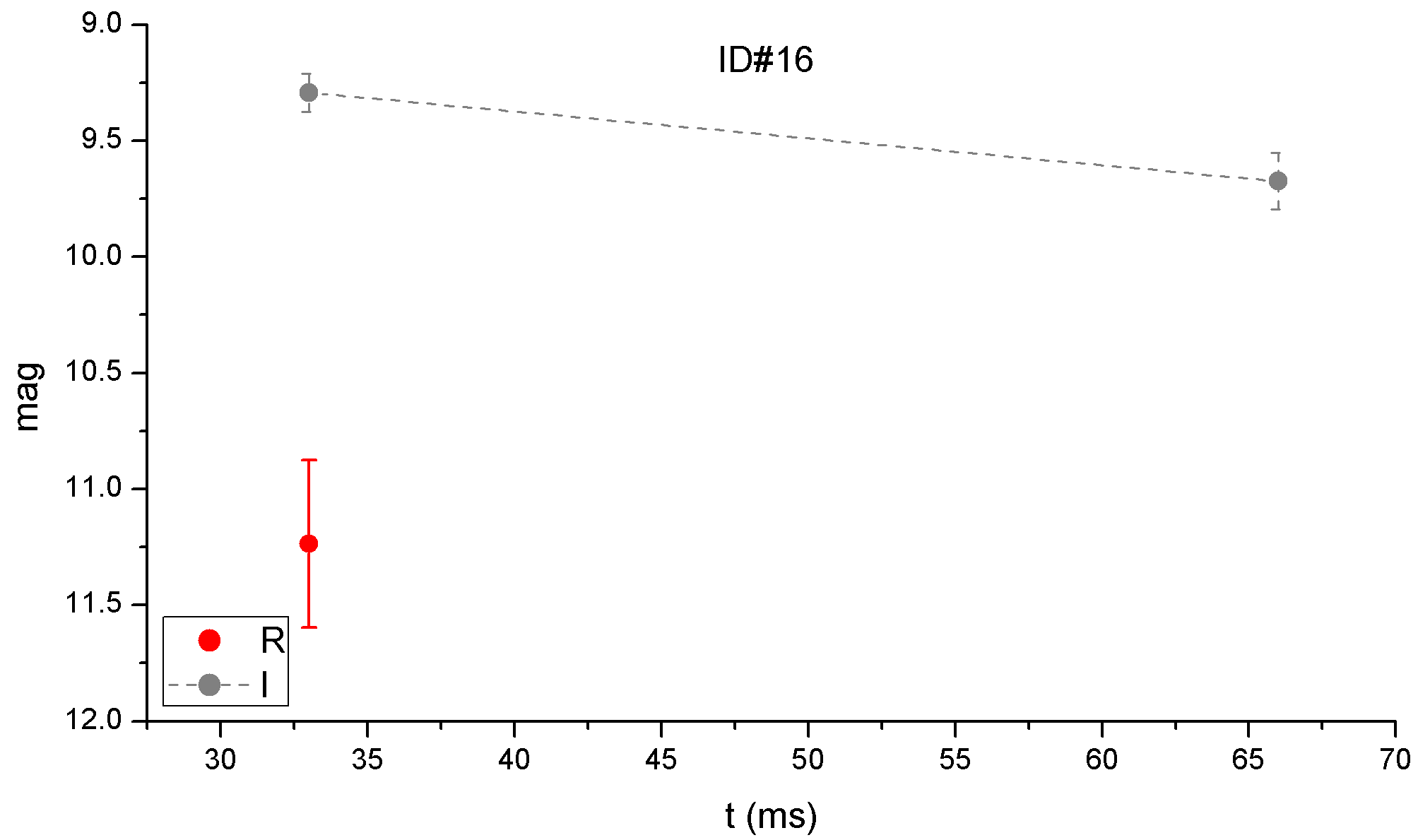}&\includegraphics[width=5.6cm]{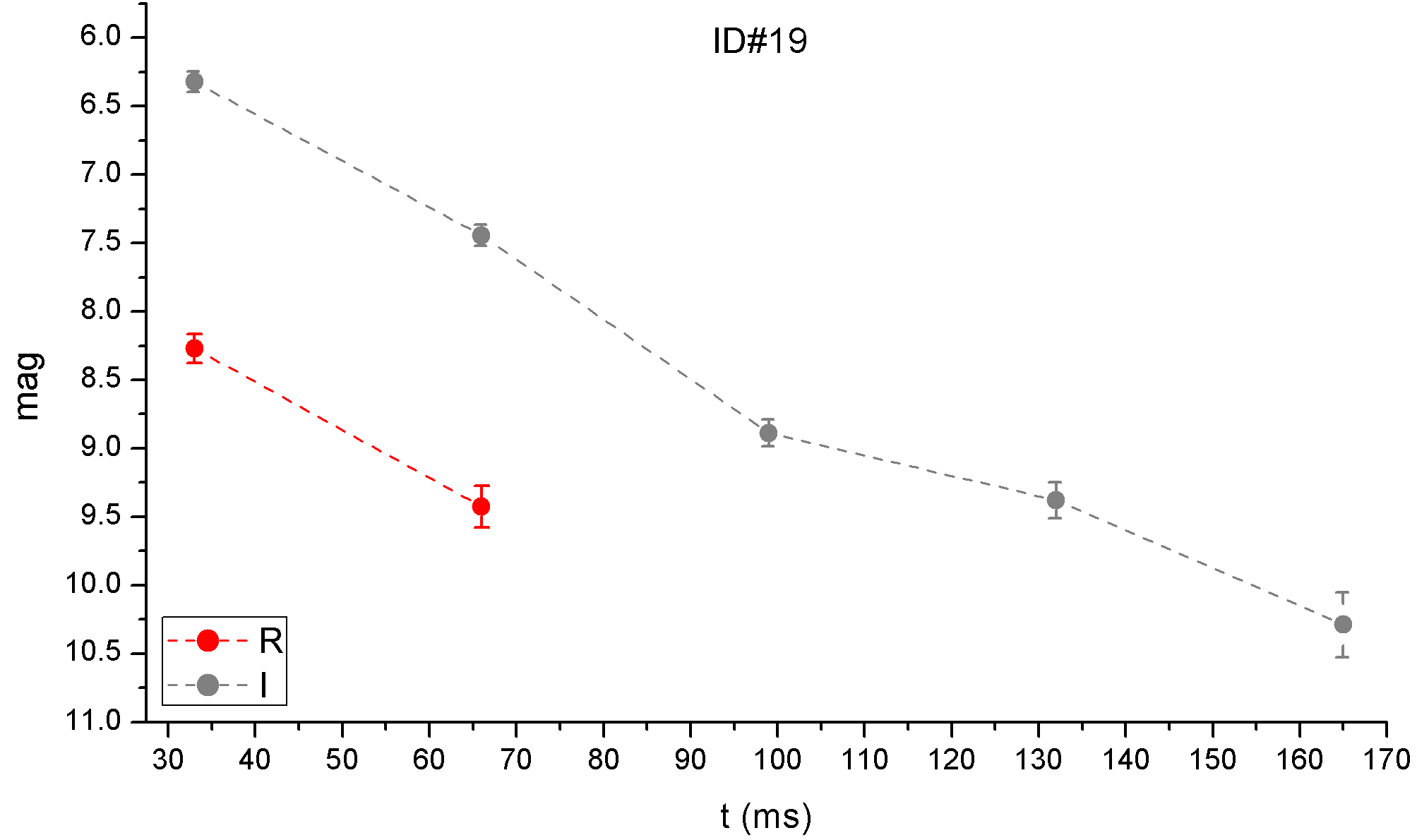}\\
\includegraphics[width=5.6cm]{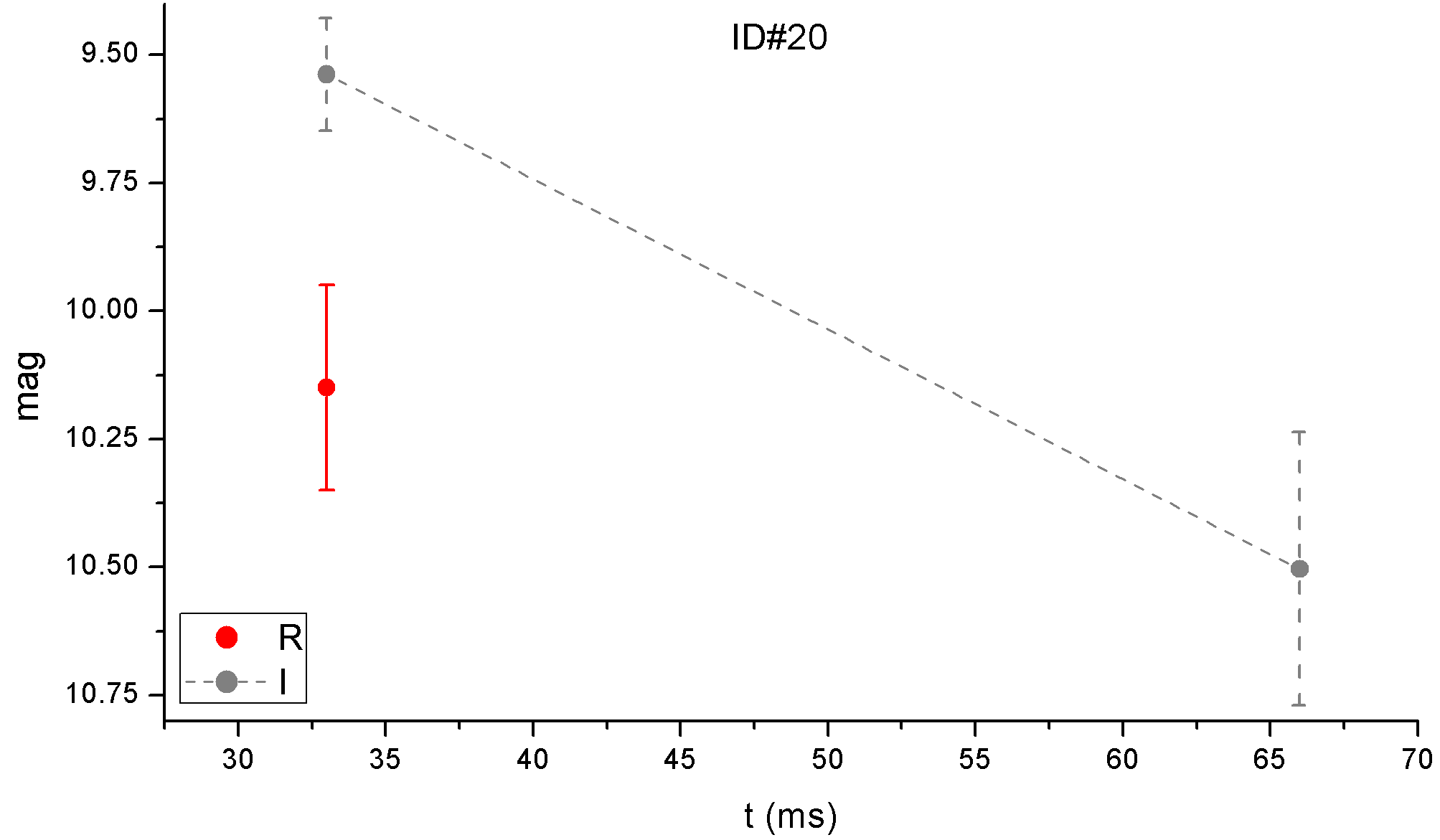}&\includegraphics[width=5.6cm]{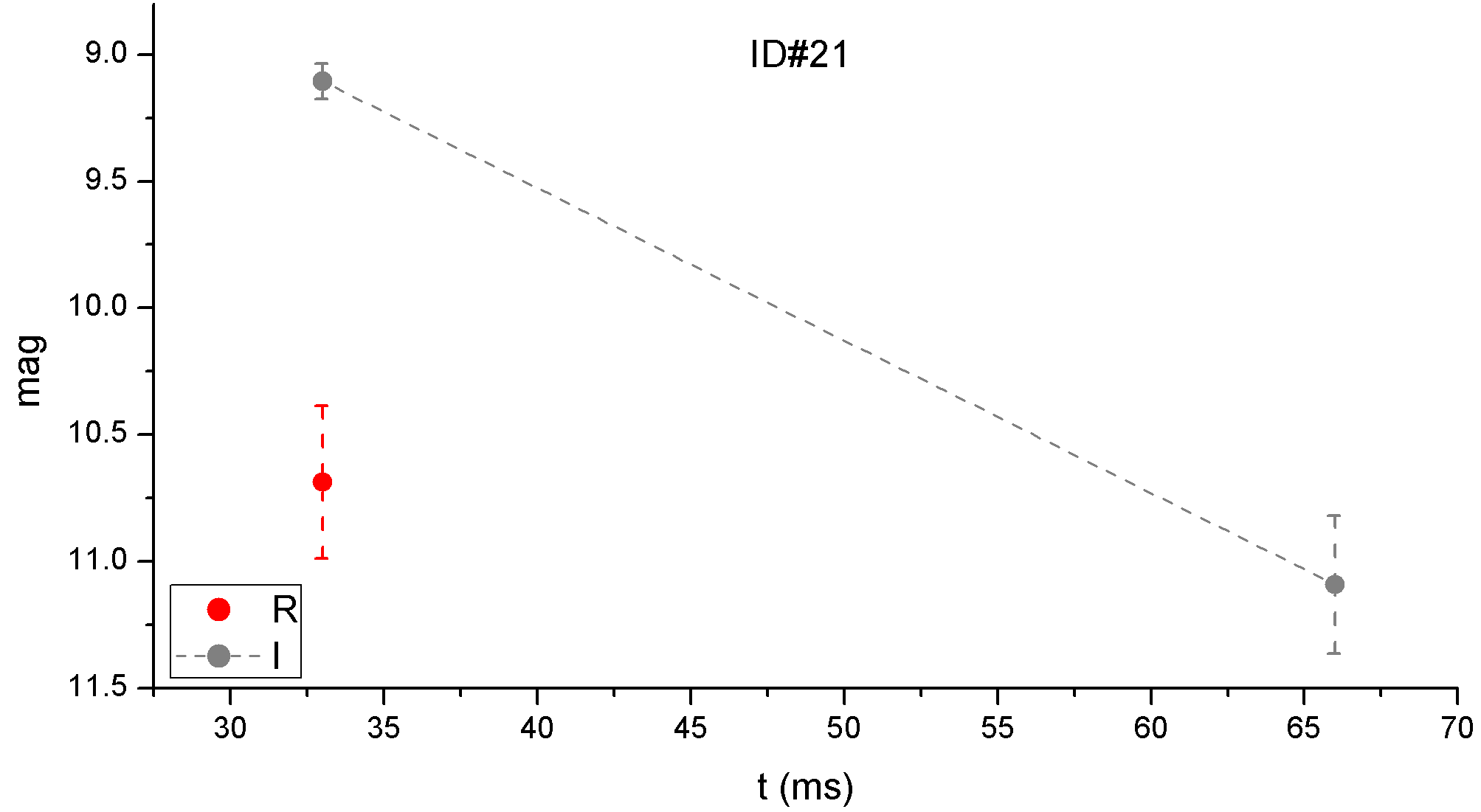}&\includegraphics[width=5.6cm]{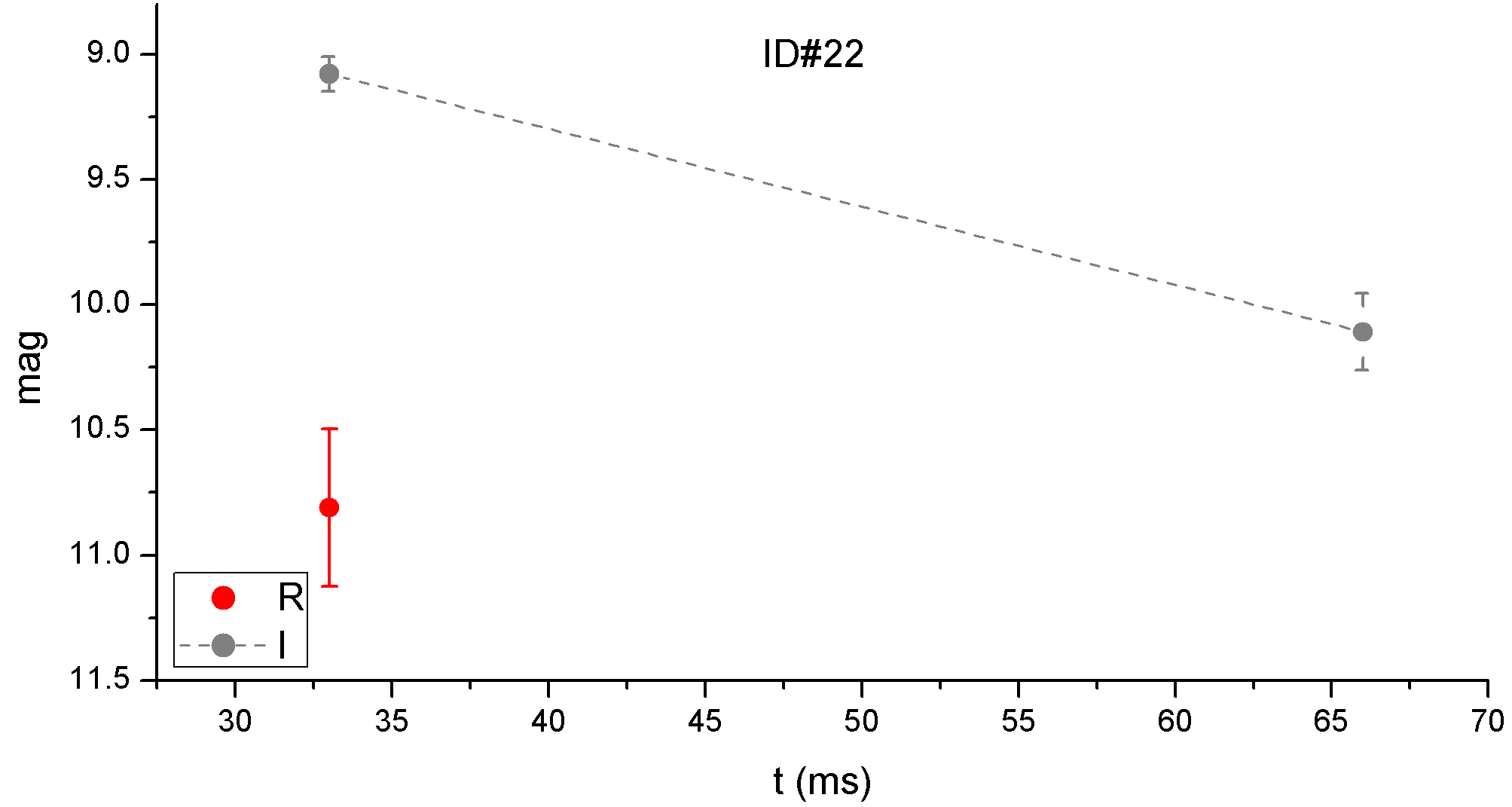}\\
\includegraphics[width=5.6cm]{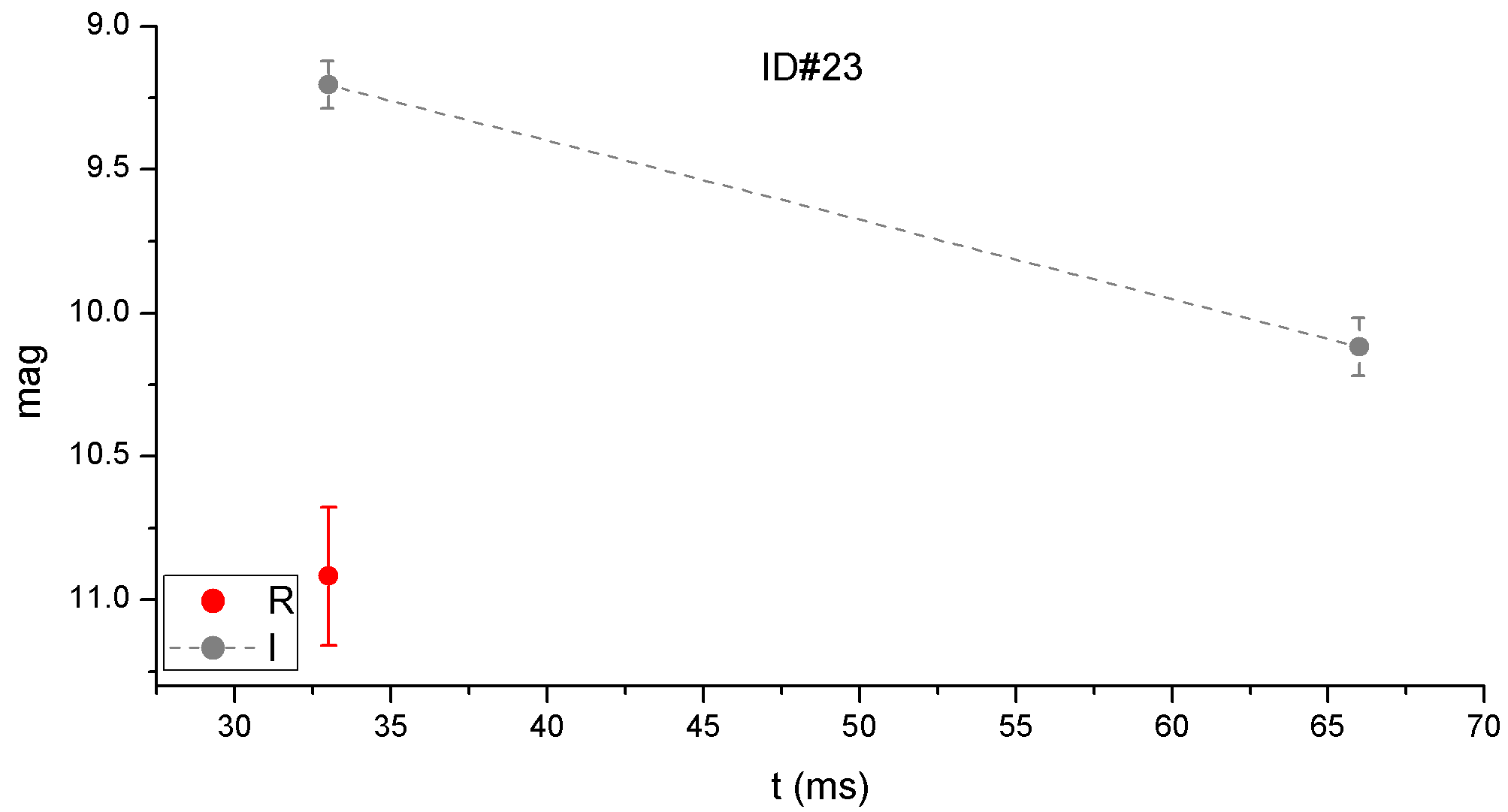}&\includegraphics[width=5.6cm]{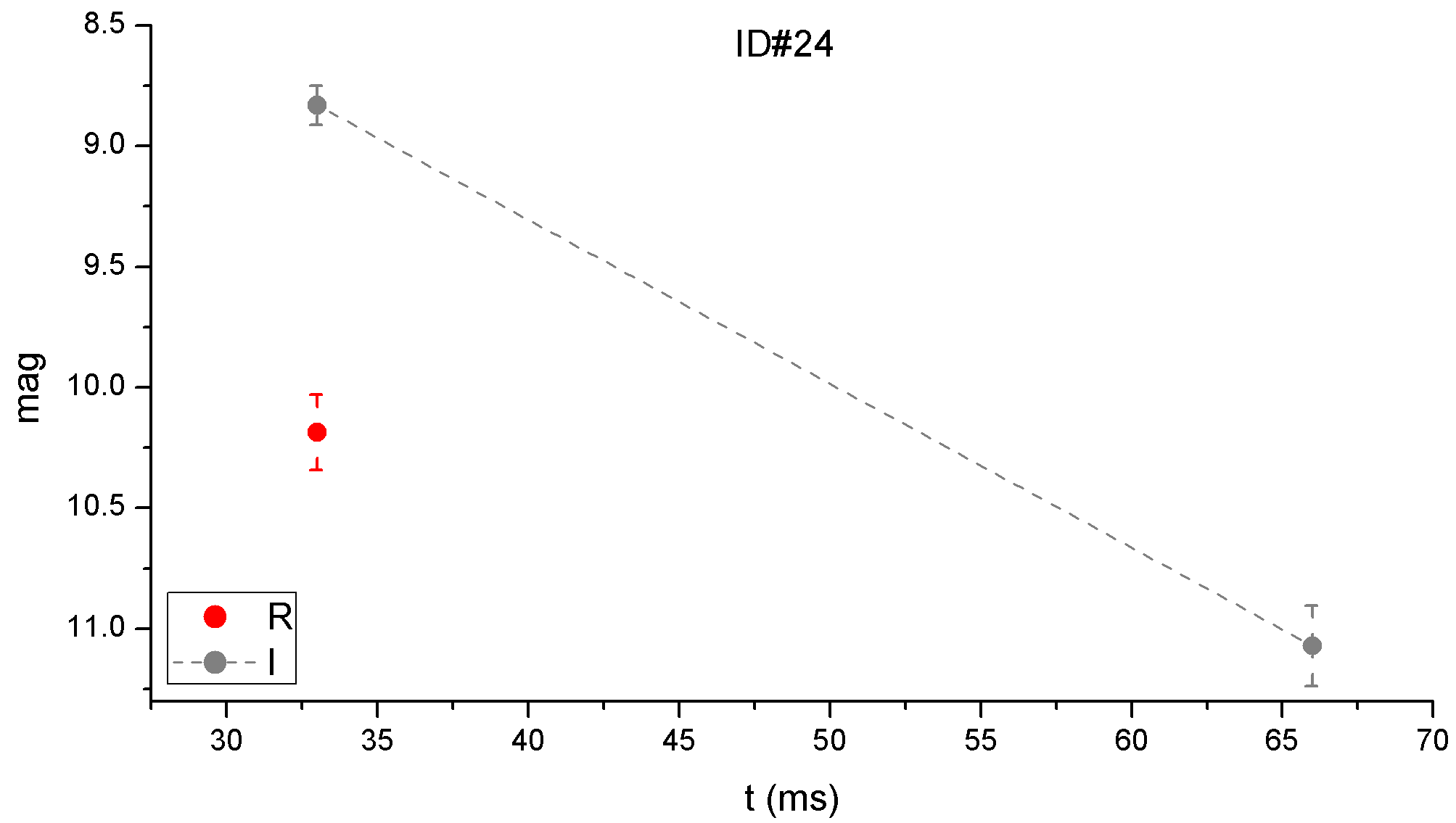}&\includegraphics[width=5.6cm]{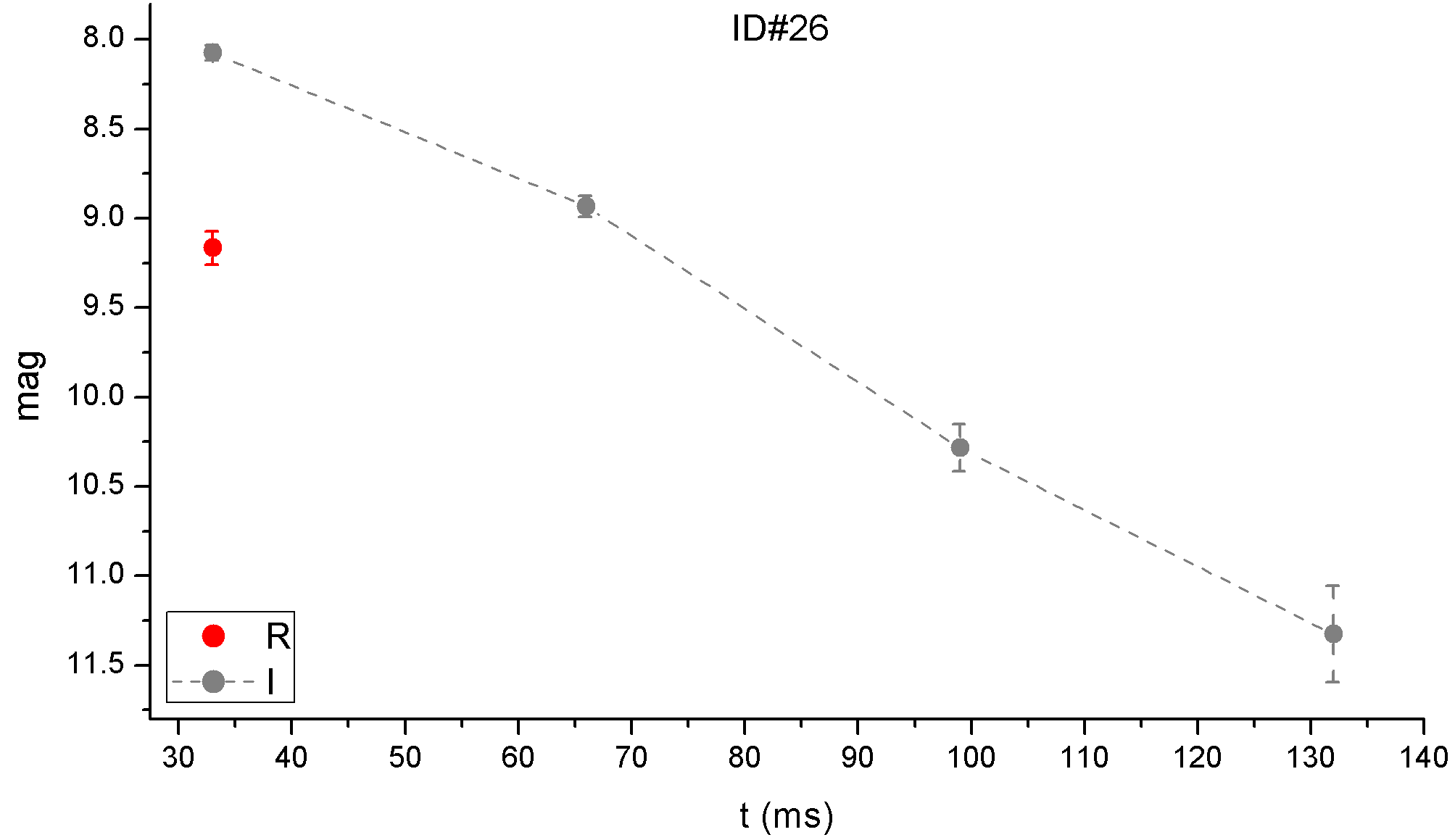}\\
\includegraphics[width=5.6cm]{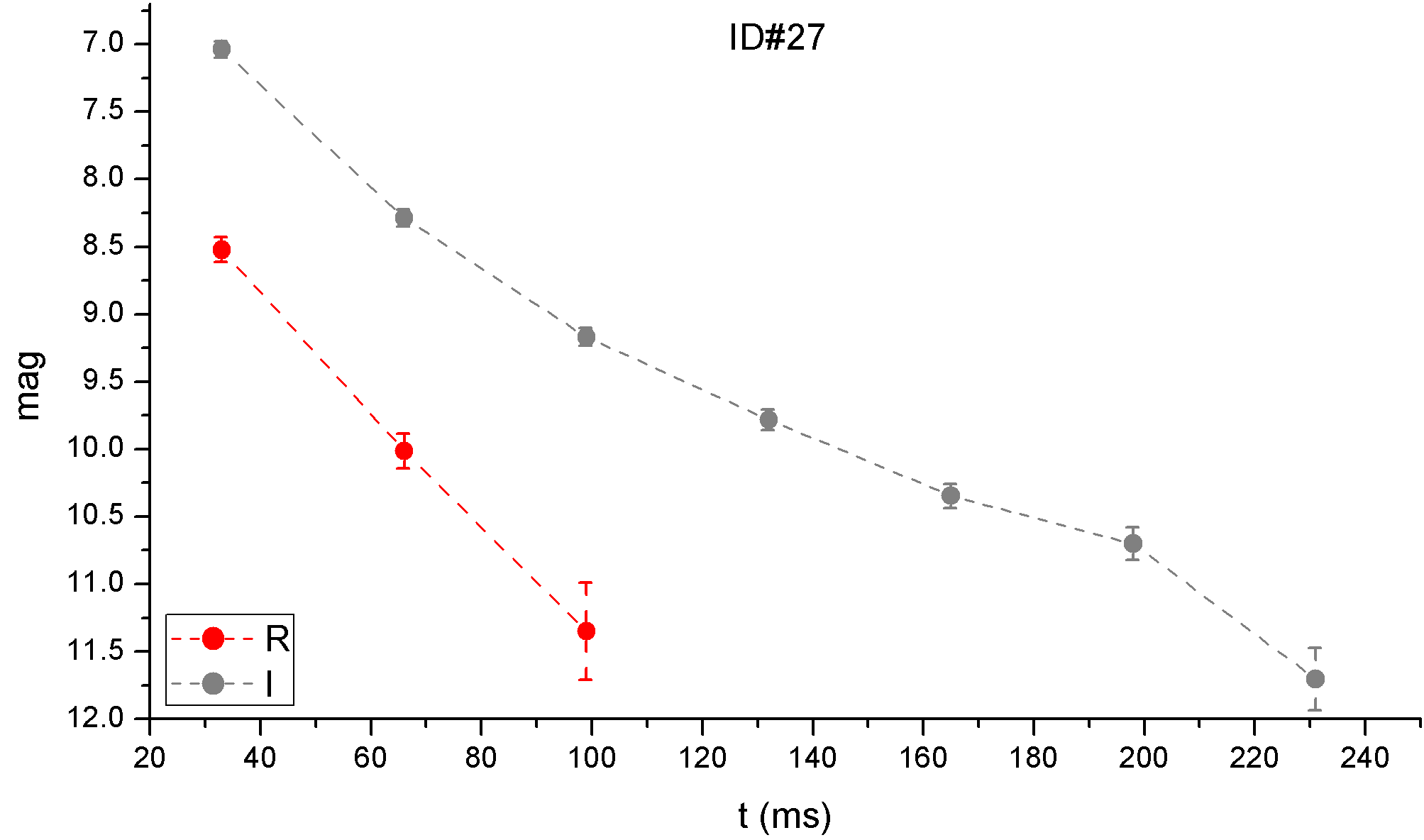}&\includegraphics[width=5.6cm]{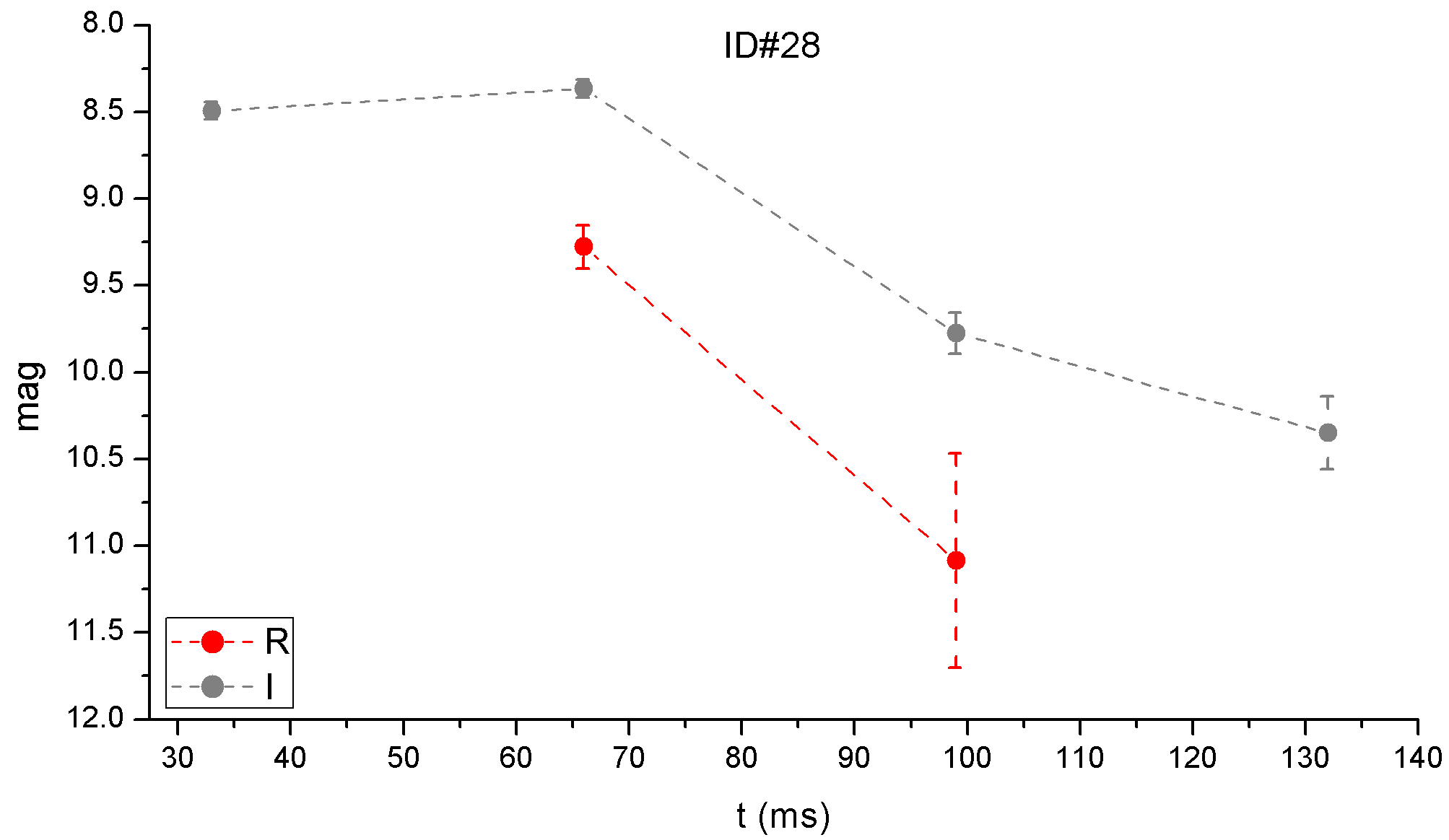}&\includegraphics[width=5.6cm]{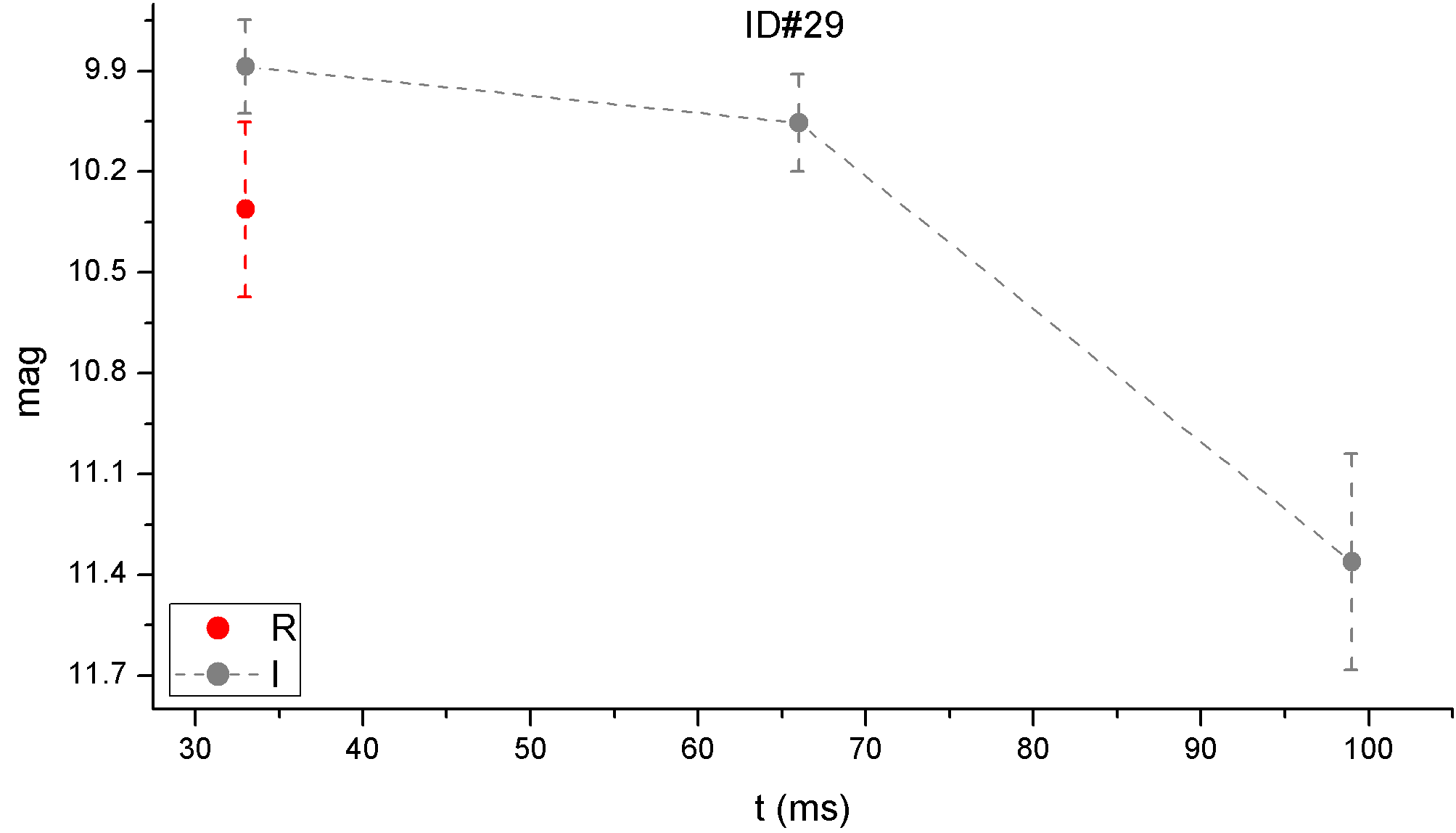}\\
\includegraphics[width=5.6cm]{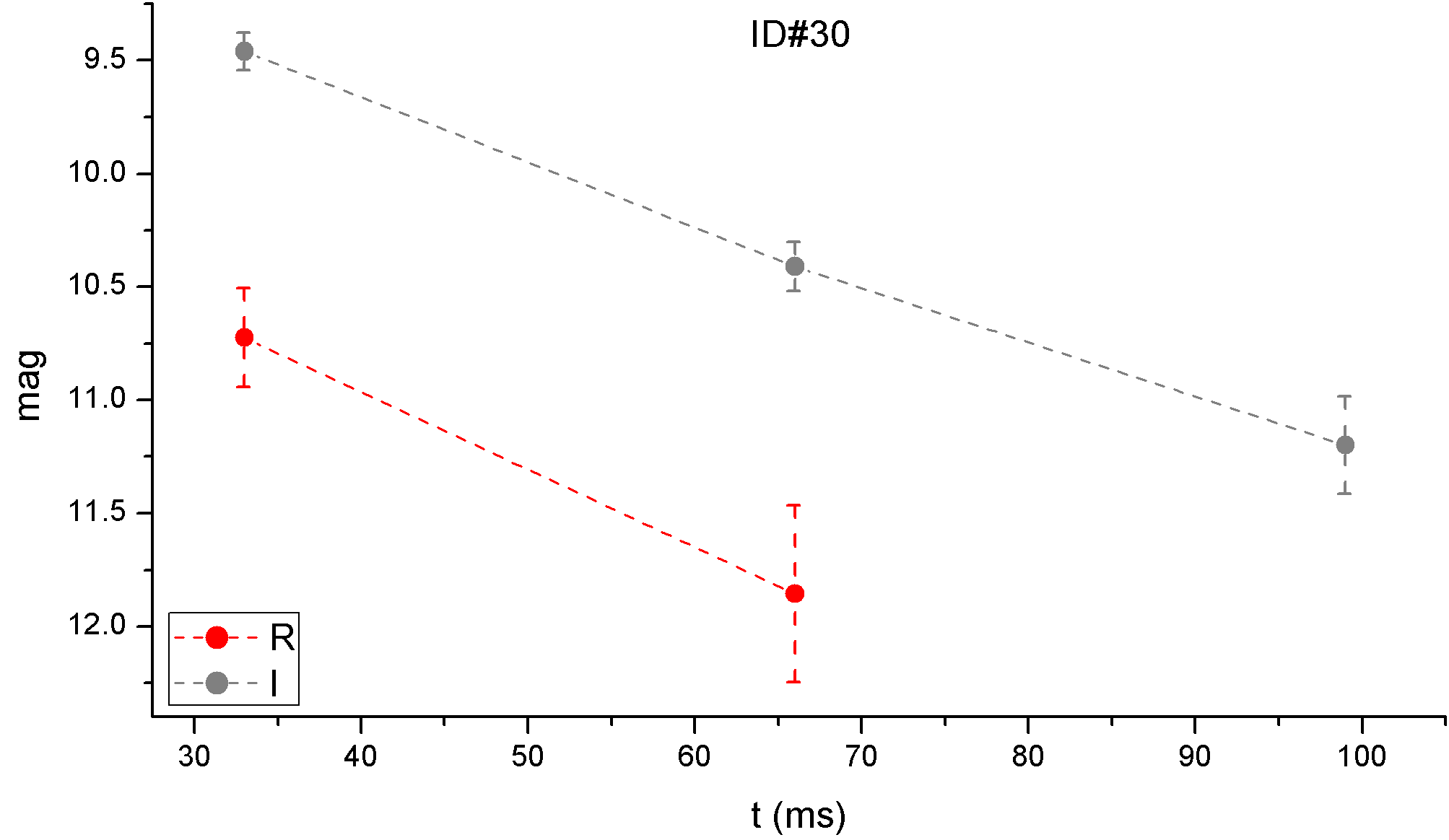}&\includegraphics[width=5.6cm]{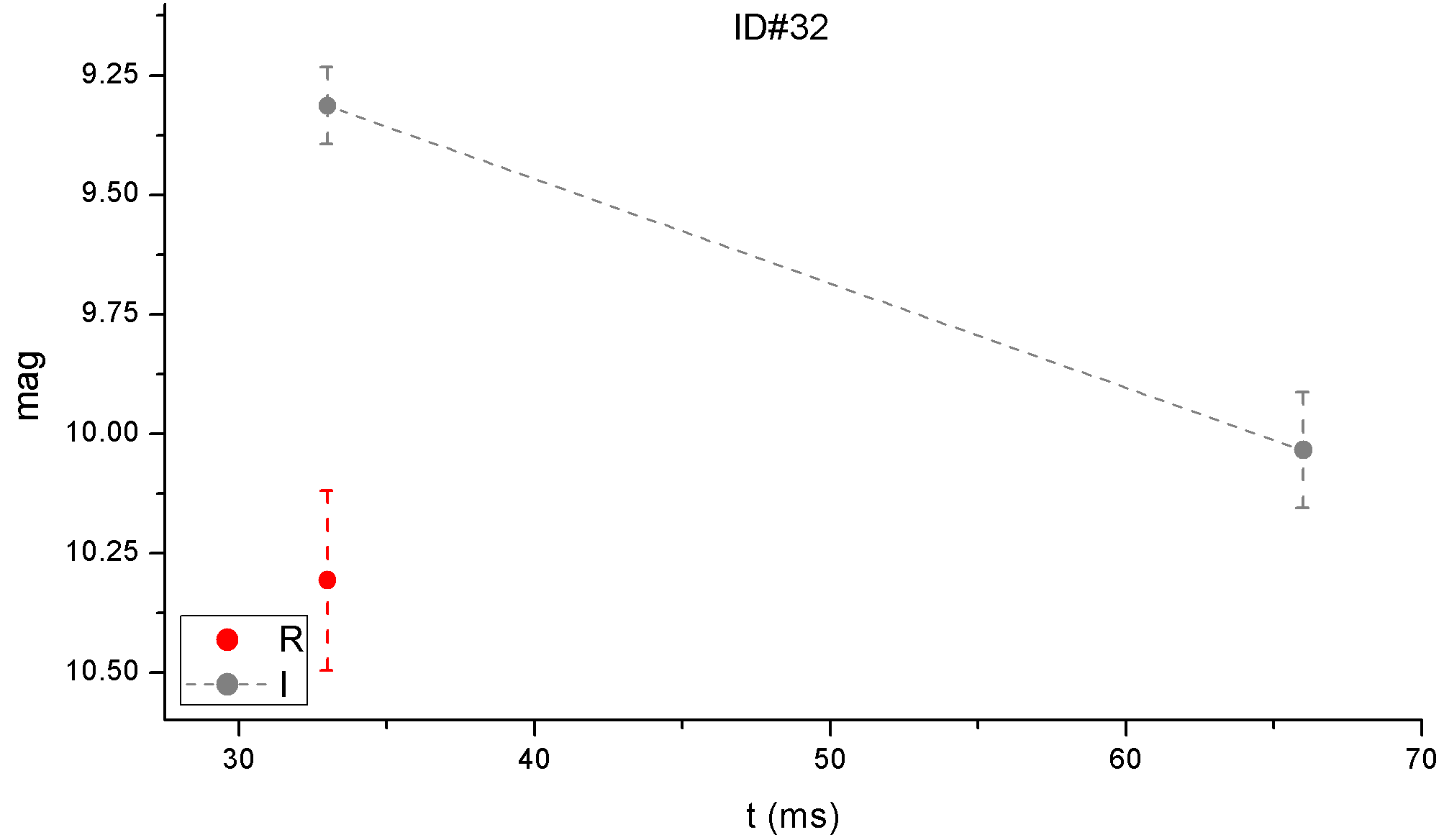}&\includegraphics[width=5.6cm]{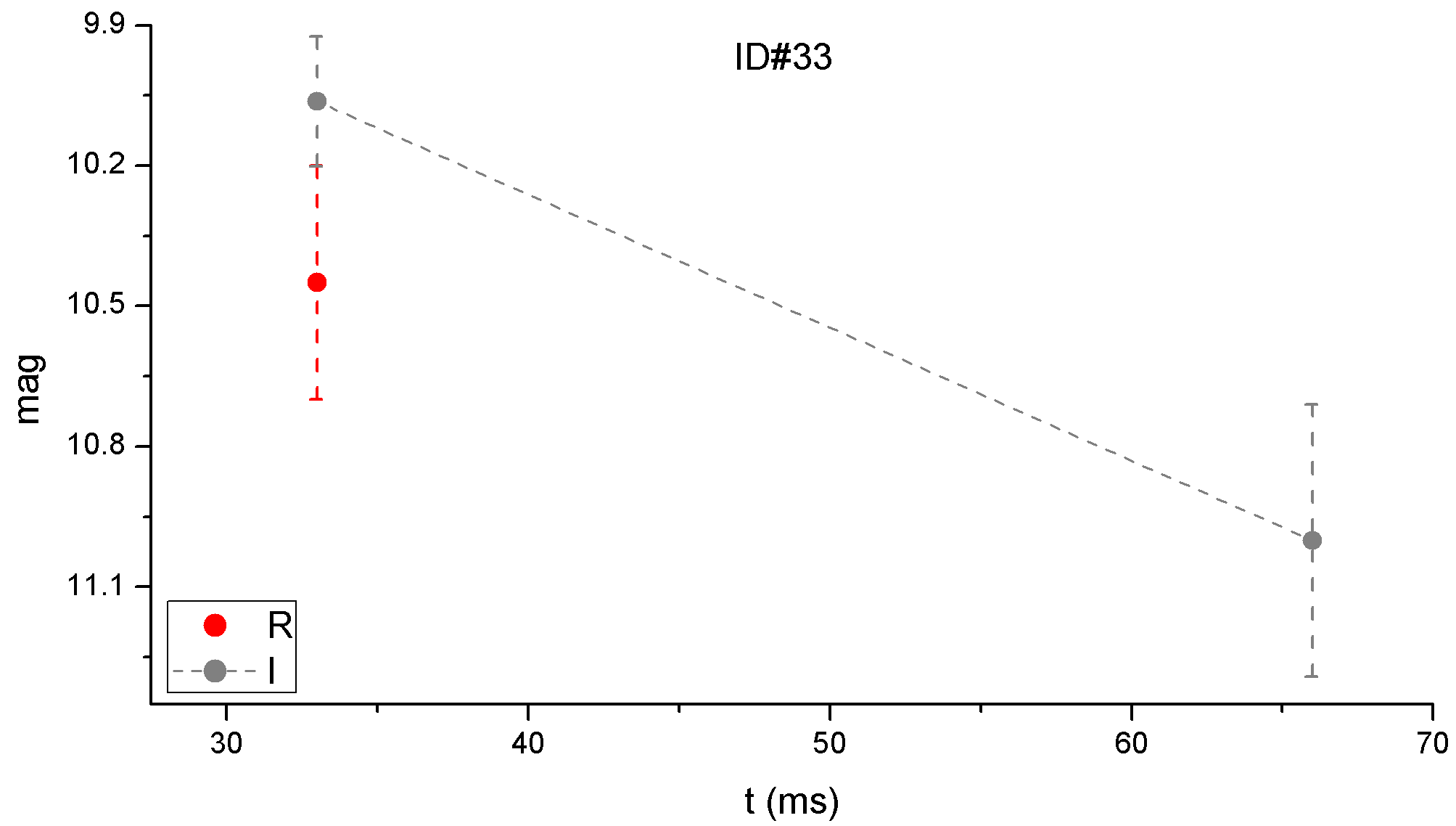}\\
\end{tabular}
\caption{Light curves of the multi-frame flashes.}
\label{fig:LCs1}
\end{figure*}	

\begin{figure*}[h]
\begin{tabular}{ccc}
\includegraphics[width=5.6cm]{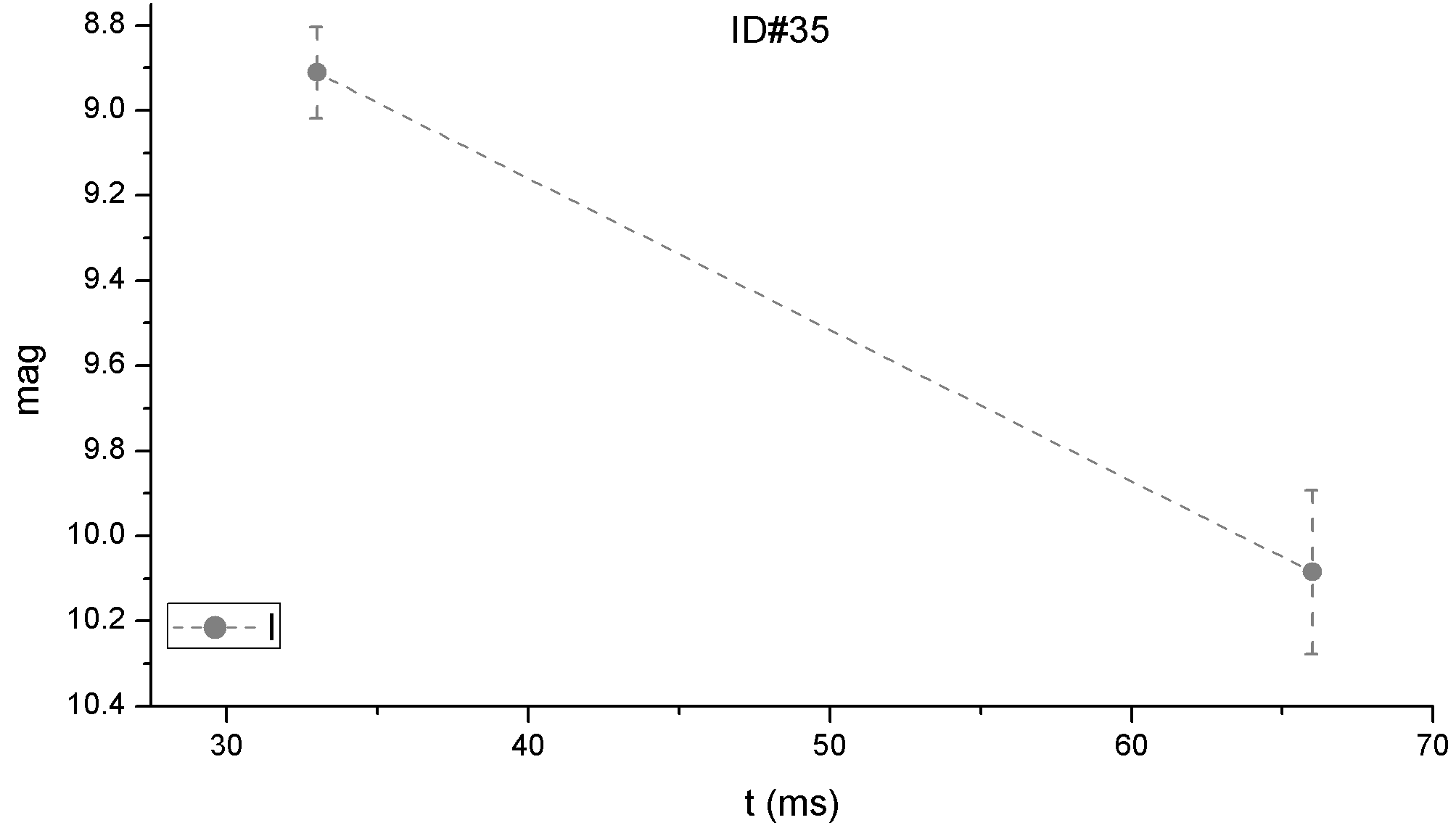}&\includegraphics[width=5.6cm]{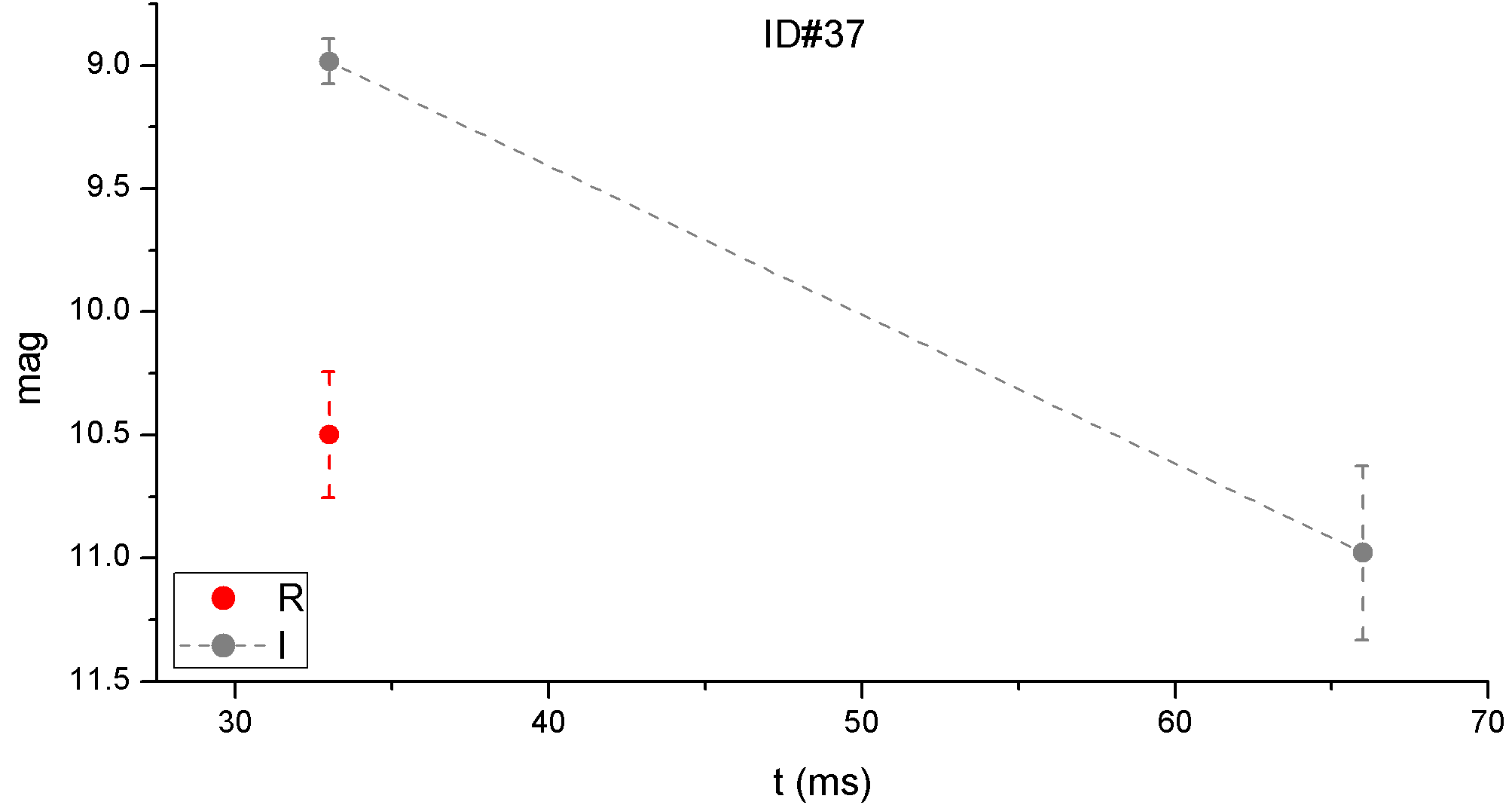}&\includegraphics[width=5.6cm]{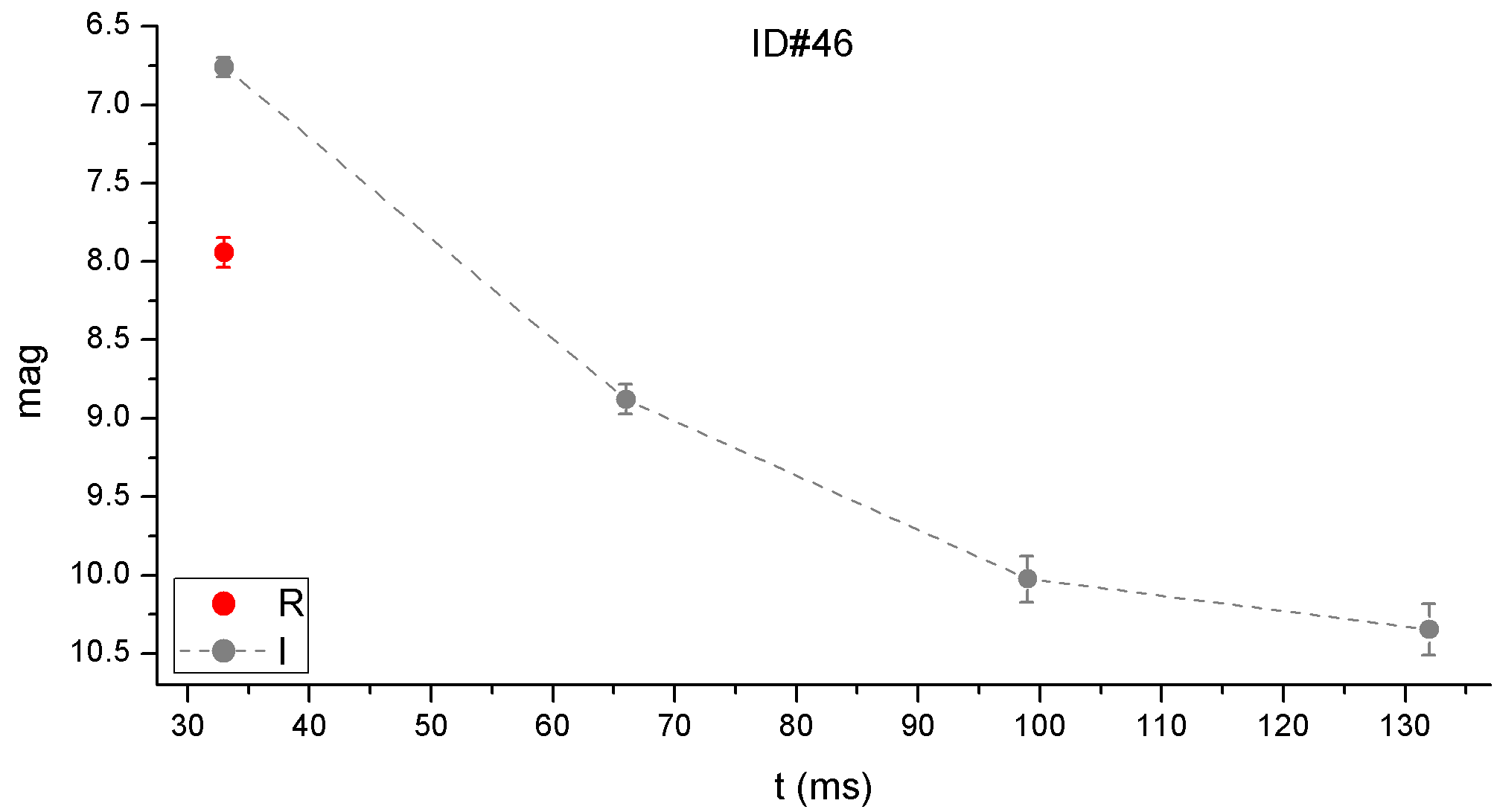}\\
\includegraphics[width=5.6cm]{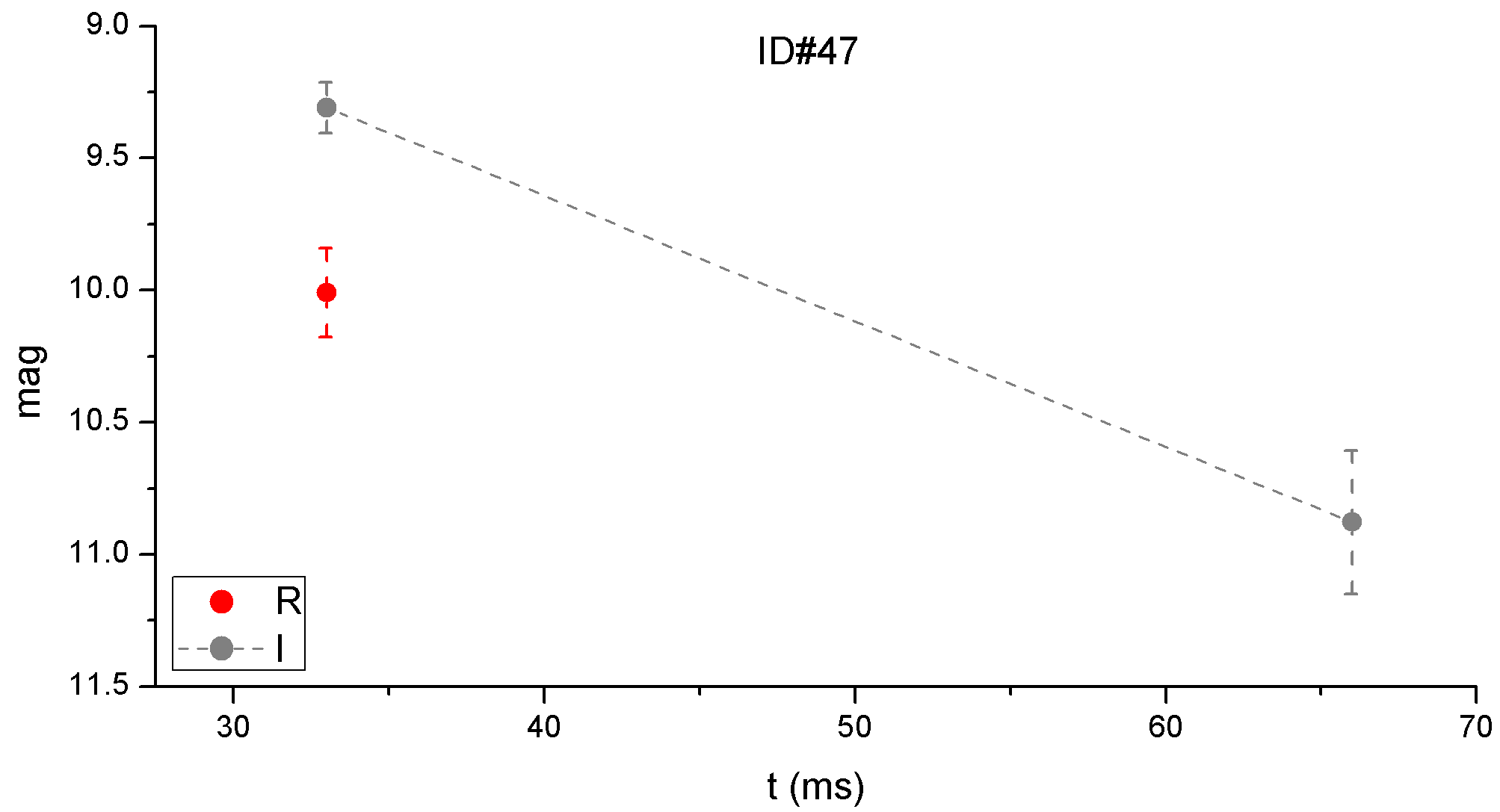}&\includegraphics[width=5.6cm]{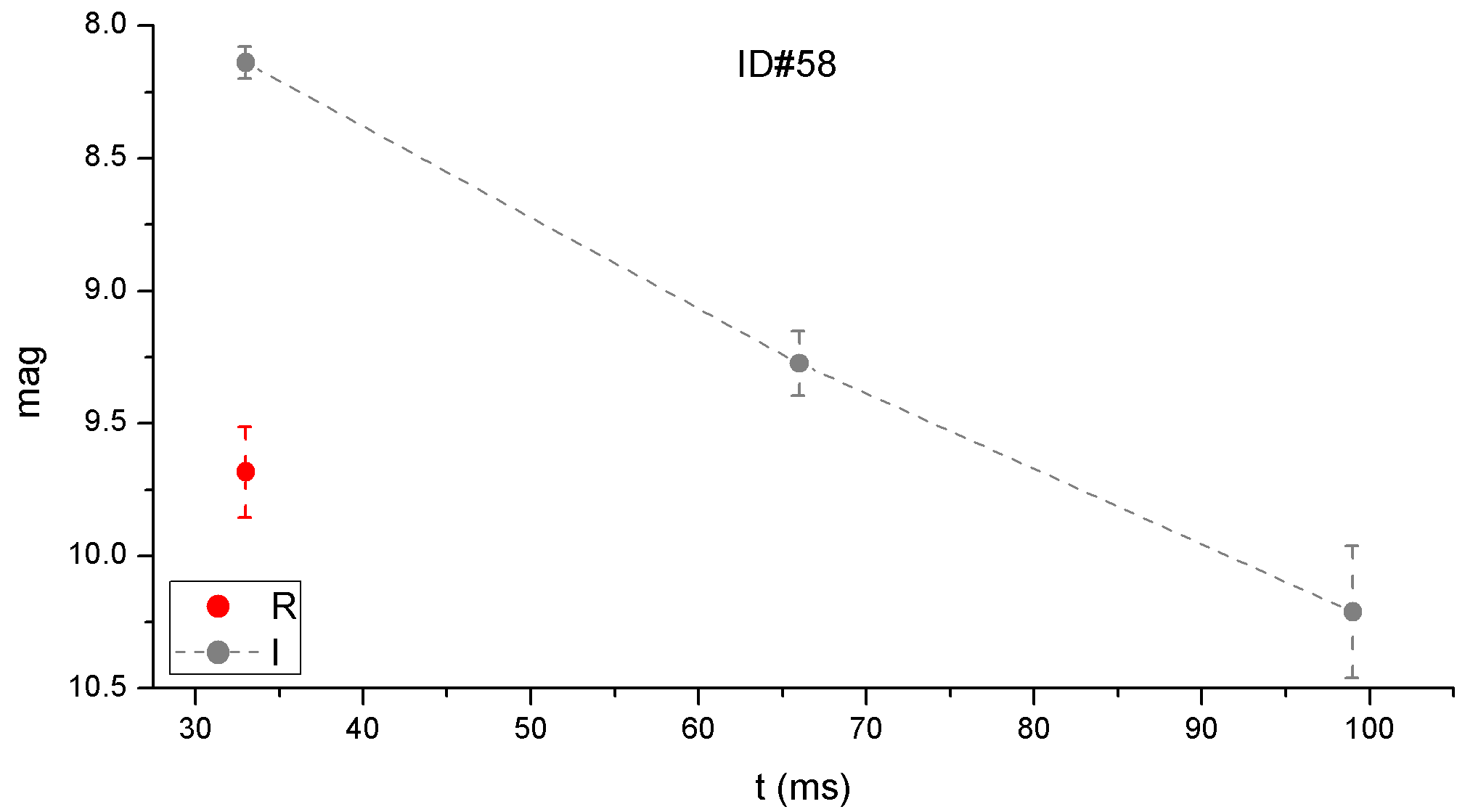}&\includegraphics[width=5.6cm]{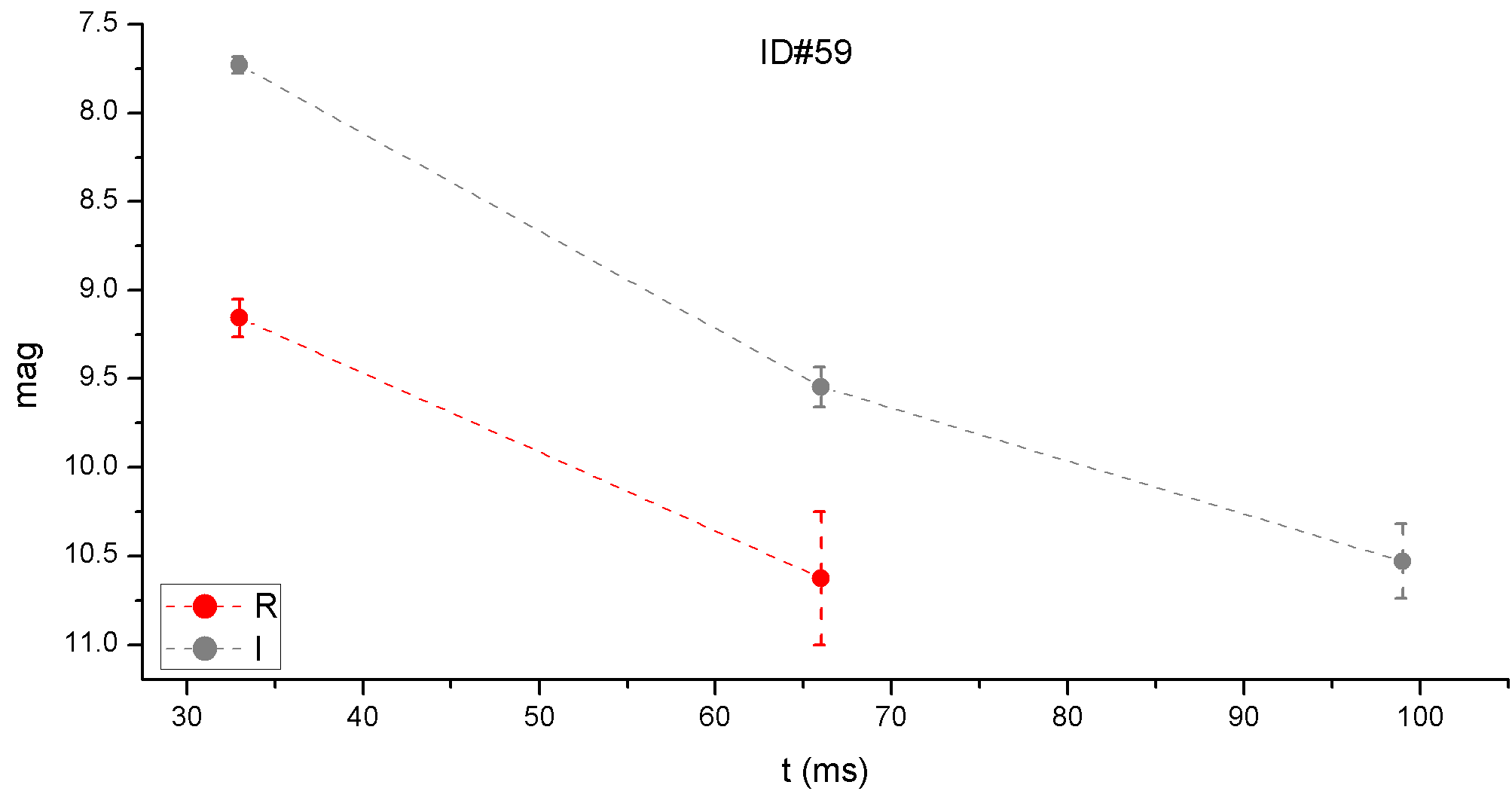}\\
\includegraphics[width=5.6cm]{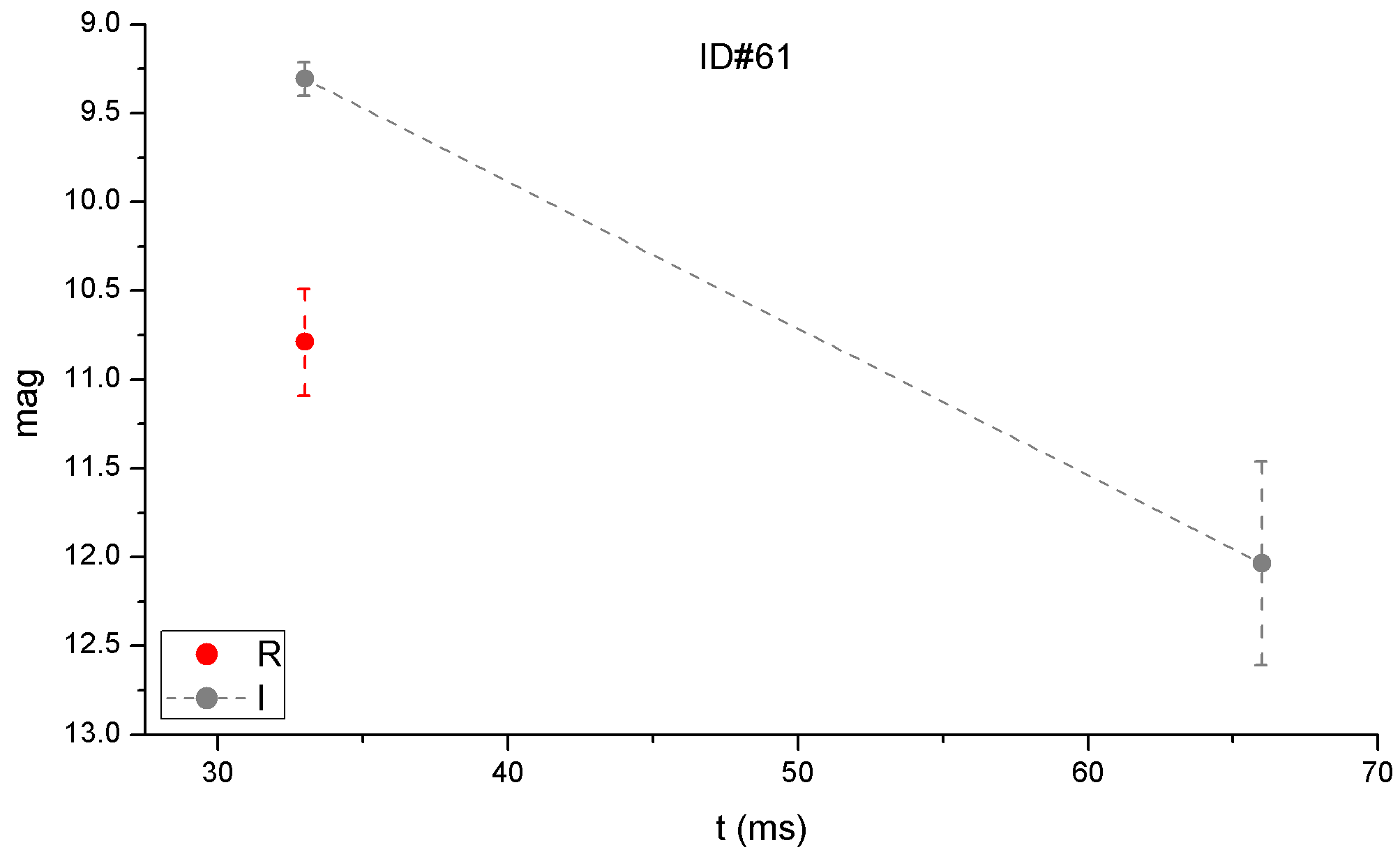}&\includegraphics[width=5.6cm]{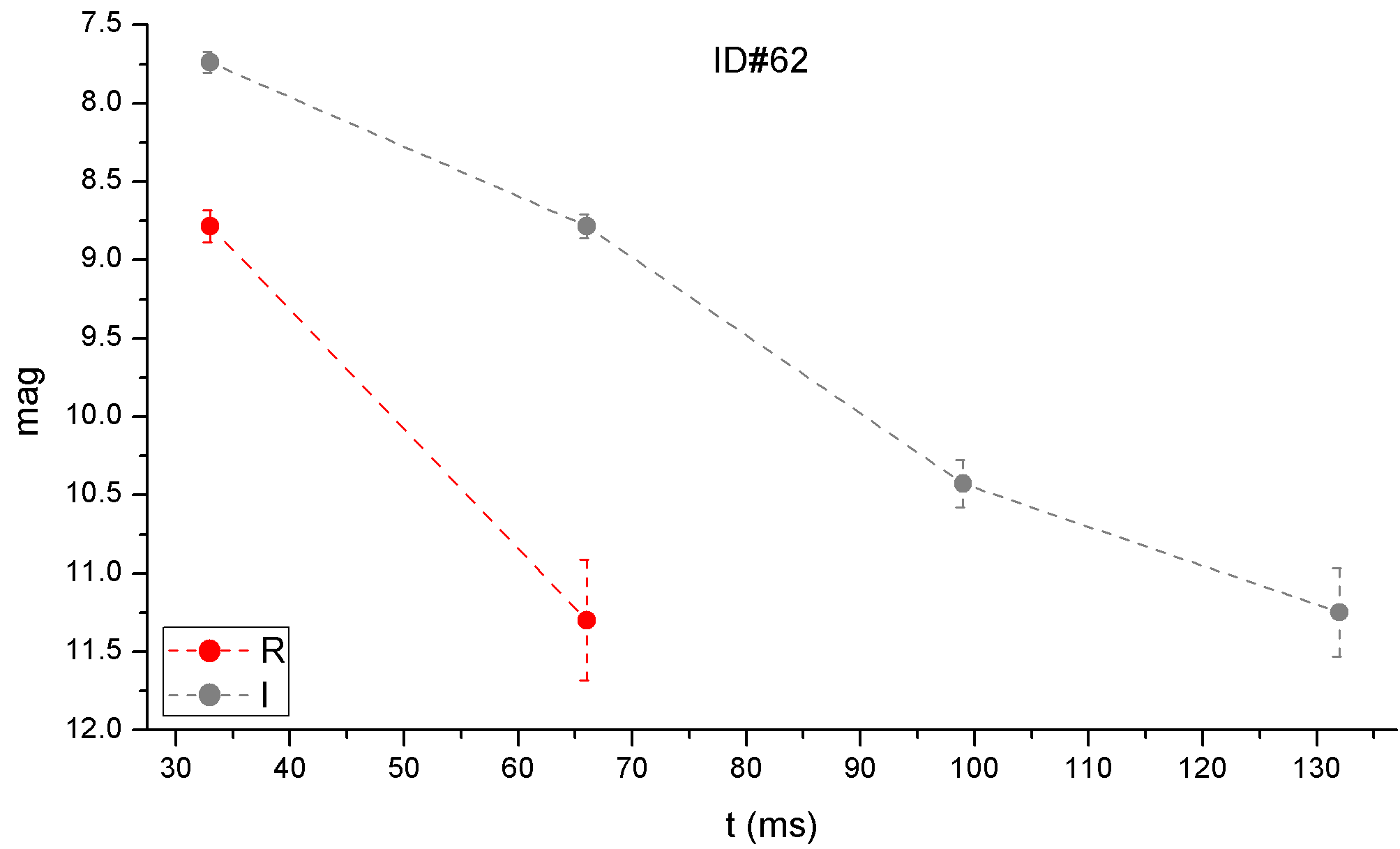}&\includegraphics[width=5.6cm]{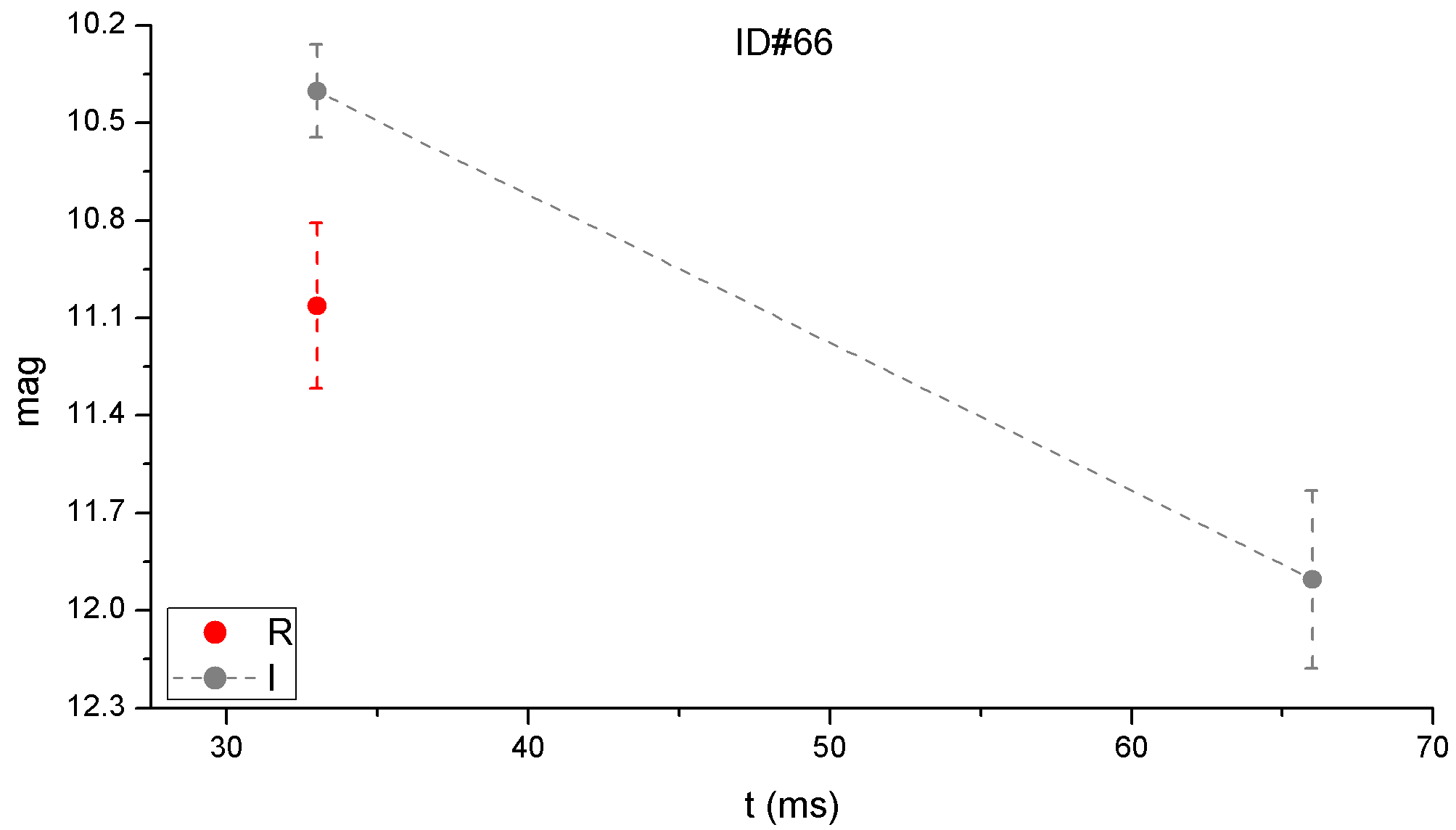}\\
\includegraphics[width=5.6cm]{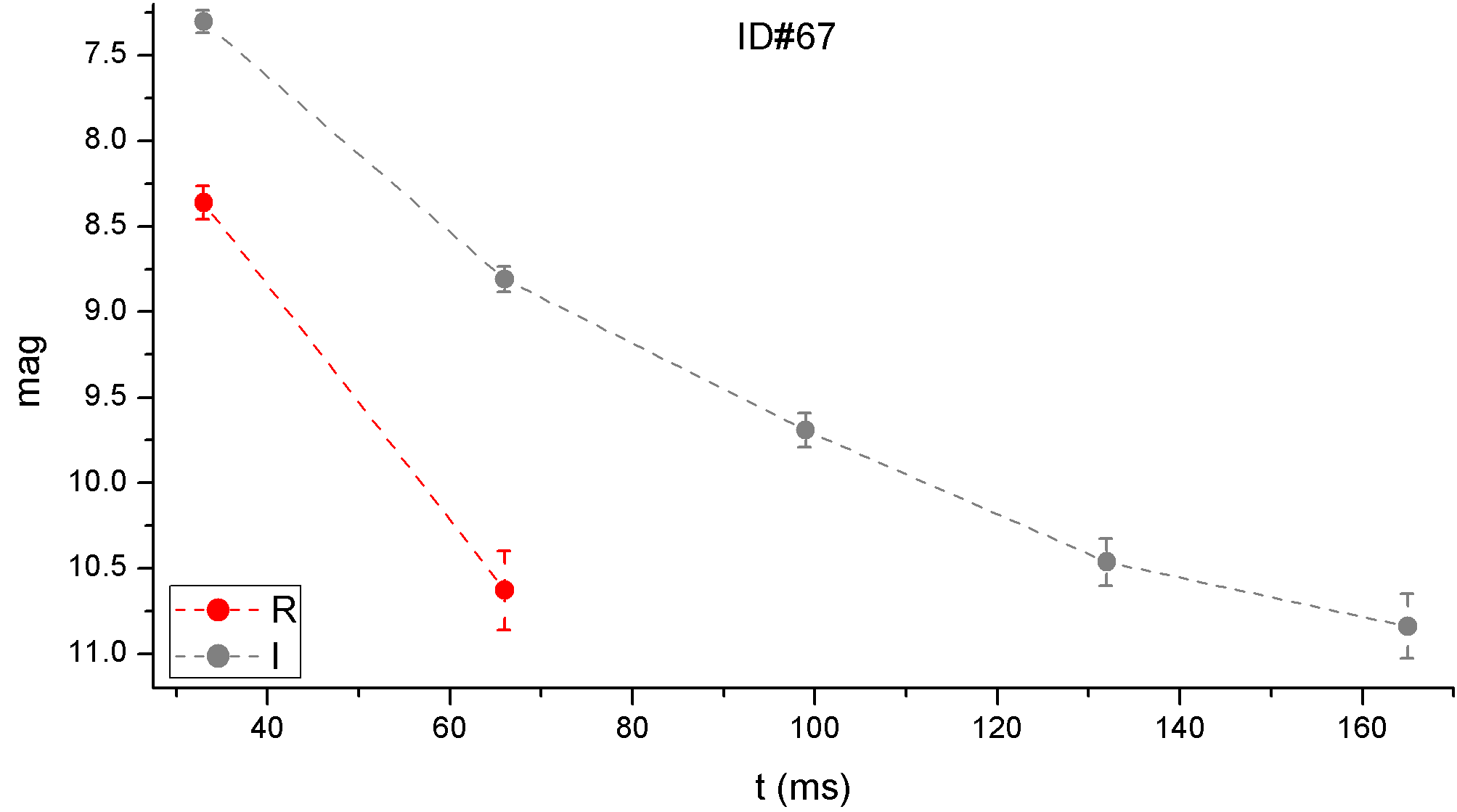}&\includegraphics[width=5.6cm]{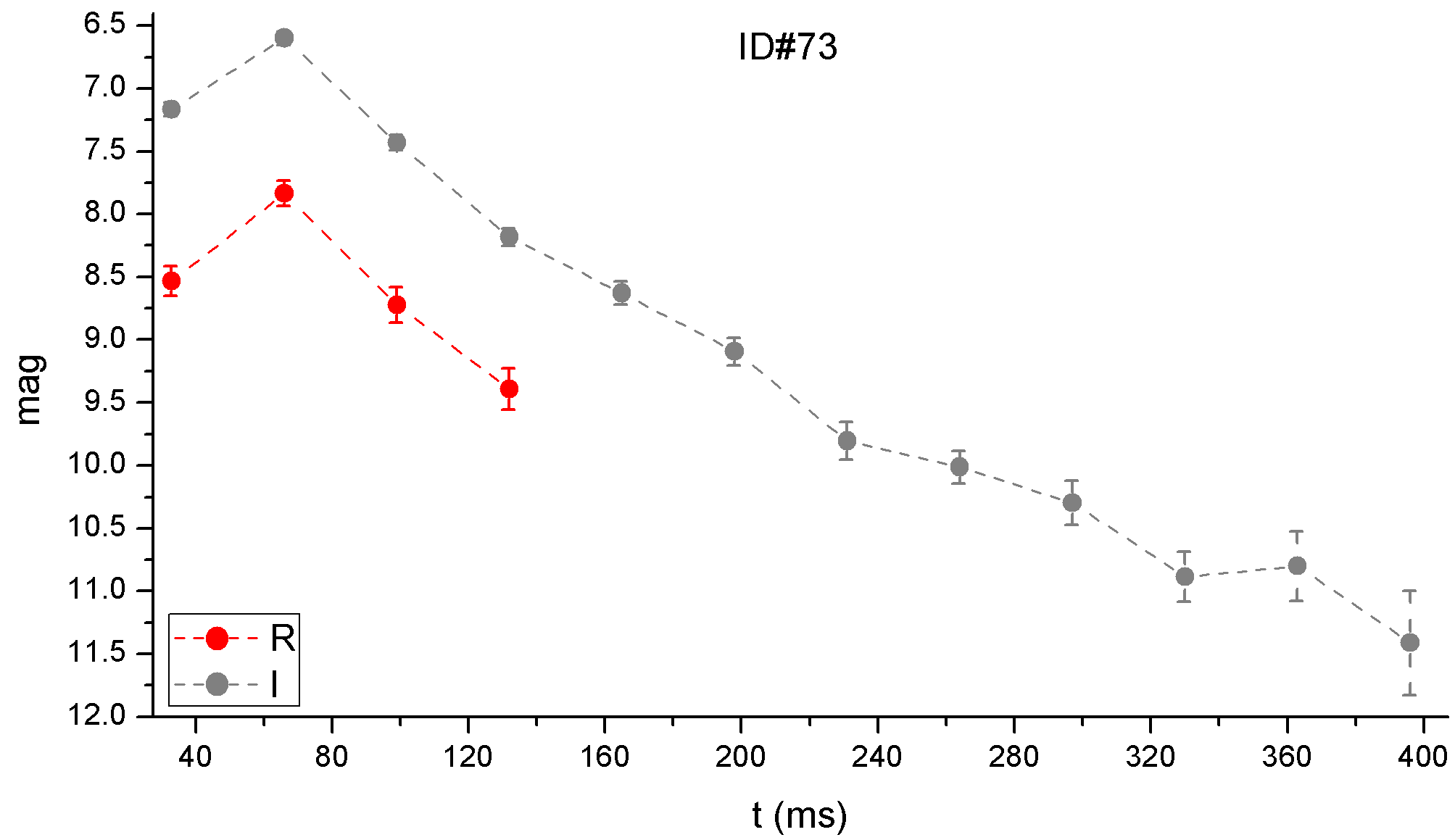}&\includegraphics[width=5.6cm]{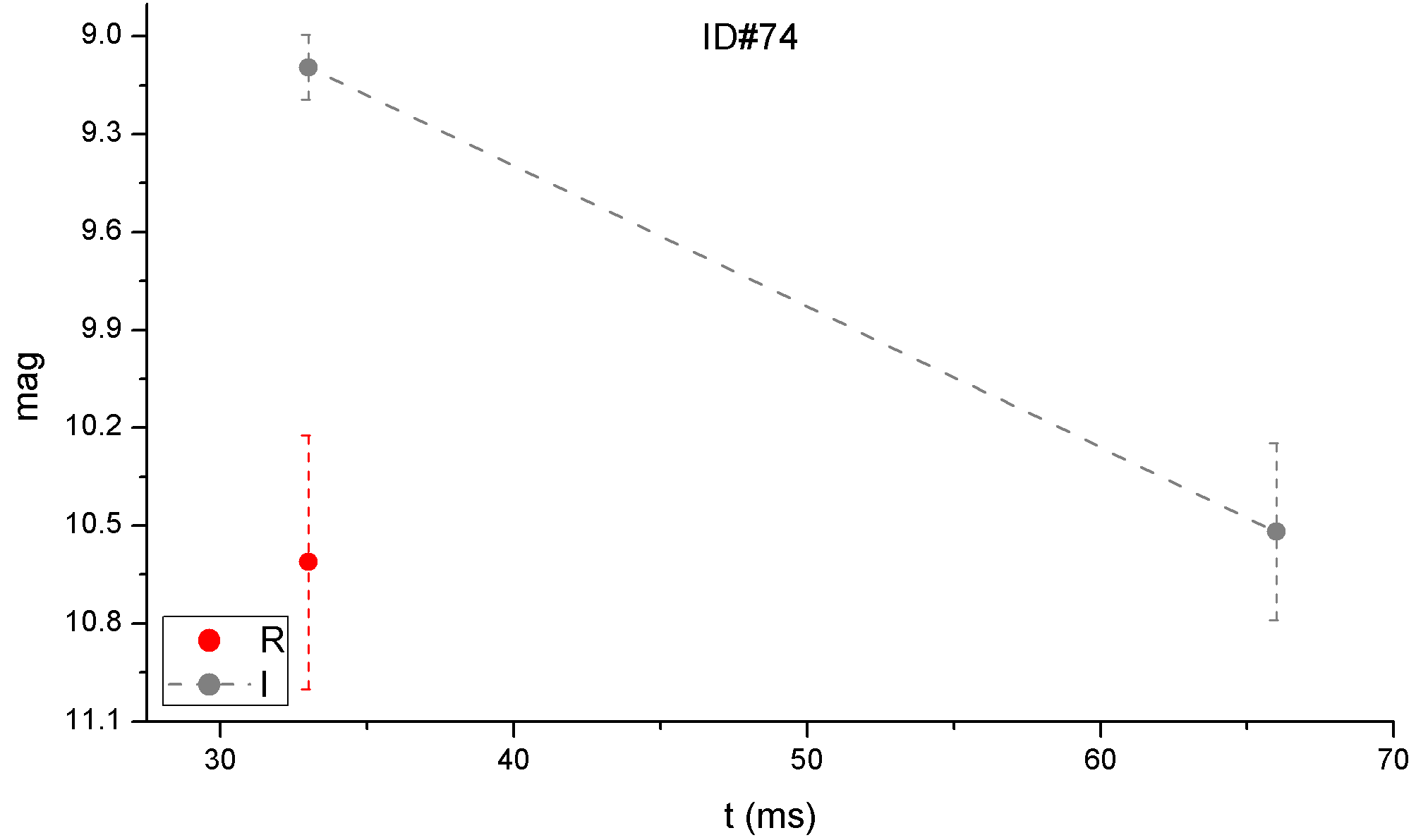}\\
\includegraphics[width=5.6cm]{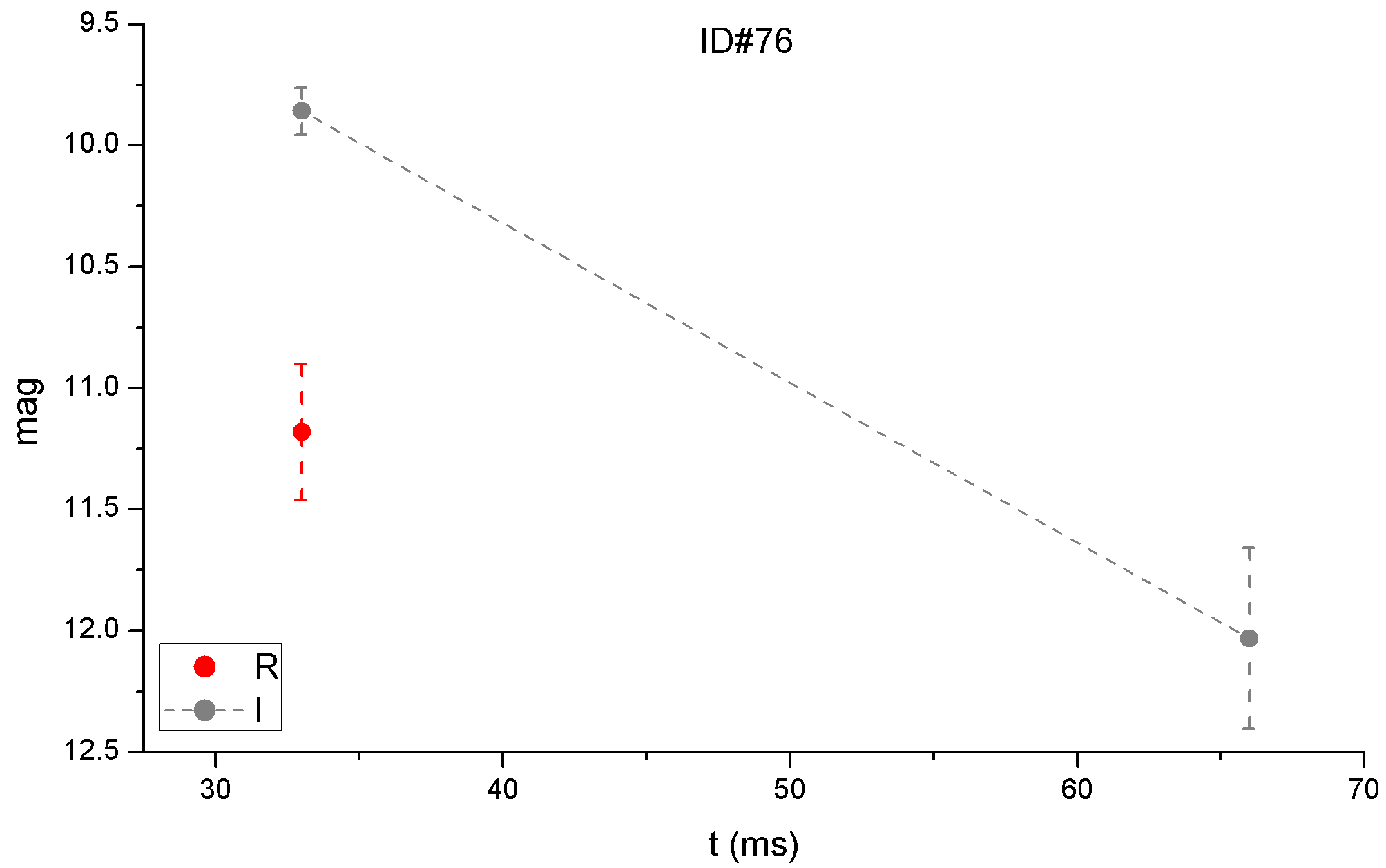}&\includegraphics[width=5.6cm]{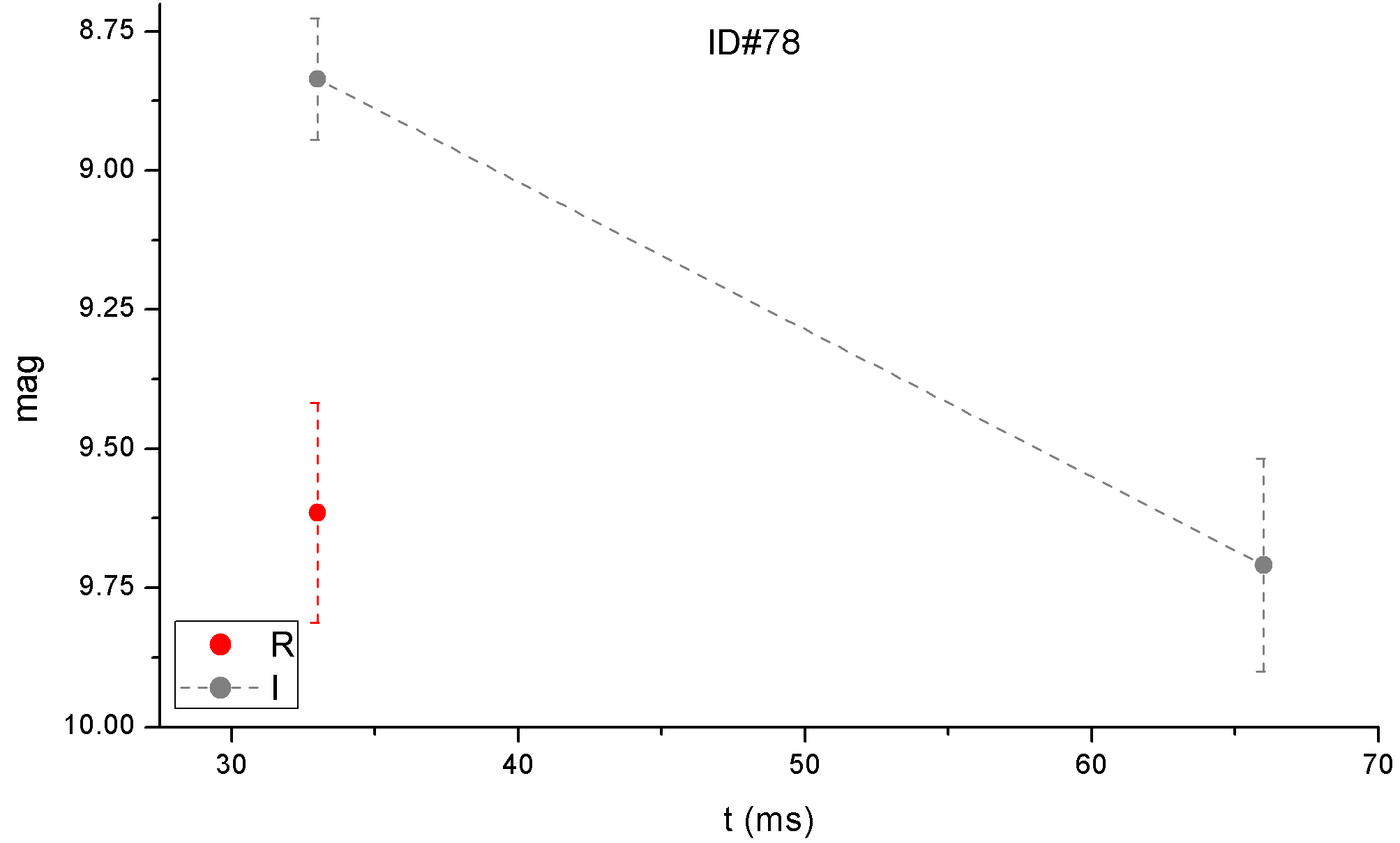}&\includegraphics[width=5.6cm]{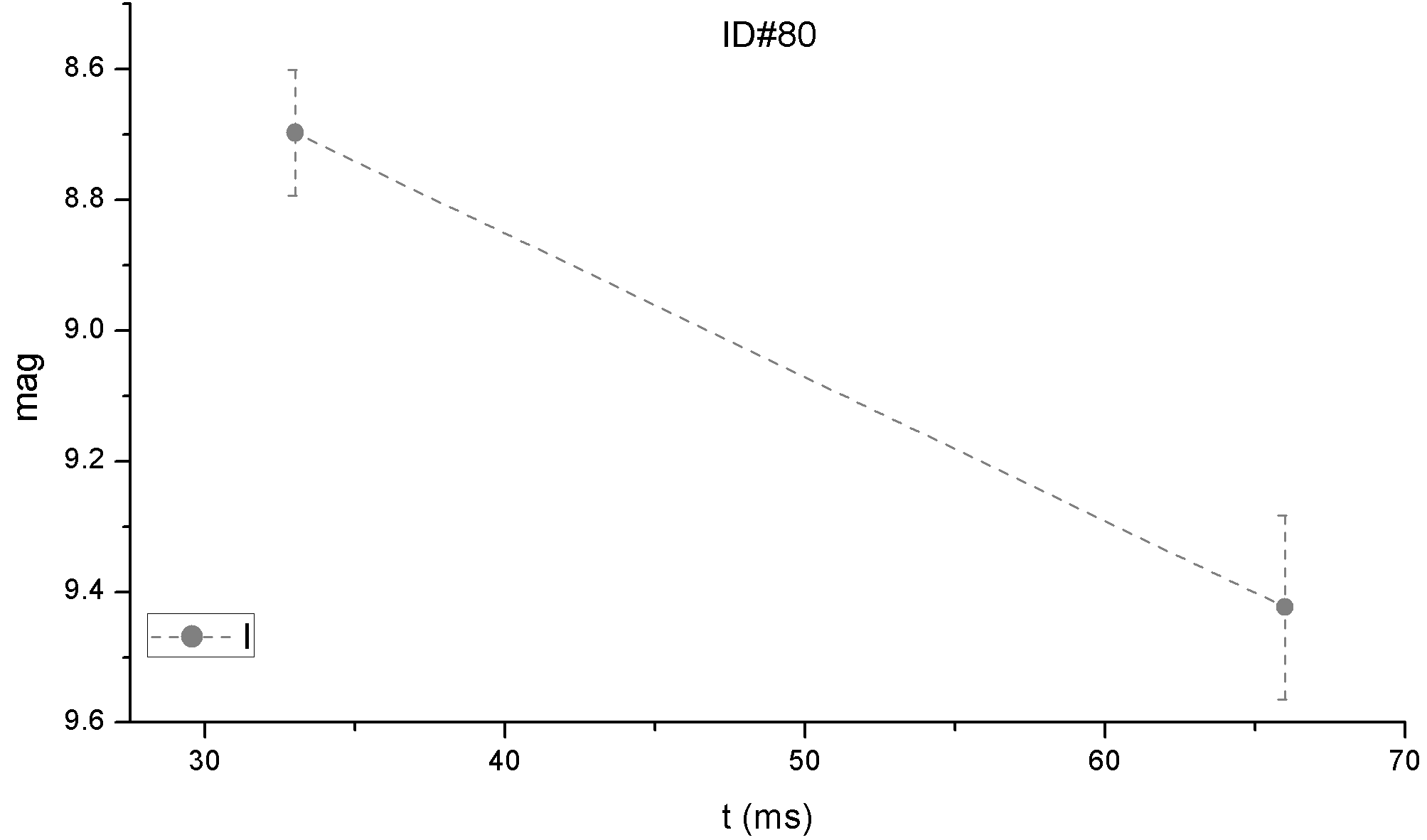}\\
\includegraphics[width=5.6cm]{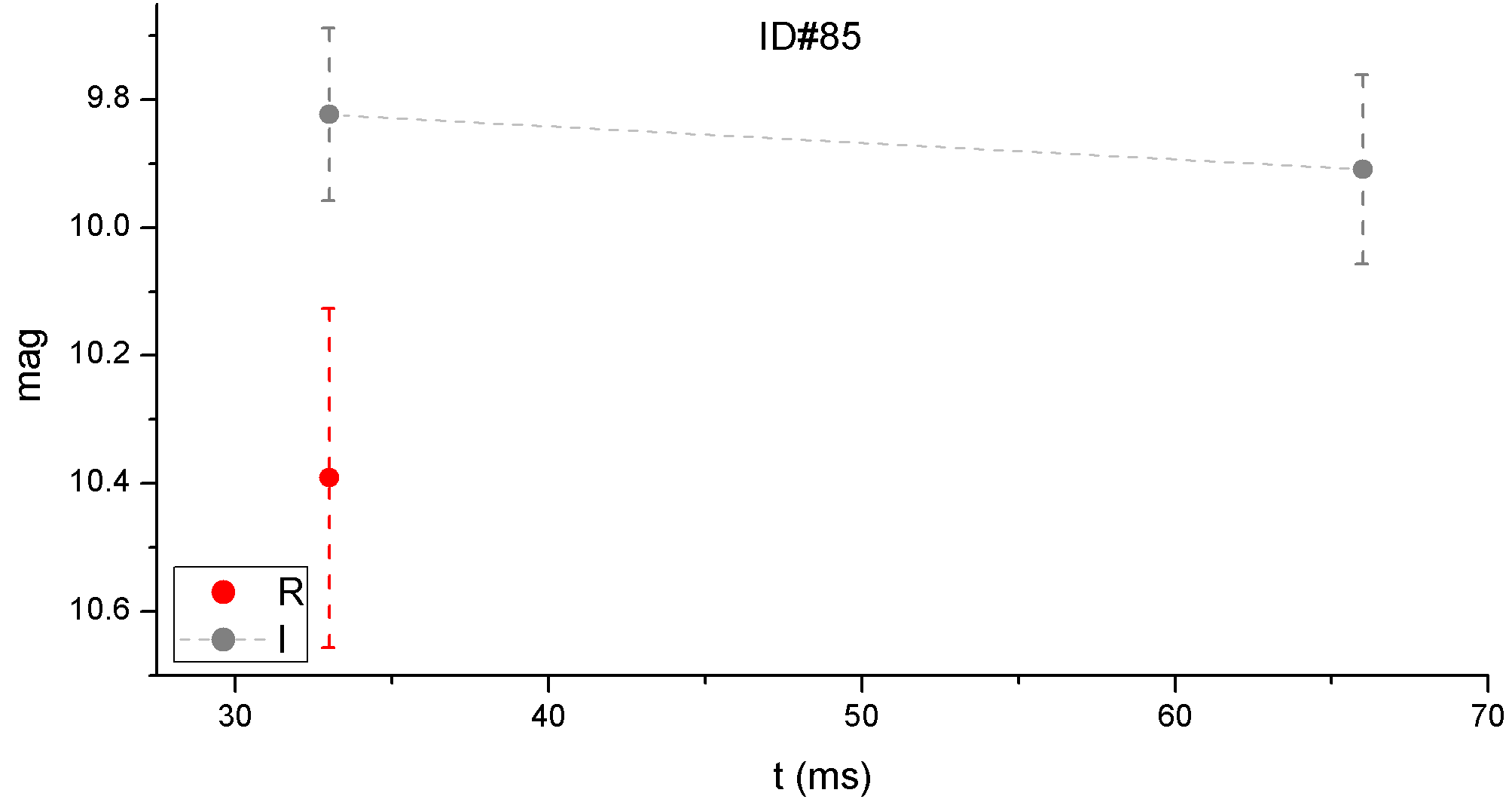}&\includegraphics[width=5.6cm]{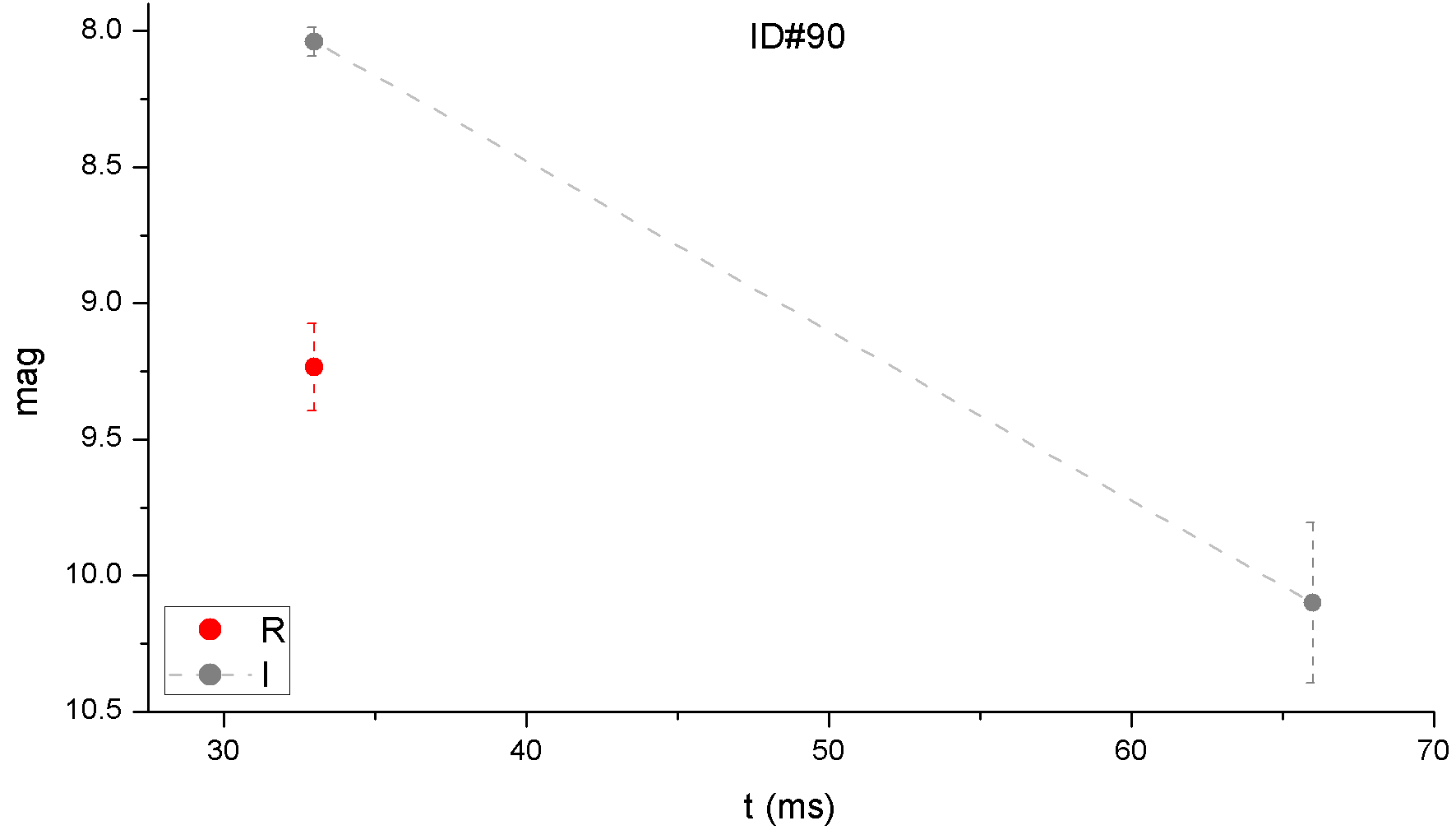}&\includegraphics[width=5.6cm]{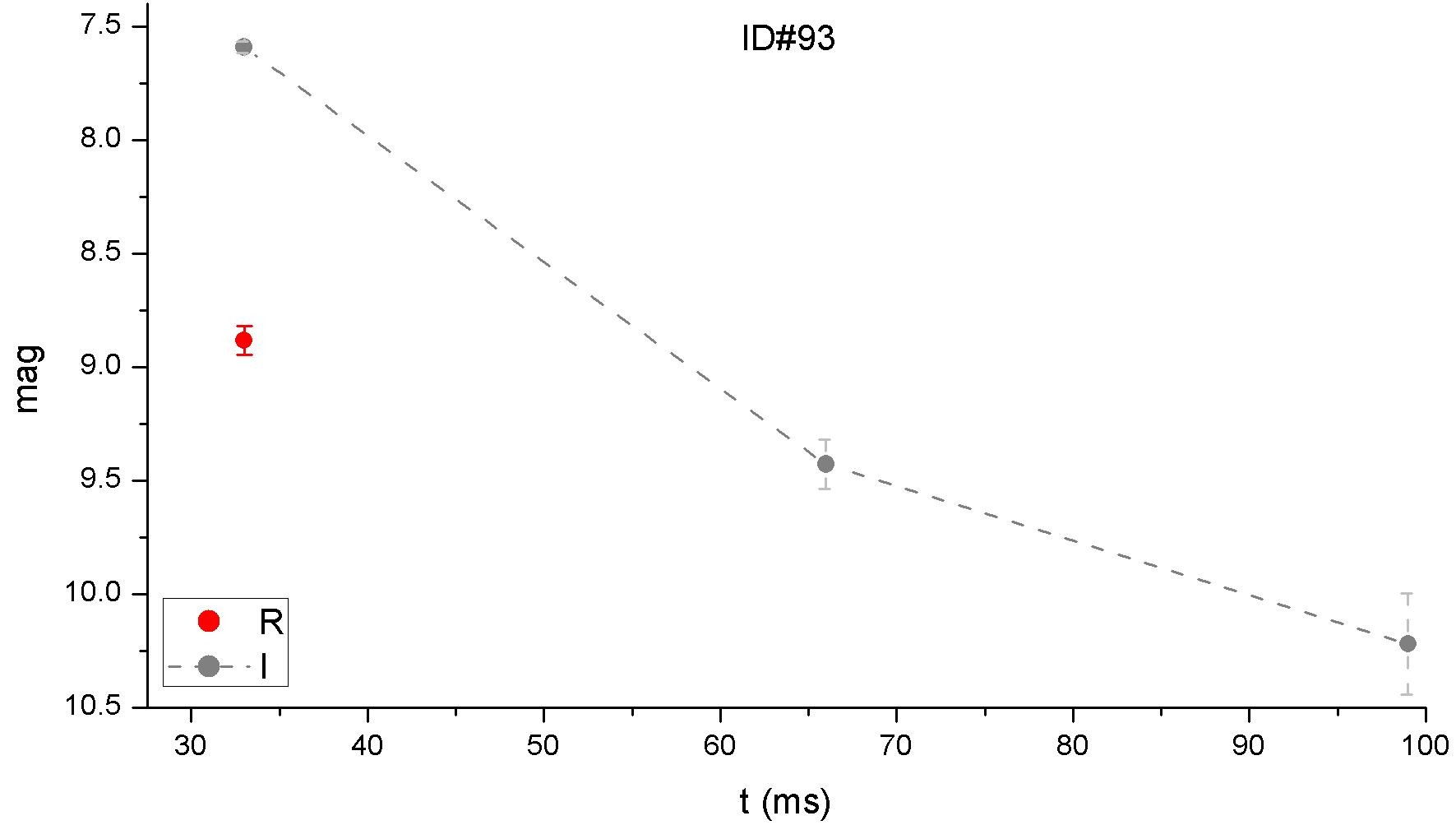}\\
\includegraphics[width=5.6cm]{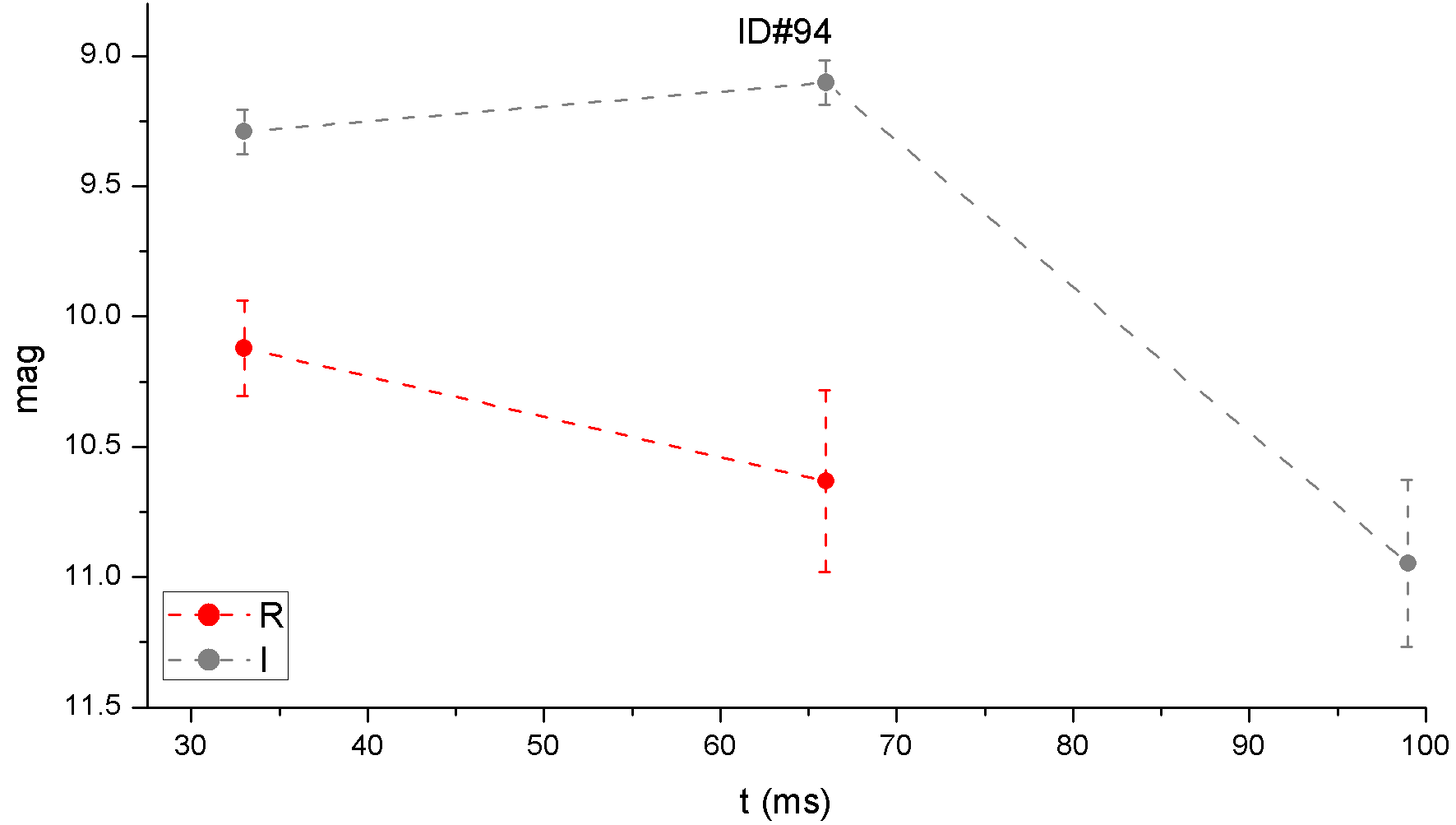}&\includegraphics[width=5.6cm]{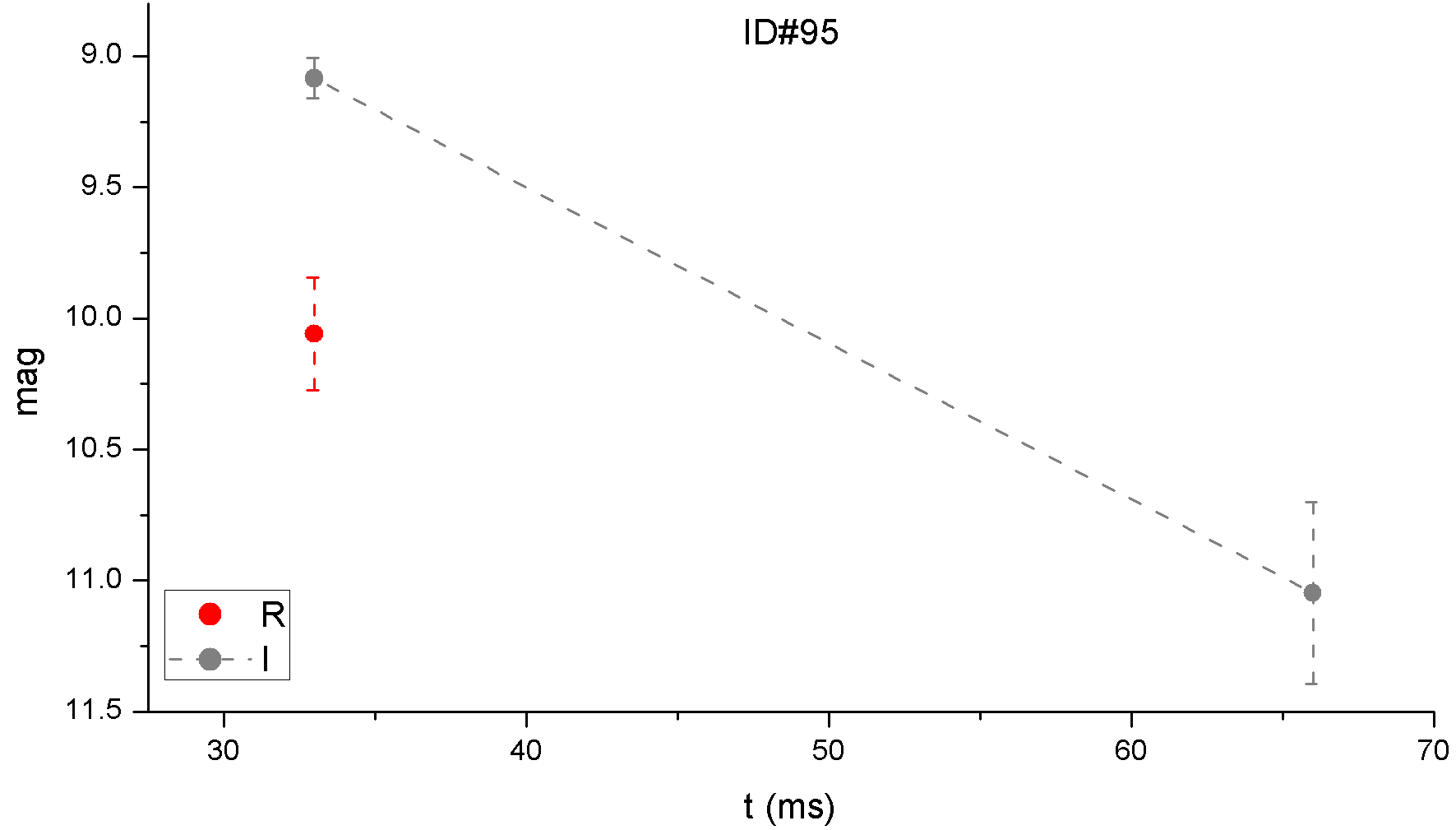}&\includegraphics[width=5.6cm]{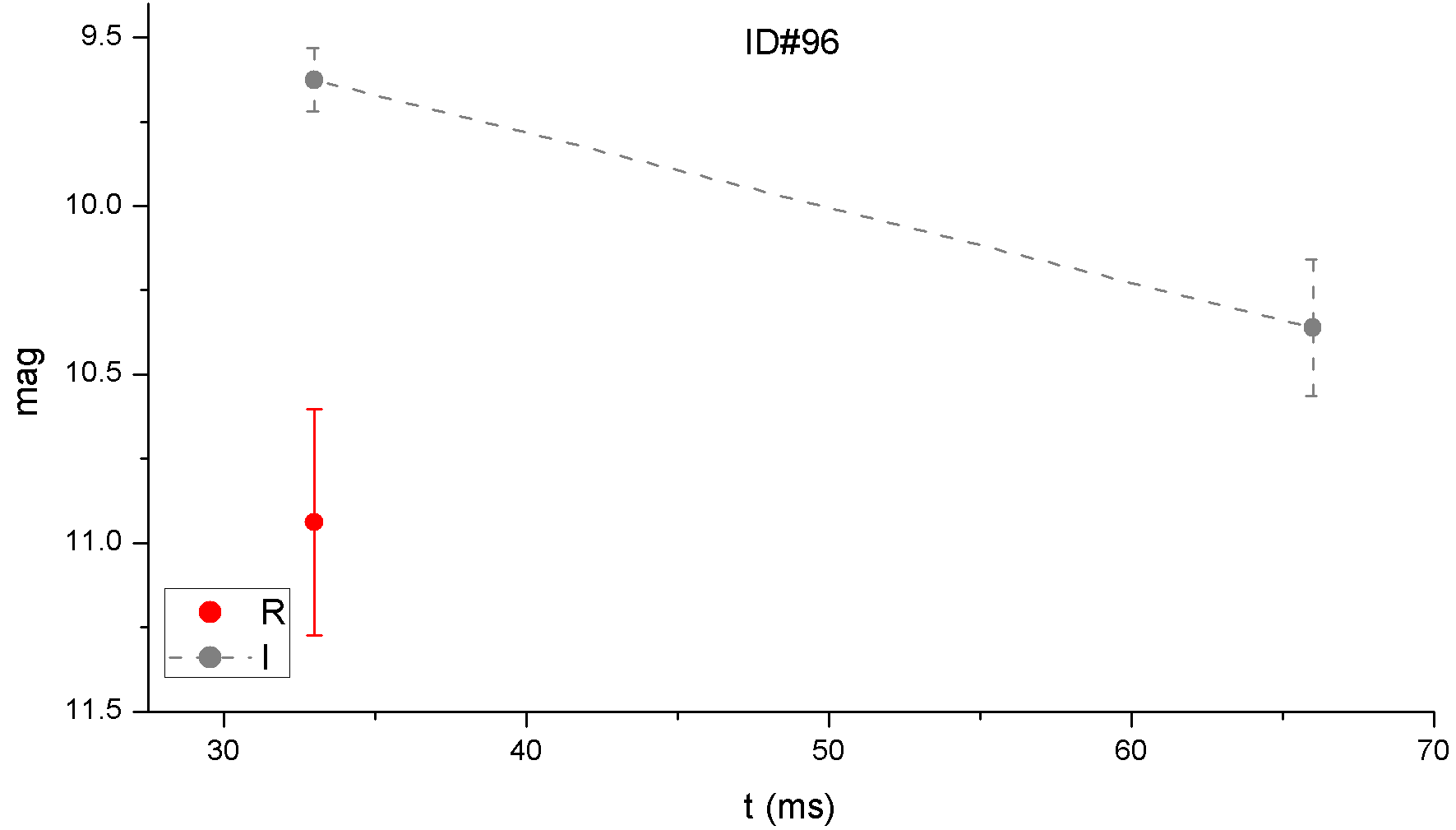}\\
\end{tabular}
\caption{Light curves of the multi-frame flashes (cont.).}
\label{fig:LCs2}
\end{figure*}	

\begin{figure*}[h]
\begin{tabular}{ccc}
\includegraphics[width=5.6cm]{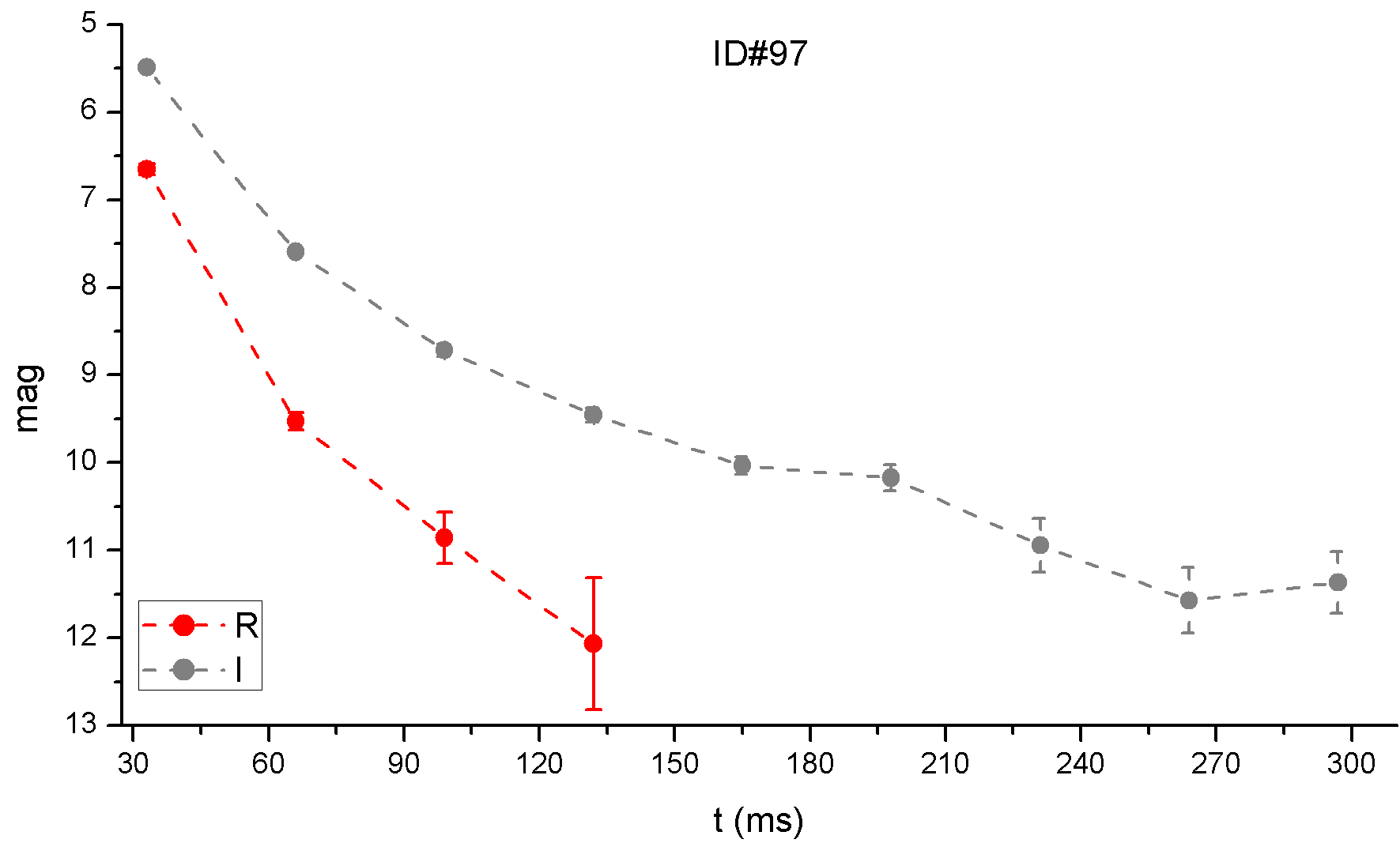}&\includegraphics[width=5.6cm]{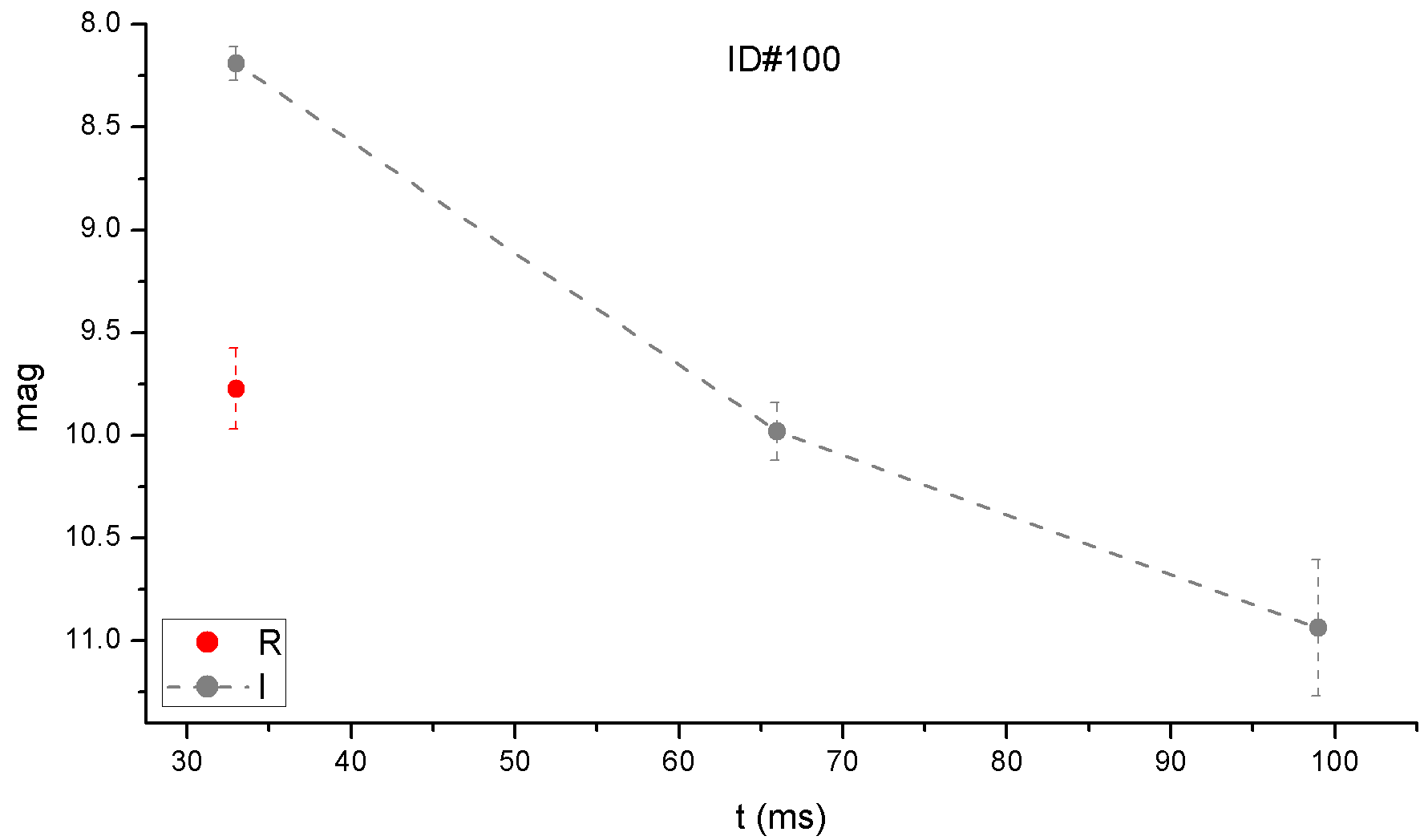}&\includegraphics[width=5.6cm]{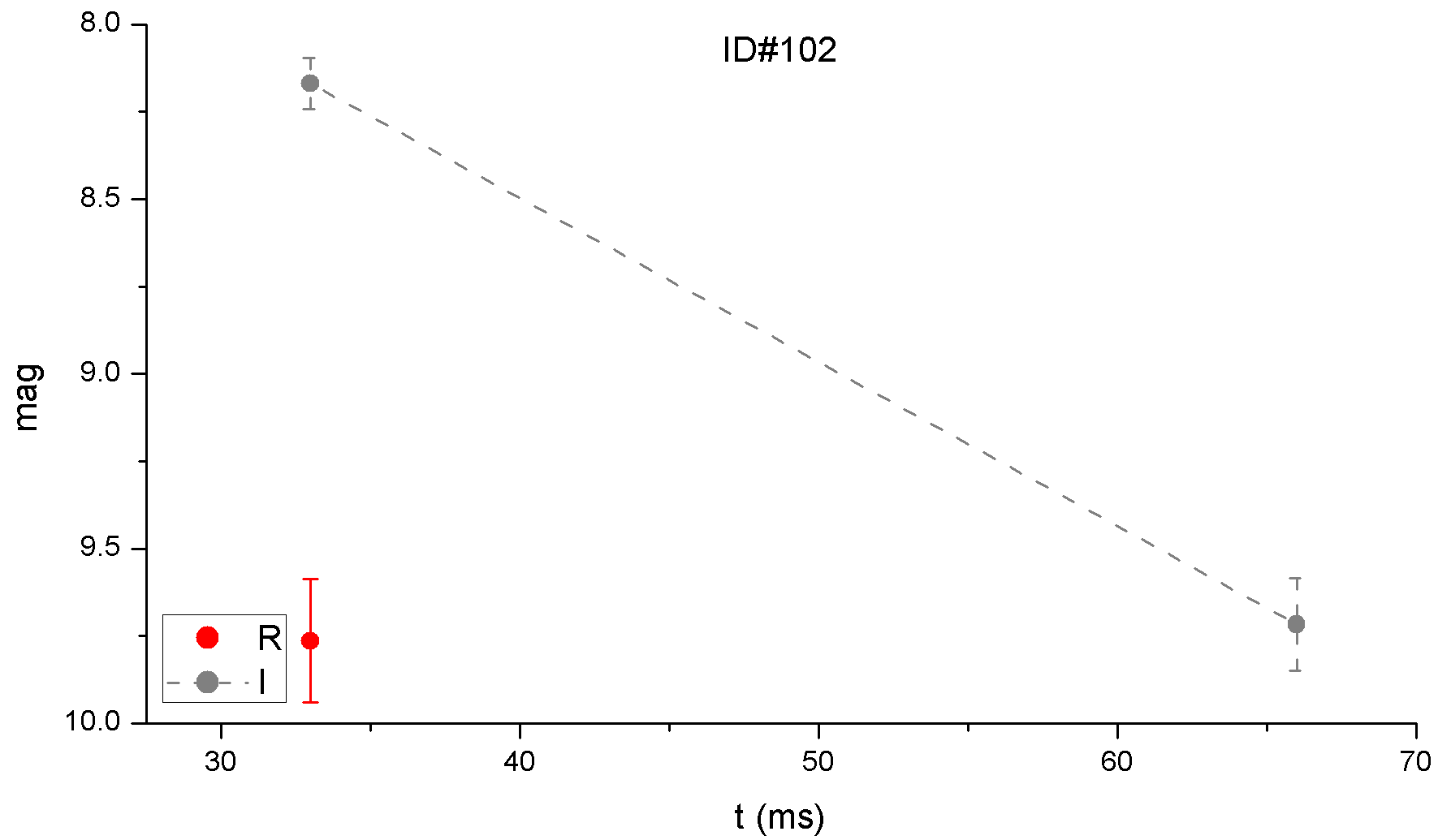}     \\
\includegraphics[width=5.6cm]{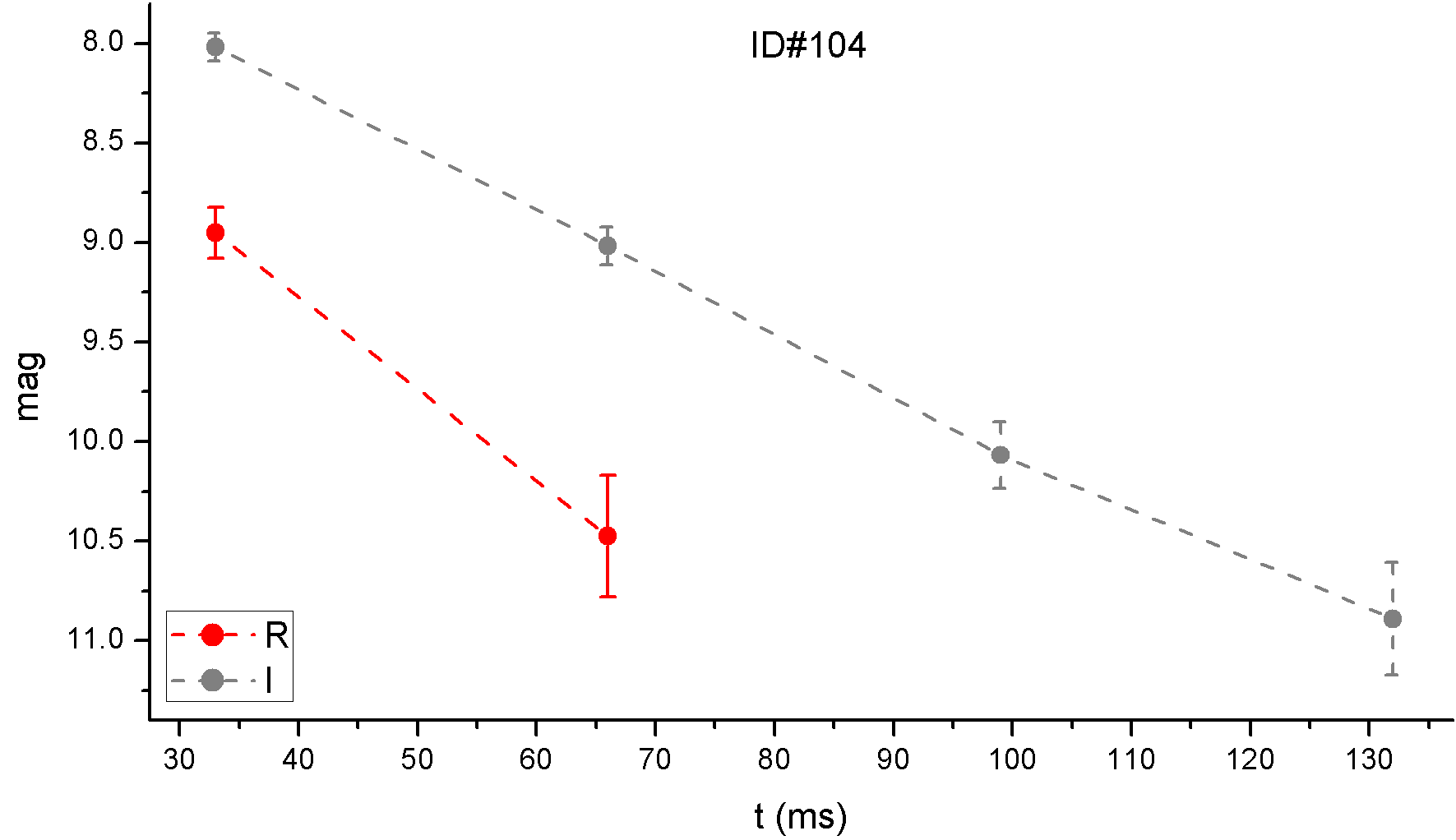}&\includegraphics[width=5.6cm]{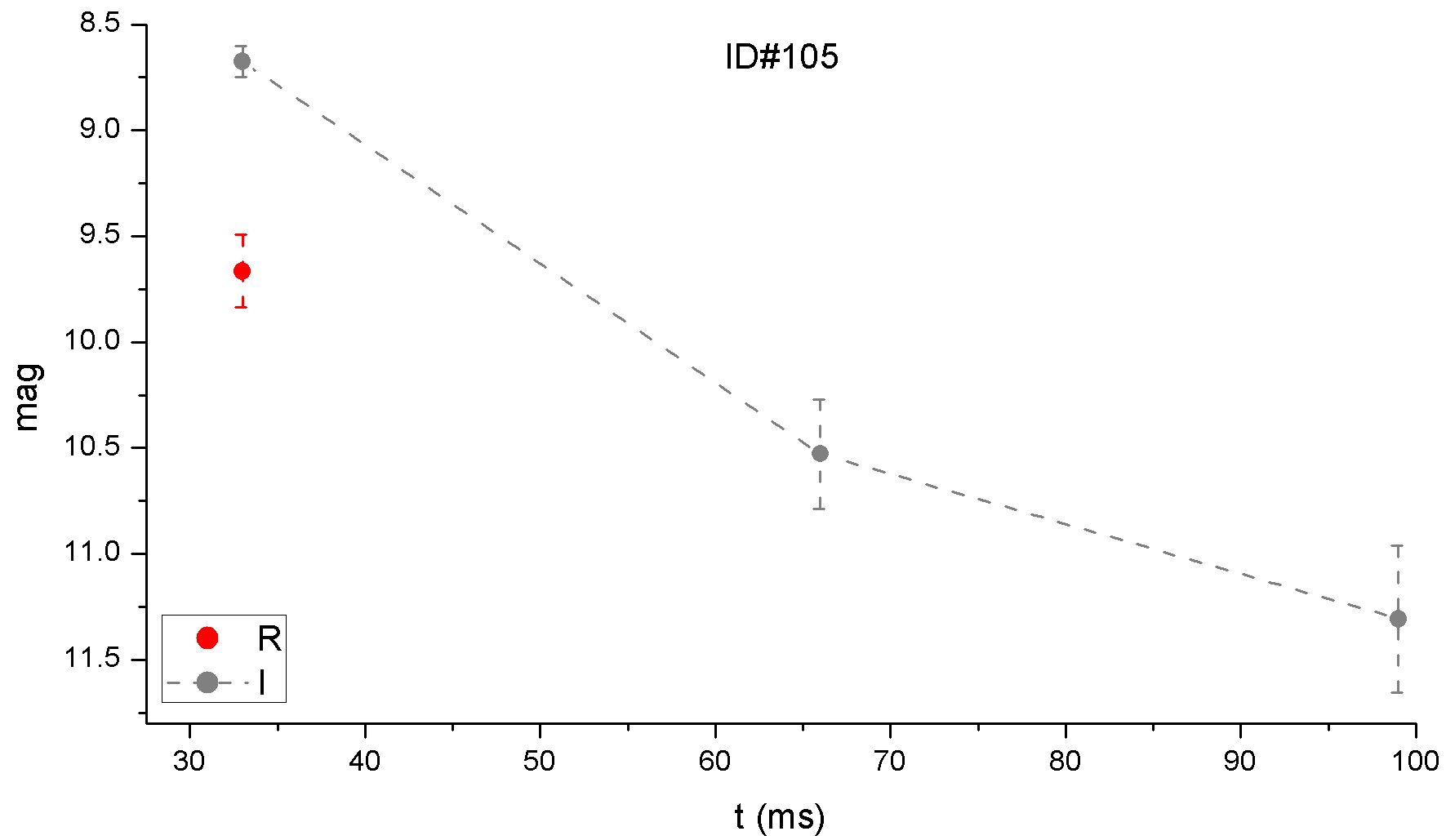}&\includegraphics[width=5.6cm]{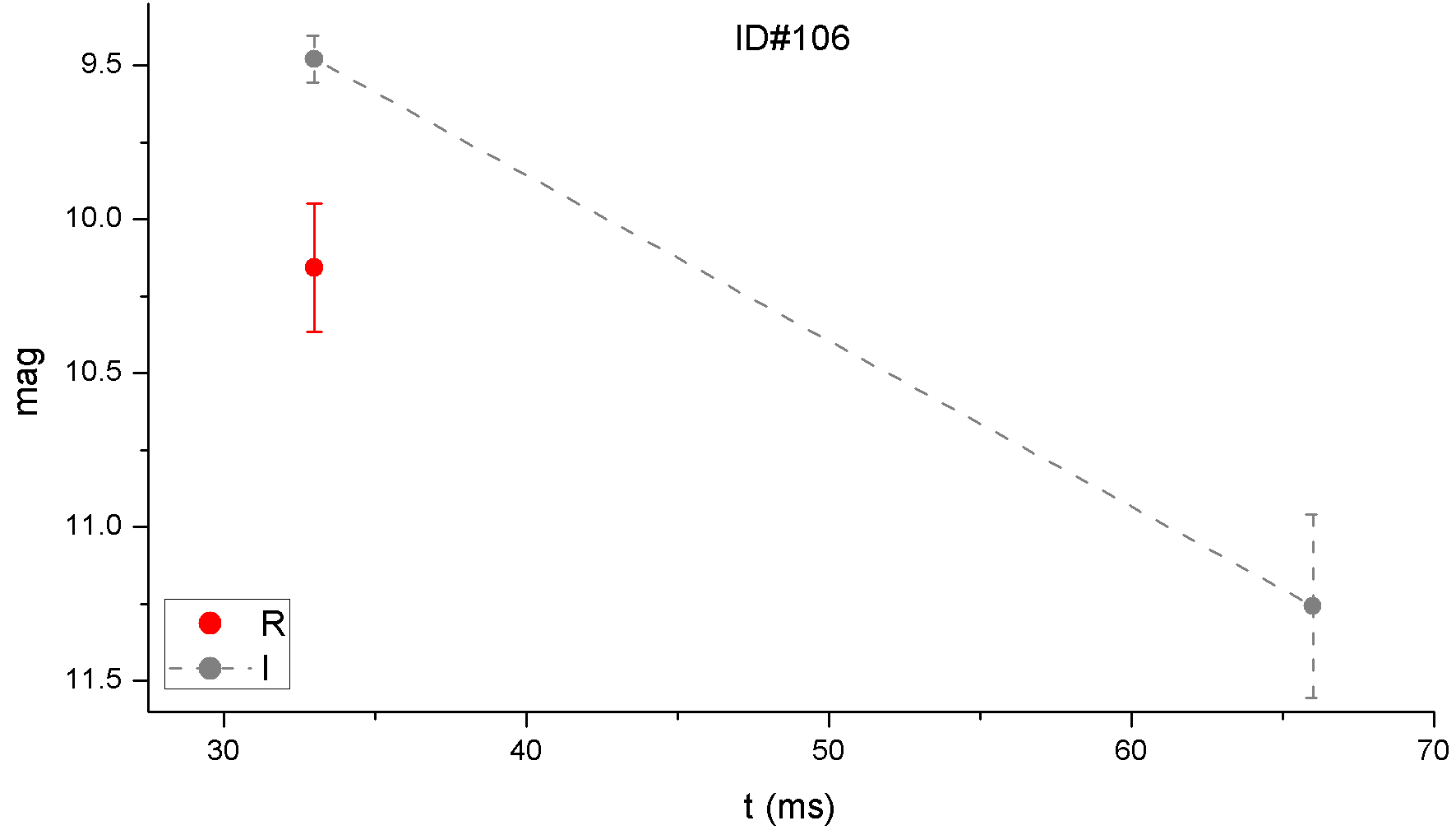}     \\
\includegraphics[width=5.6cm]{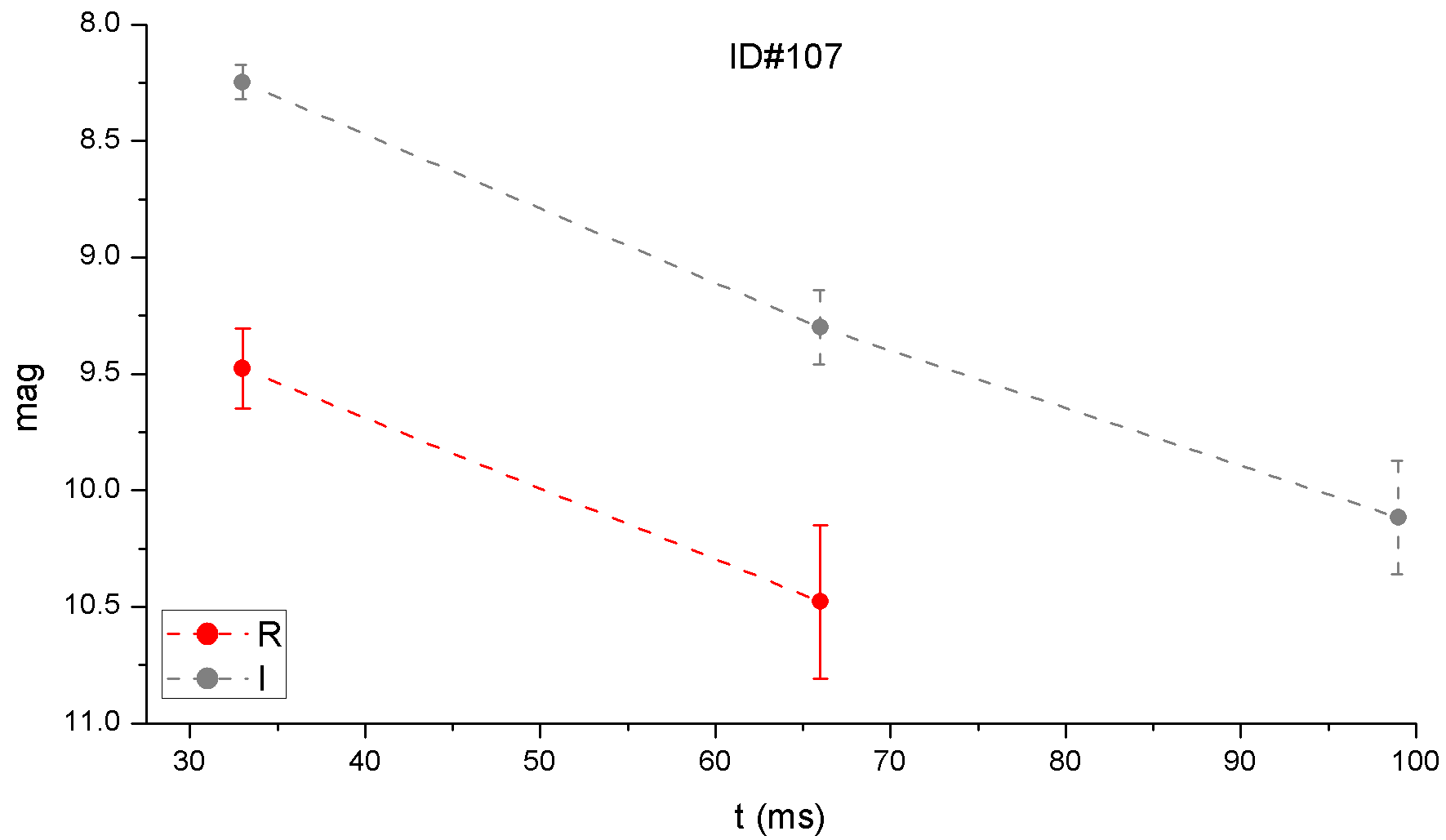}&\includegraphics[width=5.6cm]{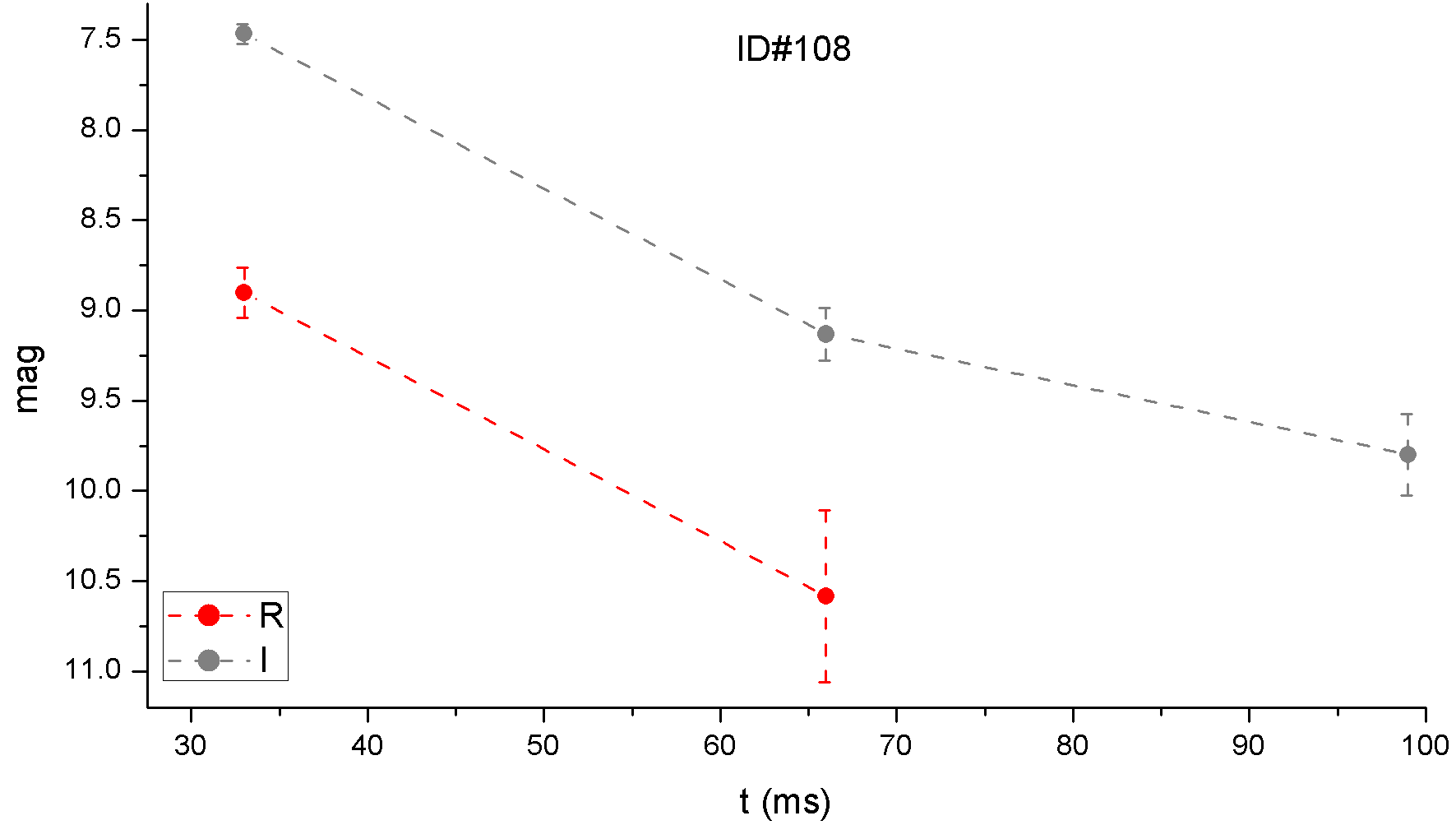}&\includegraphics[width=5.6cm]{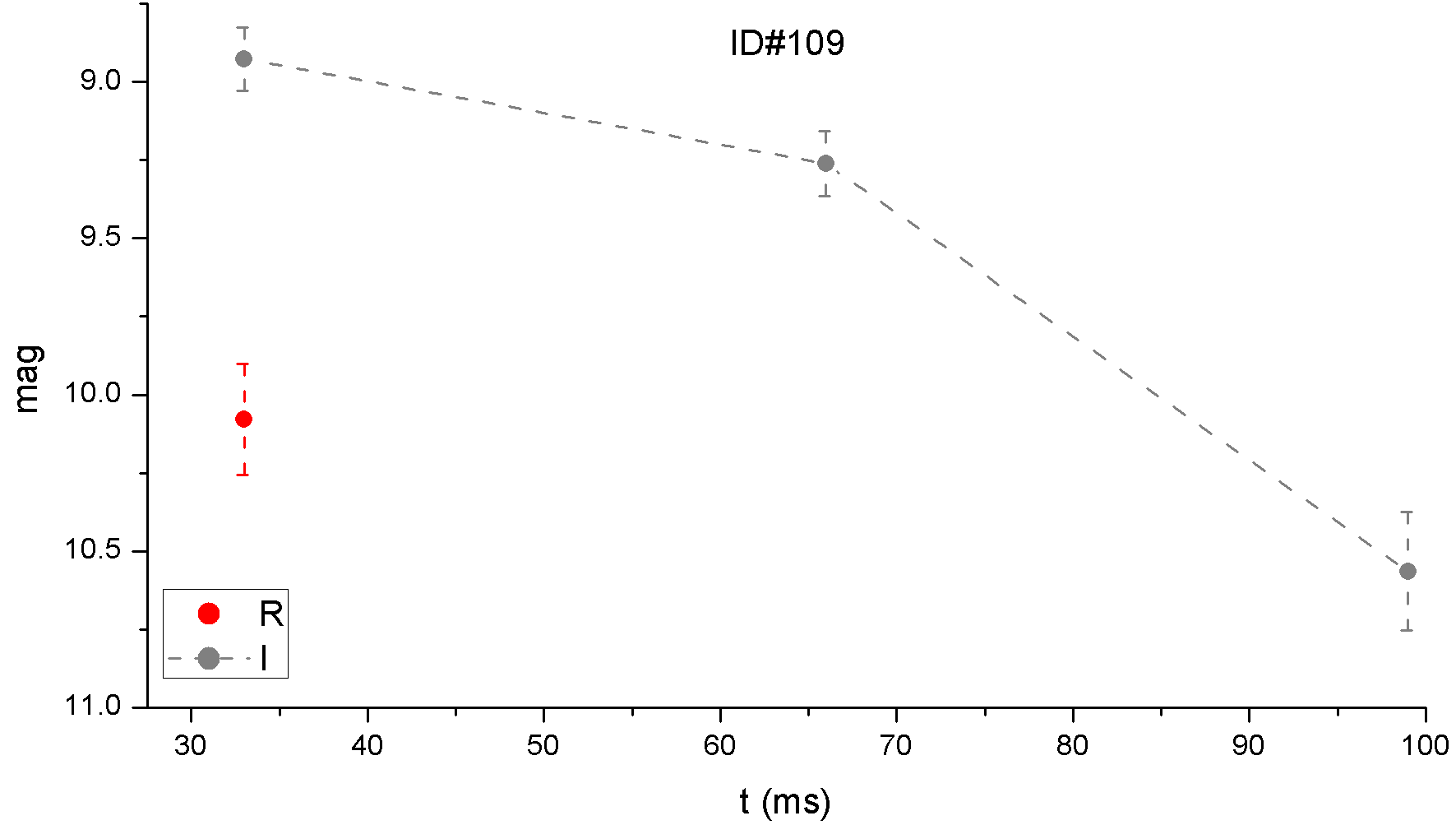}     \\
\includegraphics[width=5.6cm]{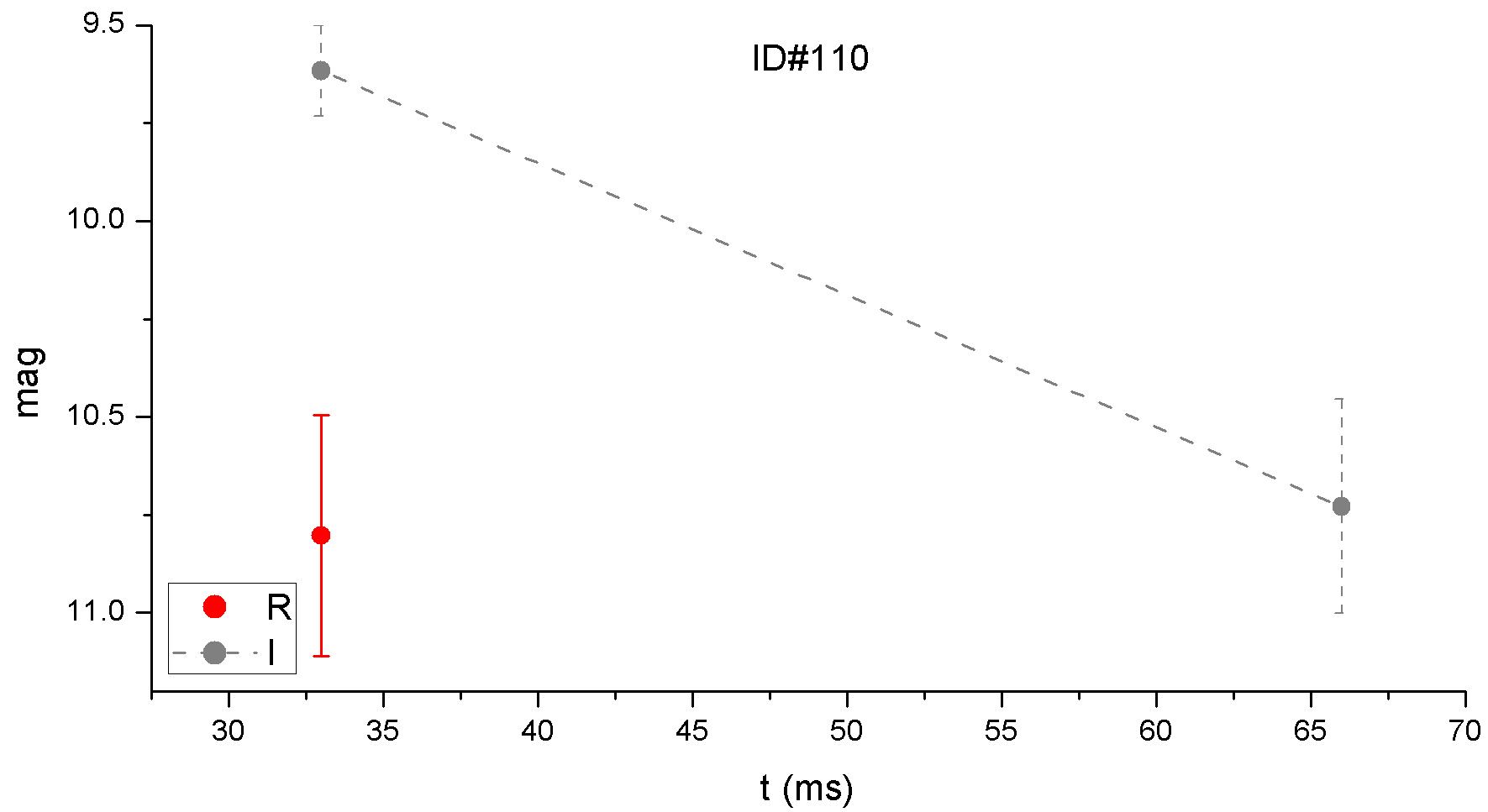}&\includegraphics[width=5.6cm]{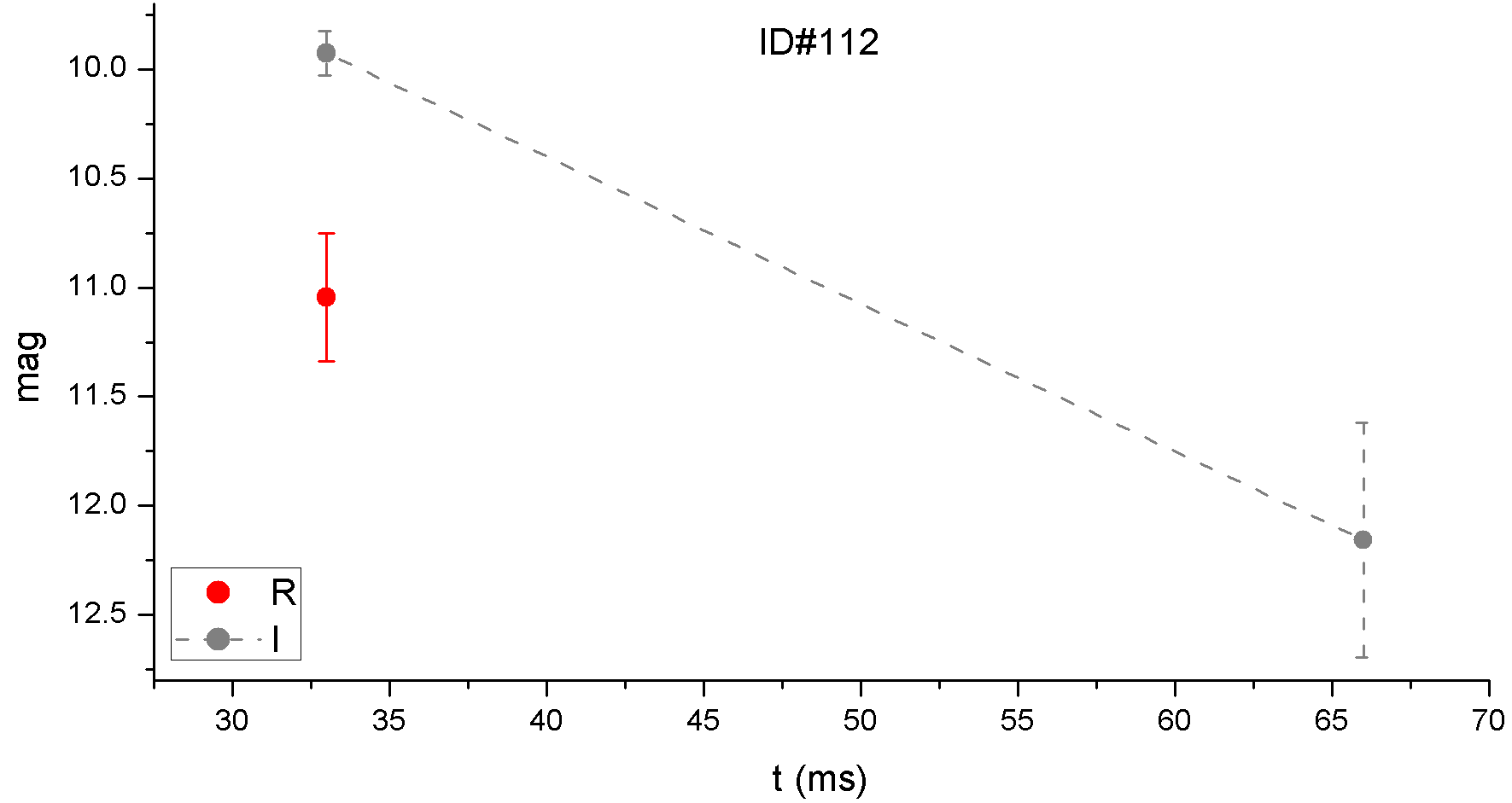}&     \\

\end{tabular}
\caption{Light curves of the multi-frame flashes (cont.).}
\label{fig:LCs3}
\end{figure*}

\section{Meteoroid stream association parameters}

\begin{table*}
\centering
\caption{Parameters of sporadic and streams meteoroids.}
\label{tab:streams}
\scalebox{0.85}{
\begin{tabular}{ccccc ccccc ccccc c}
\hline\hline																											
CODE	&	$\lambda_{\sun,\rm beg}$	&	$\lambda_{\sun,\rm max}$	&	$\lambda_{\sun,\rm end}$	&	$RA $	&	$Dec$	&	$l$	&	$b$	&	$V_{\rm p}$ 	&	$ZHR~(max)$	&	$B$	&	$m_0$	&	$s$	&	$r$	&	$\rho^a$	\\
	&	($\degr$)	&	($\degr$)	&	($\degr$)	&	($\degr$)	&	($\degr$)	&	($\degr$)	&	($\degr$)	&	(km~s$^{-1}$)	&	(hr$^{-1}$)	&	(deg$^{-1}$)	&	($\times10^{-7}$~kg)	&		&		&	(g~cm$^{-3}$)	\\
\hline																													
SPO	&		&		&		&		&		&		&		&	17	&		&		&	98.06	&	2.19	&	3	&	1.8(3)	\\
Lyr	&	29.2	&	31.7	&	34.2	&	272	&	33	&	273	&	56.4	&	49	&	18	&	0.22	&	1.09	&	1.81	&	2.1	&		\\
ETA	&	27.3	&	45.8	&	64.3	&	338	&	-1	&	339.3	&	7.64	&	66	&	50	&	0.08	&	0.31	&	1.95	&	2.4	&		\\
Ari	&	57	&	76	&	95	&	45	&	23	&	49.15	&	5.71	&	38	&	30	&	0.1	&	3.21	&	2.12	&	2.8	&		\\
SDA	&	99.4	&	124.9	&	150.4	&	339	&	-17	&	334.2	&	-7.58	&	41	&	25	&	0.091	&	2.33	&	1.99	&	2.5	&	2.4(6)	\\
Per	&	126.99	&	139.49	&	151.99	&	46	&	58	&	61.82	&	38.78	&	61	&	84	&	0.2	&	0.50	&	1.75	&	2	&	1.2(2)	\\
Ori	&	192.9	&	207.9	&	222.9	&	95	&	16	&	94.84	&	-7.36	&	67	&	20	&	0.12	&	0.31	&	1.99	&	2.5	&	0.9(5)	\\
Leo	&	230.4	&	234.4	&	238.4	&	153	&	22	&	147.06	&	10.15	&	71	&	23	&	0.39	&	0.24	&	1.92	&	2.33	&	0.4(1)	\\
Gem	&	234.4	&	261.4	&	288.4	&	112	&	32	&	108.8	&	9.99	&	36	&	88	&	0.39	&	4.56	&	1.95	&	2.4	&	2.9(6)	\\
QUA	&	253	&	283	&	313	&	231.5	&	48.5	&	203.3	&	63.3	&	41.7	&	120	&	1.8	&	2.33	&	2.30	&	1.9	&	1.9(2)	\\
Urs	&	268.3	&	270.3	&	272.3	&	223	&	78	&	120.42	&	72.5	&	35	&	11.8	&	0.61	&	4.56	&	2.19	&	3	&		\\
\hline																									
\end{tabular}}
$^a$taken from \citet{BAD09}
\end{table*}

\begin{table}
\centering
\caption{Values of $\nu^{\rm SPO}$ according to $\eta$ and $m_{\rm R,~lim}$.}
\label{tab:nu}
\begin{tabular}{ccccccc}
\hline
\hline													
$\eta$	&	\multicolumn{2}{c} {5$\times10^{-3}$}			&	\multicolumn{2}{c} {1.5$\times10^{-3}$}			&	\multicolumn{2}{c} {$5\times10^{-4}$}			\\
\hline													
$m_{\rm R,~lim}$~(mag)	&	10.5	&	11.4	&	10.5	&	11.4	&	10.5	&	11.4	\\
\hline													
$\nu^{\rm SPO} (\times 10^{-4})$ 	&	3.91	&	10.52	&	0.93	&	2.5	&	0.25	&	0.67	\\
\hline													
\end{tabular}
\end{table}





\end{appendix}

%
%



\end{document}